\documentclass[reviewcopy]{elsarticle}
\pdfoutput=1
\usepackage[reviewcopy]{adndt}
\usepackage{longtable}
\usepackage{amsmath}
\usepackage{amssymb}
\usepackage{pdfpages}

\biboptions{square,sort&compress}
\bibpunct[]{[}{]}{,}{n}{}{;}
\begin{document}
\begin{frontmatter}

\journal{Atomic Data and Nuclear Data Tables}

\title{ \normalfont\textsc{Properties of heaviest nuclei with $98\leq Z \leq 126$ and $134 \leq N \leq 192$}}


\author[One]{P. Jachimowicz}
\author[Two]{M. Kowal\corref{cor1}}
\ead{E-mail: michal.kowal@ncbj.gov.pl}
\author[Two]{J. Skalski}
\cortext[cor1]{Corresponding author: Michal Kowal }

\address[One]{Institute of Physics, University of Zielona G\'{o}ra, Szafrana 4a, 65-516 Zielona G\'{o}ra, Poland}
\address[Two]{National Centre for Nuclear Research, Pasteura 7, 02-093 Warsaw, Poland}


\begin{abstract}
 We systematically determine ground-state and saddle-point shapes and masses
 for 1305 heavy and superheavy nuclei with $Z=98-126$ and $N=134-192$,
 including odd-$A$ and odd-odd systems. From these we derive
 static fission barrier heights, one- and two-nucleon separation energies, and
 $Q_\alpha$ values for g.s. to g.s. transitions.
 Our study is performed within the microscopic-macroscopic
 method with the deformed Woods-Saxon single-particle potential and the
 Yukawa-plus-exponential macroscopic energy taken as the smooth part.
 We use parameters of the model that were fitted previously to masses
 of even-even heavy nuclei.
 For systems with odd numbers of protons, neutrons, or both, we use a standard
 BCS method with blocking.
 Ground-state shapes and energies are found by the minimization over seven
 axially-symmetric deformations. A search for saddle-points was performed
 by using the "imaginary water flow" method in three consecutive stages, using
  five- (for nonaxial shapes) and seven-dimensional (for reflection-asymmetric
 shapes) deformation spaces.
 Calculated ground-state mass excess, nucleon separation- and $Q_\alpha$
 energies, total, macroscopic
 (normalized to the macroscopic energy at the spherical shape) and
 shell corrections energies,
 and deformations are given for each nucleus in \mbox{Table 1}.
 \mbox{Table 2} contains calculated
 properties of the saddle-point configurations and the fission barrier heights.
 In \mbox{Tables 3-7}, are given calculated ground-state, inner and outer
 saddle-point and superdeformed secondary minima characteristics for
  75 actinide nuclei, from Ac to Cf, for which experimental estimates of
 fission barrier heights are known. These results are an additional test of
 our model.

\end{abstract}
\end{frontmatter}

\newpage
\tableofcontents
\listofDtables
\listofDfigures
\vskip5pc

\section{Introduction}

 On the nuclear $Z$-$N$ chart, the region of heaviest atomic nuclei
 stretches into the area of the unknown. Our knowledge on superheavy species
 is very limited by severe experimental difficulties, and our hypotheses
 rely on theoreticel predictions based on approximate models, results of which
 are often diverging with increasing $Z$ and $N$. Hence the present uncertainty
 regarding basic questions:
 How many protons and neutrons can still assemble into a bound system lasting
 long enough to be detected (today, this means $\sim 10^{-5}$ s)?
 What are their properties? What would be the largest possible atomic number
 $Z$ of a detectable atomic nucleus?
 Initial attempts to answer the last question indicated that $Z \approx 100$
  sets the limit of existence of heavy nuclei/elements
 \cite{Meitner1939,Bohr1939}.
 However, later studies, both theoretical and experimental, have shown that
 this limit can be significantly exceeded.

 In the recent experiments in Dubna, a discovery of nuclei with $Z=\:$114-118
 has been claimed
\cite{Oganessian19991,Oganessian19992,Oganessian20001,Oganessian20002,Oganessian2004,Oganessian2006,Oganessian2010,Oganessian2011,Oganessian2012},
 and was later confirmed by hot fusion cross-sections measured in GSI Darmstadt
 \cite{Dullmann2010} and LBL Berkeley \cite{Stavsetra2009}.
 Synthesis of elements: $Z=\:$107-112 was previously achieved at the GSI
 laboratory \cite{Munzenberg1984,Munzenberg1986,Munzenberg1989,Hofmann19951,Hofmann19952,Hofmann1996,Hofmann2000,Hofmann2002,Hofmann2007} while that
 of the element $Z=113$ by the Riken Lab \cite{Morita}.
 Thus, to this day all the elements with $Z\le118$ have been successfully
 produced (elements $Z=$113, 115, 117, 118 were named by IUPAC in 2016).
 However, there are many predictions that still heavier systems could be
 produced, e.g. \cite{Zagrebaev2008,Liu2011,Wang2012,SWCK2,CSWW,SWCK3,Zhu2014,Wilczynska2019,Adamian2020,Adamian2020PLB}, and there are attempts to synthesize them
\cite{Oganessian2009_Hofmann2012}. There are also ongoing attempts to
 extend the area of known isotopes of superheavy elements, e.g. \cite{Zagrebaev2011,Zagrebaev2012}.

 Theoretical predictions of the properties of these
 nuclei, like binding energies and fission barriers, provide hints
 concerning their synthesis and detection, the interpretation of present
 experiments, and ingredients for evaluating probabilities of various
 reactions and decays. In turn, they are relevant for
 scenarios of creation of elements by stellar nucleosynthesis
 and the question of how the heavy elements came to existence -
 see e.g. \cite{transnucl}.

 In this work, we present ground state (g.s. -
 \mbox{Table 1}) and saddle point (s.p. - \mbox{Table 2})
 properties of heavy and superheavy nuclei with $98 \le Z \le 126$,
 obtained within the microscopic-macroscopic (m-m) method with the
 deformed Woods-Saxon potential in a multidimensional deformation space.
 This method, relatively simpler than selfconsistent approaches,
 allows for calculating many nuclear properties for large sets of heaviest
 nuclei.
 Such systematic predictions relate to the questions specified above
 and contribute to a view on the exotic nuclear domain.

 There are relatively many published predictions of the ground state
  properties of heaviest elements, e.g.
\cite{Duflo1995,Duflo1996,Goriely2003,Goriely2007,Heenen2015,Moller2016,Pomorski2018,Stone2019,Baran2005}.
 While the experimental error in atomic masses or nuclear (g.s.) binding
 energies is smaller than 10 keV for more then half of them
  \cite{Wang20122,Wang2017},
 the best global mass calculations obtain the rms deviation from the data
 not less than 500-800~keV.
 The recent m-m calculations by P. M\"oller and co-workers \cite{Moller2016}
 achieved the rms error of 0.5595~MeV  for nuclei from
$^{16}$O to $^{265}_{106}$Sg and $^{264}_{108}$Hs (0.3549~MeV for the region
 $N \ge 65$). The selfconsistent methods, reviewed e.g. in \cite{SCM2003},
  at best approach this accuracy.
 In calculations for a series of Hartree-Fock-Bogoliubov (HFB) models with
 the Skyrme-type interactions \cite{Goriely2003,Goriely2007,Goriely2013}
 S. Goriely and collaborators reached the rms error of 0.5 MeV in the best case
 through an adjustment of a vibrational term in the phenomenological
collective correction; the mass fit with the Gogny force \cite{Goriely2009-16}
 resulted in the rms error of 0.8 MeV.
 A worse agreement with the data results from relativistic mean field models
\cite{Geng2005,Abgemava2014,Lu2015,Pena2016}, of which the last two studies
 reached the smallest error level of $\geq 1$ MeV.
 The more phenomenological formula with 10 free parameters by Duflo and Zuker
\cite{Duflo1995,Duflo1996} gives masses with the rms error of 0.574~MeV.
 A reduction of the rms deviation is possible, see e.g.
 \cite{Wang2011,Wang2011_2}, but it requires the inclusion of many corrections
 for which no consistent theoretical arguments can be produced at present.

 In spite of differences between various theoretical predictions mentioned
 above, they consistently predict prolate deformed nuclei with $Z=100-112$
 \cite{Pat1,Pat2,Pat3,Heenen2015,Dobacz2015,Pomorski2018,Stone2019}. This was confirmed
  experimentally for isotopes in the vicinity of $^{254}$No and $^{256}$Rf
by the observation of \mbox{$K$-isomers}, e.g. \cite{Herzberg2006,Tandel2006,Kondev2015}, and from rotational
bands in nuclear decay spectroscopy \cite{Reiter1999,Leino1999,Greenlees2012}.
Moreover, recently, first measurements of mean-square charge radii and/or
 electric quadrupole moments based on laser spectroscopy techniques have also
 become possible in this area of nuclei.
 So far, however, the heaviest isotopes measured in such a way are
 $^{252,253,254}$No \cite{Raeder2018}.
 On the other hand, the heavier systems with $Z=114$ and $N=174-184$ are expected to be spherical or oblate deformed
in their ground states \cite{Heenen2015,Pomorski2018}.
For obvious reasons, verification of these predictions will be much more
 difficult.

 The extensive calculations of fission saddle points and barrier heights for
 SHN like \cite{Moller2009,Abusara2012,Staszczak2013,Jachimowicz20172,
 Abgemava2017,Giuliani2018} are much fewer than those for ground states,
 although necessary to estimate the cross sections (survival probabilities)
 for the SHN synthesis.
 Especially rare are calculations for \mbox{odd-$A$} and odd-odd nuclei,
 if only because such studies involve a repetition of calculations
 for many low-lying quasiparticle states which significantly multiplies the
 numerical effort. This work brings such calculations, enlarging
 the set of existing results.



 In view of many already published predictions on SH nuclei
 one should, perhaps, emphasize specific features distinguishing
 the present results. In comparison to the global m-m model of P. M\"oller
 and coworkers \cite{Moller2016,Moller2009}, ours has a few parameters fitted
 specifically to heavy nuclei. Importantly, predictions for fission barriers
 in both m-m models are, somewhat surprisingly, {\it markedly different} in
 superheavy nuclei, in spite of a similar agreement with the data in actinides
 - cf \cite{Moller2009,Jachimowicz20172,Jachimowicz2020}.
 Selfconsistent HFB or relativistic mean-field models present a variety of
 results for fission barriers already in actinides, depending on the model
 version and applied specific corrections and approximations, see e.g.
\cite{Bonneau2004,Delaroche2006,Abusara2010,Zhou2014,Robledo2014,Lemaitre2018}.
 Comparing selfconsistent results for SHN to the present work one has to notice
 that 1) some of them do not account or account only approxiamtely for the
 triaxiality of fisssion saddles, 2) usually they cover
  only even-even nuclei, 3) they usually overestimate fission barriers in
 actinides 4) for finding fission saddles they use the
 inexact (as explained in \cite{Myers1996,Moller2000,Dubray2012}) procedure of
 minimization, as opposed to the in principle exact immersion water flow (IWF)
 method, used first in \cite{Mam1998}, then in \cite{Moller2000,Moller2009}
 and here, 5) they widely differ between themselves and from our model, both
 in g.s. and s.p. and in fission barriers,
 which is related in part to different predicted magic numbers beyond lead.
 Finally, one can mention that the nuclear properties presented here have
 recently been used to estimate cross sections for known $xn$-evaporation
 channels of the $^{48}$Ca-induced complete fusion reactions
 \cite{Adamian2020PLB}. The agreement of calculated cross-sections for the
 reactions in the $3n$ channel with the experimental ones was quite
 satisfactory.
As it turns out, in modeling of reactions leading to the SHN,
the use of a consistent i.e., coming from one source, set of nuclear data input plays a fairly important role \cite{SWCK3,Adamian2020PLB}.

As the principle of the microscopic-macroscopic method is well known, only a brief description of it is
given in \mbox{Sect. 2}, specifying involved quantities, their computation and the adopted values of parameters.
In \mbox{Sect. 3} we describe a variety of considered nuclear shapes which is an important ingredient of the model.
 A concise discussion of results for nuclei with $Z=98-126$ is given in
 \mbox{Sect. 4}, whereas results for
 selected actinide nuclei $Z=89-98$ (for which some experimental estimations
 are known), are presented in \mbox{Sect. 5}. Finally, a brief summary can be found in \mbox{Sec. 6}.

\section{Method}

 Within the microscopic-macroscopic approach, the nuclear binding energy,
 dependent on the proton number $Z$, neutron number $N$, and the nuclear
 shape/deformation, here symbolically denoted $\beta$ (see \mbox{Sect. 3}
 for more details),
 is assumed to be a sum of a macroscopic and microscopic energies:
%
\begin{eqnarray}\label{binding_E}
E_{tot}(Z,N,\beta) =  E_{mac}(Z,N,\beta)+E_{mic}(Z,N,\beta).
\end{eqnarray}
Therefore, the mass $M(Z,N,\beta)$ of a given nucleus at a given deformation
 $\beta$ can be calculated as follows:
\begin{eqnarray}
M(Z,N,\beta)= Z M_p + N M_n + E_{tot}(Z,N,\beta),
\end{eqnarray}
 with the known masses of the proton $M_{p}$ and the neutron $M_{n}$.
 For a comparison with the experimentally measured {\it atomic masses}
 $M_{at}$, one models the latter by taking
 the above nuclear mass at the ground-state deformation $\beta_{g.s.}$,
 adding the mass of $Z$ electrons, and subtracting an approximate expression
 for the binding energy of the electrons in their atomic ground state:
\begin{eqnarray}
M_{at}(Z,N)=Z M_e + M(Z,N,\beta_{g.s.}) - a_{el}Z^{2.39} ,
\end{eqnarray}
 where $M_{e}$ is the electron mass, and $a_{el}$ is the parameter
 of the model.

In general, $E_{mac}$ is the dominant part of $E_{tot}$, varying smoothly
 as a function of $Z$ and $N$. $E_{mic}$ is a small - typically a few MeV -
 but very important energy correction that shows oscillations as a function of
 the number of nucleons, resulting from the quantum nature of the nucleus.

\subsection{Macroscopic energy}

 In this work, the macroscopic part of binding energy (\ref{binding_E})
is taken in the liquid-drop form \cite{Krappe1979,Muntian2001}:
\begin{eqnarray} \label{mmacr}
 E_{mac}(Z,N,\beta)&=&
-a_{v}(1-\kappa_{v} I^2)A + a_{s}(1-\kappa_{s}
I^2)A^{2/3}B_S(\{\beta \}) \nonumber\\
&& + a_0 A^0 + c_1 Z^2 A^{-1/3} B_C(\{\beta \}) - c_4 Z^{4/3}
A^{-1/3} \\
&&- f(k_{F} r_{p})Z^2 A^{-1}
+ \bar{\Delta}_{mac}, \nonumber
\end{eqnarray}
where $I=(N-Z)/A$ is the relative neutron excess and $A=Z+N$ is
the mass number of a nucleus.

 The volume term - the first one in the above expression - is the largest,
 but shape-independent. The whole shape dependence of $E_{mac}$ resides in
 the functions $B_{S}(\beta)$ and $B_{C}(\beta)$, describing the
 dependence of the surface and Coulomb energies, respectively, on deformation
 $\beta$.
 We adopted these functions in the form given by the Yukawa-plus-exponential
 model
formulated by Krappe and Nix \cite{Krappe1979}. They read
\cite{Moller1981,Moller1995}:
\begin{equation}
B_S = \frac{A^{-2/3}}{8\pi^2r_0^2a^4}\int\int_V \left(
2-\frac{r_{12}}{a} \right) \frac{e^{-r_{12}/a}} {r_{12}/a} d^3r_1
d^3r_2,
\end{equation}
\begin{equation}
B_C = \frac{15}{32\pi^2} \frac{A^{-5/3}}{r_0^5}\int\int_V
\frac{1}{r_{12}} \left[ 1- \left(
1+\frac{1}{2}\frac{r_{12}}{a_{ den}} \right) e^{-r_{12}/a_{den}} \right] d^3r_1 d^3r_2,
\end{equation}
where $r_{12}$=$|\overrightarrow{r_1}-\overrightarrow{r_2}|$, with
$\overrightarrow{r_1}$ and $\overrightarrow{r_2}$ describing the
positions of two interacting volume elements, $a$ is the range of
the Yukawa interaction on which the model is based, $a_{den}$
is the range of the Yukawa function used to generate nuclear
charge distribution. The functions are normalized in such a way
that they are equal 1 for a spherical nucleus in the limit
case of $a$=0 (for $B_S$) and $a_{den}$=0 (for $B_C$),
corresponding to the traditional liquid-drop model with a sharp
surface. The integrations are over the volume of a nucleus.
After turning them into surface integrals, $B_S$ and $B_C$ were calculated
by using a four-fold (or three-fold, for the axially symmetric shape) 64-point
 Gauss quadrature.

The quantities $c_1$ and $c_4$ appearing in the direct Coulomb energy and the
Coulomb exchange correction, respectively, are:
\begin{equation}
c_1 = \frac{3}{5} \frac{e^2}{r_0},
\qquad c_4 = \frac{5}{4} \left( \frac{3}{2\pi} \right)^{2/3} c_1,
\end{equation}
where $e$ is the elementary electric charge and $r_0$ is the
nuclear-radius parameter.
The quantity $f(k_{F} r_{p})$
appearing in the proton form-factor correction to the Coulomb
energy in Eq.~(\ref{mmacr}) has the form:
\begin{equation}
f(k_{F} r_{p}) = -\frac{1}{8} \frac{e^2r_{p}^2}{r_0^3}
\left[ \frac{145}{48} - \frac{327}{2880}(k_{F} r_{p})^2 +
\frac{1527}{1\,209\,600}(k_{F} r_{p})^4 \right],
\end{equation}
where the Fermi wave number is taken as:
\begin{equation}
k_{F} = \left( \frac{9\pi Z}{4A} \right)^{1/3} r_0^{-1},
\end{equation}
and $r_{p}$ is the proton root-mean-square radius. The term by $A^0$ is just
 an overall constant.
 The average pairing energy contribution $\bar{\Delta}_{mac}$ takes
 different values for each $Z$- and $N$-parity, as specified
   in \mbox{Ref. \cite{Jachimowicz2014}}. For even-even systems
 $\bar{\Delta}_{mac}=0$, while it modifies binding of odd-$A$ and odd-odd
 systems: $\bar{\Delta}_{mac}=0.824$ MeV, $1.013$ MeV and $1.703$ MeV for
($Z$-even, $N$-odd), ($Z$-odd, $N$-even) and ($Z$-odd, $N$-odd) nuclei,
 respectively.

 The values of the following parameters are adopted after \cite{Moller1981}:
\begin{eqnarray}\label{det_pars3}
&&a_{s} = 21.13 \; {\rm  MeV}, \qquad \kappa_{s}=2.30, \qquad a = 0.68 \; {\rm fm},  \qquad a_{ den}=0.70\;{\rm fm}, \\
&&r_0=1.16\;{\rm fm}, \qquad  r_{ p}=0.80\;{\rm fm}, \qquad a_{el} = 1.433\cdot 10^{-5}\;{\rm MeV}.  \nonumber
\end{eqnarray}
 The remaining parameters of the formula (\ref{mmacr}),
 which stand by the terms independent of the nuclear shape,
  are: $a_{v}$, $\kappa_{v}$, and $a_0$.
 These three parameters were fitted to experimental
 binding energies (masses) of heavy even-even nuclei with $Z \geqslant 84$
 in \cite{Muntian2001}. Since then, we adopted in (\ref{mmacr}) the parameter
 values resulting from this fit:
\begin{equation}\label{det_pars1}
a_{v} = 16.0643, \qquad \kappa_{v} = 1.9261, \qquad a_0 = 17.926.
\end{equation}
Following \cite{Muntian2001}, we omitt here two terms
considered in \cite{Moller1981}: the charge-asymmetry term $c_{a}(N-Z)$
and the Wigner term (with the coefficient $W$),
as they do not significantly influence the precision of calculated binding
 energies (masses) of heaviest nuclei.

\subsection{Microscopic energy}

 For the microscopic energy we calculate the Strutinsky shell correction
 \cite{Strutinsky1967,Strutinsky1968}, based on the deformed Woods-Saxon
 single-particle potentiali, in the form:
 :
\begin{eqnarray}  \label{mic}
E_{mic}(Z,N,\beta) &=&  E^{sh}_{corr}(Z,N,\beta) +  E^{pair}_{ corr}(Z,N,\beta),
\end{eqnarray}
where $E^{sh}_{corr}$ and $E^{pair}_{corr}$ are
the shell and pairing corrections, respectively.
For systems with an odd number of protons, neutrons, or both, these
corrections are calculated by blocking the lowest-lying quasiparticle states
on levels from the tenth below to the tenth above the Fermi level, and
 choosing at each deformation the configuration with the lowest $E_{tot}$.

\subsubsection{Woods-Saxon potential}

 The Woods-Saxon potential consists of the central and spin-orbit parts.
The central part of the  Woods-Saxon potential $V_{WS}$,
which depends on the deformation $\beta$, is given by:
\begin{equation}\label{ws_pot}
V_{WS}(\vec{r})=-\frac{V}{1+e^{d(\vec{r},{\beta})/a_{ws}}},
\end{equation}
where $V$ determines the potential depth, $d(\vec{r},{\beta})$ is
the distance from the point $\vec{r}$ to the surface of the
nucleus, $a_{ws}$ is the diffuseness of the nuclear surface.
In the case of a spherical shape ($\beta=0$), the function $d(\vec{r},{\beta})$
reduces to: $r-R_0$ with a constant radius of a nucleus $R_0 = r_0 A^{1/3}$.
The depth of the potential is
\begin{equation}\label{ws_pot_depth}
V = V_0 (1 \pm \kappa I),
\end{equation}
where $I=(N-Z)/A$, as in \mbox{Eq. (\ref{mmacr})}, is the relative neutron excess and $V_0$ and
$\kappa$ are adjustable parameters. The sign ($+$) is for protons
and ($-$) for neutrons.

The full single-particle potential has the form (e.g. \cite{Cwiok1987}):
\begin{equation}\label{micr_pot}
V_{micr}=V_{WS}+\lambda\left(\frac{\hbar}{2mc}\right)^2
\left( \nabla {\tilde V}_{WS}\right) \cdot
\left(\vec{\sigma}\times\vec{p}/\hbar\right)+V_{C},
\end{equation}
where the second term is the spin-orbit potential, with ${\tilde V}_{WS}$ of
 the same form as $V_{WS}$, but with a different radius parameter. The third
 term is the Coulomb potential that occurs only for protons, of the following
 form:
\begin{equation}\label{coul_pot}
V_{C}(\vec{r})=\rho_{c}\int\frac{d^3r'}{|\vec{r}-\vec{r'}|},
\end{equation}
where $\rho_{c}=(3Ze)/(4\pi R_0^3)$ is the uniform charge density and the
integration extends over the volume enclosed by the nuclear surface.

In our calculations we use the ''universal'' set of parameters of
the potential given in \cite{Cwiok1987}:
\begin{eqnarray}
V_0 = 49.6  \;{\rm MeV}, & a_{\rm ws} = 0.70 \;{\rm fm}, & \kappa = 0.86,
\end{eqnarray}
and separately for protons and neutrons:
\begin{eqnarray}
r_0 = 1.275 \;{\rm fm},  & (r_0)_{ so} = 1.32 \;{\rm fm}, & \lambda = 36.0, \qquad  ({\rm for \; protons}) \\
r_0 = 1.347 \;{\rm fm},  & (r_0)_{ so} = 1.31 \;{\rm fm}, & \lambda = 35.0, \qquad  ({\rm for \; neutrons}) \nonumber
\end{eqnarray}
where $r_0$ and $(r_0)_{so}$ are the radius
parameters for the central and spin-orbit parts of the potential,
respectively.

The single-particle potential is diagonalized in the
deformed harmonic-oscillator basis. The $n_{p}=450$ lowest proton levels
and $n_{n}=550$ lowest neutron levels from the $N_{max}=19$ lowest
 harmonic oscillator shells are taken into account in the
diagonalization procedure. We have determined the single-particle
spectra for every investigated nucleus. These calculations
therefore do not include any scaling relation to the \emph{central} nucleus.
A standard value of \mbox{$\hbar\omega_{0}=41/A^{1/3}$ MeV} is taken for
the oscillator energy.

\subsubsection{Shell correction}

The shell correction energy is calculated as proposed by
Strutinsky \cite{Strutinsky1967,Strutinsky1968}:
\begin{equation}
E^{sh}_{corr}=E_{shell}-\widetilde{E}_{shell},
\end{equation}
where $E_{shell}$ is the sum of single-particle energies $\varepsilon_\nu$ over
all occupied energy levels $\nu$,
\begin{equation}
E_{shell} = \sum_{\nu} \varepsilon_\nu
=\int^{\varepsilon_{ F}}_{-\infty} \rho
(\varepsilon )\varepsilon d \varepsilon,
\end{equation}
while
\begin{equation}\label{sp_dens}
\rho(\varepsilon)=\sum_\nu \delta(\varepsilon-\varepsilon_\nu)
\end{equation}
is the density of the single-particle levels per energy
unit and $\varepsilon_{F}$ is the Fermi energy.

The "smooth" microscopic energy
$\widetilde{E}_{shell}$ is defined by means of the "smooth" density of
the single-particle levels $\widetilde{\rho}(\varepsilon)$:
\begin{equation}\label{sm_sp_en}
\widetilde{E}_{shell}=\int^{\widetilde{\varepsilon}_{F}}_{-\infty}
\widetilde{\rho} (\varepsilon )\varepsilon d \varepsilon.
\end{equation}
The value of $\widetilde{\varepsilon}_{F}$,
found from the following condition for the particle number $N$:
\begin{equation}\label{sh_cor1}
N=\int^{\varepsilon_{F}}_{-\infty} \rho (\varepsilon ) d
\varepsilon=\int^{\widetilde{\varepsilon}_{F}}_{-\infty}
\widetilde{\rho} (\varepsilon) d\varepsilon ,
\end{equation}
is in general different from the Fermi energy $\varepsilon_{F}$.

The "smooth" density, appearing in (\ref{sm_sp_en}) and
(\ref{sh_cor1}), is obtained as
\begin{equation}\label{smooth_den_def}
\widetilde{\rho}(\varepsilon)= \frac{1}{\gamma}
\int^{\infty}_{-\infty} \rho (\varepsilon^{\prime} )
f_p\left(\frac{\varepsilon^{\prime}-\varepsilon}{\gamma}\right)
d \varepsilon^{\prime},
\end{equation}
where $f_p$ is a folding function of the Gaussian type,
taken as the formal expansion of the $\delta$-function,
truncated to the first $2p$ terms:
\begin{equation}\label{delta_expand_approx}
f_p(x) = \frac{1}{\sqrt{\pi}} \sum_{n=0}^{2p} C_n H_n(x)
e^{-x^2},
\end{equation}
with
\begin{equation}\label{C_n}
C_n= \frac{1}{2^n n!} H_n(0) = \left\{
\begin{array}{ll}
\frac{(-1)^\frac{n}{2}}{2^n(\frac{n}{2})!} & \textrm{for even $n$} \\
0 &  \textrm{for odd $n$.} \\
\end{array} \right .
\end{equation}
The width $\gamma$ is of the order of shell energy gaps.
Using $f_p$ one obtains the averaged density $\widetilde{\rho}$:
\begin{equation}
\widetilde{\rho}(\varepsilon
)=\frac{1}{\gamma\sqrt{\pi}}\sum_{\nu=1} e^{-u_\nu^2}
\sum_{n=0}^{2p}C_n H_n(u_\nu),
\end{equation}
where $u_\nu=(\varepsilon-\varepsilon_\nu)/\gamma$.

 In the smoothing procedure we use all energy levels up to
 energy 3.2 $\hbar \omega_0$ above the Fermi level.
 The energy (\ref{sm_sp_en}) in
 general depends on the parameters $\gamma$ and $p$. The method is meaningful,
if there is a certain interval of $\gamma$ and corresponding $p$,
for which the energy does not practically depend on them (so called "plateau condition").
Here, we use $\gamma=1.2 \hbar\omega_{0}$ for the Strutinski smearing parameter and the sixth-order correction polynomial ($p=6$) for $f_p$.

\subsubsection{Pairing correlations}

In this work, pairing is included within the
Bardeen-Cooper-Schrieffer (BCS) theory \cite{Bardeen1957}. We assume a constant
matrix element $G$ of the (short-range) monopole pairing interaction.
The pairing hamiltonian, treated separately for neutrons and for protons,
 may be written as:
\begin{equation}\label{bcs_ham}
H = \sum_{\nu}\varepsilon_\nu a_{\nu}^{+} a_{\nu} -
G \sum_{\nu,\nu^{'}>0}  a_{\nu}^+ a_{\nu^{'}}^{+}
a_{\bar{\nu}^{'}}  a_{\bar{\nu}},
\end{equation}
where, as before, $\varepsilon_{\nu}$ denotes the energy of a single-particle
state $\nu$. Each state $\nu$ has its time-reversal-conjugate
$\bar{\nu}$ with the same energy (Kramers degeneration).
As the BCS wave function is a superposition of components with
different numbers of particles, one requires that the expectation
value of the particle number has a definite value $n$ ($n=N$ for neutrons and
 $n=Z$ for protons):
\begin{equation}
\left<\hat{n}\right>=2\sum_{\nu>0}v_{\nu}^2 = n.
\end{equation}
The occupation numbers are given by
\begin{equation}
v_{\nu}^2=\frac{1}{2}\left[ 1-\frac{(\varepsilon_{\nu}-\varepsilon_{ F})}
{\sqrt{\left(\varepsilon_{\nu}-\varepsilon_{ F}\right)^2+\Delta^2}} \right],
\end{equation}
where the parameters $\varepsilon_{ F}$ and $\Delta$ are solutions of the system
of two equations, for the average particle number:
\begin{equation}
\qquad\quad  n = \sum_{\nu>0} \left[1-
\frac{\varepsilon_{\nu}-\varepsilon_{ F}}
{\sqrt{\left(\varepsilon_{\nu}-\varepsilon_{ F}\right)^2+\Delta^2}} \right],
\end{equation}
and for the pairing gap:
\begin{equation}
\frac{2}{G} = \sum_{\nu>0}
\frac{1}{\sqrt{\left(\varepsilon_{\nu}-\varepsilon_{ F}\right)^2+\Delta^2}}.
\end{equation}
 Energy of the system in the BCS state reads:
\begin{equation}\label{bcs_ener}
E_{ BCS} = 2\sum_{\nu>0} \varepsilon_{\nu} v_{\nu}^2
- \frac{\Delta^2}{G} - G\sum_{\nu>0} v_{\nu}^4.
\end{equation}
 Within our model, the BCS equations are solved on $N$ (for neutrons) and $Z$
 (for protons) lowest doubly degenerate levels.

\subsubsection{Pairing correction}

Pairing correction energy $E^{pair}_{corr}$ of
Eq.~(\ref{mic}) is usually constructed in analogy to the shell
correction energy $E^{sh}_{corr}$,
\begin{equation}\label{bcs_E_pair}
E^{pair}_{ corr} = E_{pair} - \widetilde{E}_{pair},
\end{equation}
where $E_{pair}$ is the pairing energy corresponding to real
single-particle level distribution $\rho(\varepsilon)$,
Eq.~(\ref{sp_dens}), and $\widetilde{E}_{pair}$ is this
energy for the smoothed single-particle level distribution,
$\widetilde{\rho}(\varepsilon)$, Eq.~(\ref{smooth_den_def}).

The $E_{pair}$ is
\begin{equation}
E_{pair} = E_{ BCS} - E_{BCS}^{\Delta=0},
\end{equation}
where $E_{BCS}^{\Delta=0}$ is the $E_{BCS}$ energy in the
limit of disappearing pairing correlations ($\Delta = 0$). Thus,
using Eq.~(\ref{bcs_ener}),
\begin{equation}\label{bcs_ener_limit}
E_{BCS}^{\Delta=0} =
2\sum_{\nu=1}^{n/2} \varepsilon_{\nu} - \frac{Gn}{2}.
\end{equation}
because for $\Delta=0$, the probability $v_{\nu}^2$ of
the occupation of any state $\nu$ is either 0 or 1.

The smoothed pairing energy term  is included  in a schematic form, resulting
 from a model with the constant density of pairs, or doubly degenerate
 levels, $\bar{\rho}$, taken equal to
 ${\widetilde {\rho}} (\widetilde\varepsilon_{F})/2$:
\begin{equation}
\widetilde{E}_{pair}= - \frac{N_p^2}{\bar{\rho}}(\sqrt{1+x^2}-1)
+ \frac{\bar{G}N_p x}{2} \arctan (1/x).
\end{equation}
In the above expression, $x=\bar{\rho}\bar{\Delta}/N_p$, $N_p=n/2$ is a number of pairs,
$\bar{\Delta}$ is an average value of the pairing gap in the neighbourhood of a studied nucleus,
related to the average pairing strength $\bar{G}$ via the BCS formula for the
constant level density:
\begin{equation}
\frac{1}{\bar{\rho}\bar{G}} = \ln \left ( \frac{\sqrt{1+x^2}+1}{x}  \right ).
\end{equation}
The values of $\bar{\Delta}$ are taken from the fit \cite{Madlannd1998} as:
\begin{equation}
\bar{\Delta}=\frac{5.72}{N^{1/3}} \exp {(-0.119I - 7.89I^2)},
\end{equation}
for neutrons and
\begin{equation}
\bar{\Delta}=\frac{5.72}{Z^{1/3}} \exp {(0.119I - 7.89I^2)},
\end{equation}
for protons, with $I=(N-Z)/A$. The smoothed pairing energy term
calculated in this way shows nearly no deformation dependence. For example,
it varies by about $50$ keV over the whole deformation range in actinides.
Thus, it could be omitted in energy landscapes, while it shows up in binding
 energies.

The pairing interaction strengths in \mbox{Eq. (\ref{bcs_ham})},
 $G_p$ for protons and $G_n$ for neutrons, are taken according to
 \cite{Nilsson1969} as:
\begin{equation}\label{G_param}
G_p = (g_{0p}+g_{1p}I)/A,  \quad  \quad  G_n = (g_{0n}+g_{1n}I)/A ,
\end{equation}
with the constants:
\begin{eqnarray}
g_{0p}  = 13.40 \; {\rm MeV},  & g_{1p} =   44.89 \; {\rm MeV}, \\
g_{0n}  = 17.67 \; {\rm MeV},  & g_{1n} =  -13.11 \; {\rm MeV}, \nonumber
\end{eqnarray}
 fixed in \cite{Muntian2001} by adjusting the calculated proton $\Delta_p$
and neutron $\Delta_n$ gap parameters to the three-point odd-even mass
 differences:
\begin{eqnarray}
\Delta_p = (-1)^Z \left\{ \frac{1}{2} \left[M(Z+1,N)+M(Z-1,N)\right] - M(Z,N)\right\}, \\
\Delta_n = (-1)^N \left\{ \frac{1}{2} \left[M(Z,N+1)+M(Z,N-1)\right] - M(Z,N)\right\}, \nonumber
\end{eqnarray}
 obtained using all measured masses of nuclei with $Z \geqslant 88$.

\section{Shape parametrization}

The essential point of any microscopic-macroscopic study is the kind and
 number of independent deformations used to describe a variety of nuclear
 shapes.
Obviously, there is no ideal shape parametrization which would allow for an
effective and simple description of different nuclear shapes in different
regions of nuclei.
 However, as far as we are interested in
superheavy nuclei, which have relatively short fission barriers and rather
 compact shapes at the saddle, a traditional expansion of the nuclear radius
 $R(\vartheta ,\varphi)$ onto spherical harmonics
 ${Y}_{\lambda \mu}(\vartheta ,\varphi)$ \cite{Hasse1988} can be used:
\begin{eqnarray}
R(\vartheta ,\varphi)= c R_0\{ 1+\sum_{\lambda=1}^{\infty}
\sum_{\mu=-\lambda}^{+\lambda} \beta_{\lambda \mu} {Y}_{\lambda \mu}
 (\vartheta ,\varphi) \},
\end{eqnarray}
 where $c$ is the volume-fixing factor depending on deformation and
 $R_0 = 1.16 \cdot A^{1/3}$ fm is the radius of a spherical nucleus.

\subsection{Deformation space in the analysis of the ground states}

After many tests, we confined analysis for nuclear ground states
to axially-symmetric shapes, with the expansion of the nuclear radius
truncated at $\beta_{80}$ (for brevity, arguments of $Y_{\lambda \mu}$ are
 suppressed below):
\begin{eqnarray} \label{Rgs}
R(\vartheta ,\varphi)= c R_0\{
1 \kern-1em & \:+\:& \kern-1em \beta_{20}{Y}_{20}+\beta_{30}{Y}_{30}+\beta_{40}{Y}_{40}+\beta_{50}{Y}_{50} \\
\kern-1em &+& \kern-1em \beta_{60}{Y}_{60}+\beta_{70}{Y}_{70}+\beta_{80}{Y}_{80}\}. \nonumber
\end{eqnarray}
In general, there is no physical principle which would forbid a nonaxial shape
 of a nuclear ground-state in its intrinsic frame of reference. However,
 systematic calculations made by
P. M\"{o}ller et al. \cite{Moller2006} and our studies
\cite{Jachimowicz2009,Jachimowicz2010,Jachimowicz2011,Jachimowicz2017}
suggest that in the investigated nuclei the effect of nonaxiality in the
 ground states
(including non-axial octupole deformation parameterized by $\beta_{32}$)
is either small or non-existent. On the other hand, competing axially symmetric
minima are frequent \cite{Jachimowicz2010_2}. Therefore, we assumed here
the axial symmetry of the ground states. The total energy was minimized
simultaneously in 7 degrees of freedom present in (\ref{Rgs}), by using
 the conjugate gradient method. To locate the relevant global minima, the
 minimization was repeated at least 30 times for a given nucleus with randomly selected (within reasonable limits) starting
values of $\beta_{\lambda 0}$.
In order to gain some confidence in our results we also performed additional
 tests, described in \cite{Jachimowicz2017}, where the ground states of about
 $3000$ heavy and superheavy nuclei, $82\le Z \le126$, were found by the
simultaneous minimization over ten deformation parameters:
the axial ones from (\ref{Rgs}) plus non-axial $\beta_{22}$, $\beta_{32}$ and $\beta_{42}$.

It is worth mentioning that for odd systems, the additional minimization over
configurations was performed at every step of
the gradient procedure. Considered configurations consist of the unpaired
  particle occupying one of the levels close to the Fermi level -
 from the tenth state above to the tenth below the Fermi level -
 and the rest of the particles forming a paired BCS state on the remaining
 levels.

\subsection{Deformation space in the analysis of the saddle points}

  Saddle points were searched in consecutive stages by using the Imaginary
Water Flow (IWF) technique, see e.g. \cite{Luc1991,Moller2009,Jachimowicz2017},
 at each of them. At the first stage, the search was performed
 in a five-dimensional deformation space including triaxial shapes, spanned by
$\beta_{20},\beta_{22},\beta_{40},\beta_{60},\beta_{80}$.
 The appropriate nuclear radius expansion had the form:
\begin{eqnarray}
\label{deformation_space}
R(\vartheta ,\varphi)= c R_0\{ 1 \kern-1em & \:+\: & \kern-1em \beta_{20}{Y}_{20}+
\frac{\beta_{22}}{\sqrt{2}} \left[{Y}_{22}+{Y}_{2-2} \right] \\
\kern-1em &+& \kern-1em \beta_{40}{Y}_{40}+\beta_{60}{Y}_{60}+\beta_{80}{Y}_{80}  \},   \nonumber
\end{eqnarray}
and the potential energy was calculated in the following basic grid
points:
\begin{eqnarray}\label{basicgrid}
\beta_{20}\: \kern-1em   & = & \kern-1em   \;\;\: 0.00 \: (0.05) \: 0.60, \nonumber\\
\beta_{22}\: \kern-1em   & = & \kern-1em   \;\;\: 0.00 \: (0.05) \: 0.45, \nonumber\\
\beta_{40}\: \kern-1em   & = & \kern-1em  -0.20 \: (0.05) \: 0.20, \\
\beta_{60}\: \kern-1em   & = & \kern-1em  -0.10 \: (0.05) \: 0.10, \nonumber\\
\beta_{80}\: \kern-1em   & = & \kern-1em  -0.10 \: (0.05) \: 0.10, \nonumber
\end{eqnarray}
where the numbers in the parentheses specify the step with which the
calculation was done for a given variable.
This made a grid of 29 250 points for given nucleus. As for the ground states,
 here and in the following stages, for odd-$A$ and odd-odd nuclei
 the minimization over configurations was performed at each grid point
 among those with the odd particle blocking one
 of levels from the 10-th below to the 10-th above the Fermi level.
 Then, our primary grid (\ref{basicgrid}) was subject to the fivefold
 interpolation in all directions. Thus, we
finally had energy values at a total of 50 735 286 grid points with the step
0.01 in each dimension. To find the saddles on such
a giant grid we used the IWF method. In addition, for more elongated
fission barriers (few cases), grid (\ref{basicgrid}) was further
extended to $\beta_{20} > 0.60$
For nuclei that had two or even more saddles of a comparable height
we found all of them within the $0.5$ MeV energy window.

In the second stage, we checked the stability of all those saddles
(the first, the second, axially
symmetric or triaxial), against the mass-asymmetry.
This was done by an additional 3-dimensional energy minimization with respect
to $\beta_{30},\beta_{50},\beta_{70} $ around each saddle that was previously
determined on the mesh (\ref{basicgrid}). One has to emphasize that to
perform this minimization we used the Woods-Saxon code in which
both nonaxial and reflection-asymmetric deformations are included
simultaneously. This more general code should be, in principle, used in
a search for saddle points (which would be then a one-stage task) but:
  1) at present, a large size of the resulting mesh of deformations prohibits
  such calculations for many nuclei, 2) the results of the performed
 minimization give a hope that such an effort would be unnecessary.

It turned out that the minimization lowers energy
of only those saddles in which there is no triaxiality,
and the saddle deformation is around $\beta_{20} \approx 0.3$. Therefore,
we could find all mass-asymmetric fission saddle points
by using (in principle exact) IWF method on the following 7-dimensional grid
of axially symmetric deformations:
\begin{eqnarray}
\label{7Dgrid}
\beta_{20}\: \kern-1em & = & \kern-1em  \;\;\: 0.25 \: (0.05) \: 0.40, \nonumber\\
\beta_{30}\: \kern-1em & = & \kern-1em  \;\;\: 0.00 \: (0.05) \: 0.25, \nonumber\\
\beta_{40}\: \kern-1em & = & \kern-1em        -0.15 \: (0.05) \: 0.20, \nonumber\\
\beta_{50}\: \kern-1em & = & \kern-1em  \;\;\: 0.00 \: (0.05) \: 0.15, \\
\beta_{60}\: \kern-1em & = & \kern-1em        -0.10 \: (0.05) \: 0.10, \nonumber\\
\beta_{70}\: \kern-1em & = & \kern-1em  \;\;\: 0.00 \: (0.05) \: 0.15, \nonumber\\
\beta_{80}\: \kern-1em & = & \kern-1em        -0.10 \: (0.05) \: 0.10, \nonumber
\end{eqnarray}
containing 1 690 730 496 points (after the fivefold interpolation in all
directions).

In the third stage of our calculations we included the effect of hexadecapole
nonaxiality on the saddles. To evaluate this effect without
increasing the basic grid dimensions (\ref{basicgrid})
we constrained $\beta_{42}$ and $\beta_{44}$ to be functions
of $\beta_{22}$ and $\beta_{40}$, according to \cite{Hasse1988}.
Now, the equation of the nuclear surface had the following form:
\begin{eqnarray}\label{Rhexadeca}
R(\vartheta ,\varphi)= c R_0\{ \nonumber
1\kern-1em & \:+\:& \kern-1em \beta \cos (\gamma){ Y}_{20}+ \frac {\beta \sin(\gamma)}{\sqrt{2}} \left [{Y}_{22} + {Y}_{2-2} \right ] \\
 \kern-1em & \:+\:& \kern-1em \beta_{40}\frac{1}{6}\left [ 5 \cos^2(\gamma)+1 \right ] {Y}_{40} - \beta_{40}\frac{1}{6} \sqrt{\frac{15}{2}} \sin (2\gamma) \left [{Y}_{42} + {Y}_{4-2} \right ] \\
 \kern-1em & \:+\:& \kern-1em \beta_{40}\frac{1}{6} \sqrt{\frac{35}{2}} \sin^2 (\gamma) \left [{Y}_{44} + {Y}_{4-4} \right ] + \beta_{60}{Y}_{60} +  \beta_{80}{Y}_{80} \}, \nonumber
\end{eqnarray}
with $\beta=\sqrt{\beta_{20}^2+\beta_{22}^2}$ and
$\gamma= {\arctan}(\beta_{22} / \beta_{20})$ - for details, see
\cite{Jachimowicz2017}.
The further method of proceeding was analogous to that
used in the study of the saddle mass-asymmetry, namely:
initial checking the effect of the non-axial hexadecapole deformation by a
 2-dimensional minimization with respect to $\beta_{42}$, $\beta_{44}$ at each
 saddle found on the basic grid (\ref{basicgrid}), and
 then finding the "exact" location of each saddle that was significantly
 lowered by the minimization. The latter step was performed on the
 following grid:
\begin{eqnarray}
 \beta_{20}\: \kern-1em & = & \kern-1em \;\;\:0.00 \: (0.05) \: 0.60, \nonumber\\
 \beta_{22}\: \kern-1em & = & \kern-1em \;\;\:0.00 \: (0.05) \: 0.45, \nonumber\\
 \beta_{40}\: \kern-1em & = & \kern-1em      -0.20 \: (0.05) \: 0.20, \\
 \beta_{60}\: \kern-1em & = & \kern-1em      -0.10 \: (0.05) \: 0.10, \nonumber\\
 \beta_{80}\: \kern-1em & = & \kern-1em      -0.10 \: (0.05) \: 0.10, \nonumber
\end{eqnarray}
where, according to (\ref{Rhexadeca}), the parameter $\beta_{40}$ appears
as the coefficient by the real combinations of the spherical harmonics
 ${ Y}_{4\pm 2}$ and ${ Y}_{4\pm 4}$.
This five-dimensional grid, composed of 29 250 deformations, was subject to
 the fivefold interpolation in all directions before it was used in the IWF
 procedure.
Thus, in this case we finally had the values of the potential energy at a
 total of \mbox{50 735 286} grid points.

\section{Discussion of the results}

The calculated properties of ground-states of
 nuclei in the range \mbox{$Z=98-126$}, \mbox{$N=134-192$}
 are given in \mbox{Table 1} and those of the saddle points in \mbox{Table 2}.
Total energy $E_{tot}$ given in those tables and depicted on the following
  energy maps is normalized in such a way that its macroscopic part is equal
 zero at the spherical shape.
 \mbox{Tables 3-6} which contain i.a. the same ground-state and saddle point
 quantities calculated for the selected $75$ actinides
 are discussed in Sect. 5.

\subsection{Ground state properties}

\begin{figure}[h]
\centering
\includegraphics[scale=0.26]{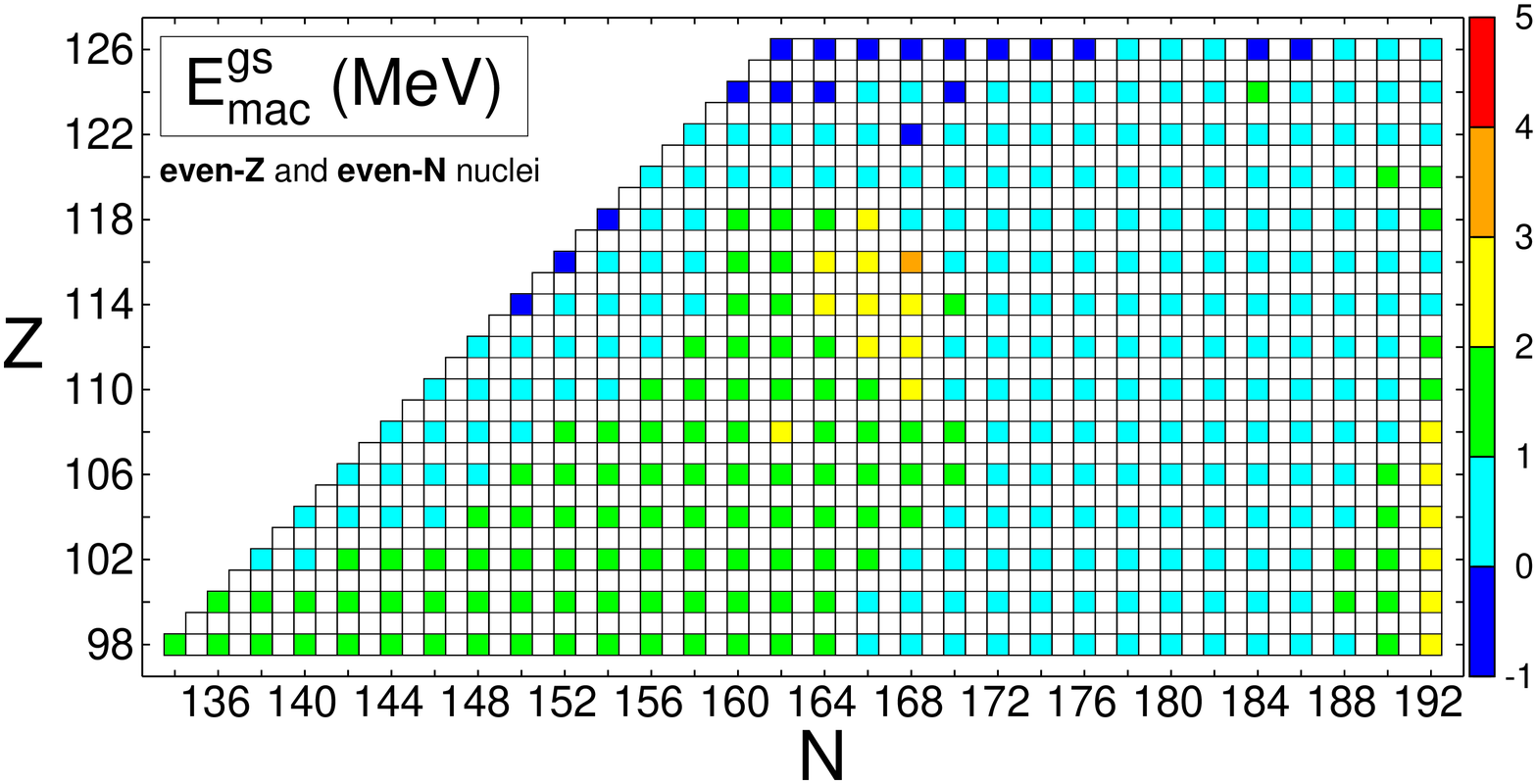}
\includegraphics[scale=0.26]{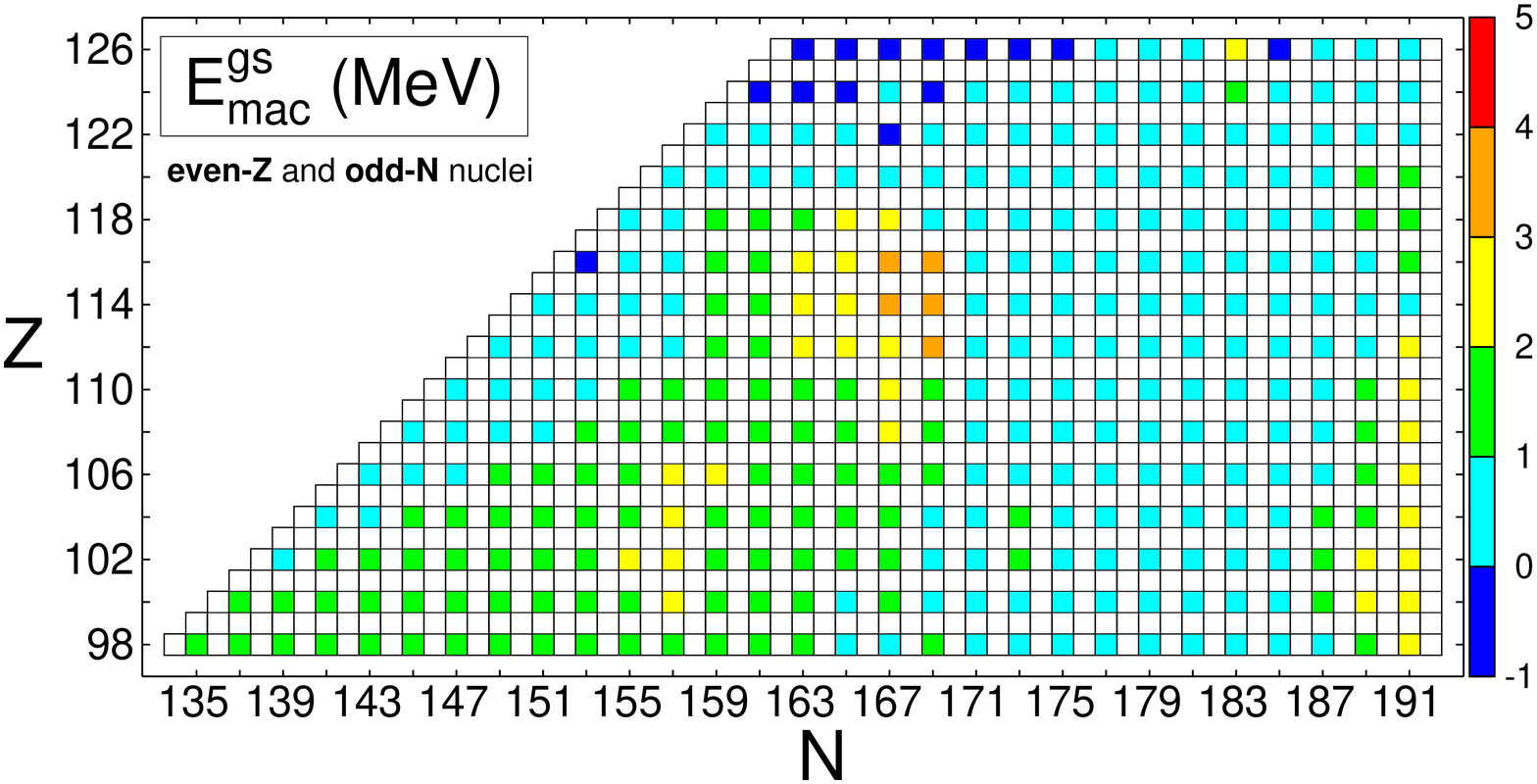}
\includegraphics[scale=0.26]{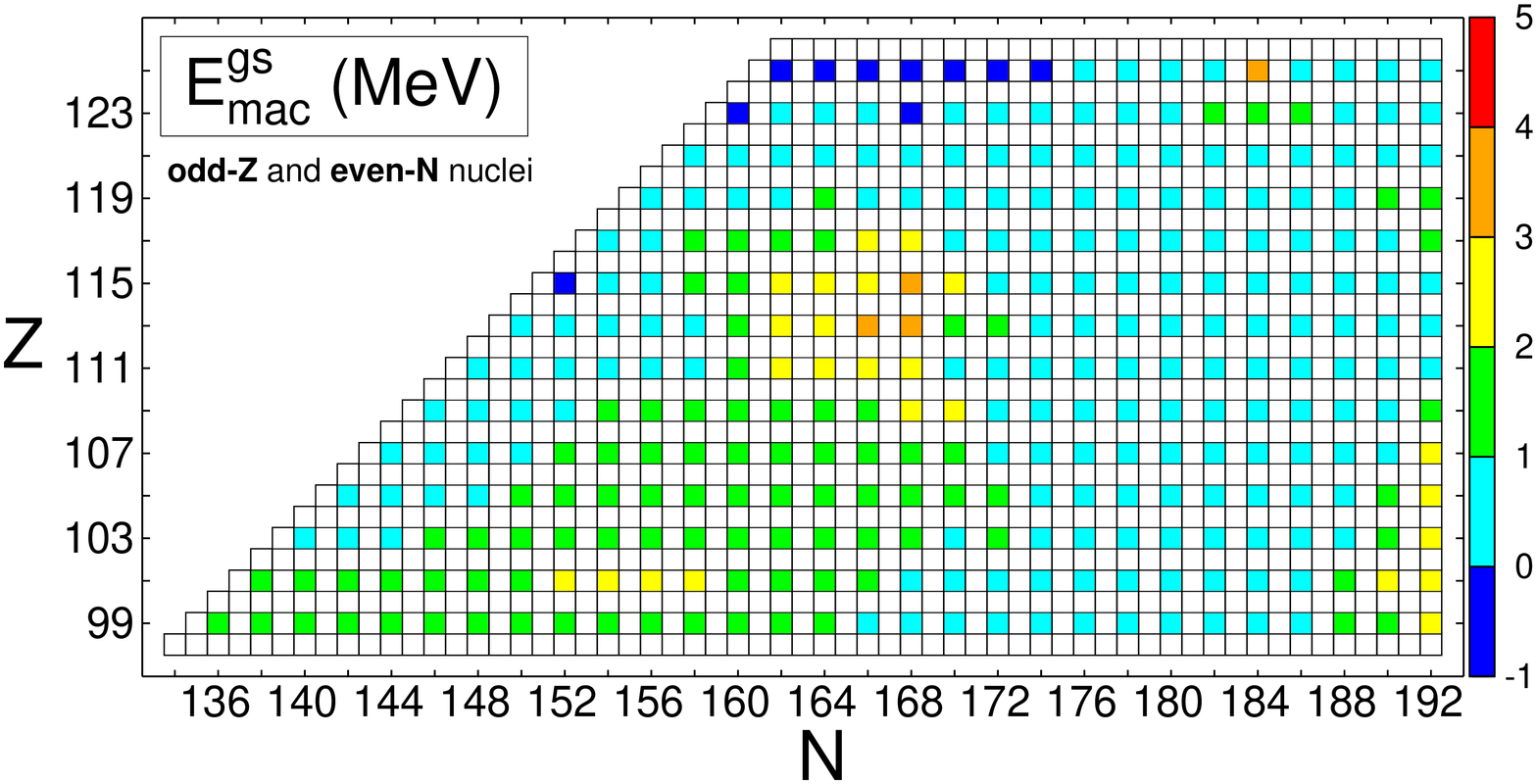}
\includegraphics[scale=0.26]{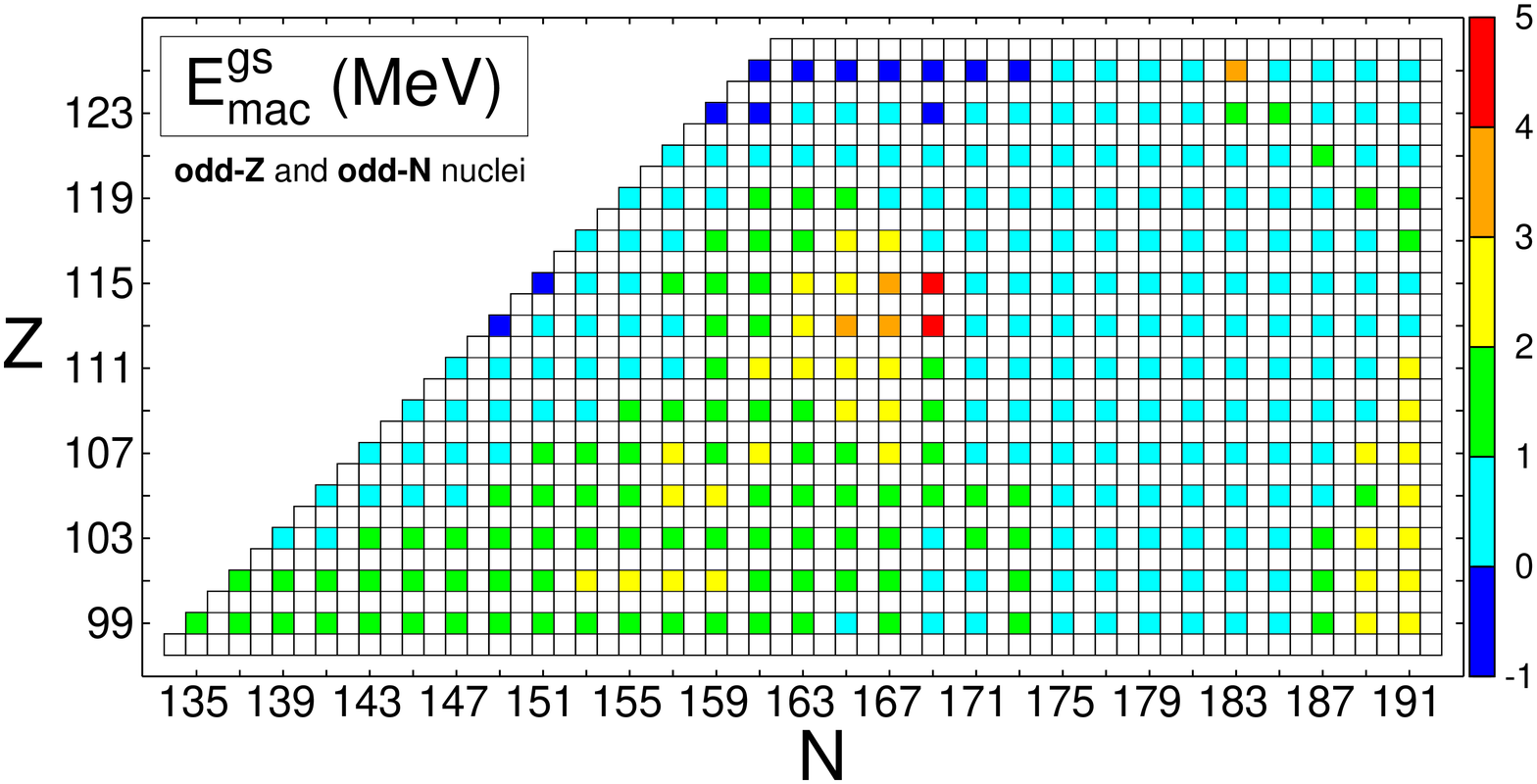}
\caption{Calculated macroscopic component of the
ground state binding energy (in 4 separate groups of nuclei), normalized to
the macroscopic energy at the spherical shape,
$E^{gs}_{mac}=E_{mac}^{gs}(deformation)-E_{mac}^{gs}(sphere)$.}
\label{fig1}
\end{figure}

\begin{figure}[h]
\centering
\includegraphics[scale=0.26]{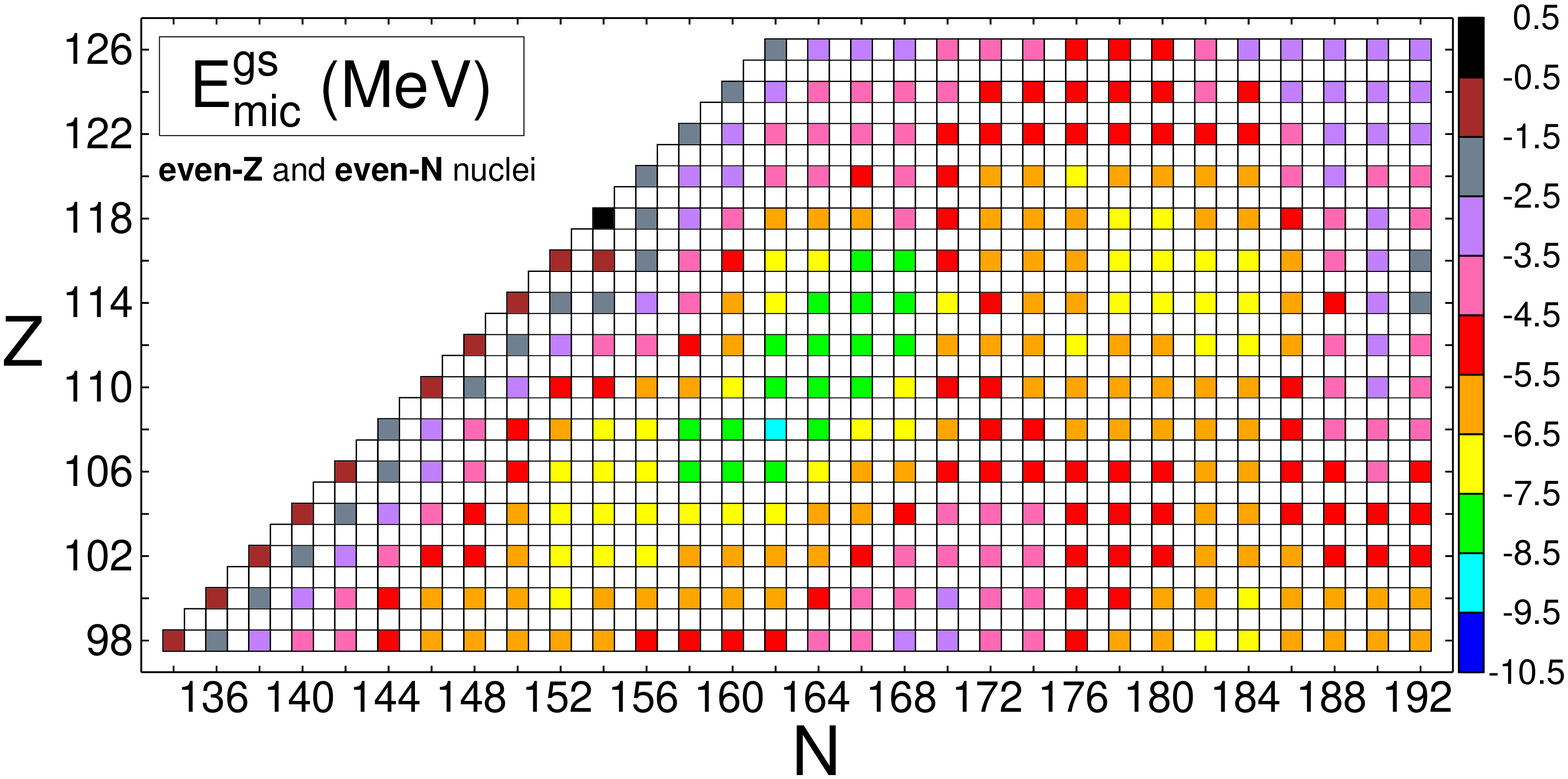}
\includegraphics[scale=0.26]{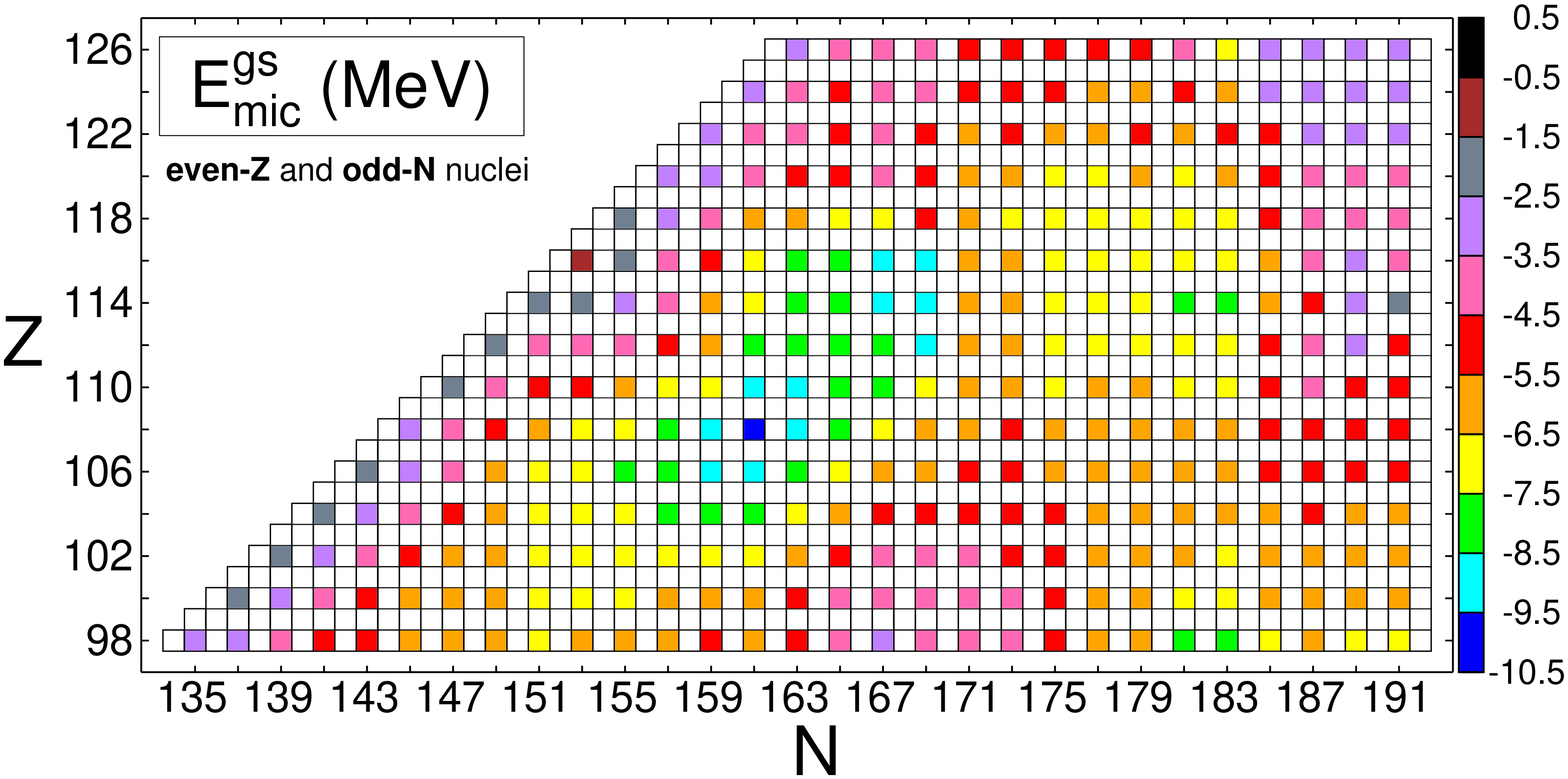}
\includegraphics[scale=0.26]{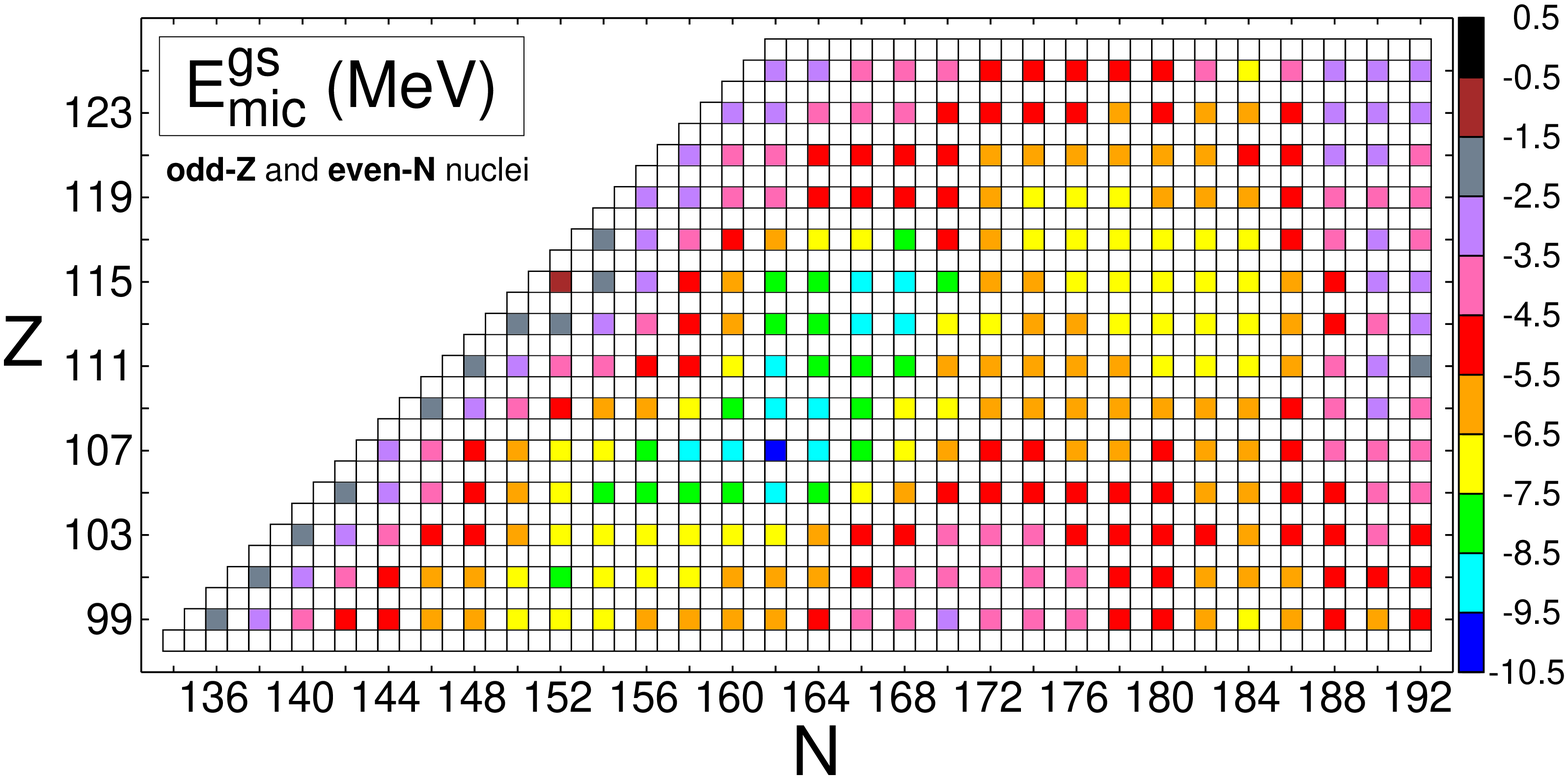}
\includegraphics[scale=0.26]{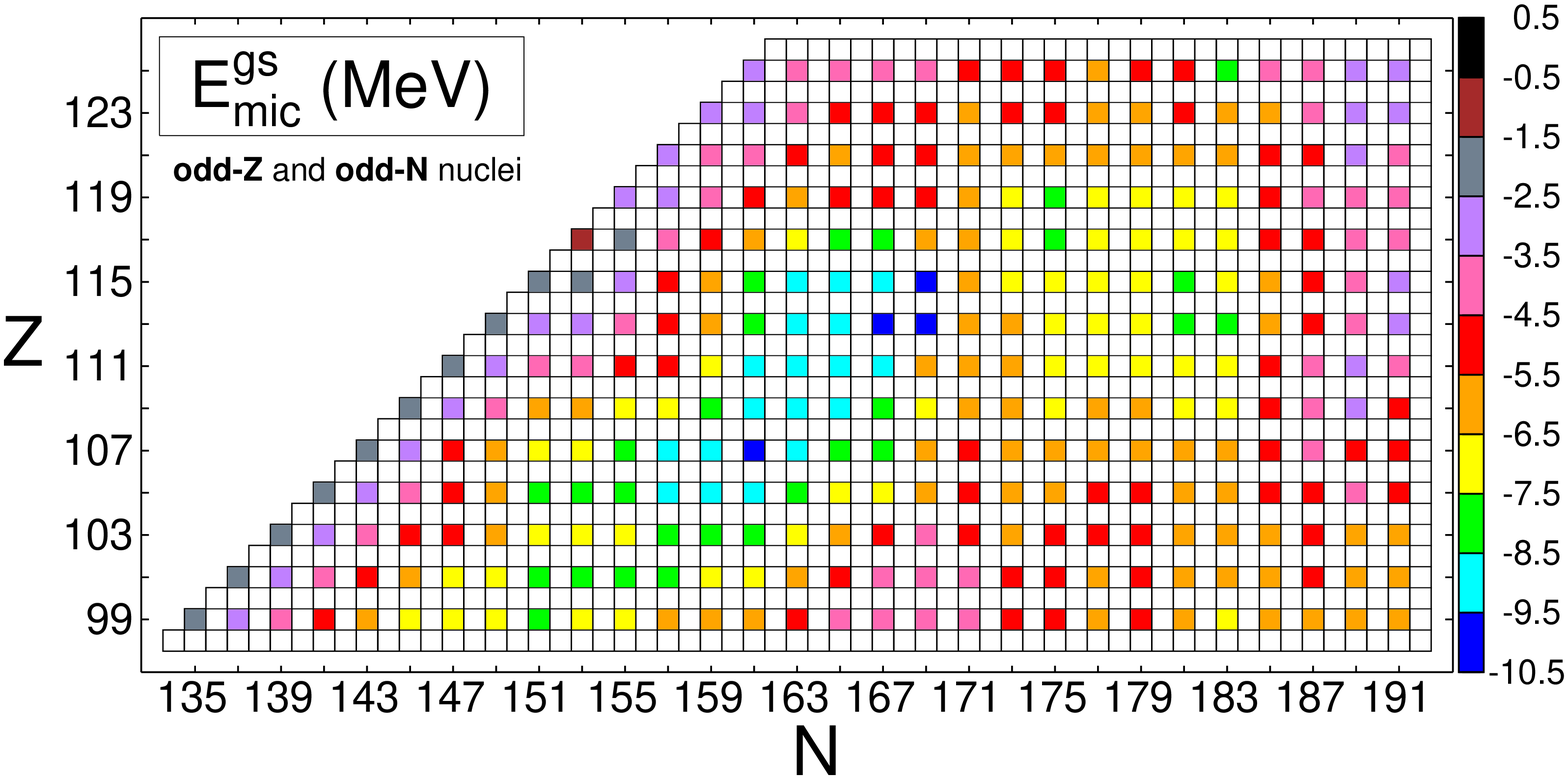}
\caption{Calculated microscopic component $E^{gs}_{mic}$ of the
ground state binding energy in 4 separate groups of nuclei.}
\label{fig2}
\end{figure}

 Calculated ground-state masses and shapes are the result of the energy
 minimization and reflect an interplay of the macroscopic and microscopic
 energies.
 These macroscopic and microscopic components of $E_{ tot}^{ gs}$ are shown in
\mbox{Fig. \ref{fig1}} and \mbox{Fig. \ref{fig2}}, respectively.
 In these figures, as in the following ones pertaining to all nuclei, the
 quantities are plotted separately for each of the groups of nuclei:
($Z$-even, $N$-even), ($Z$-even, $N$-odd), ($Z$-odd, $N$-even) and
 ($Z$-odd, $N$-odd).
 The biggest microscopic correction, up to \mbox{$\sim 10$ MeV}, is
observed around $Z=108$, $N=162$ and $Z=114$, $N=170$,
 i.e. close to the magic or semi-magic numbers: $Z=114$ \cite{Sob1966}
 and $N=162$ \cite{Patyk1991}.
As in other microscopic-macroscopic calculations, e.g.
 \cite{Moller2016,Wang2013},
the vast area of large negative microscopic corrections
\mbox{$\mid E_{mic}^{gs} \mid \geqslant 6.5$ MeV} lies also
within the range: $Z\approx109-120$ and $N \approx 174-184$.
When superposed with weakly deformation-dependent macroscopic part \mbox{(Fig. \ref{fig1})},
this component is largely responsible for the emergence of global minima in
superheavy nuclei.

\subsubsection{Ground state shapes (deformations)}

\begin{figure}[h]
\centering
\includegraphics[height=6cm]{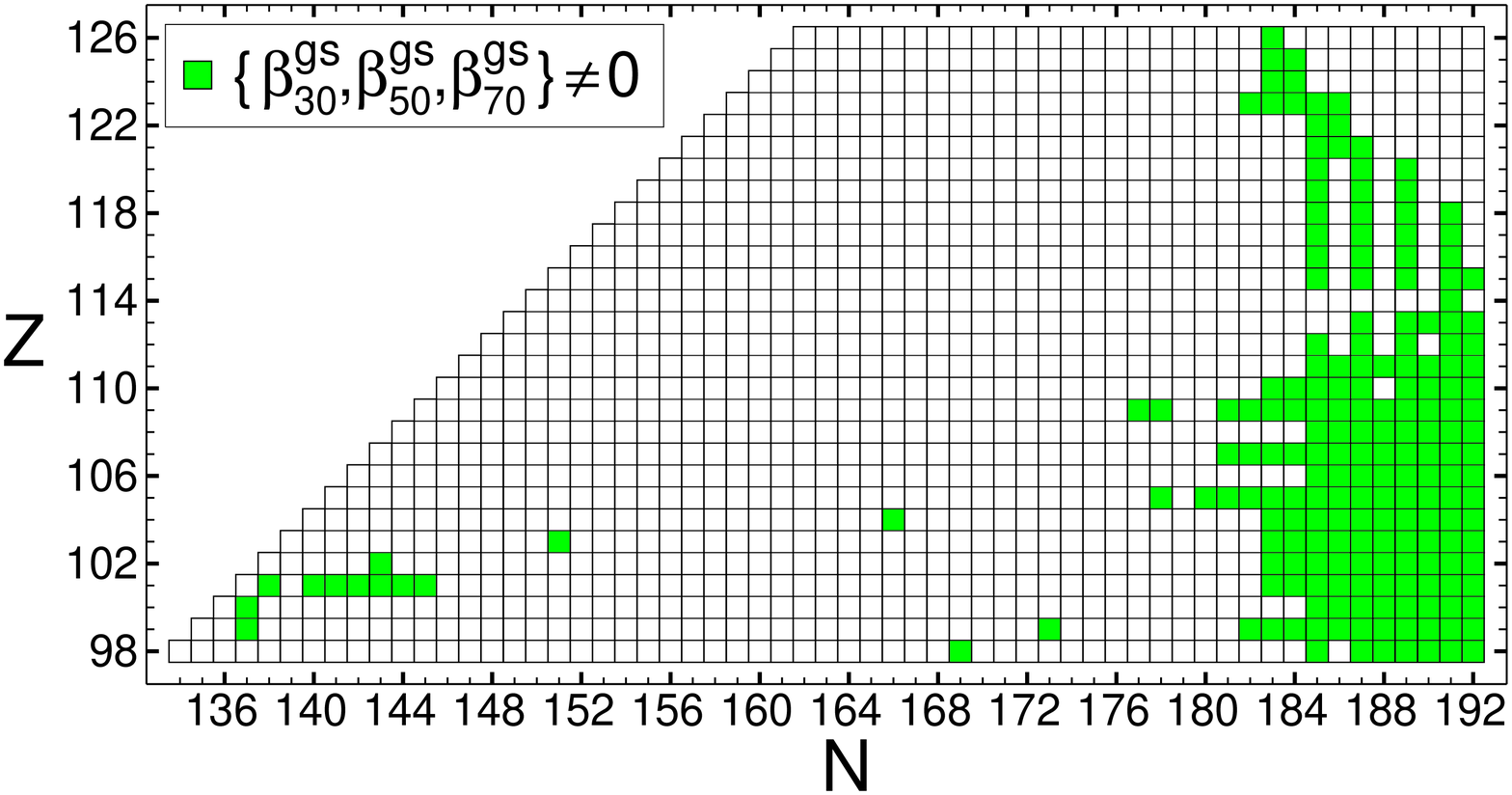}
\caption{Nuclei with a nonzero values of deformation parameters:
$\beta_{30}, \beta_{50}, \beta_{70}$ in their ground states,
marked with green squares.
}
\label{fig3}
\end{figure}

The calculated ground-state shapes are mostly reflection-symmetric.
 For about \mbox{$15$\%} of the considered nuclei with $Z \geqslant 98$ we obtained
 nonzero values of deformation parameters: $\beta_{30}, \beta_{50}, \beta_{70}$
 and these are given in \mbox{Table 1}. Additionally, the nuclei with
 reflection-asymmetric ground states are marked in \mbox{Fig. \ref{fig3}}
 with green squares.
As may be seen, those deformations occur practically only
in ultra-neutron-rich nuclei with $N \geqslant 182$. Therefore, a vast majority
  of them are very exotic and probably nonexistent.
Additionally, for selected superheavy nuclei with $Z > 118$
we have found g.s. minima exhibiting more exotic triaxial shapes \cite{Jachimowicz2010_2,Jachimowicz2017}.
In most cases those minima are degenerate with axially symmetric ones, specified in \mbox{Table 1}.
However, for about $40$ superheavy systems in the area of
$Z = 119-126$, $N = 173-188$, their ground states exhibit a combined slight oblate
and octupole distortions which lead to the shape of the $\beta_{33}$-type
with respect to the axis of the oblate deformation. In the extreme case,
for the nucleus $^{296}123$,  the oblate-$\beta_{33}$ minimum is lower by
\mbox{$\sim 0.7$ MeV} \cite{Jachimowicz2017} than the axially symmetric one.
This effect {\it is not included} in \mbox{Table 1}.

\begin{figure}[h]
\centering
\includegraphics[scale=0.26]{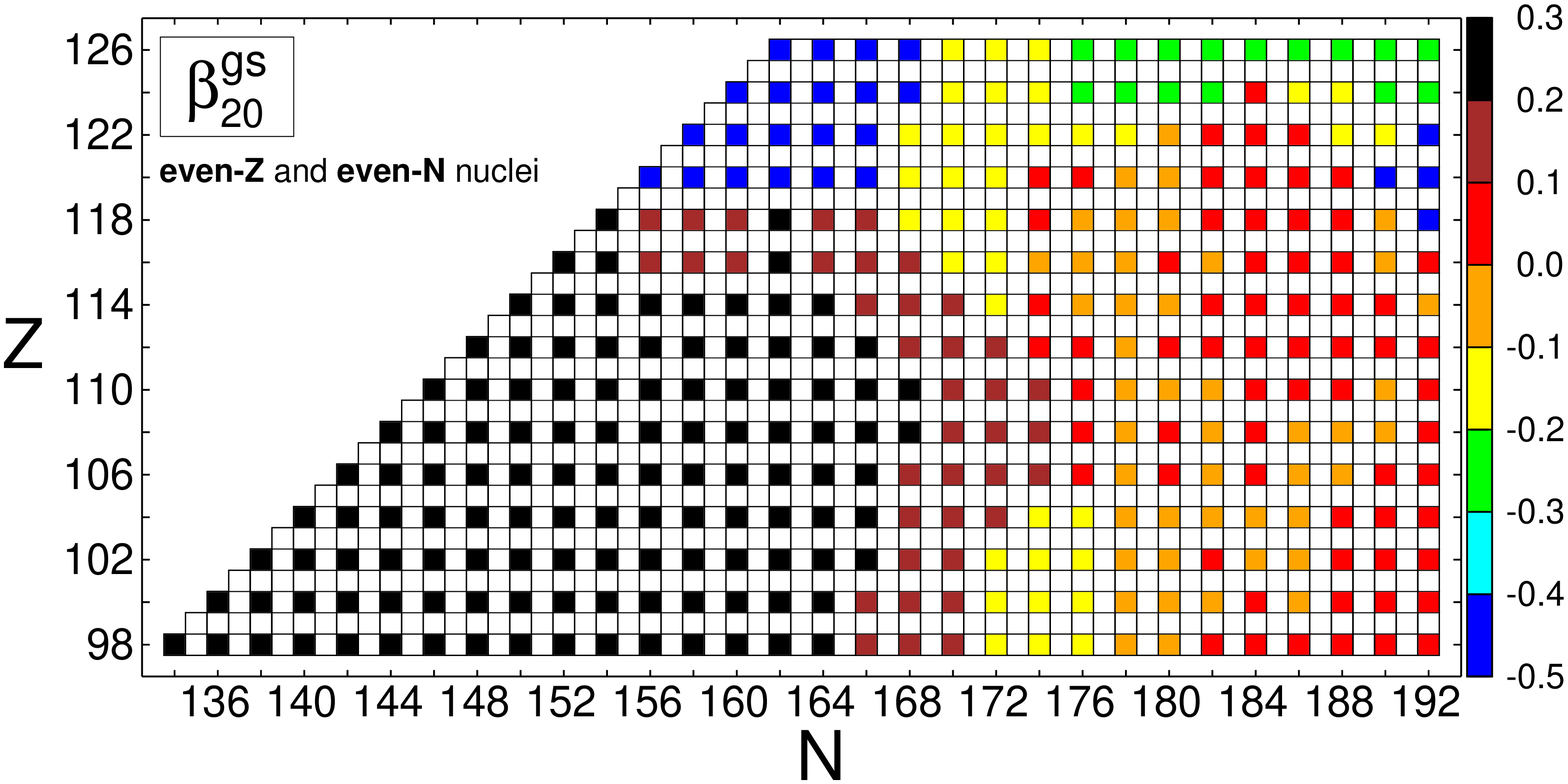}
\includegraphics[scale=0.26]{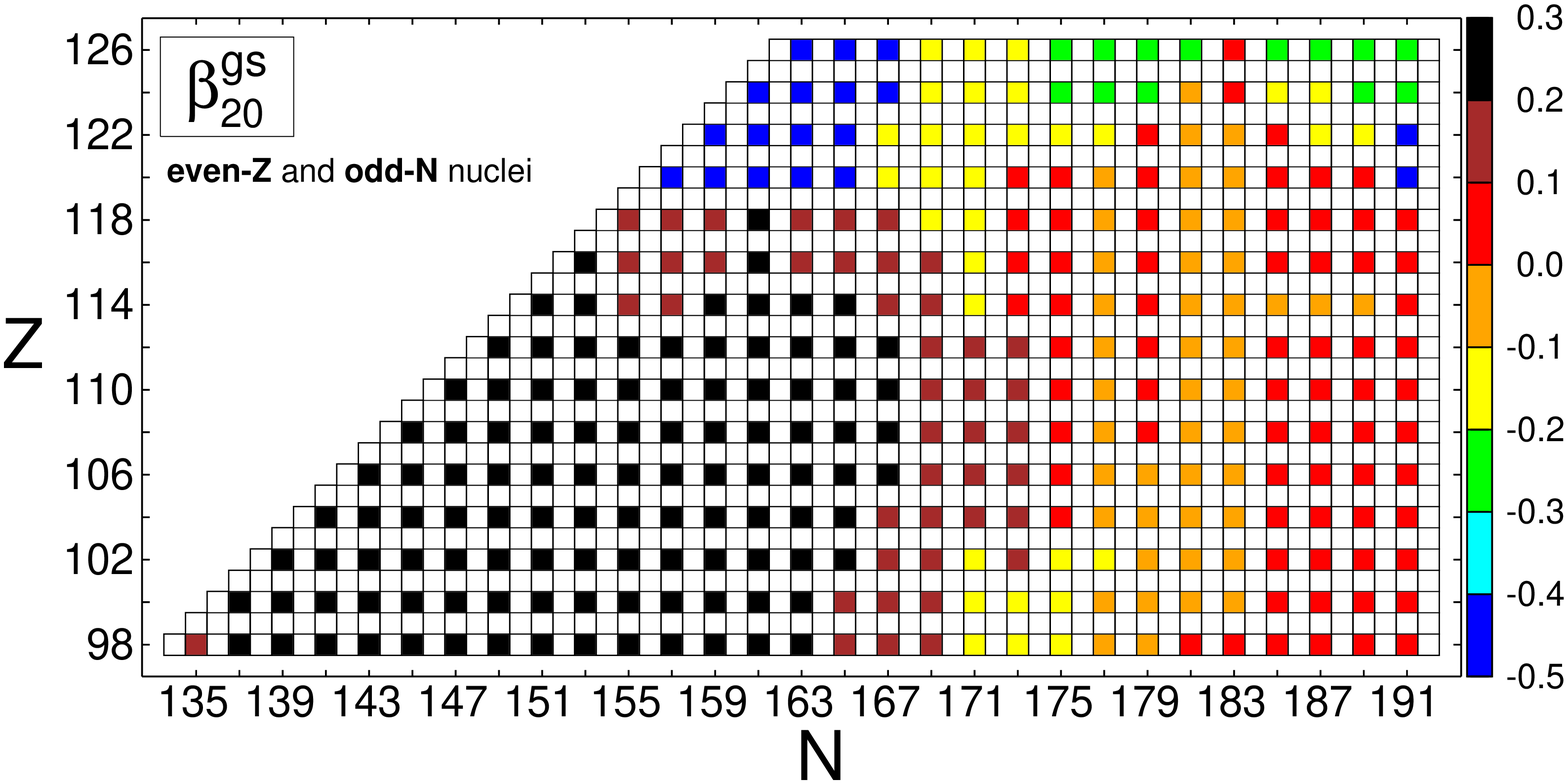}
\includegraphics[scale=0.26]{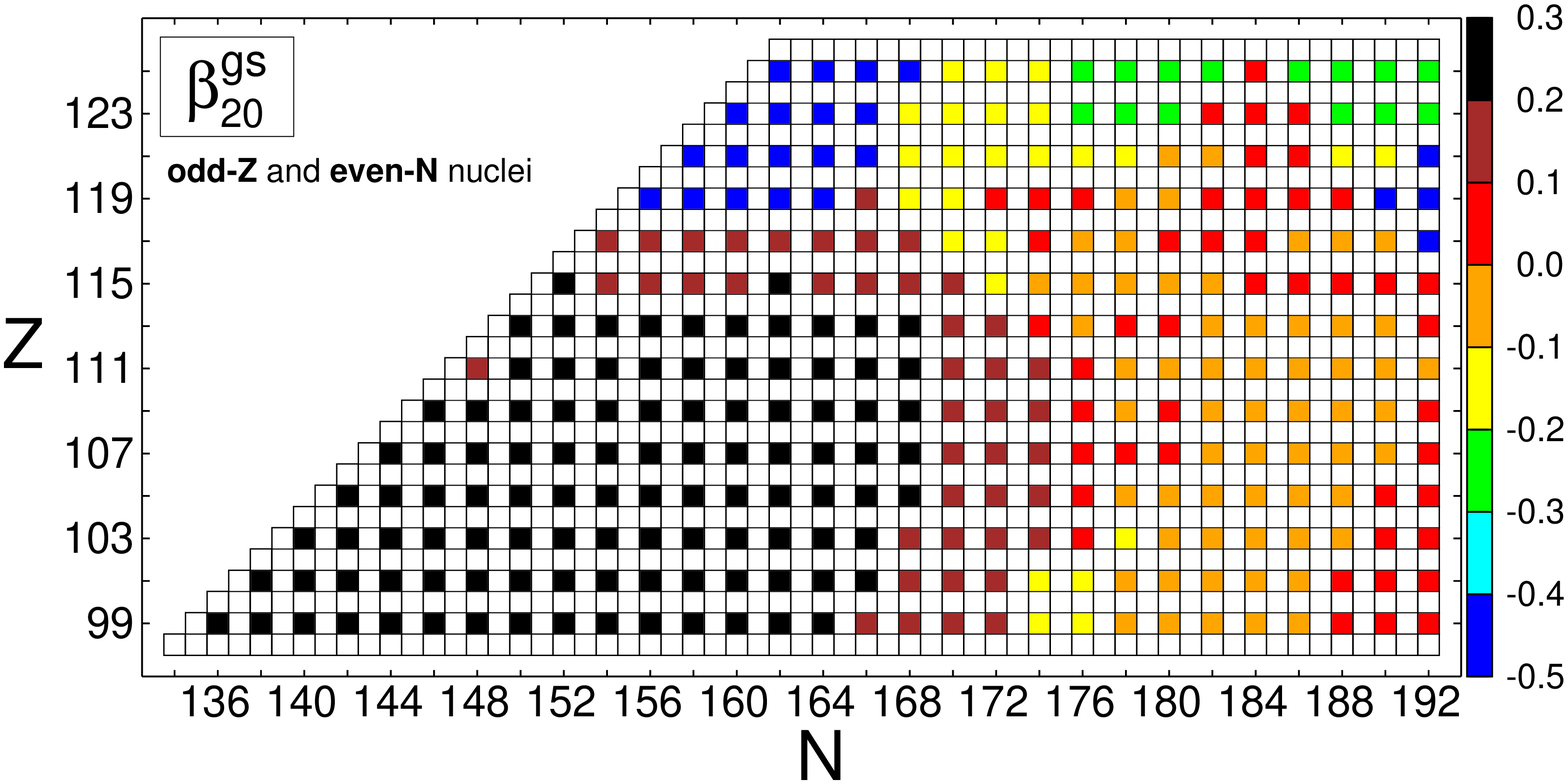}
\includegraphics[scale=0.26]{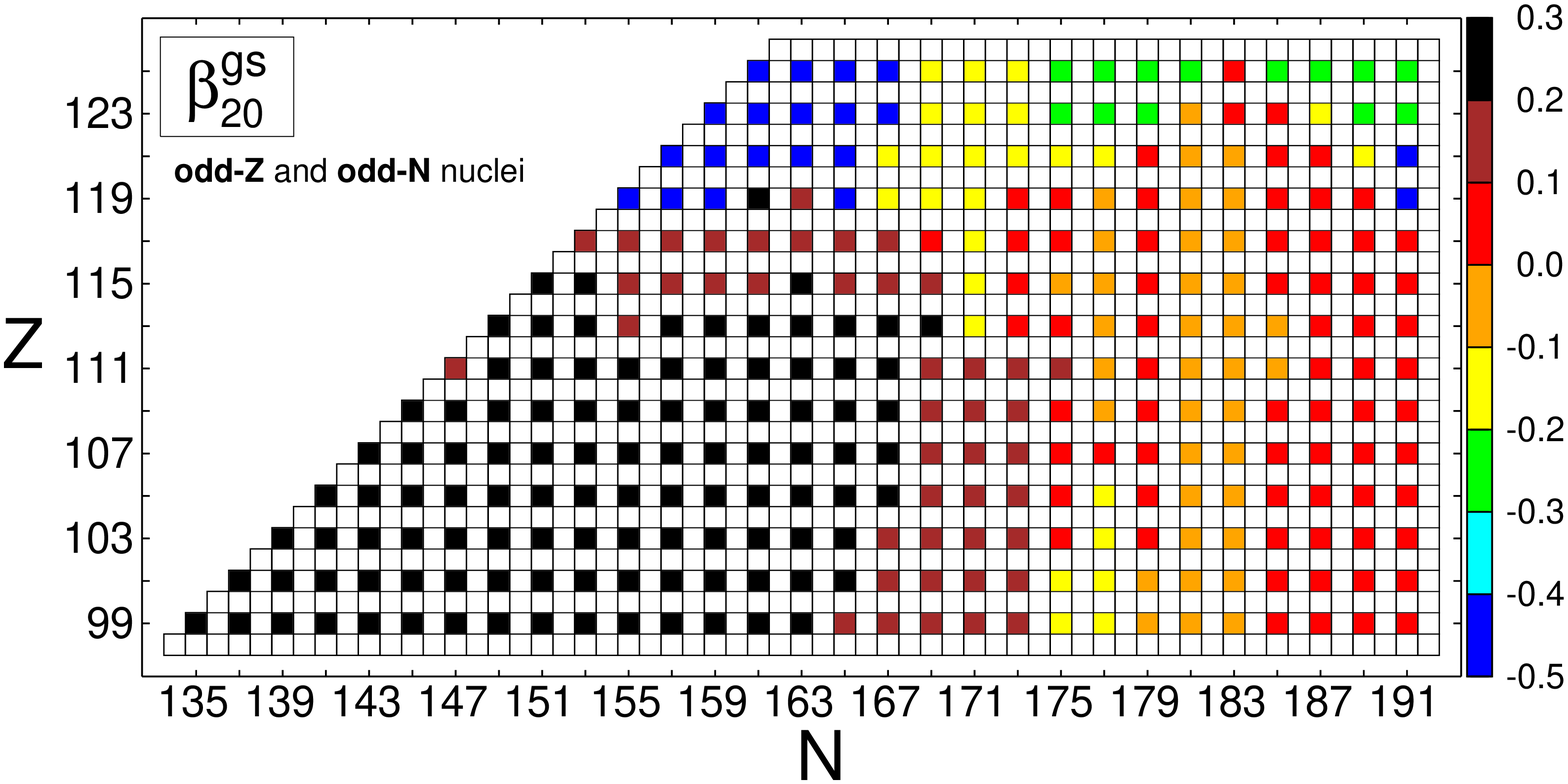}
\caption{Calculated ground-state quadrupole deformations $\beta^{gs}_{20}$
in 4 separate groups of nuclei.}
\label{fig4}
\end{figure}

The calculated ground-state equilibrium quadrupole deformations
$\beta^{gs}_{20}$ are shown in a $Z$ vs. $N$ plot in \mbox{Fig. \ref{fig4}}.
 Clearly visible is a vast region of prolate-deformed nuclei with
$Z \approx 98-112$ and $N\lesssim 164$, denoted by black squares.
The largest quadrupole deformations in this region
are close to $\beta^{gs}_{20} \approx 0.25$ and occur
 in the vicinity of $^{254}$No. With increasing neutron number, deformation
$\beta^{gs}_{20}$ decreases, ground states become gradually weakly deformed
 and then spherical, as the predicted spherical magic number $N=184$ is
 approached.
 According to our results, almost all of the heaviest elements
 $Z=123-126$ prefer oblate deformations in their ground state.

 As an illustration, in \mbox{Fig. \ref{fig5}} we present a sample of energy
 landscapes in ($\beta_{20}$,$\beta_{22}$) plane,
 obtained from the five - dimesional deformation space
 (\ref{deformation_space}, \ref{basicgrid}) by the minimization over the
 remaining
  $\beta_{40}, \beta_{60}, \beta_{80}$. Included are maps for selected
 experimentally known and hypothetical [denoted (h)] nuclei:
 $^{245}$Md, $^{252}$Md, $^{259}$Md,
  $^{256}$Rf, $^{265}$Rf, $^{270}$Rf (h),
  $^{266}$Mt, $^{270}$Mt, $^{277}$Mt,
  $^{280}$Fl (h), $^{286}$Fl, $^{292}$Fl (h),
    $^{294}$Og, $^{298}$Og (h),
    $^{296}$124 (h), $^{304}$124 (h).
 Except for the last two, they show trends within or close to the
 area of known isotopes. There, Md and Rf nuclei
 are well deformed, in Mt isotopes deformation is smaller, it becomes small
 and $\gamma$-soft or vanishes in Fl and Og, while the isotopes of $Z=124$ show
 oblate ground states. Besides minima, the maps in \mbox{Fig. \ref{fig5}}
 also show fission saddles and allow to appreciate fission barrier heights
 which will be discussed in the next section.


\begin{figure}[h]
\centering
\includegraphics[scale=0.45]{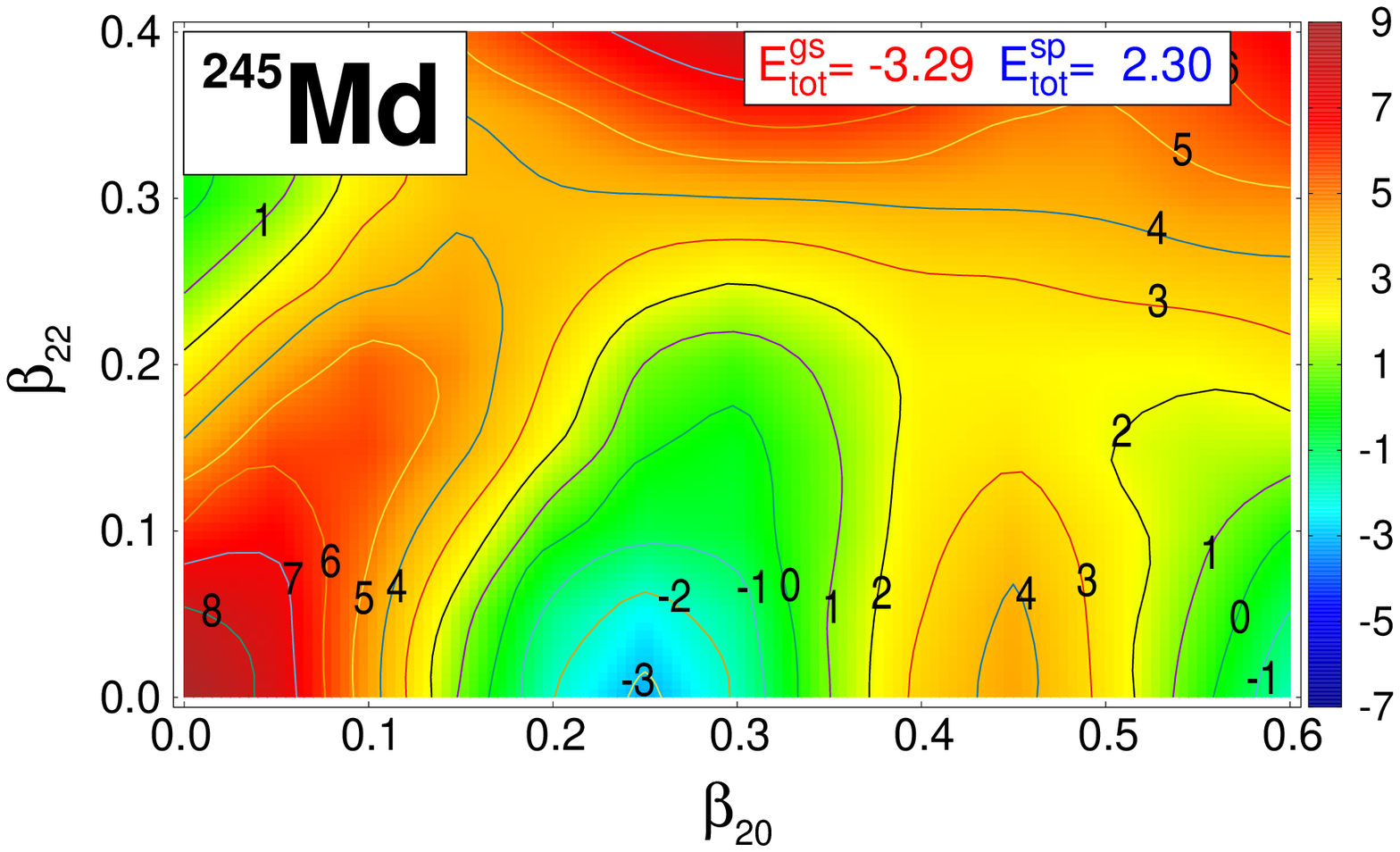}
\includegraphics[scale=0.45]{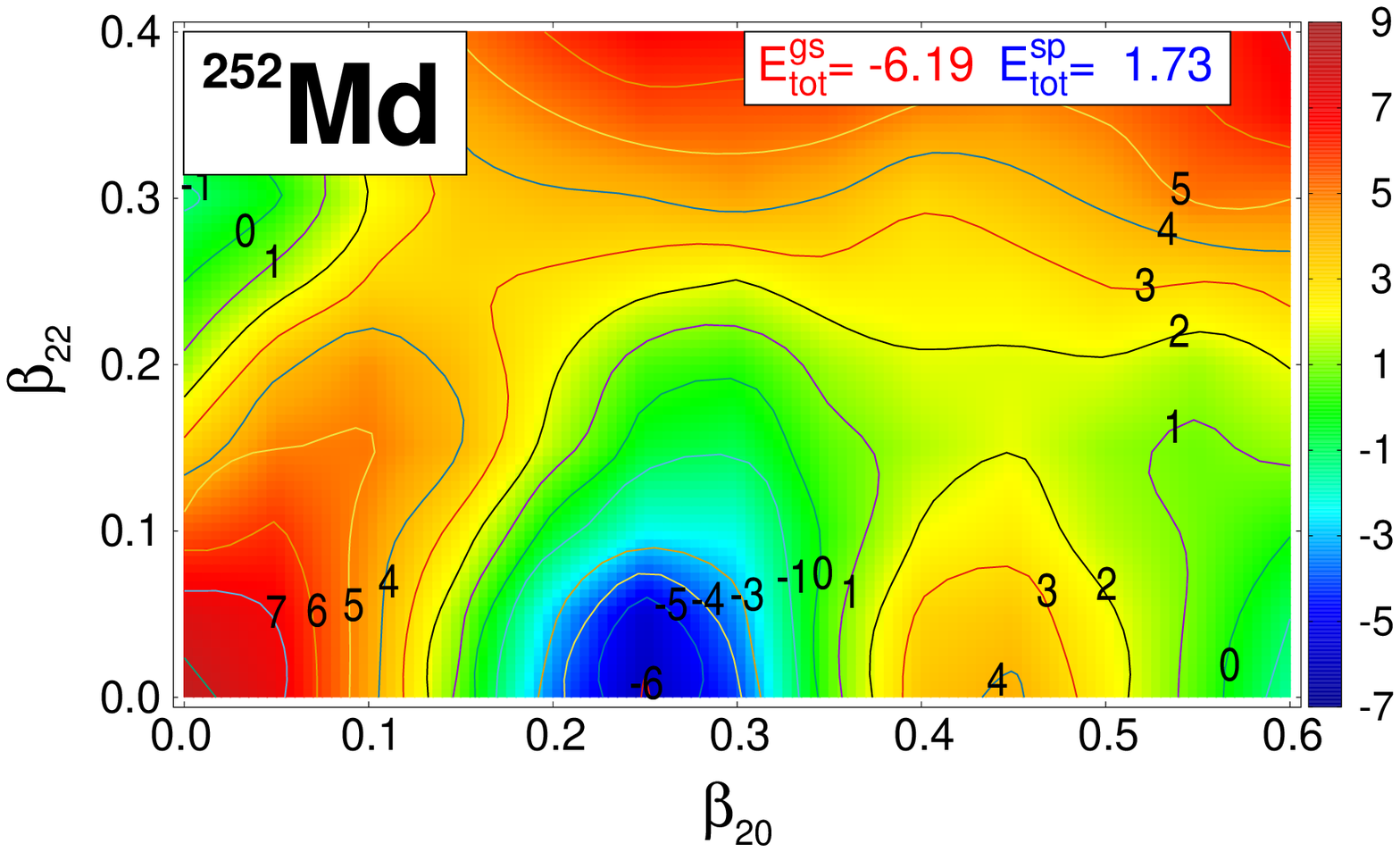}
\includegraphics[scale=0.45]{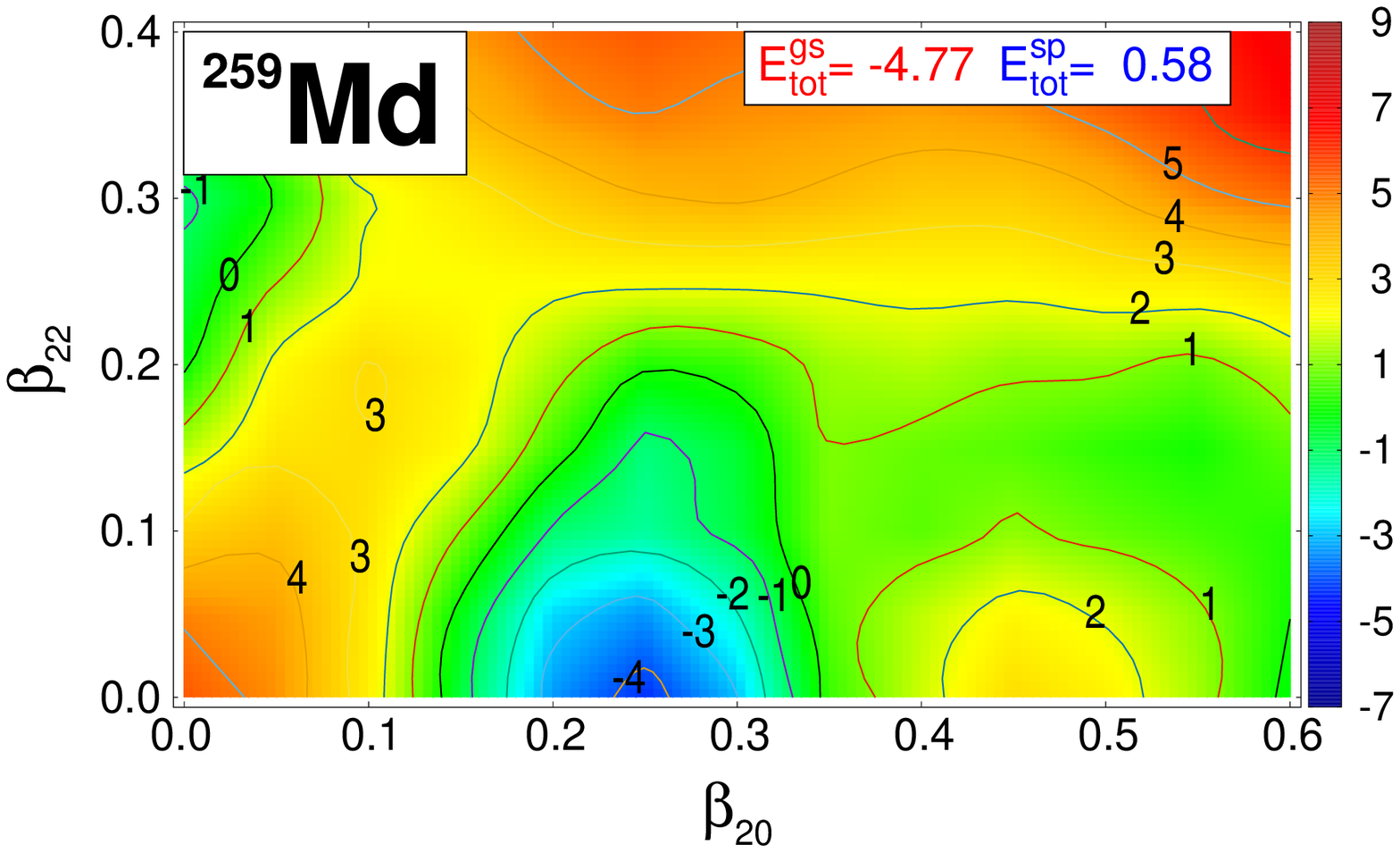}
\includegraphics[scale=0.45]{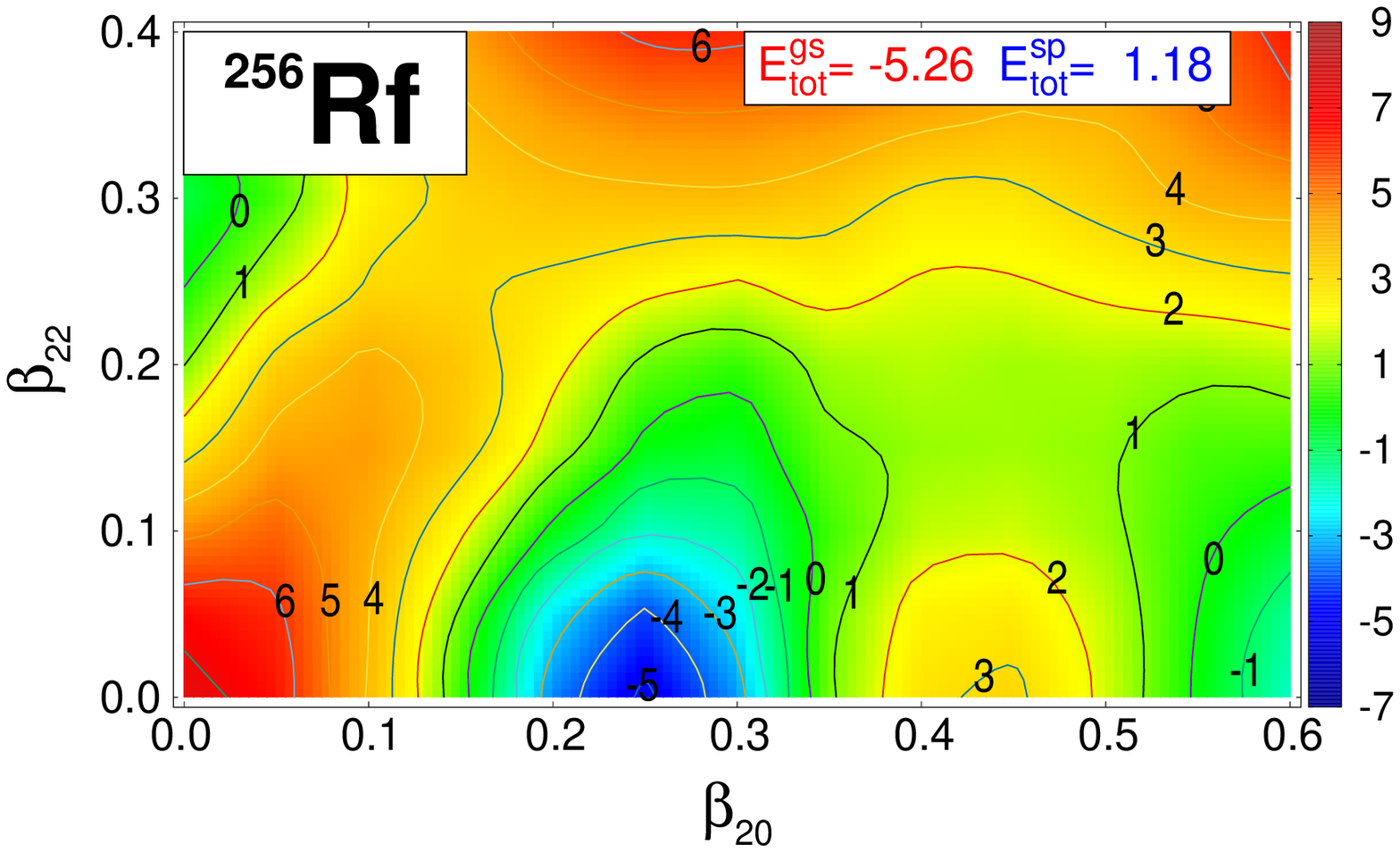}
\includegraphics[scale=0.45]{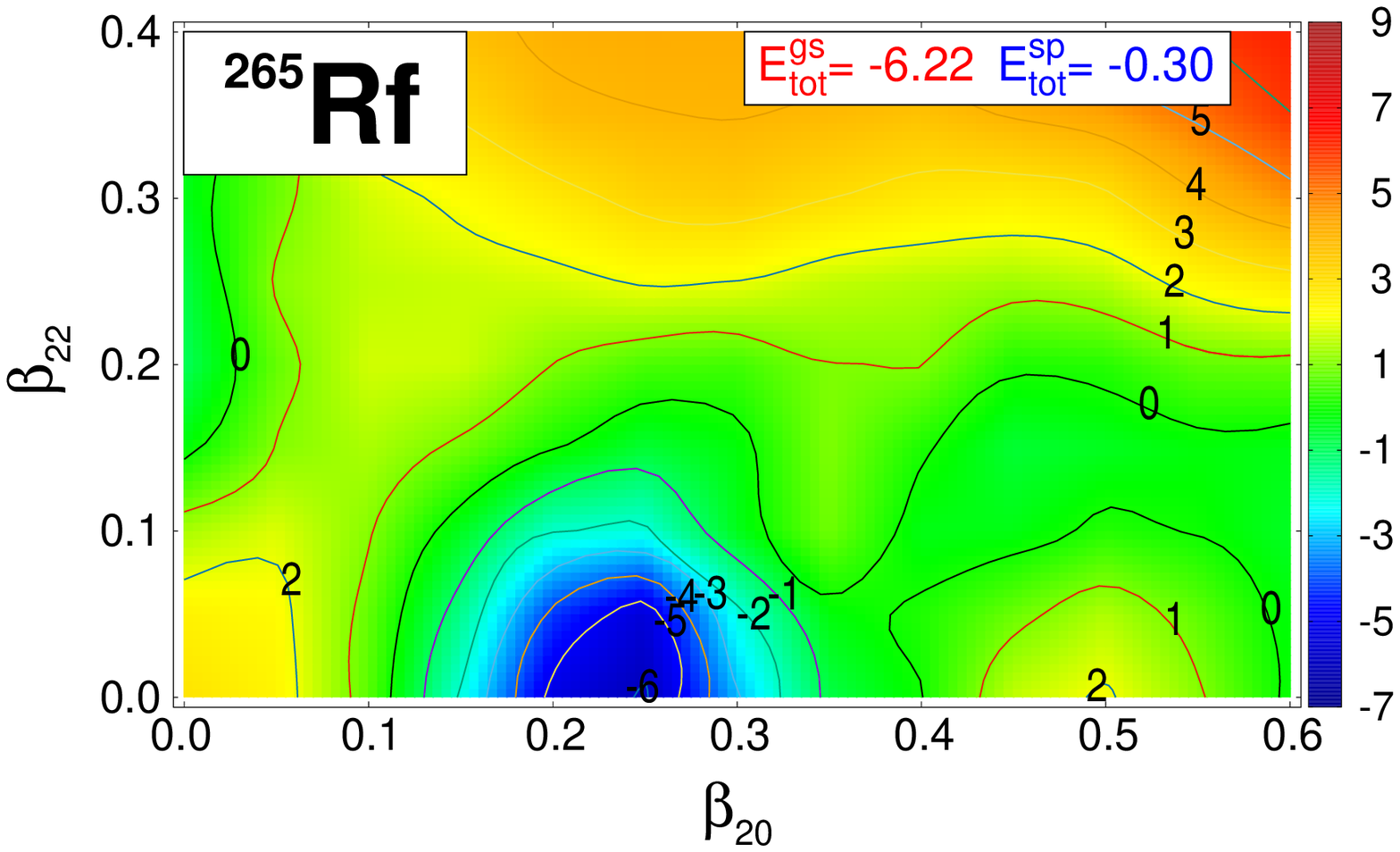}
\includegraphics[scale=0.45]{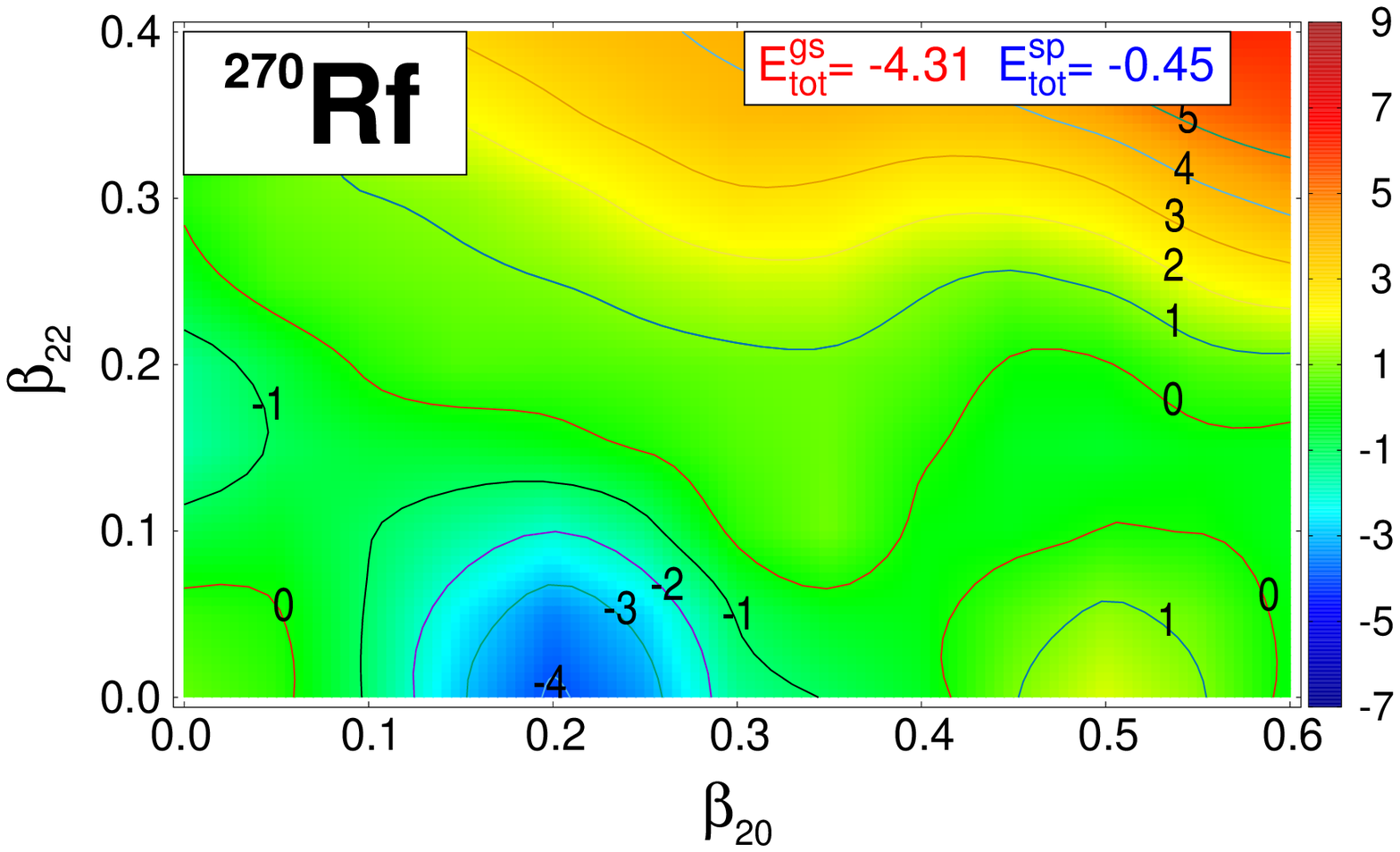}
\includegraphics[scale=0.45]{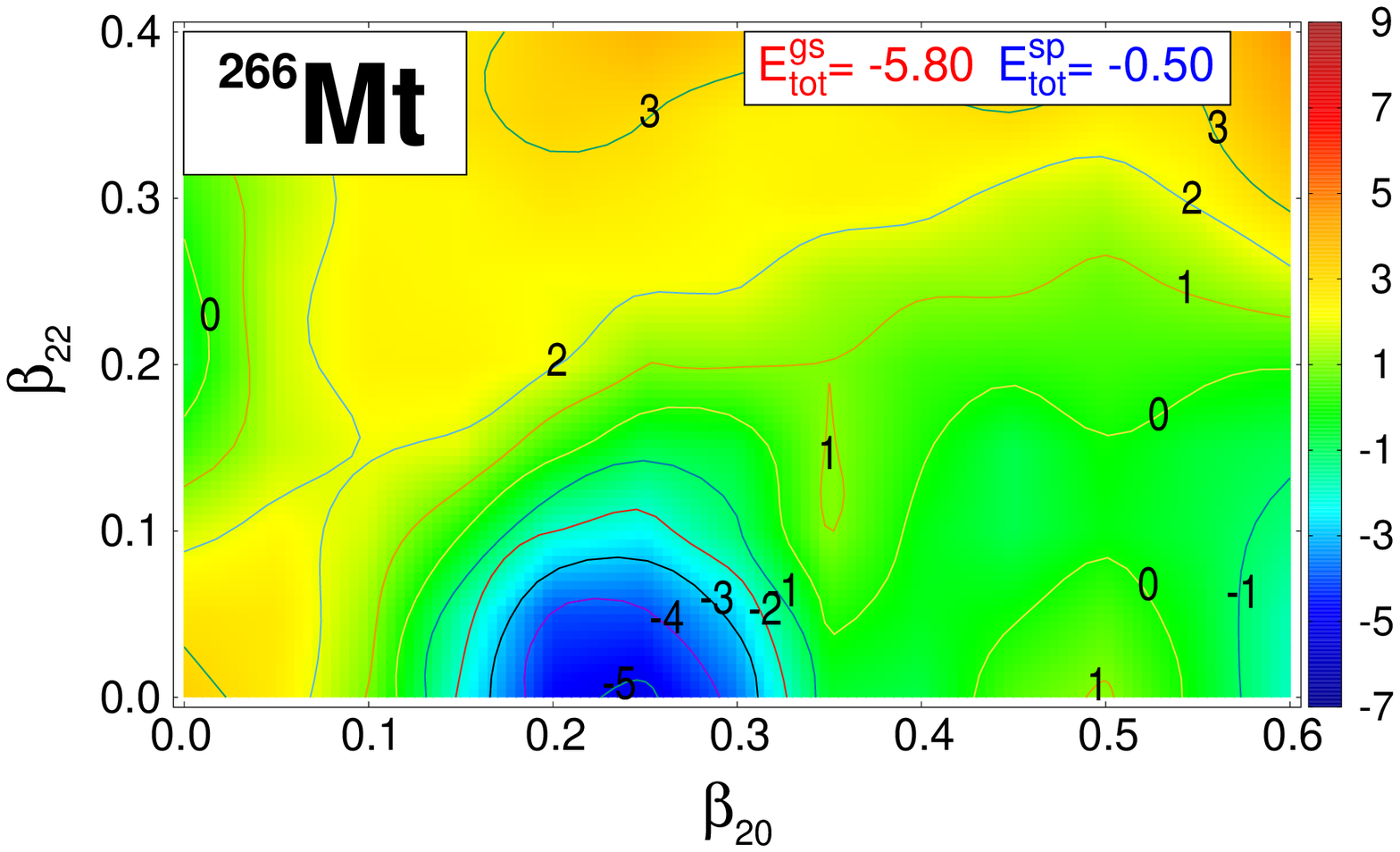}
\includegraphics[scale=0.45]{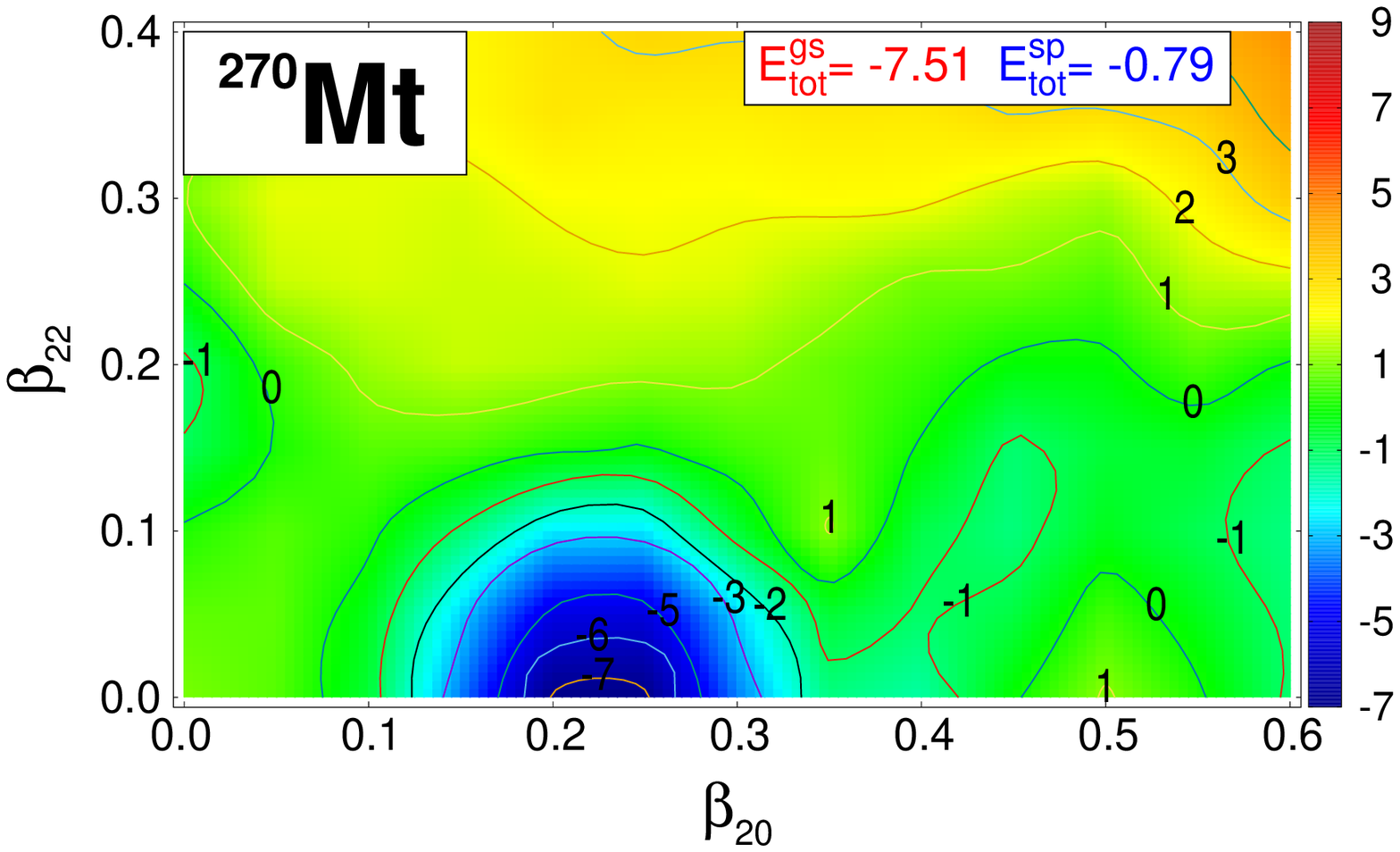}
\end{figure}
\begin{figure}[h]
\centering
\includegraphics[scale=0.45]{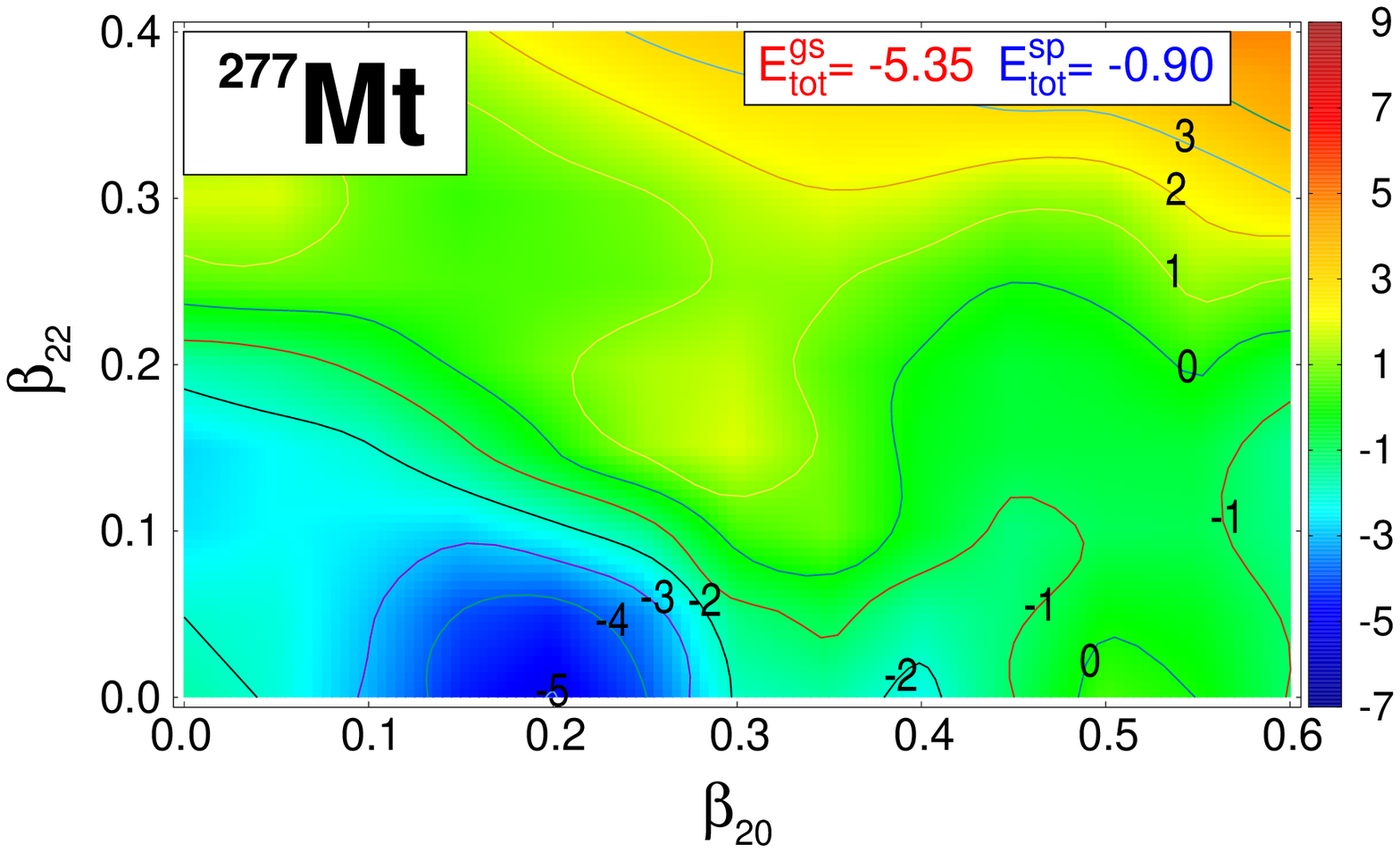}
\includegraphics[scale=0.45]{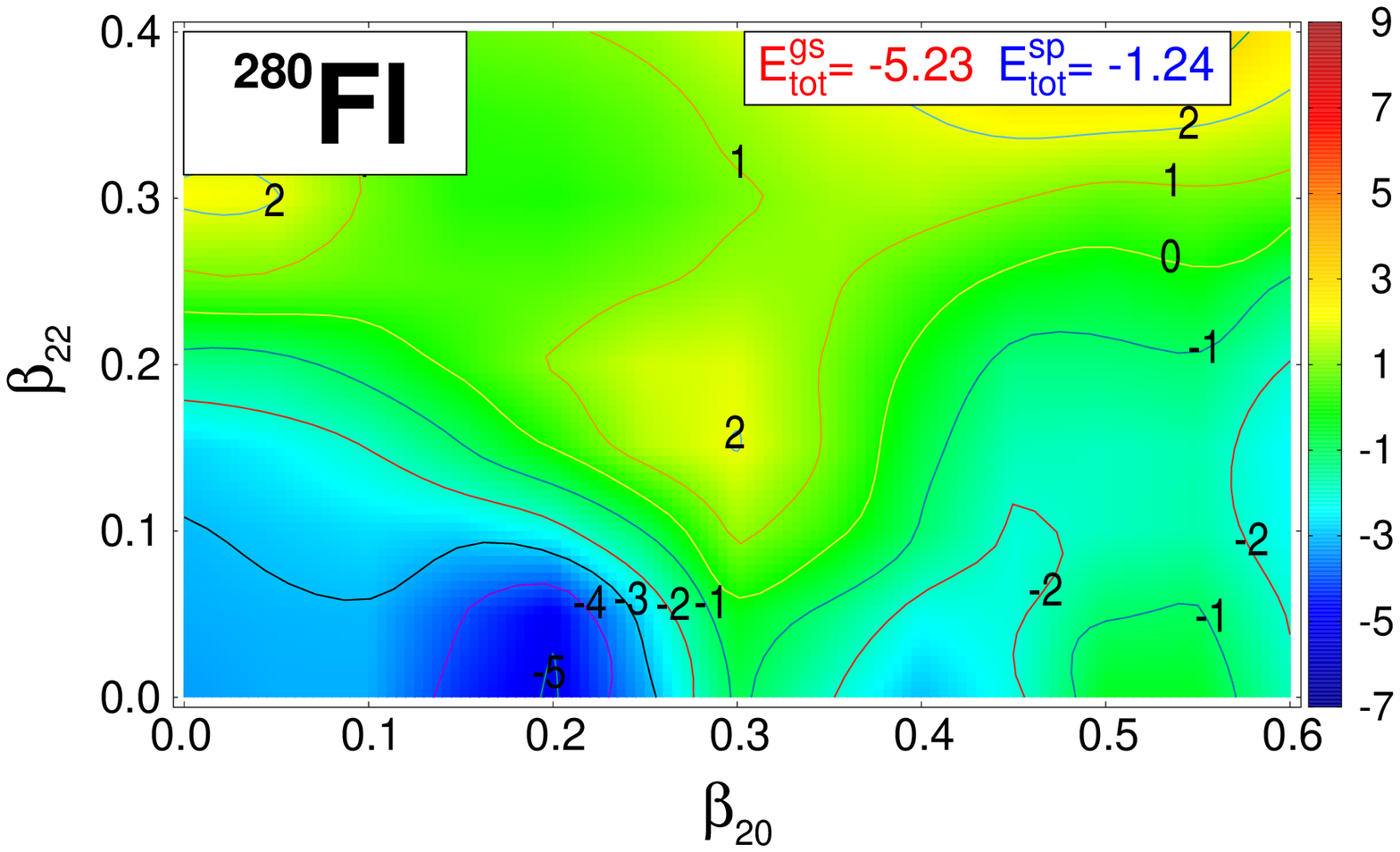}
\includegraphics[scale=0.45]{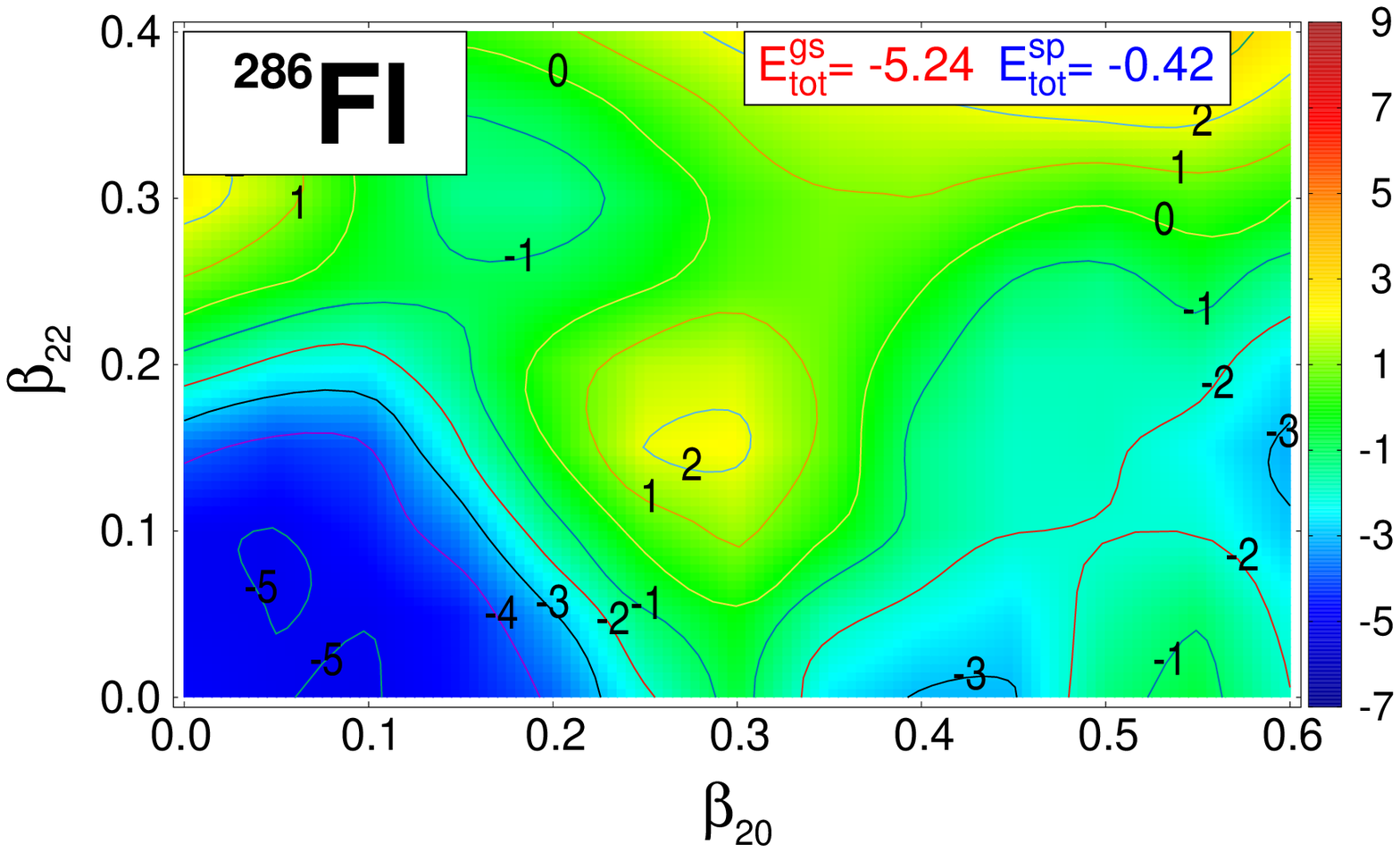}
\includegraphics[scale=0.45]{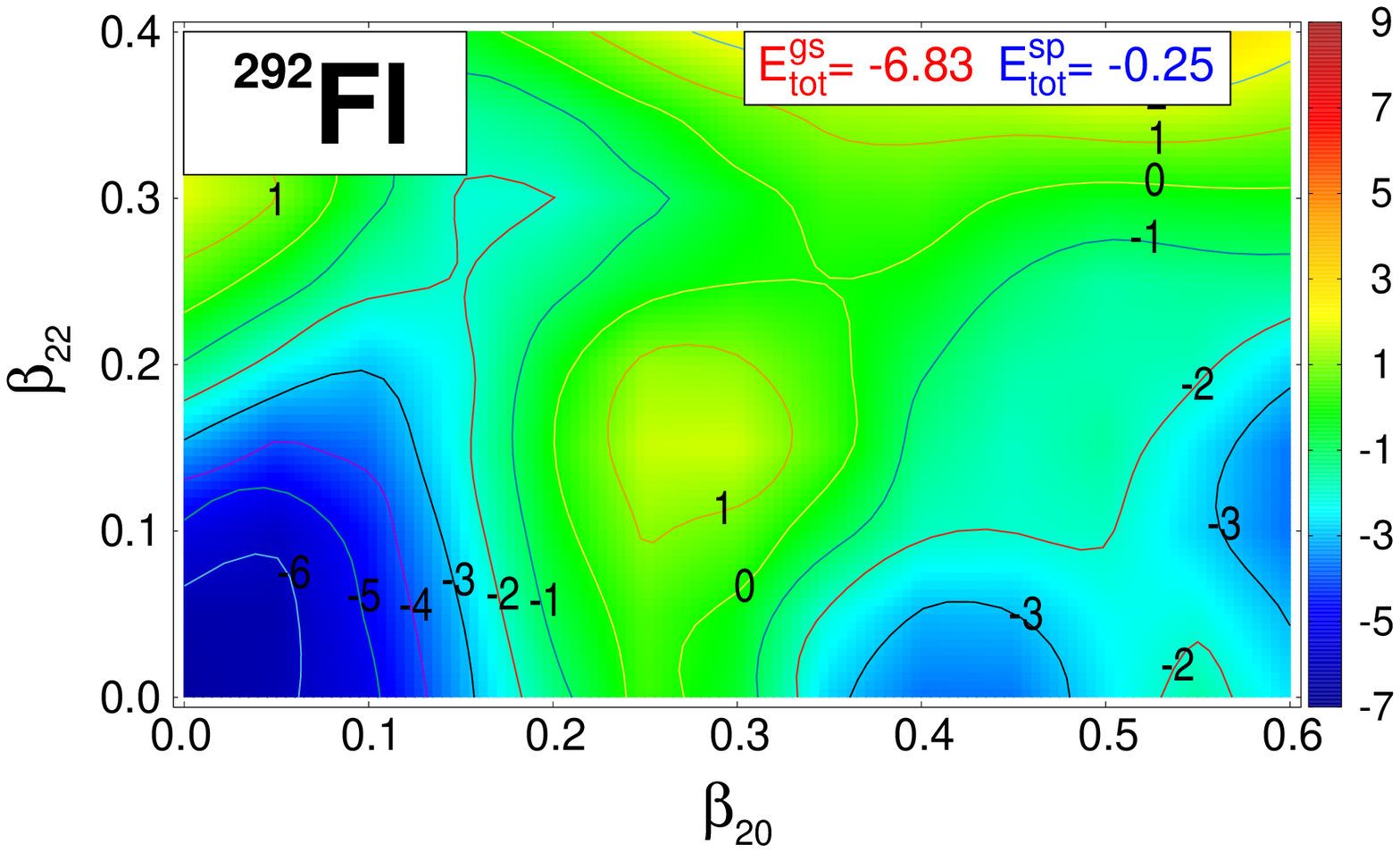}
\includegraphics[scale=0.45]{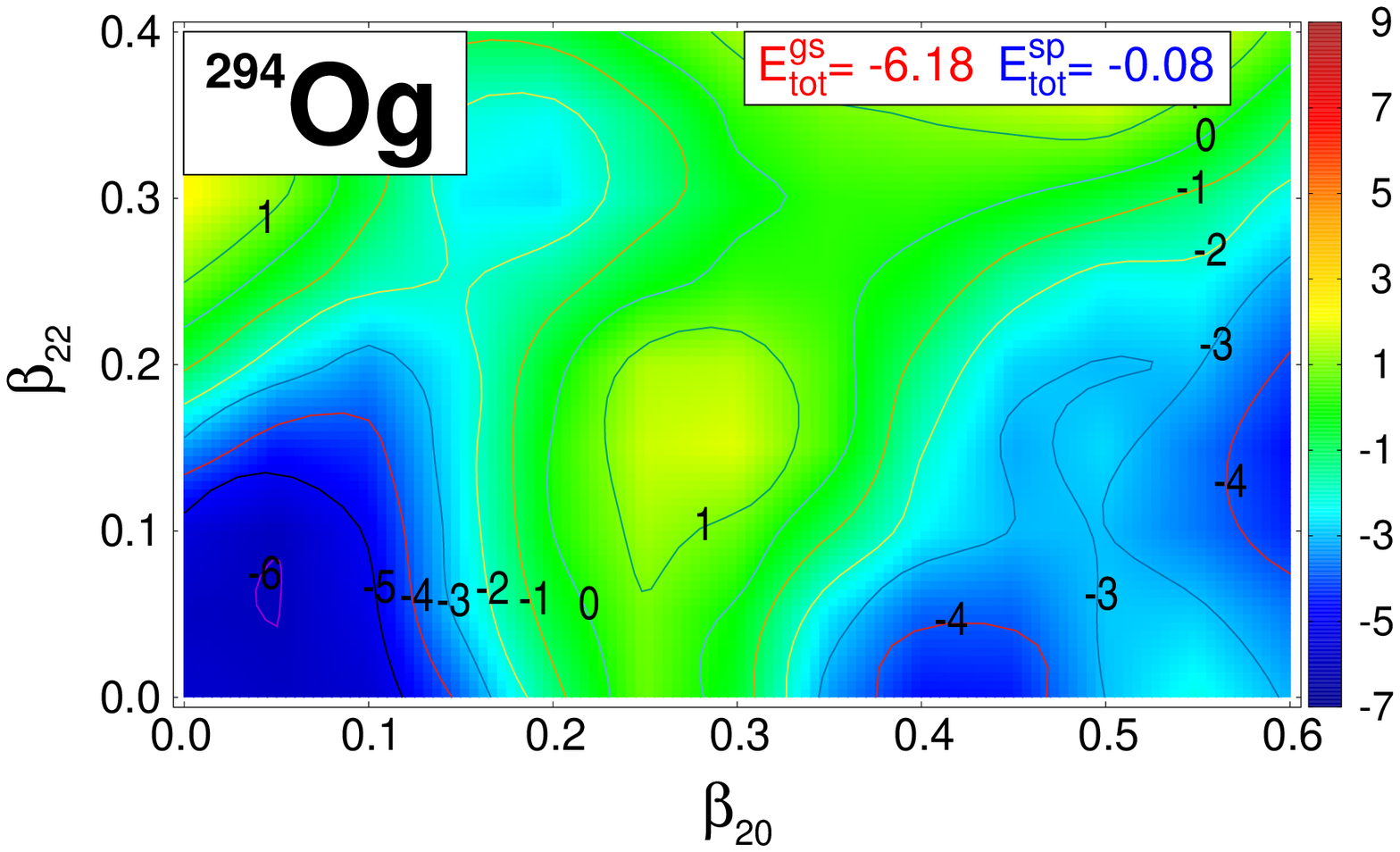}
\includegraphics[scale=0.45]{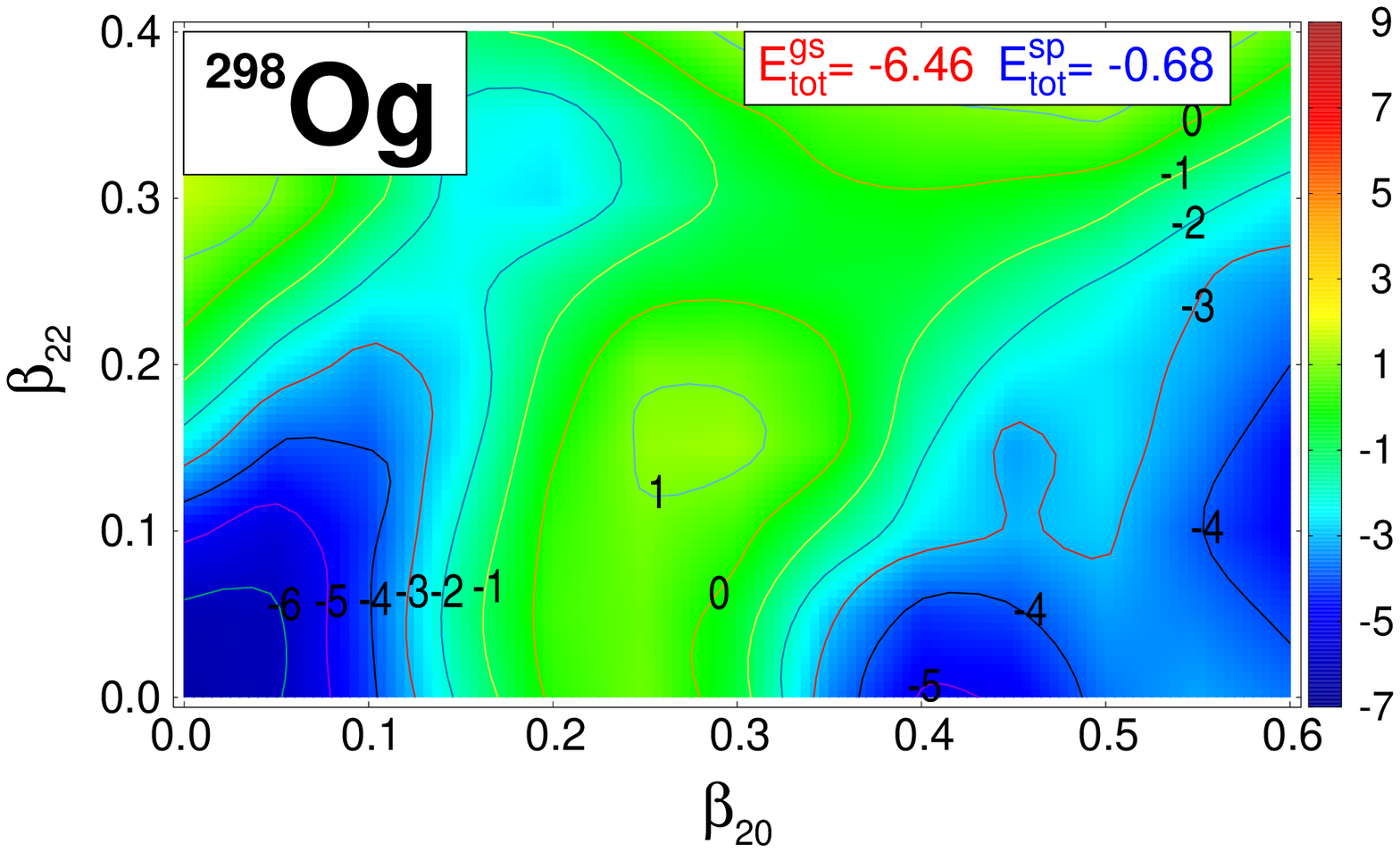}
\includegraphics[scale=0.45]{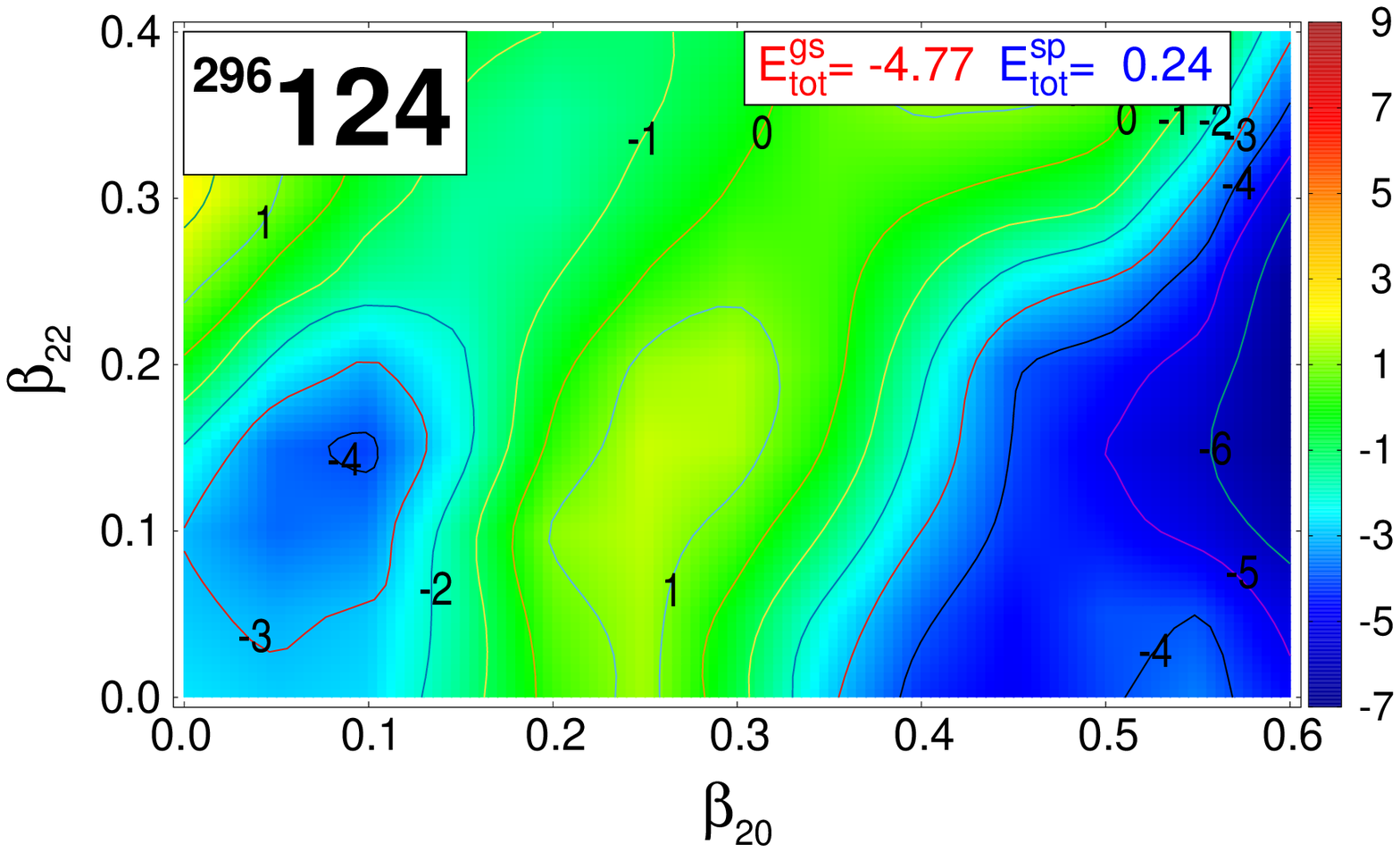}
\includegraphics[scale=0.45]{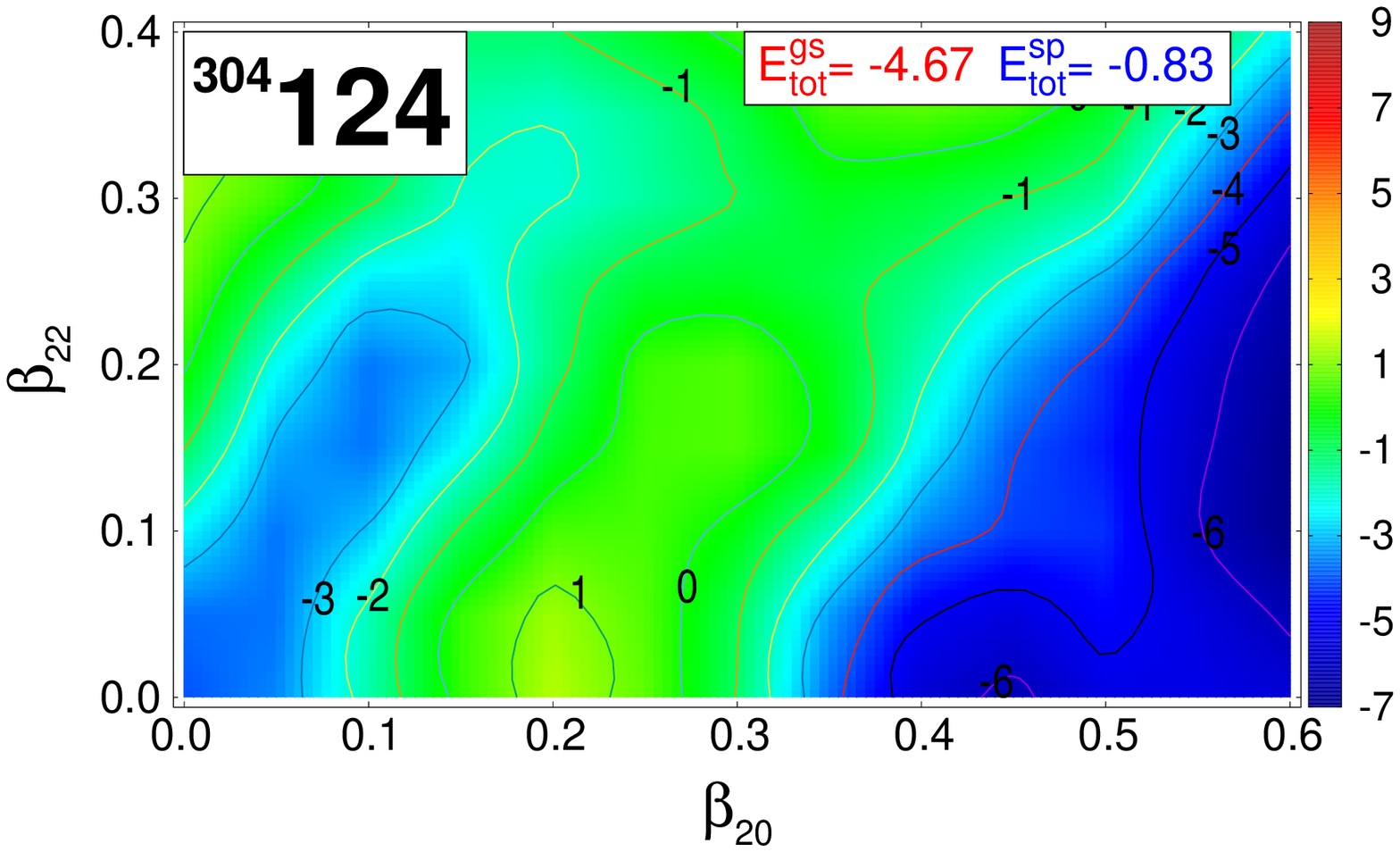}
\caption{Examples of potential energy landscapes in $\beta_{20}, \beta_{22}$
 plane from the minimization over the remaining three
 deformation parameters in (\ref{deformation_space}): $\beta_{40}, \beta_{60},
 \beta_{80}$.
Visible energy $E_{tot}$ (in MeV) is calculated relative to the macroscopic energy at the spherical shape:
\mbox{$E_{mac}=E_{mac}(deformation)-E_{mac}(sphere)$}.
The red cross on a given map indicates the location of the ground state,
 the blue one - of the saddle point.}
\label{fig5}
\end{figure}

\begin{figure}[h]
\centering
\includegraphics[scale=0.45]{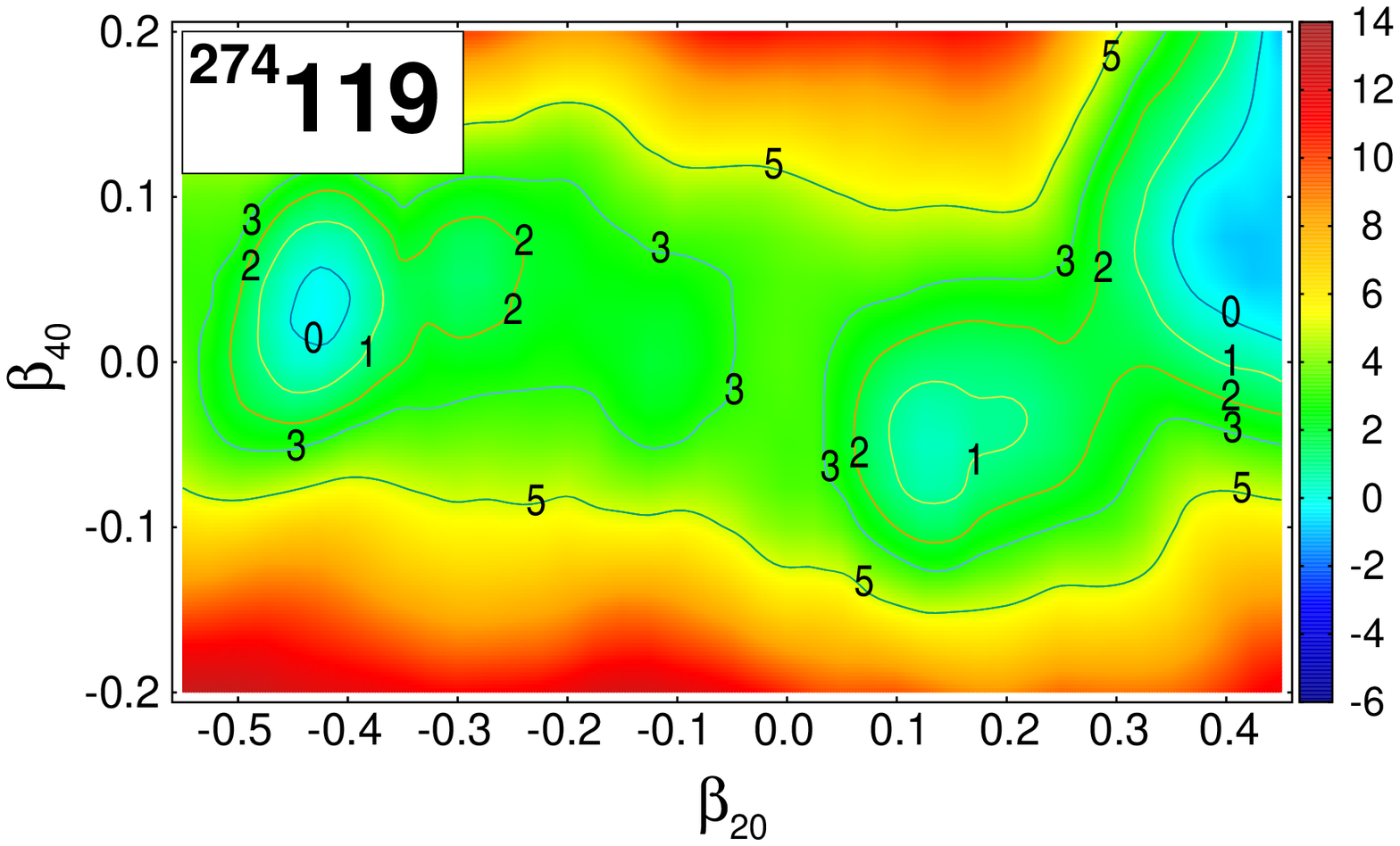}
\includegraphics[scale=0.45]{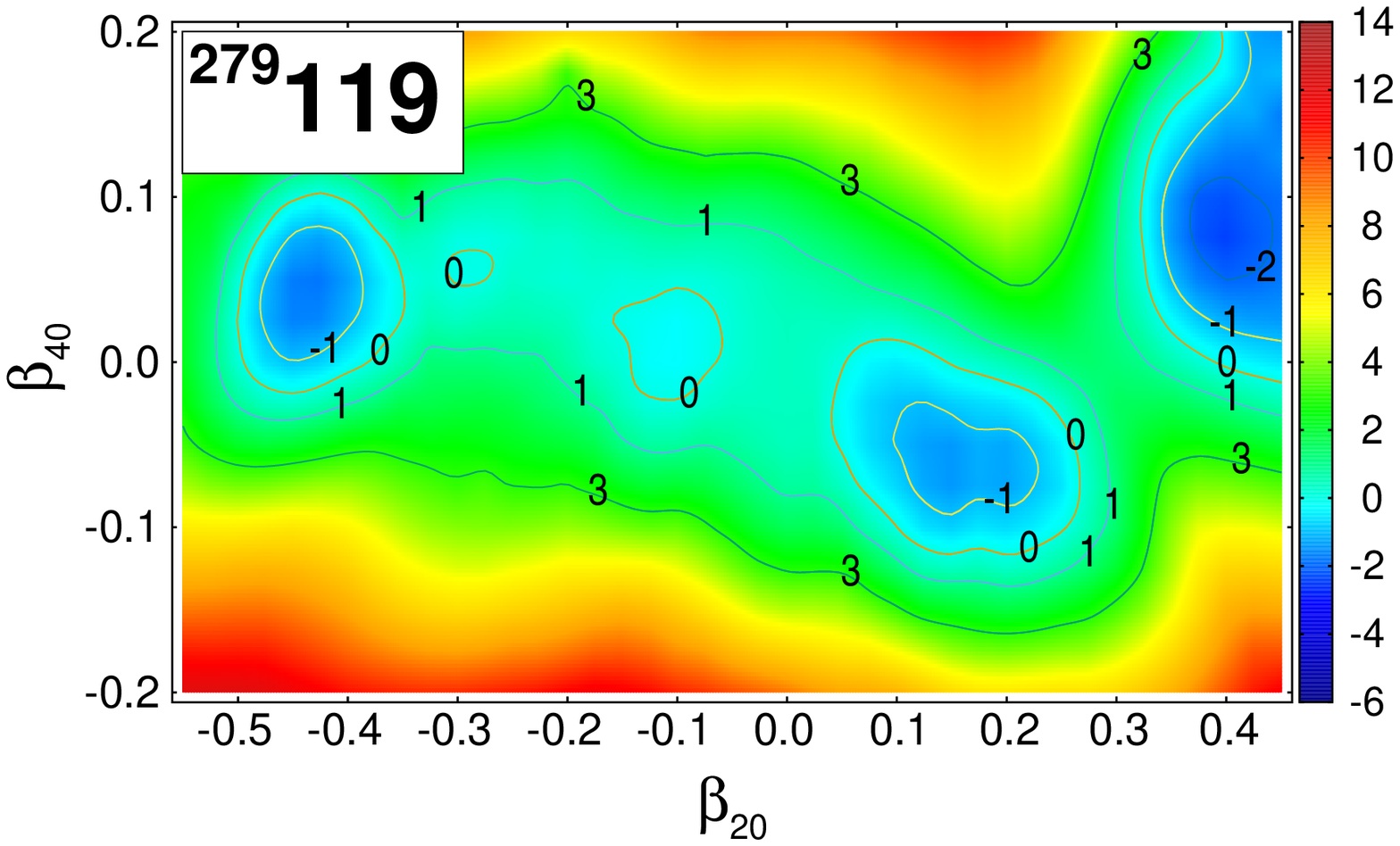}
\includegraphics[scale=0.45]{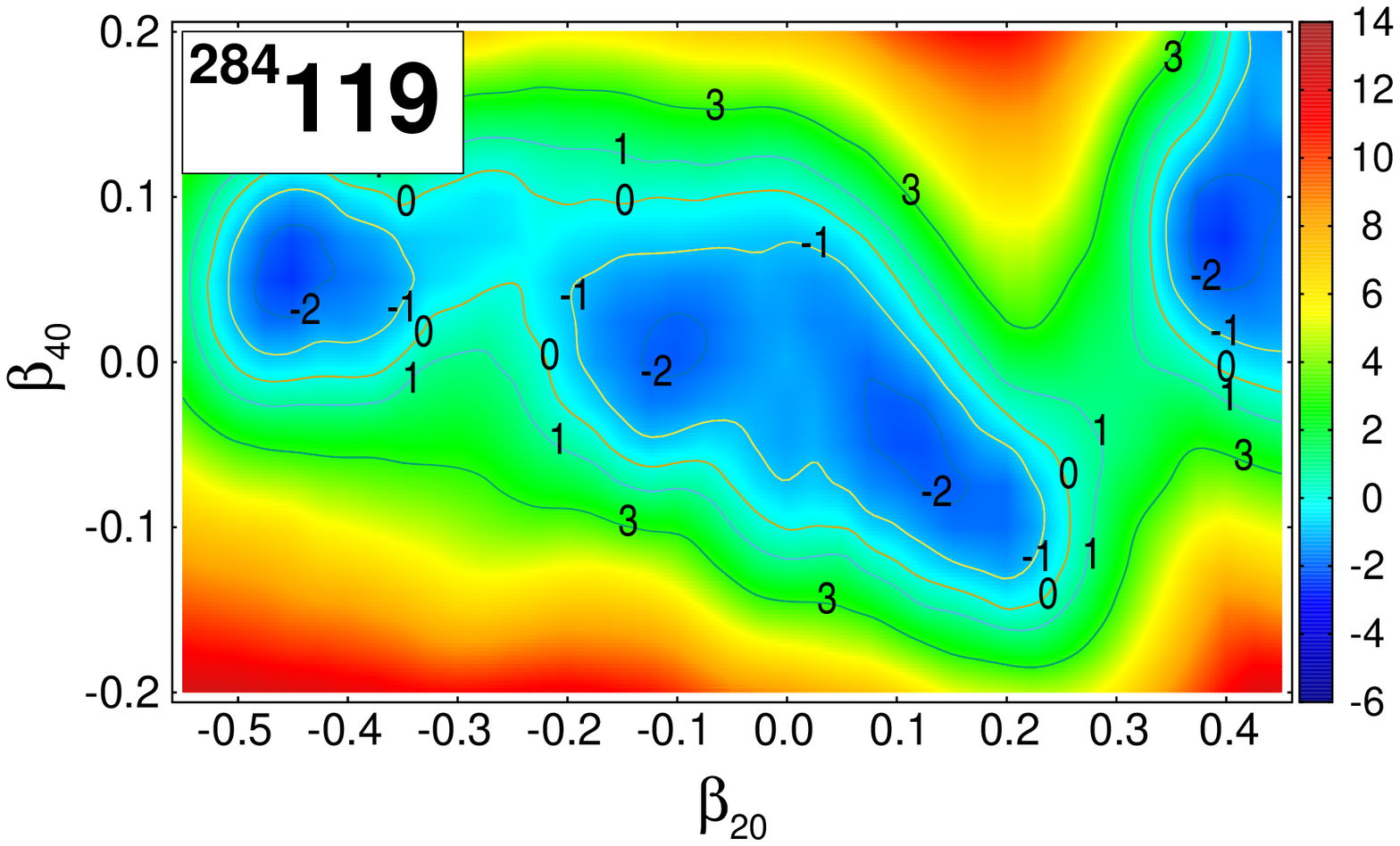}
\includegraphics[scale=0.45]{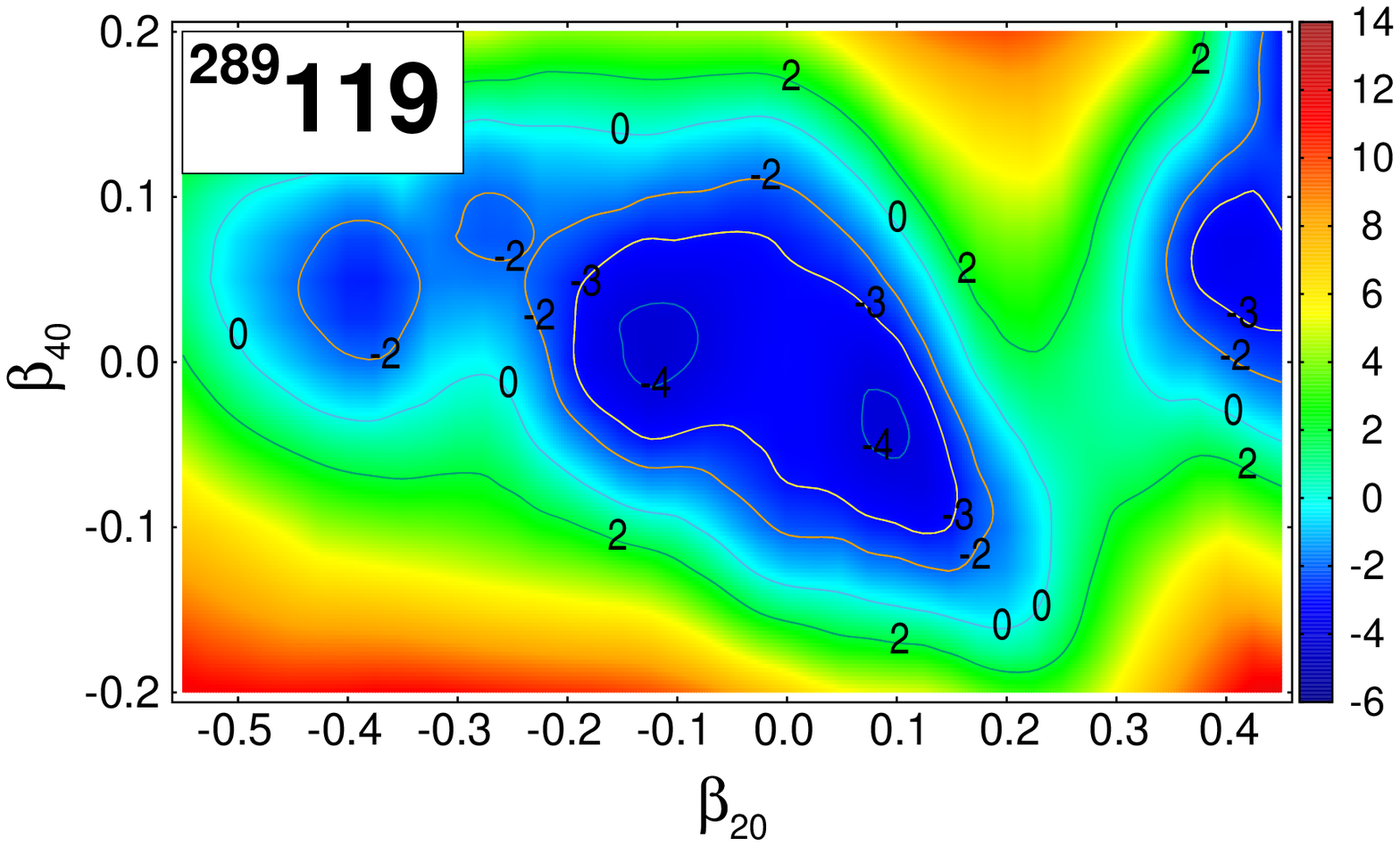}
\includegraphics[scale=0.45]{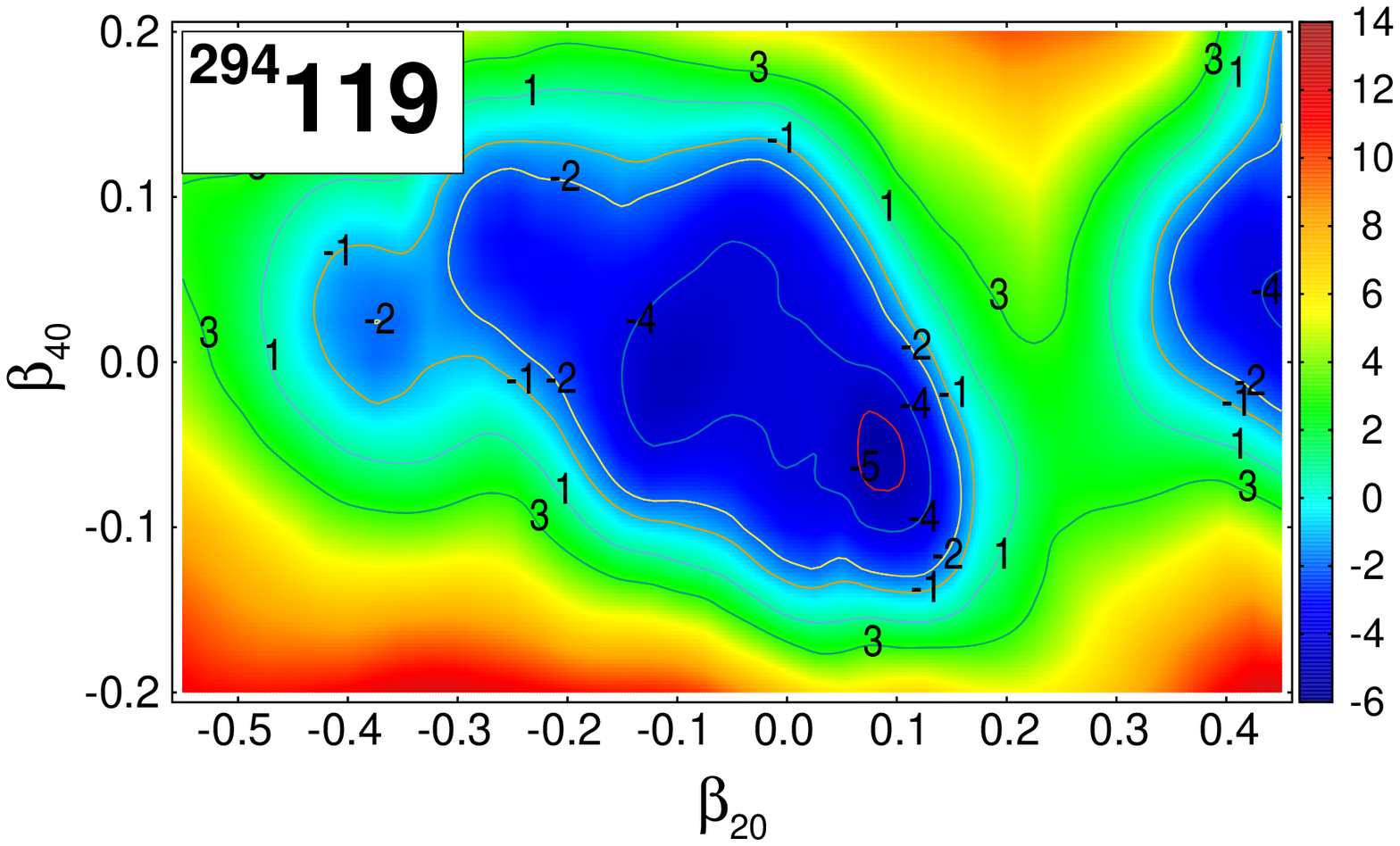}
\includegraphics[scale=0.45]{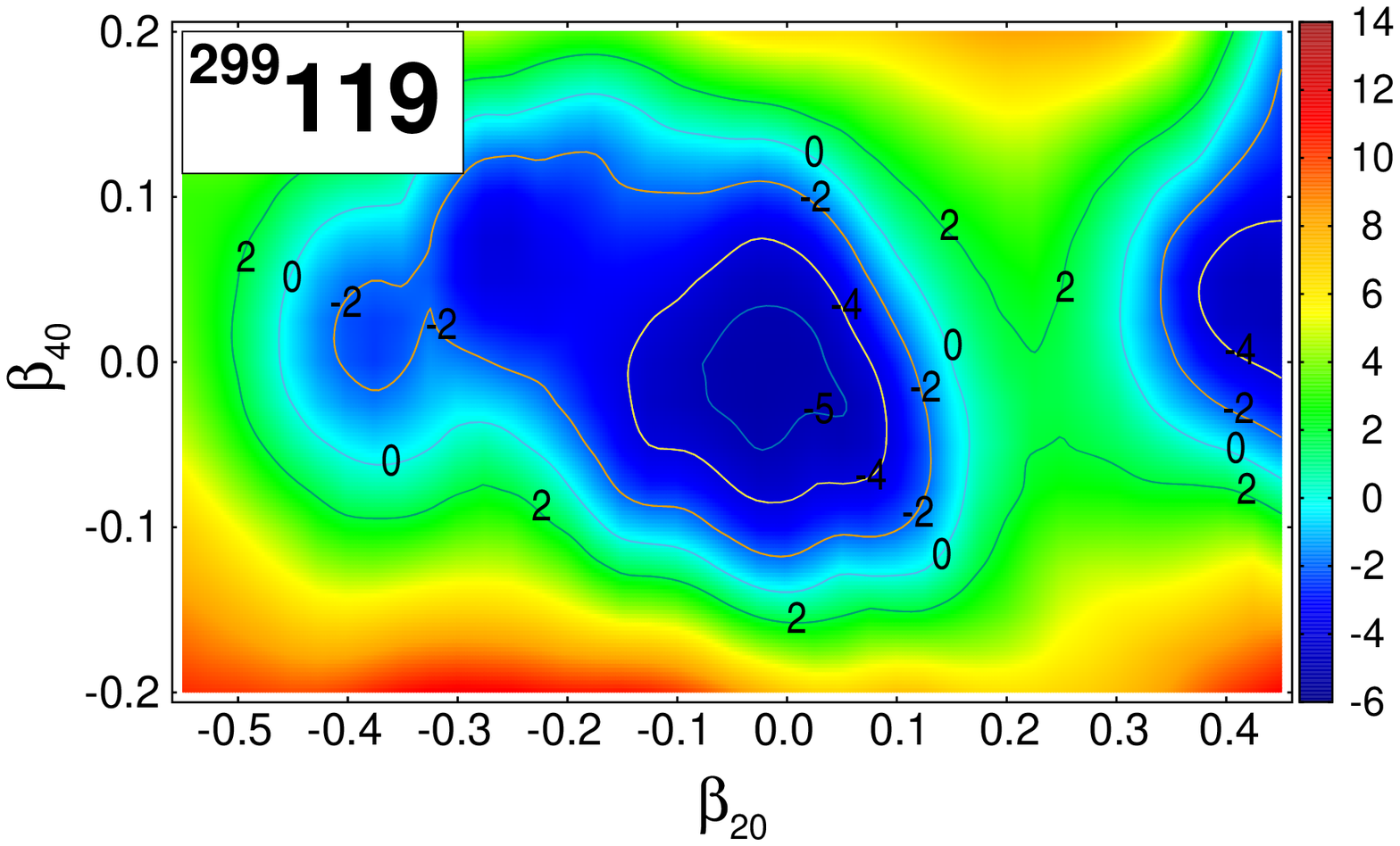}
\includegraphics[scale=0.45]{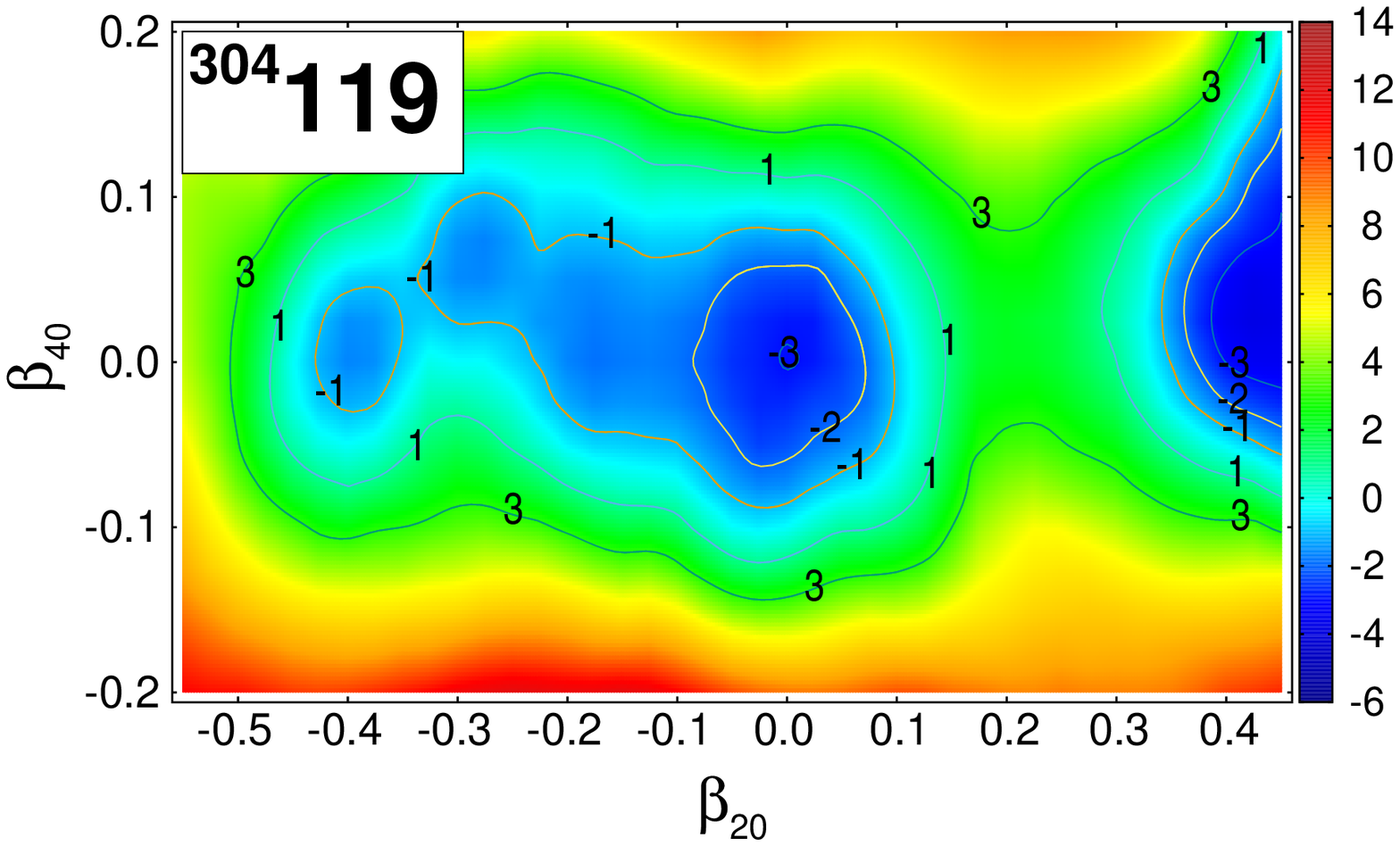}
\includegraphics[scale=0.45]{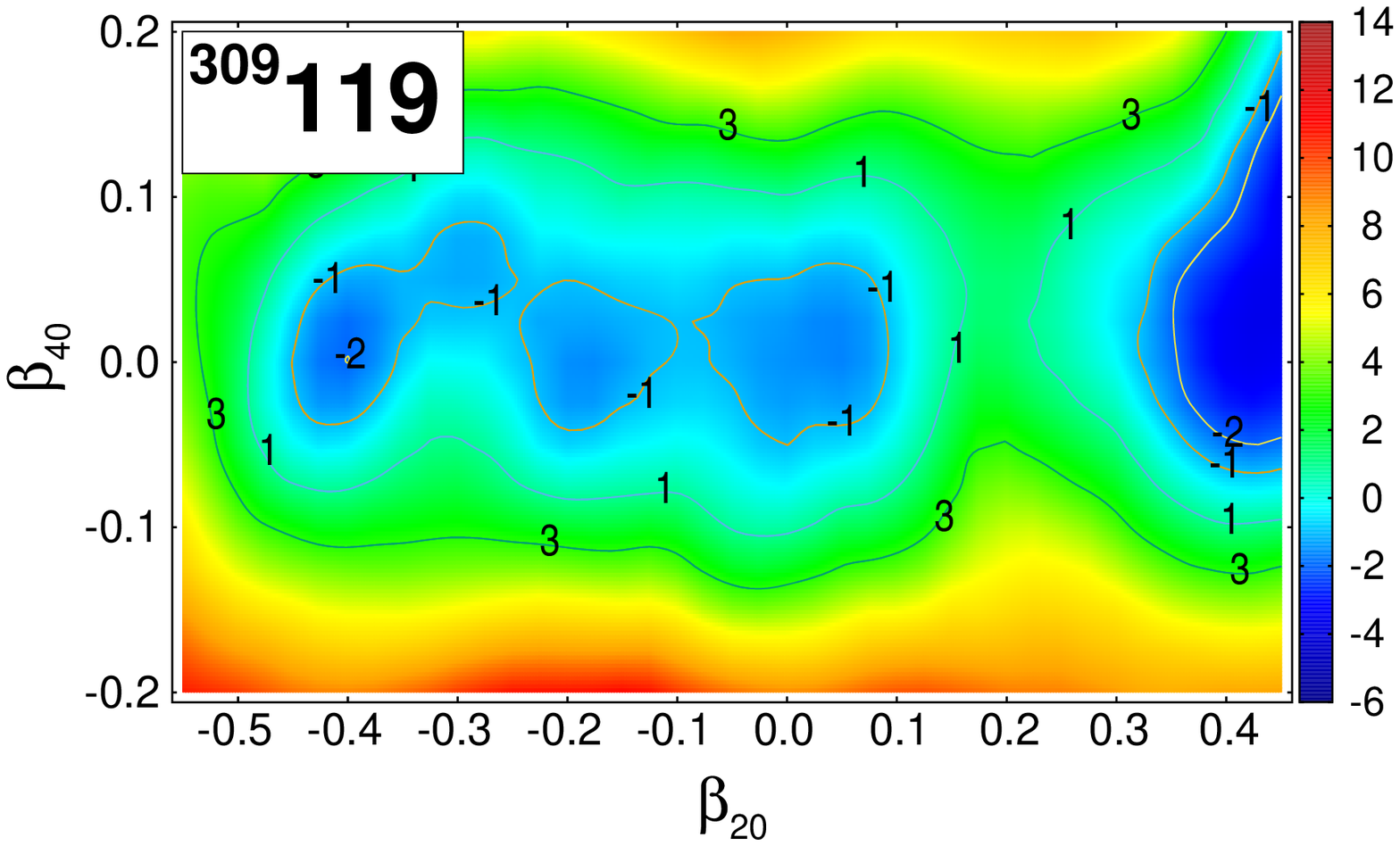}
\caption{Potential energy landscapes in $\beta_{20}, \beta_{40}$ plane for selected
$8$ isotopes of $Z=119$, calculated from the minimization over the remaining five
deformation parameters in (\ref{Rgs}): $\beta_{30}, \beta_{50}, \beta_{60}, \beta_{70}, \beta_{80}$.
Visible energy $E_{tot}$ (in MeV) is calculated relative to the macroscopic energy at the spherical shape:
$E_{mac}=E_{mac}(deformation)-E_{mac}(sphere)$.
The red cross on a given map indicates determined location of the ground-state.}
\label{fig6}
\end{figure}

 In addition to typical prolate, spherical and oblate shapes,
the superdeformed oblate (SDO) shapes with $\beta_{20}^{ gs}\approx-0.45$
 (axes ratio 3:2) appear in some $Z \ge 119$ nuclei, as initially discussed
 (for the case of even-even systems) in \cite{Jachimowicz20112}.
 The evolution of g.s. deformation in these nuclei is illustrated by a series
 of ($\beta_{20}$,$\beta_{40}$) maps for $Z=119$ isotopes in
 \mbox{Fig. \ref{fig6}}. The g.s. minimum on each map is marked by the red
 cross.
 As one can see, in neutron deficient isotopes the prolate shapes compete with
 the SDO. For larger $N$,
 the prolate minimum disappears and "normal" oblate minimum becomes lower
 than the SDO minimum.
 The spherical configuration becomes the g.s. in the close neighborhood of
 $N \approx 184$.
 For still heavier isotopes, i.e. $N\geqslant190$, the SDO minimum reappears
 as the g.s. Deformations $\beta_{20} > 0.45$ (i.e. larger elongations) are not
 shown in the maps, but let us mention that the apparent secondary prolate
 minima at $\beta_{20}\approx 0.40$ are so shallow that they cannot be
 seriously considered as candidates for the configurations of equilibrium.

\subsubsection{Ground state mass excess}

The accuracy of our approach may be partially assessed by comparing
calculated and experimental (ground state) masses.
We have done such comparison in \cite{Jachimowicz2014} for separate groups of
($Z$-even, $N$-even), ($Z$-odd, $N$-even), ($Z$-even, $N$-odd) and
 ($Z$-odd, $N$-odd) heavy nuclei.
Obtained accuracy in the description of the ground-state masses of $252$ heavy
nuclei with $Z\ge82$, for which experimental data \cite{Audi2003} were
 available is summarized in \mbox{Table \ref{1}} below, taken from
 \cite{Jachimowicz2014}.
One can notice that our results are worse for ($Z$-odd, $N$-even) and ($Z$-even, $N$-odd) systems
than for the even-even ones. Moreover, the quality deteriorates further in the
 group of odd-odd nuclei. However, the largest $\delta_{rms}$ deviation of the
 order of 0.67 MeV, obtained for odd-odd systems, is comparable with the
 results of the best available theoretical global mass calculations -
 see Introduction.

\begin{table}
\centering
\caption{Statistical parameters of calculated mass excess
in $4$ separate groups of heavy nuclei with $Z\ge82$, in relation to
experimental data taken from \cite{Audi2003}. All quantities are in MeV:
the average discrepancy \mbox{$\langle \mid M_{th}^{gs}- M_{exp}^{gs}\mid \rangle$},
the maximal difference \mbox{${\max} \mid M_{th}^{gs}- M_{exp}^{gs}\mid$},
and the  rms  deviation $\delta_{rms}$, where $n_g$ is the number
of considered nuclei in a given group.} \label{1}

\begin{tabular}{|c|c|c|c|c|}
\hline                                                &   ($Z$-even, $N$-even)    &   ($Z$-odd, $N$-even)    &
($Z$-even, $N$-odd)    &   ($Z$-odd, $N$-odd)     \\
\hline
$n_g$                                                 &   74     &   56     &   69     &   53       \\
$\langle \mid M_{th}^{gs}- M_{exp}^{gs}\mid \rangle$  &  0.212   &  0.340   & 0.356    &  0.566     \\
${\max} \mid M_{th}^{gs}- M_{exp}^{gs}\mid$           &  0.833   &  0.836   & 1.124    &  1.387     \\
$\delta_{ rms}$                                       &  0.284   &  0.425   & 0.435    &  0.666     \\
\hline
\end{tabular}
\end{table}

\subsubsection{Q-alpha energies}

\begin{figure}[h]
\centering
\includegraphics[scale=0.26]{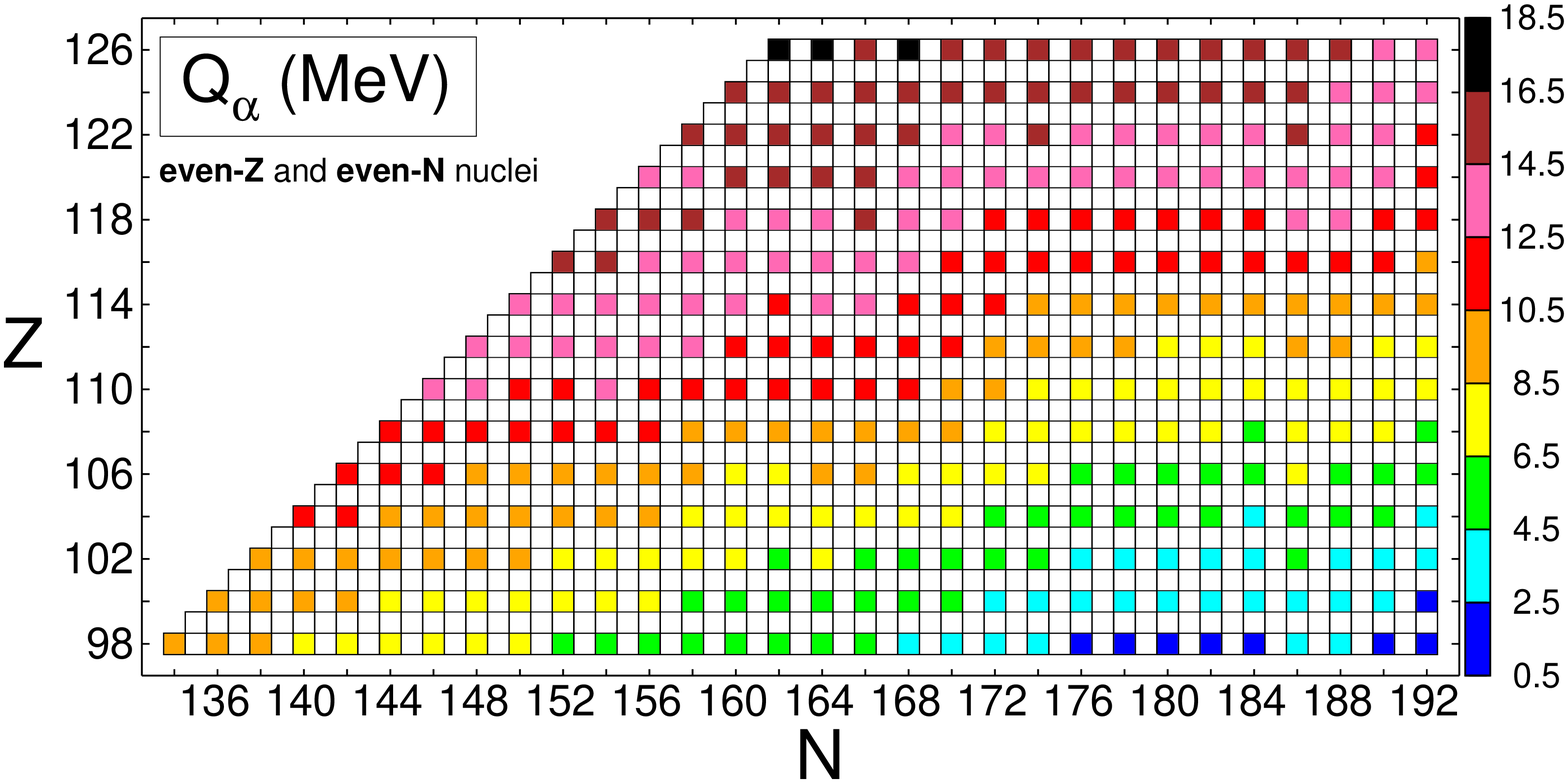}
\includegraphics[scale=0.26]{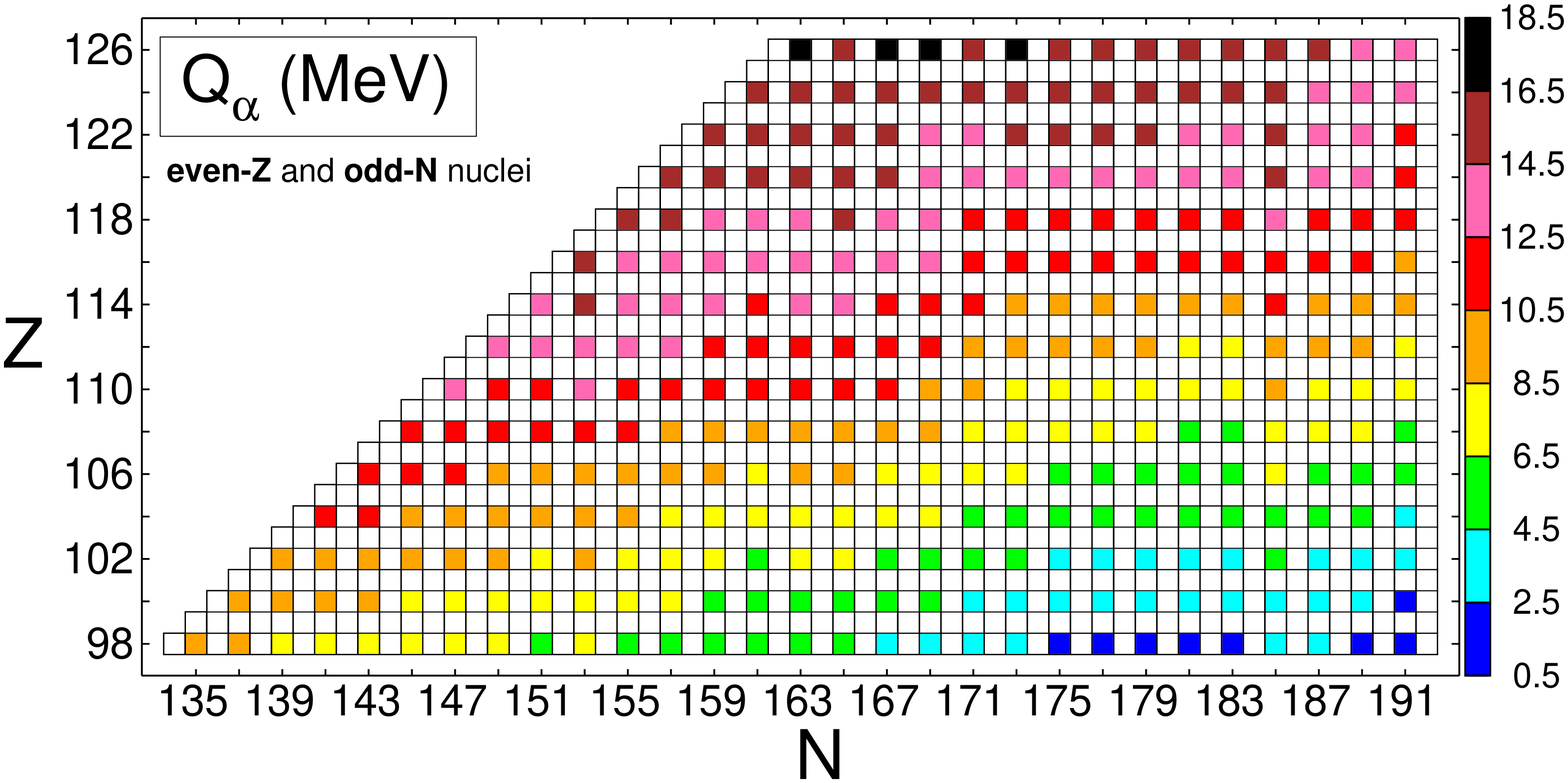}
\includegraphics[scale=0.26]{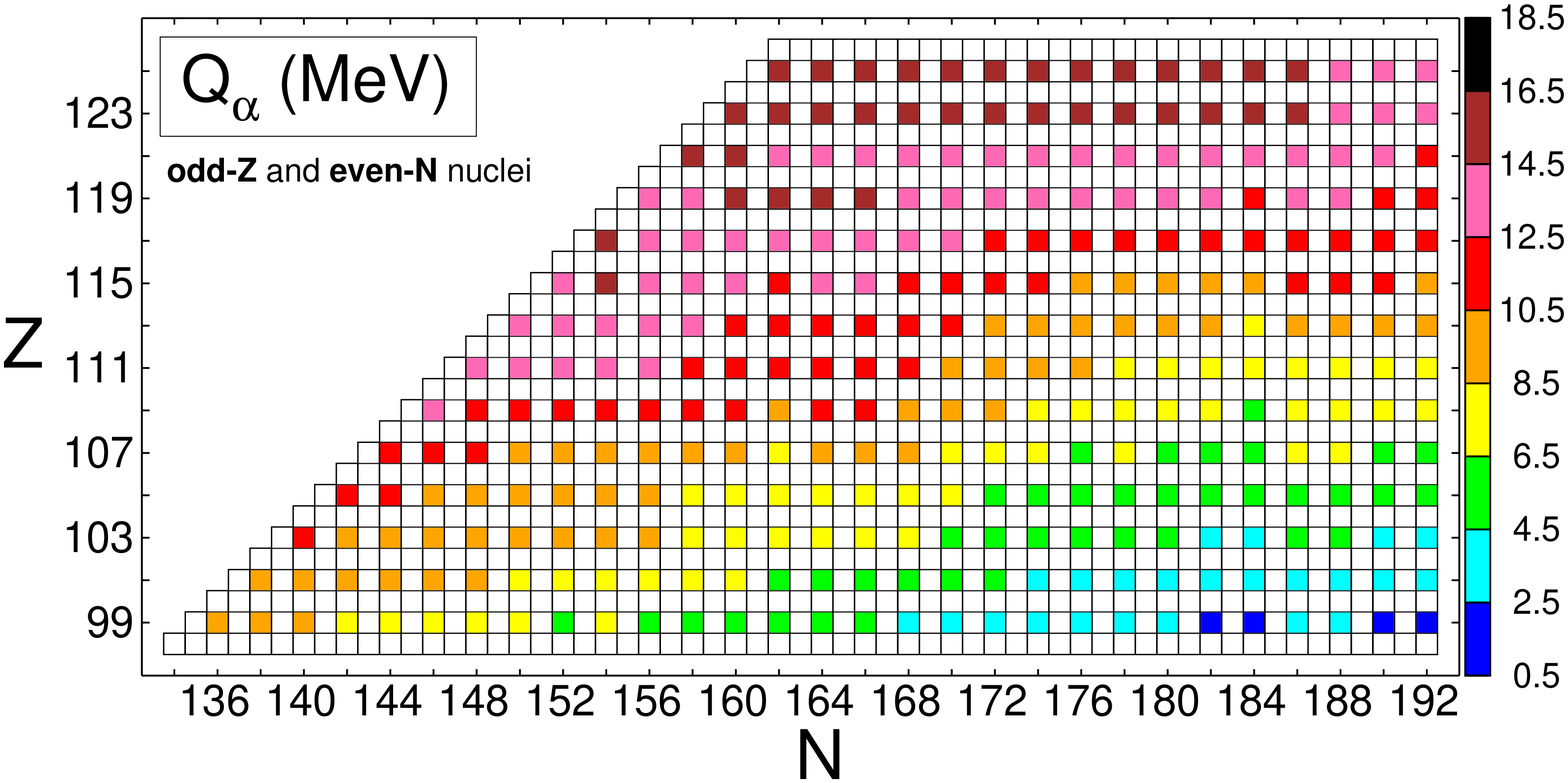}
\includegraphics[scale=0.26]{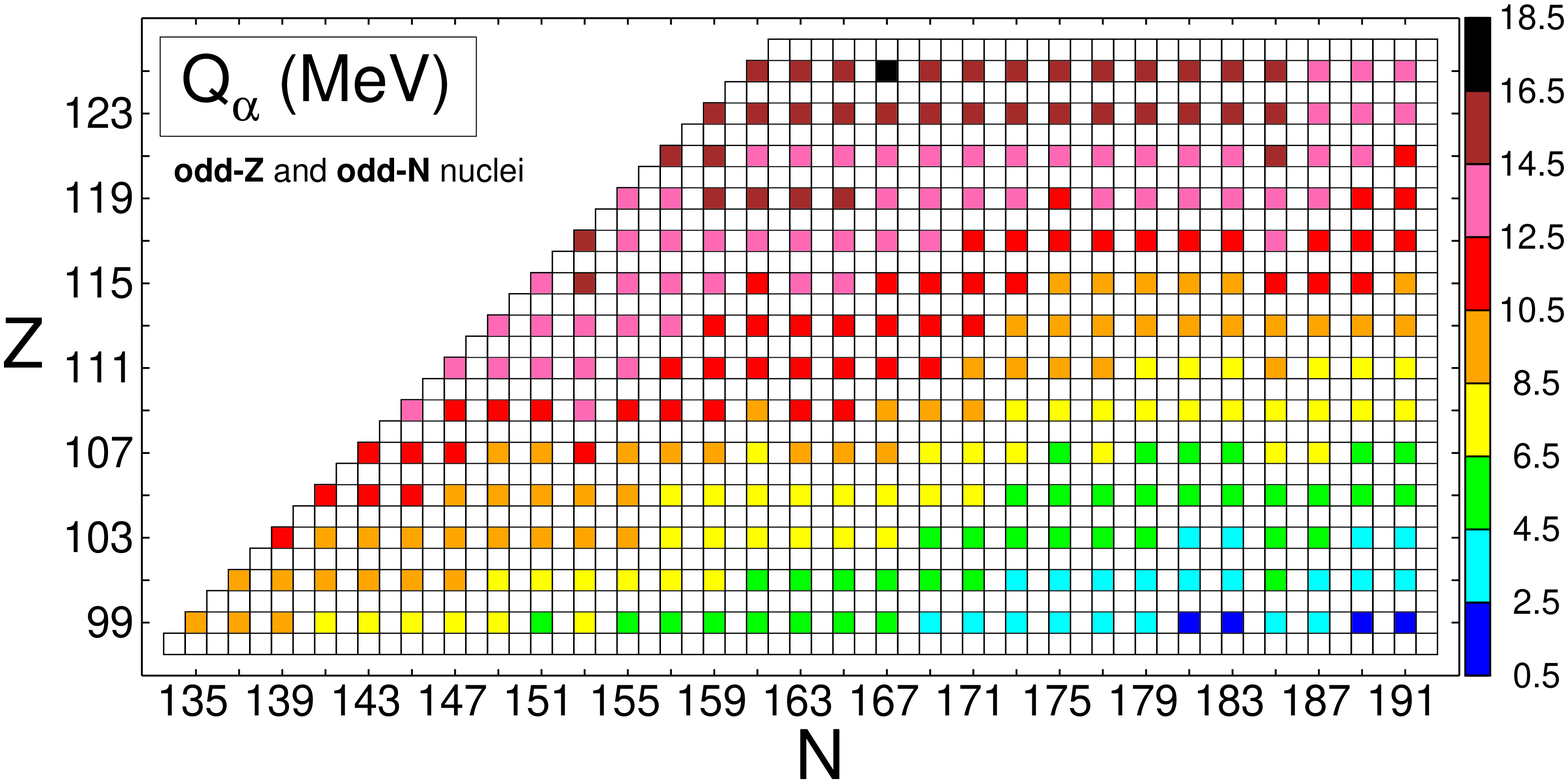}
\caption{Calculated $\alpha$-decay energies for g.s.$\rightarrow$g.s. transitions as a function
of proton $Z$ and neutron $N$ numbers in 4 separate groups of nuclei.}
\label{fig7}
\end{figure}

$Q_{\alpha}^{th}$ values, given in \mbox{Table 1} and
shown in \mbox{Fig. \ref{fig7}} as a function of proton and neutron numbers,
are calculated always for the g.s. to g.s transitions, even if the
 corresponding parent-daughter deformations/configurations differ widely and
 hence one can expect a substantial decay hindrance, see e.g.
\cite{Jachimowicz2014,Jachimowicz2015,Jachimowicz2018}. $Q_{\alpha}$ energy for
a nucleus with $Z$ protons and $N$ neutrons is directly obtained from masses
 (here and below given in MeV):
\begin{equation}
Q_{\alpha}^{th} (Z,N) = M_{th}^{gs}(Z,N)-M_{th}^{gs}(Z-2,N-2)-M(2,2).
\end{equation}
One can notice a sudden decrease in our $Q_{\alpha}^{th}$ values along $N=152$
and $162$, and a more pronounced one, at $N=184$. They signal particularly well
 bound systems at these neutron numbers.
The ones at $N=152$ and $162$ are connected with particularly stable prolate-
 deformed configurations,
 whereas the one at $N=184$ with the predicted "magic" spherical configuration,
 not yet tested in experiment.
The comparison between calculated $\alpha$-decay energies and those
measured in experiments was done in \cite{Jachimowicz2014} for a set
of $204$ nuclei with $Z \ge 82$. It is worth mentioning that in most cases
the experimental $Q_{\alpha}^{exp}$ values there were taken from
 \cite{Wang20122}.
 Average discrepancy
$\langle \mid Q_{\alpha}^{th}- Q_{\alpha}^{exp}\mid \rangle$
obtained for all those nuclei was 326 keV, whereas the rms deviation
 $\delta_{ rms}=426$ keV.
The quality of our model's predictions can be also inferred from
 the camparison of the predicted vs. experimental $\alpha$ -  decay chains,
 starting at isotopes: $^{294}$Ts, $^{295}$Og, or the hypothetical $^{297}119$,
  shown in \cite{Jachimowicz2014}.

\subsubsection{Neutron and proton separation energies}

\begin{figure}[h]
\centering
\includegraphics[scale=0.26]{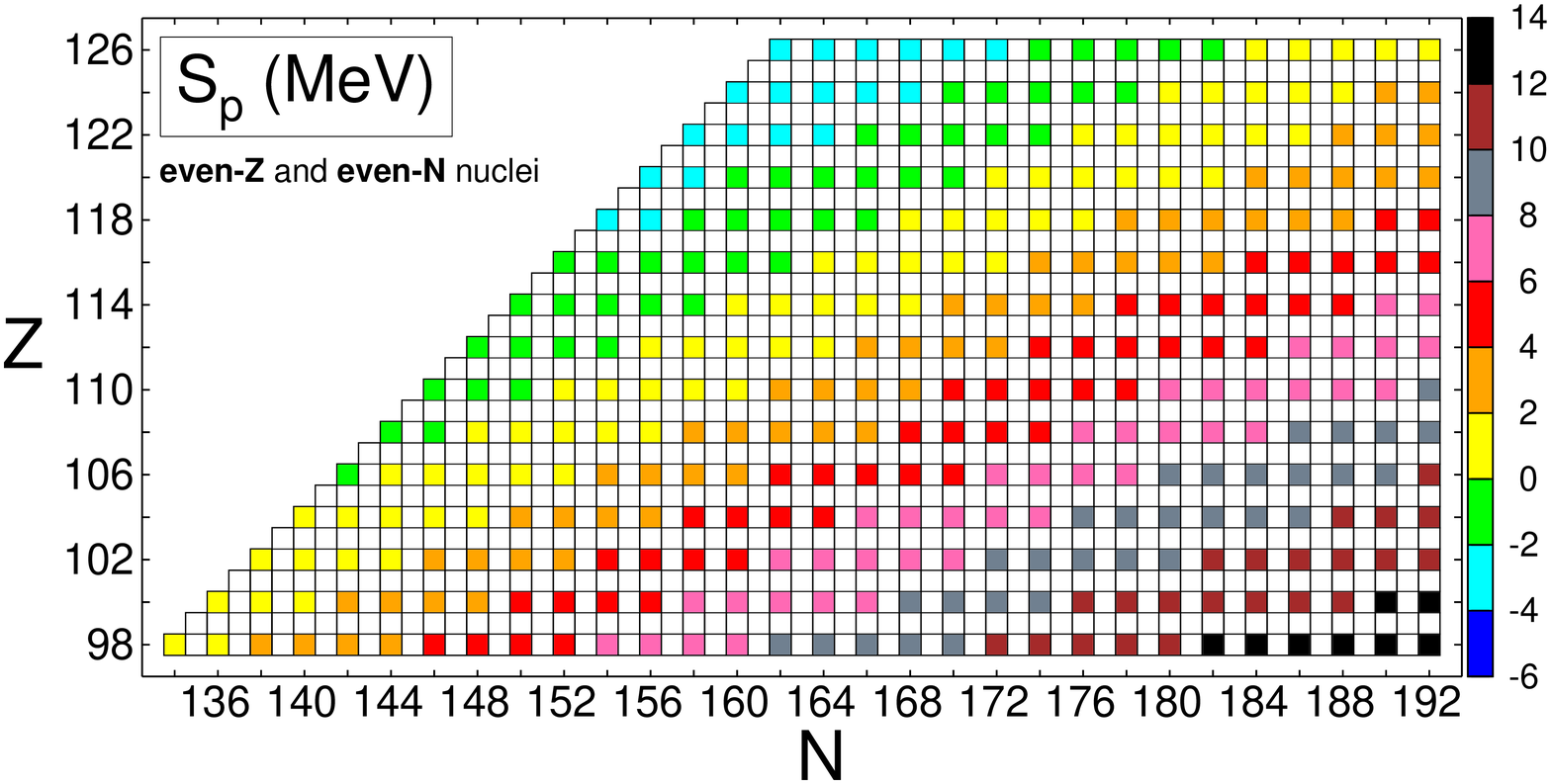}
\includegraphics[scale=0.26]{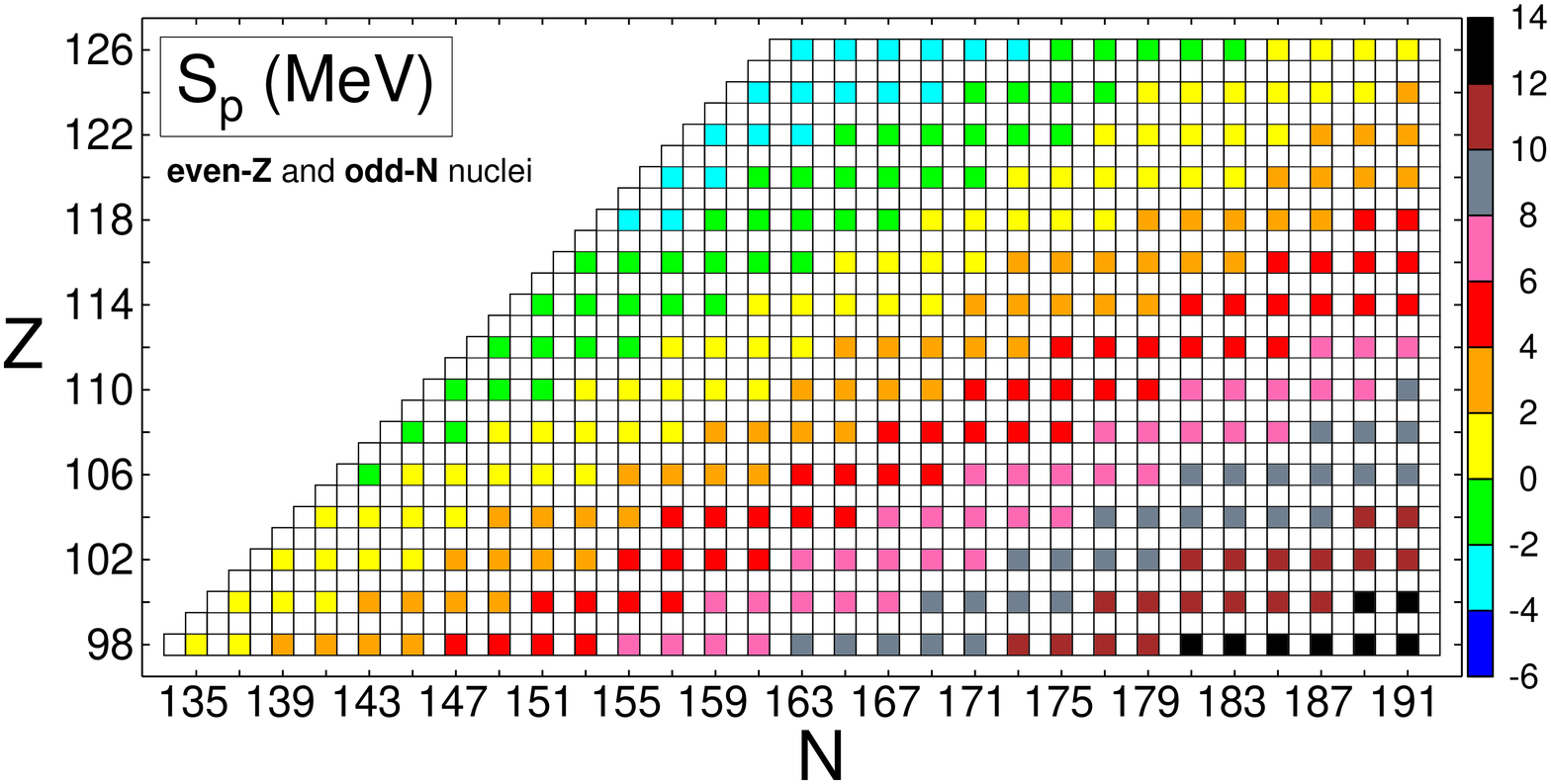}
\includegraphics[scale=0.26]{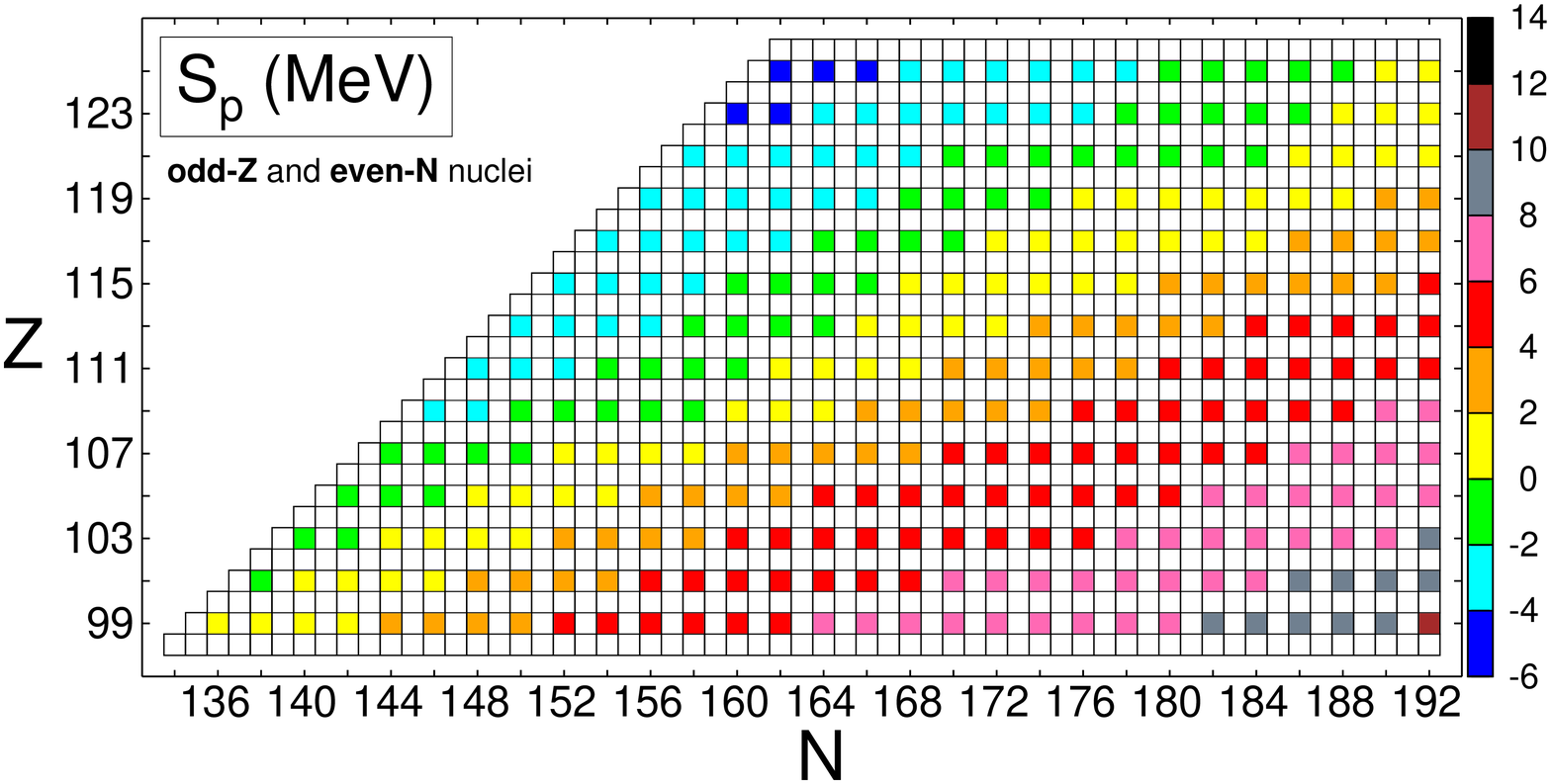}
\includegraphics[scale=0.26]{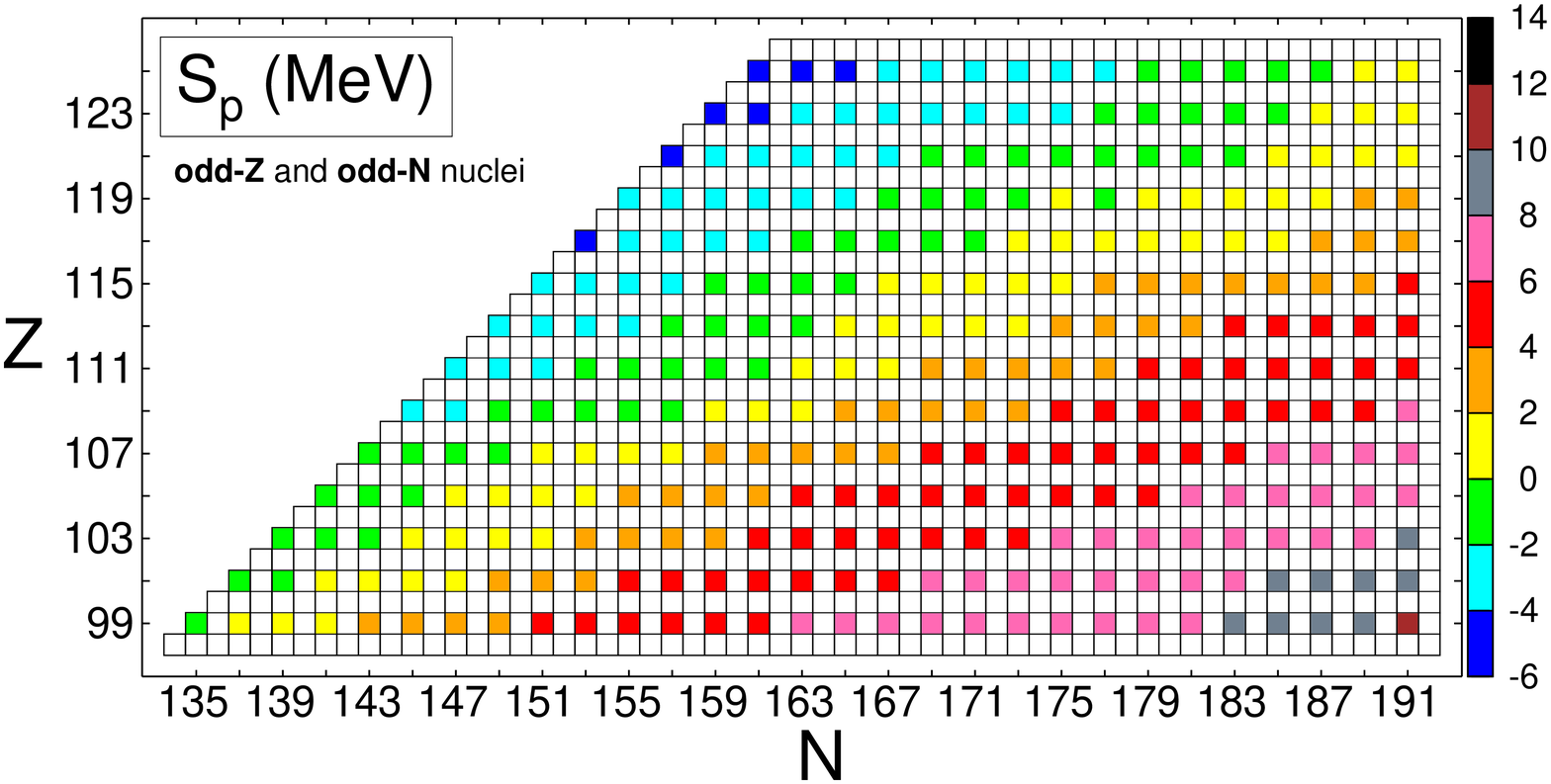}
\caption{Proton separation energy $S_p$ (in MeV) in 4 separate groups of nuclei
as a function of proton $Z$ and neutron $N$ numbers.}
\label{SP}
\end{figure}

\begin{figure}[h]
\centering
\includegraphics[scale=0.26]{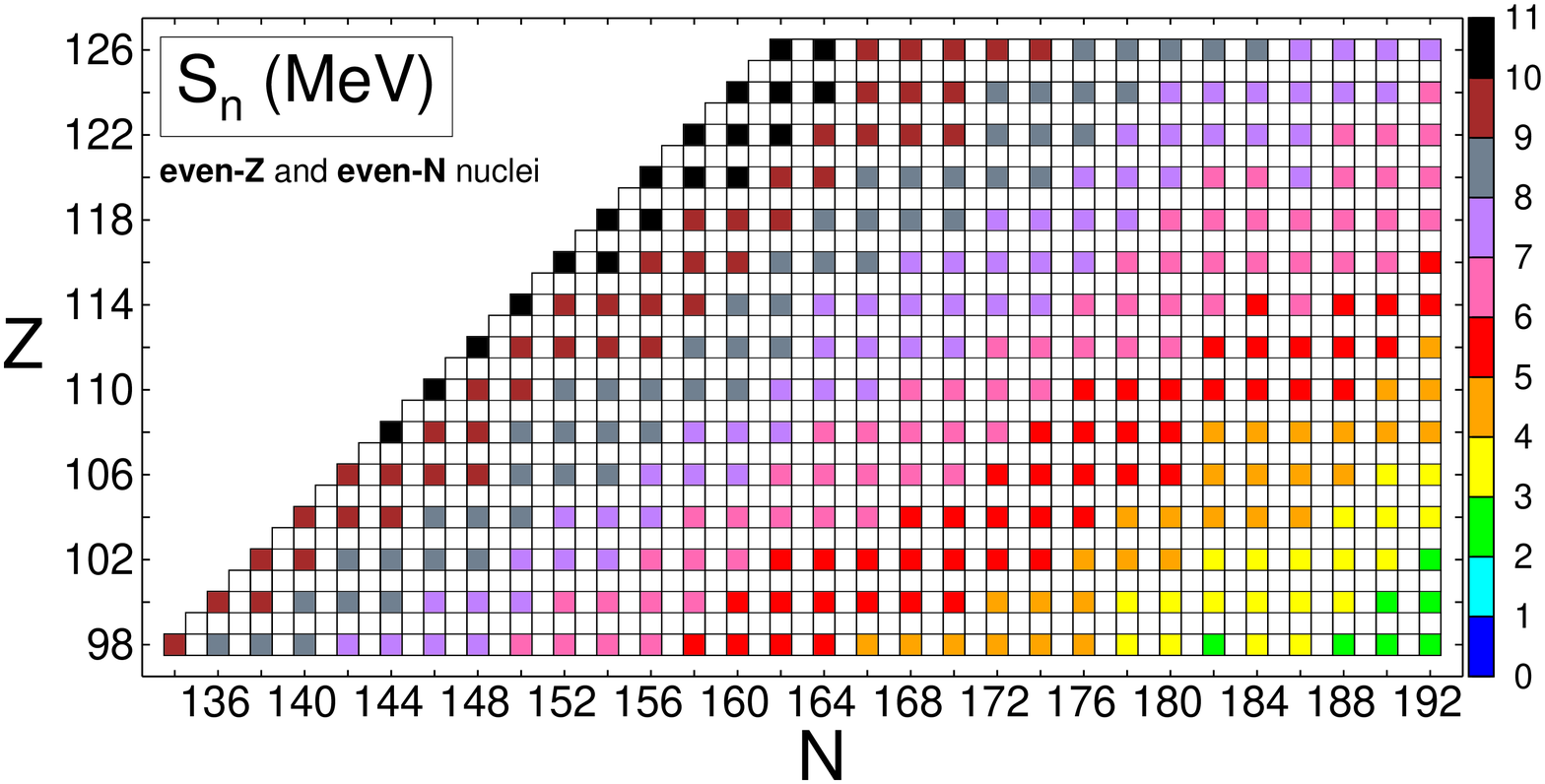}
\includegraphics[scale=0.26]{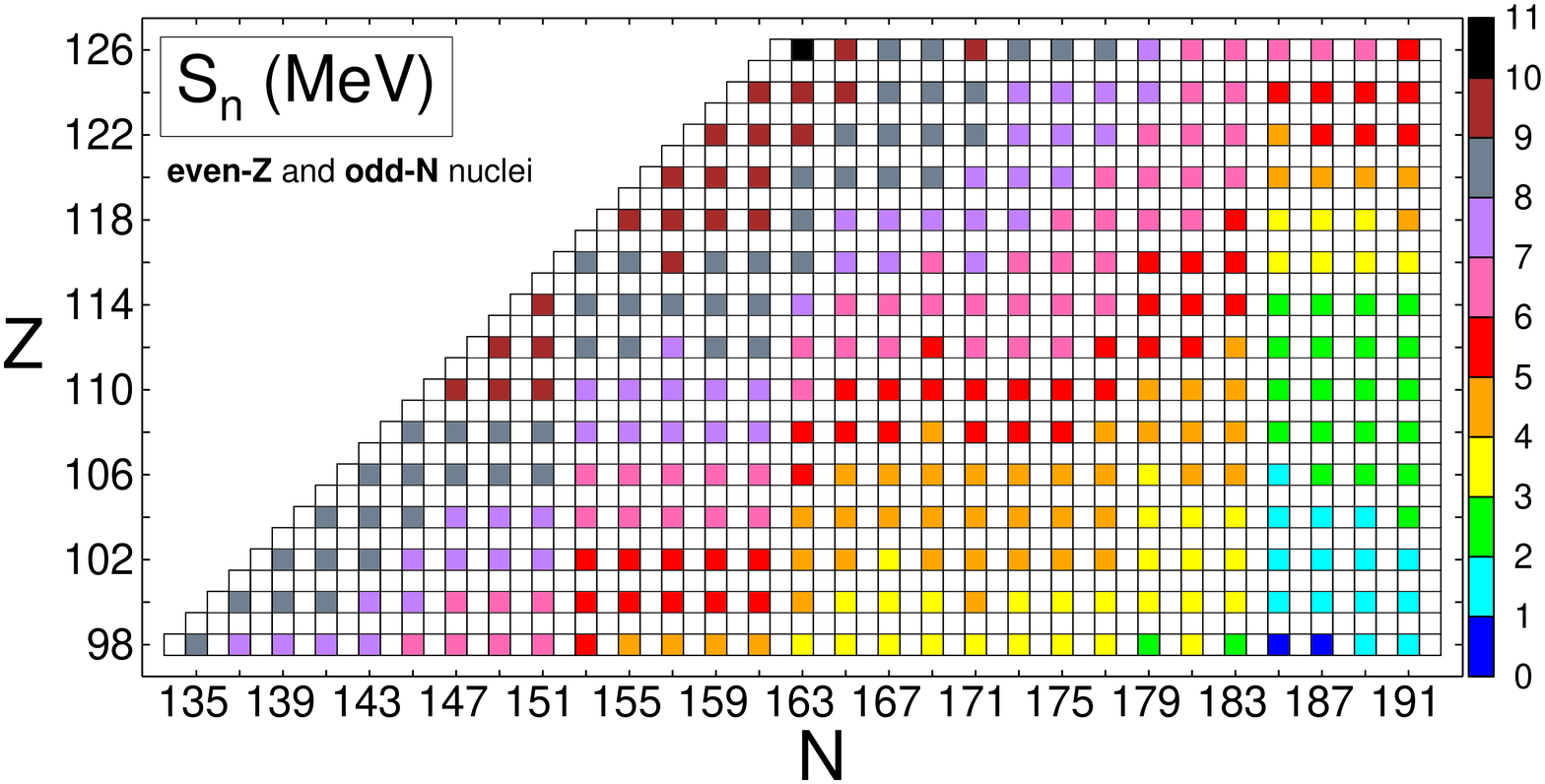}
\includegraphics[scale=0.26]{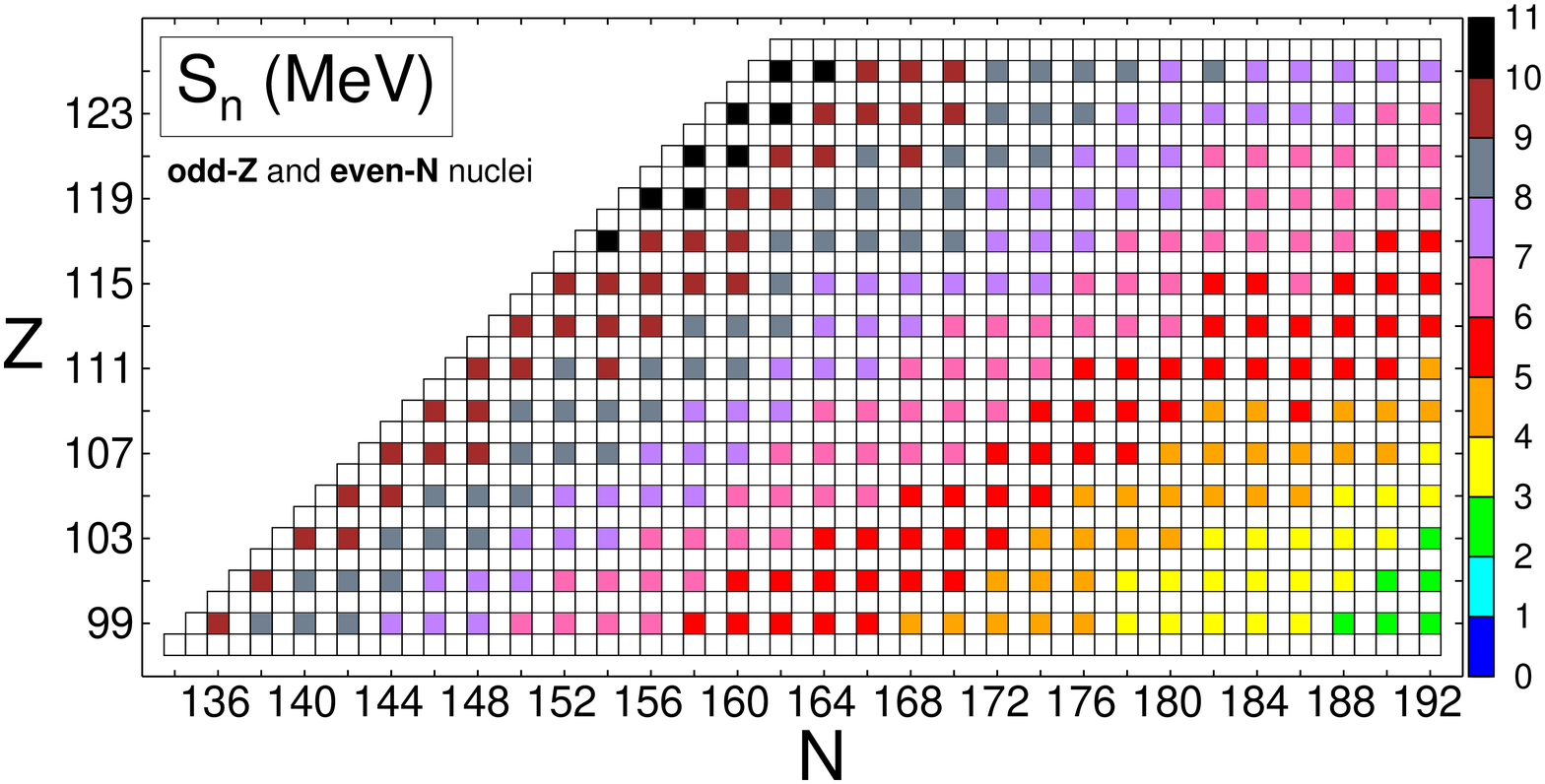}
\includegraphics[scale=0.26]{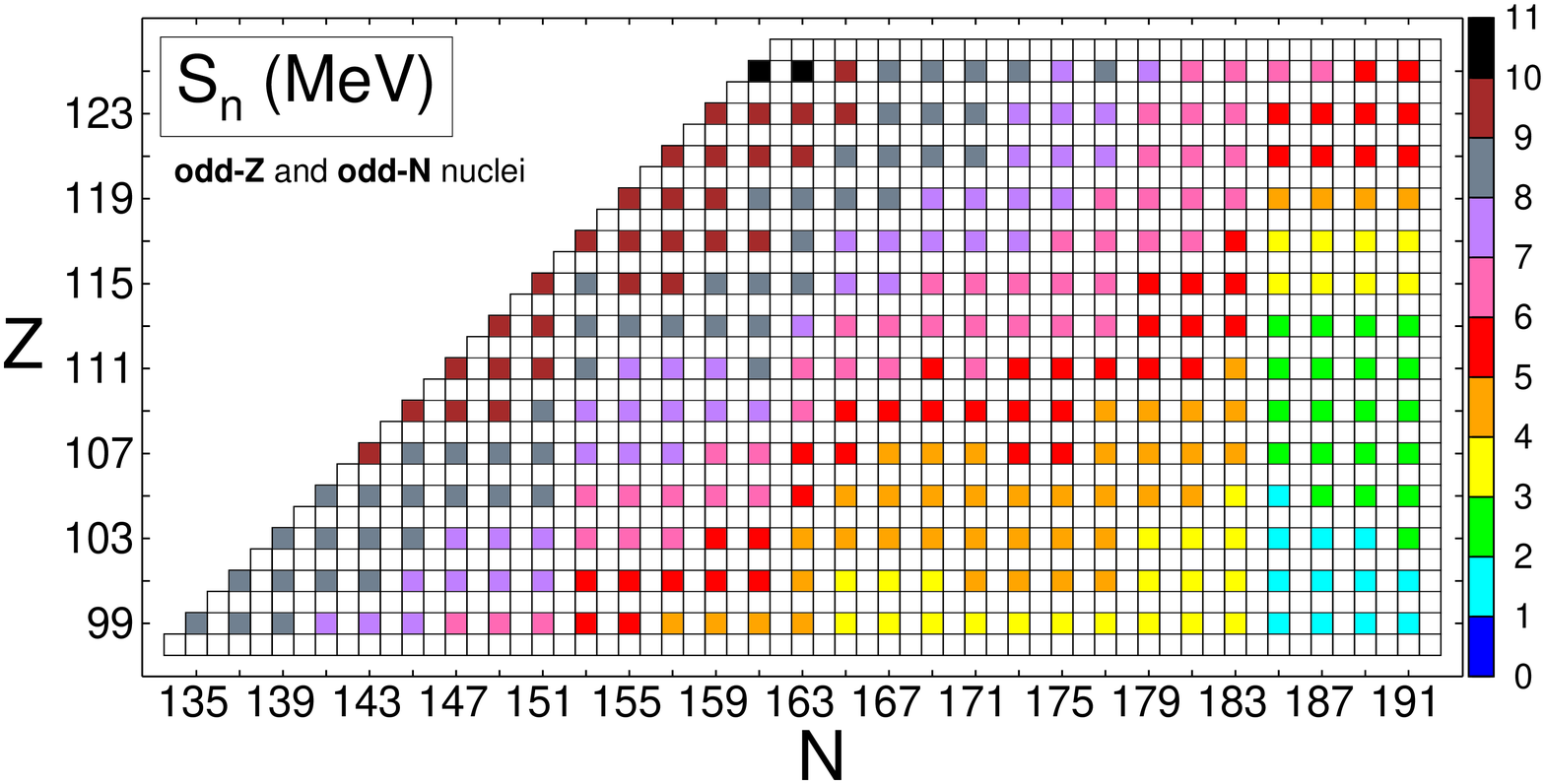}
\caption{Neutron separation energy $S_n$ (in MeV) in 4 separate groups of nuclei
as a function of proton $Z$ and neutron $N$ numbers.}
\label{SN}
\end{figure}

\begin{figure}[h]
\centering
\includegraphics[scale=0.26]{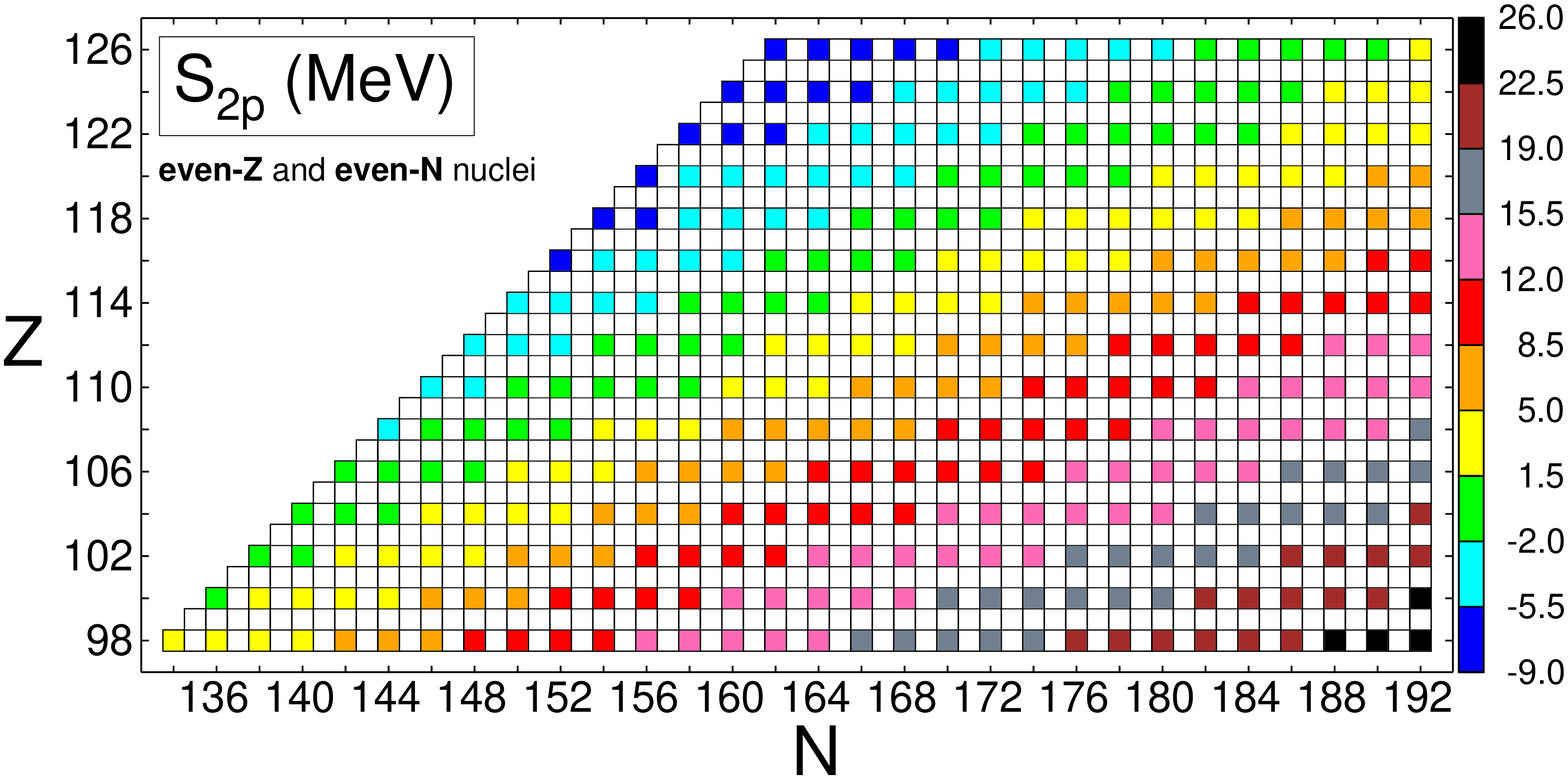}
\includegraphics[scale=0.26]{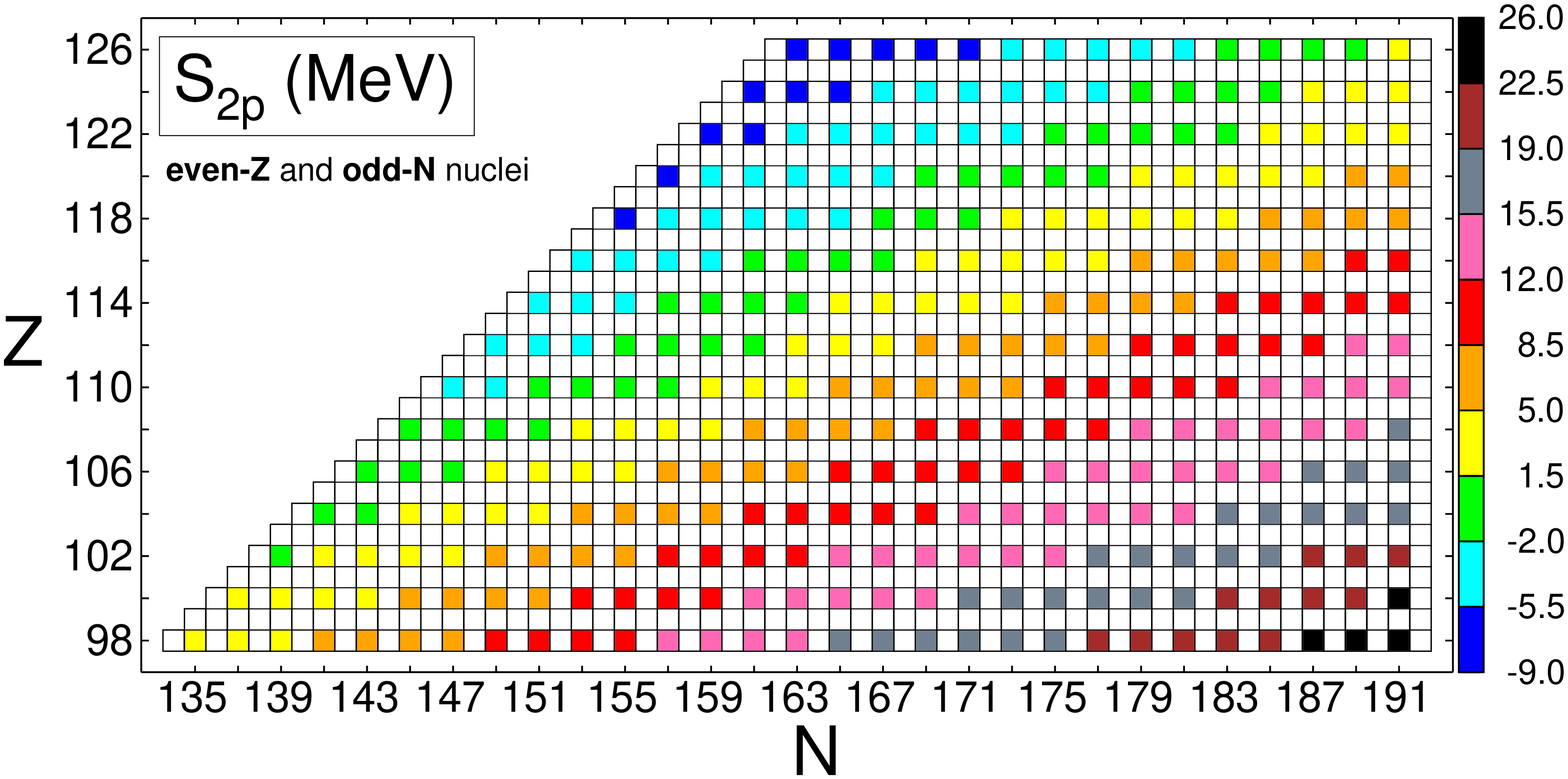}
\includegraphics[scale=0.26]{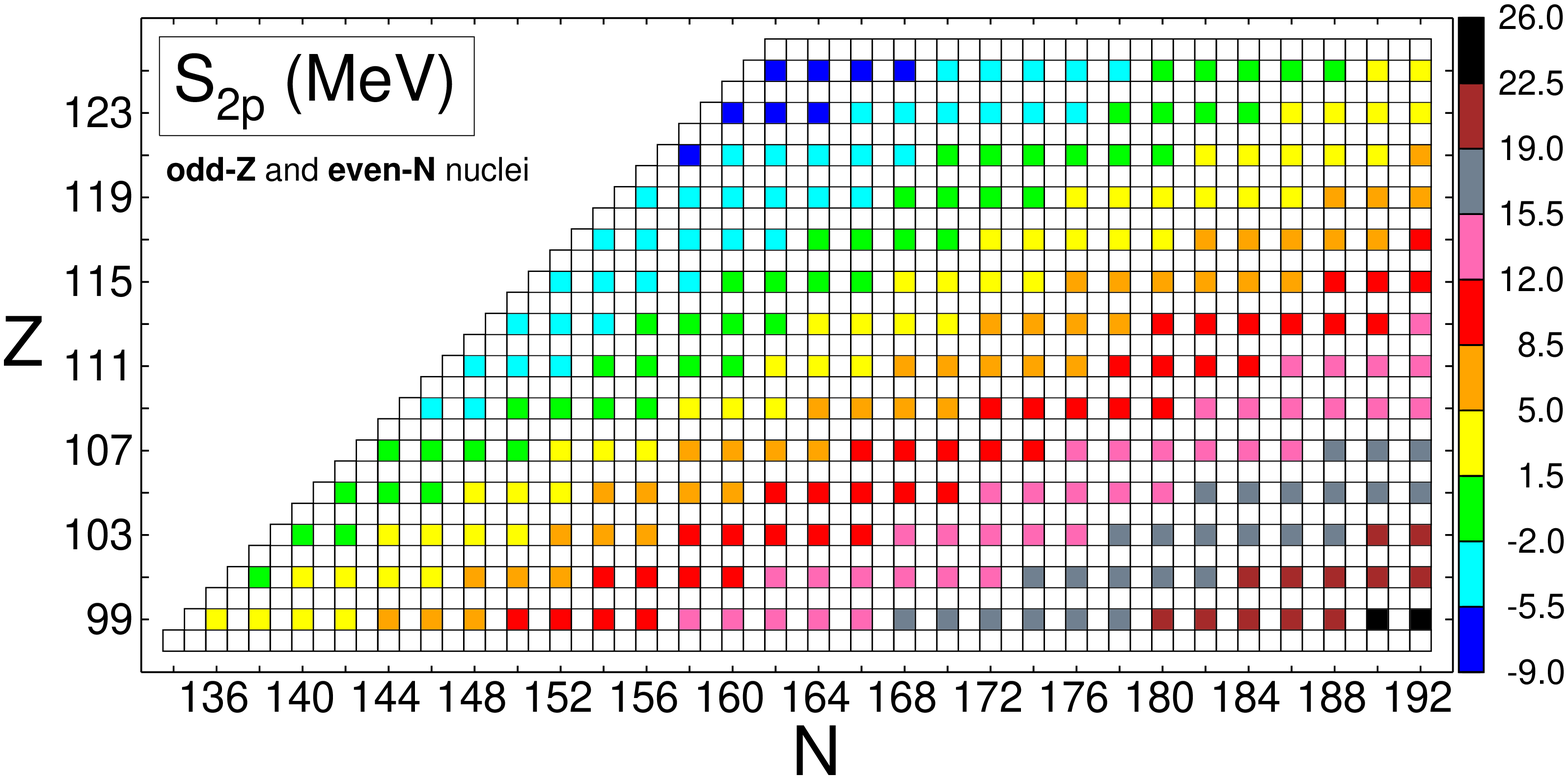}
\includegraphics[scale=0.26]{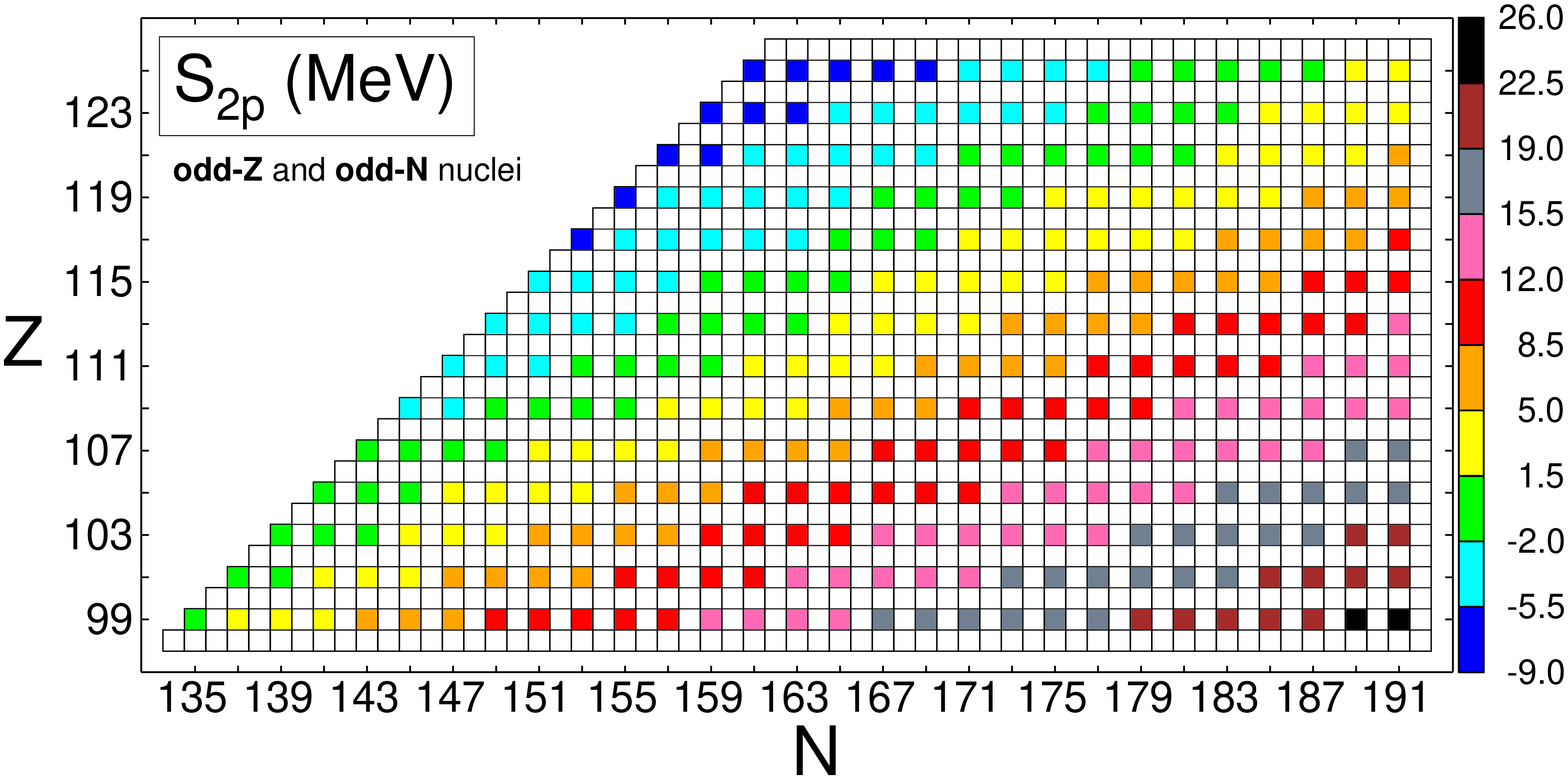}
\caption{Two proton separation energy $S_{2p}$ (in MeV) in 4 separate groups of nuclei
as a function of proton $Z$ and neutron $N$ numbers.}
\label{S2P}
\end{figure}

\begin{figure}[h]
\centering
\includegraphics[scale=0.26]{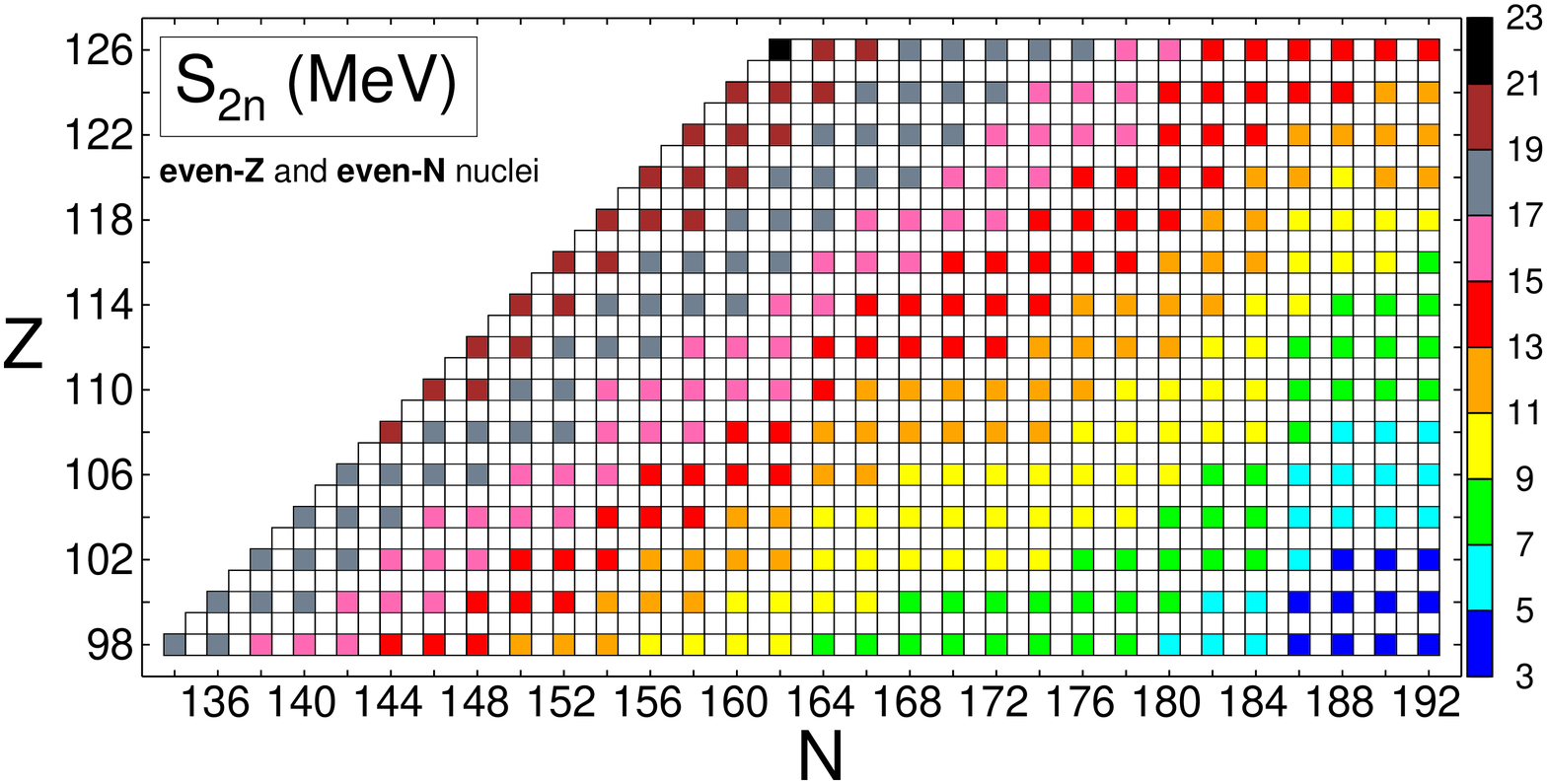}
\includegraphics[scale=0.26]{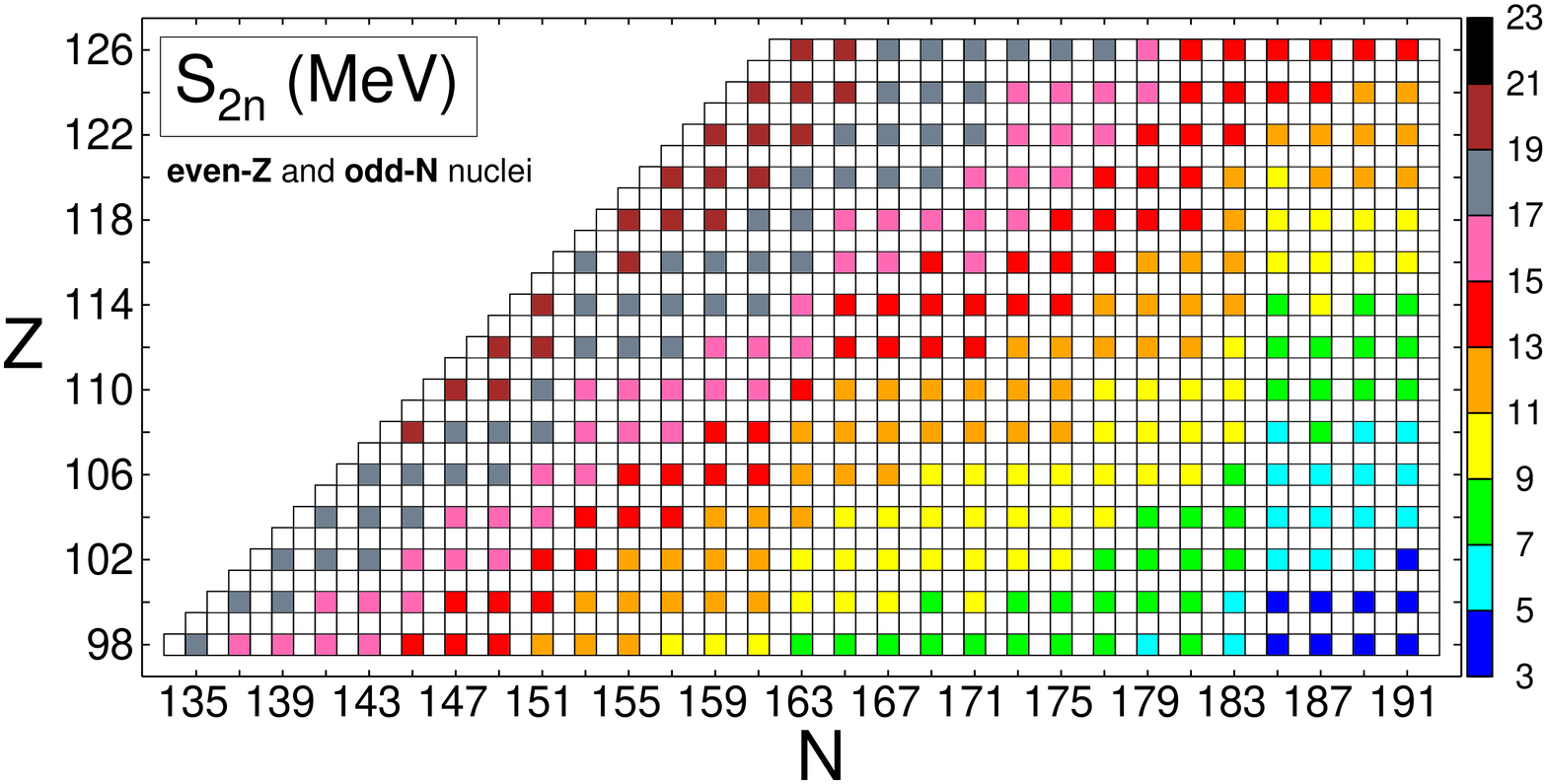}
\includegraphics[scale=0.26]{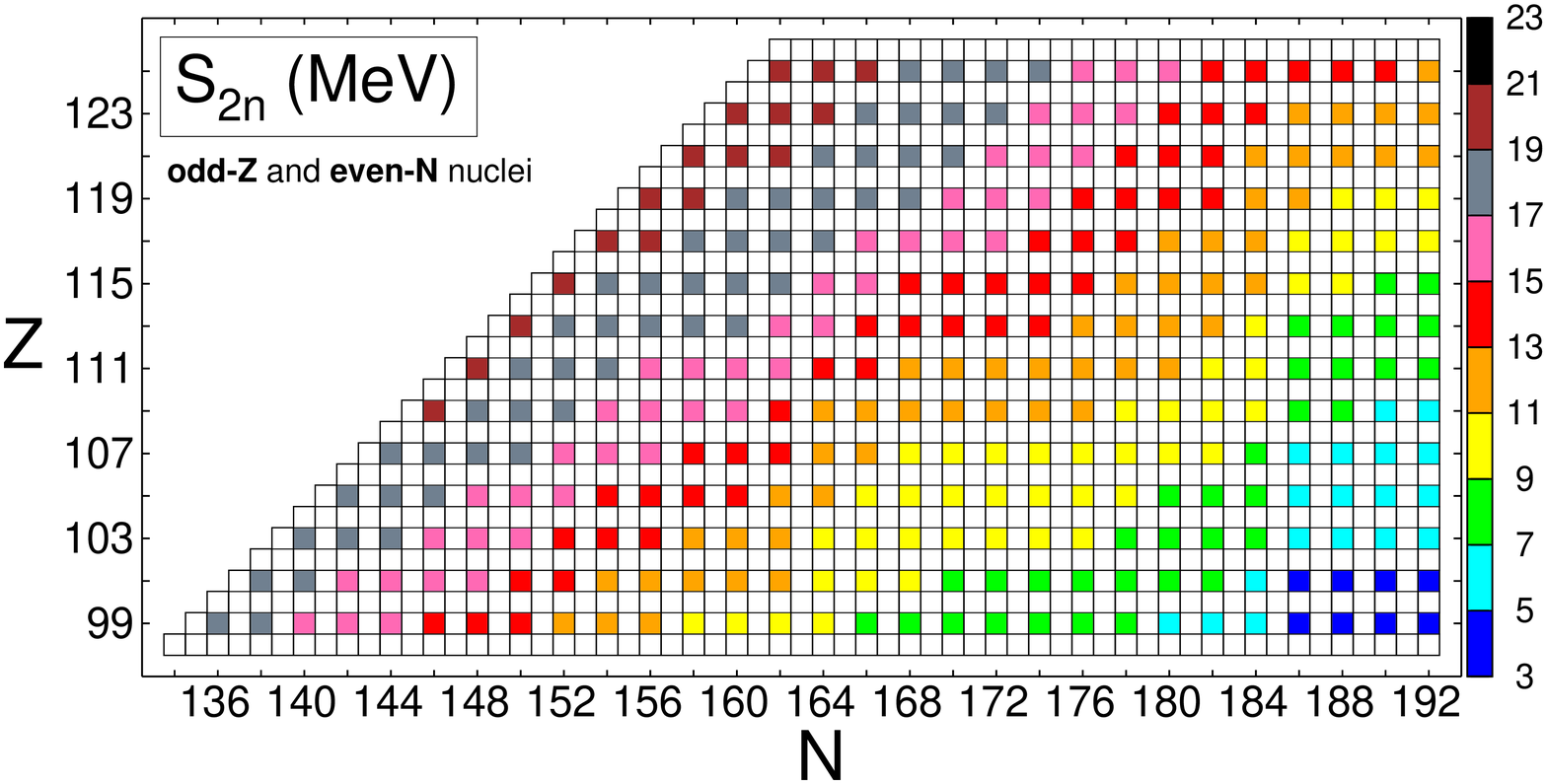}
\includegraphics[scale=0.26]{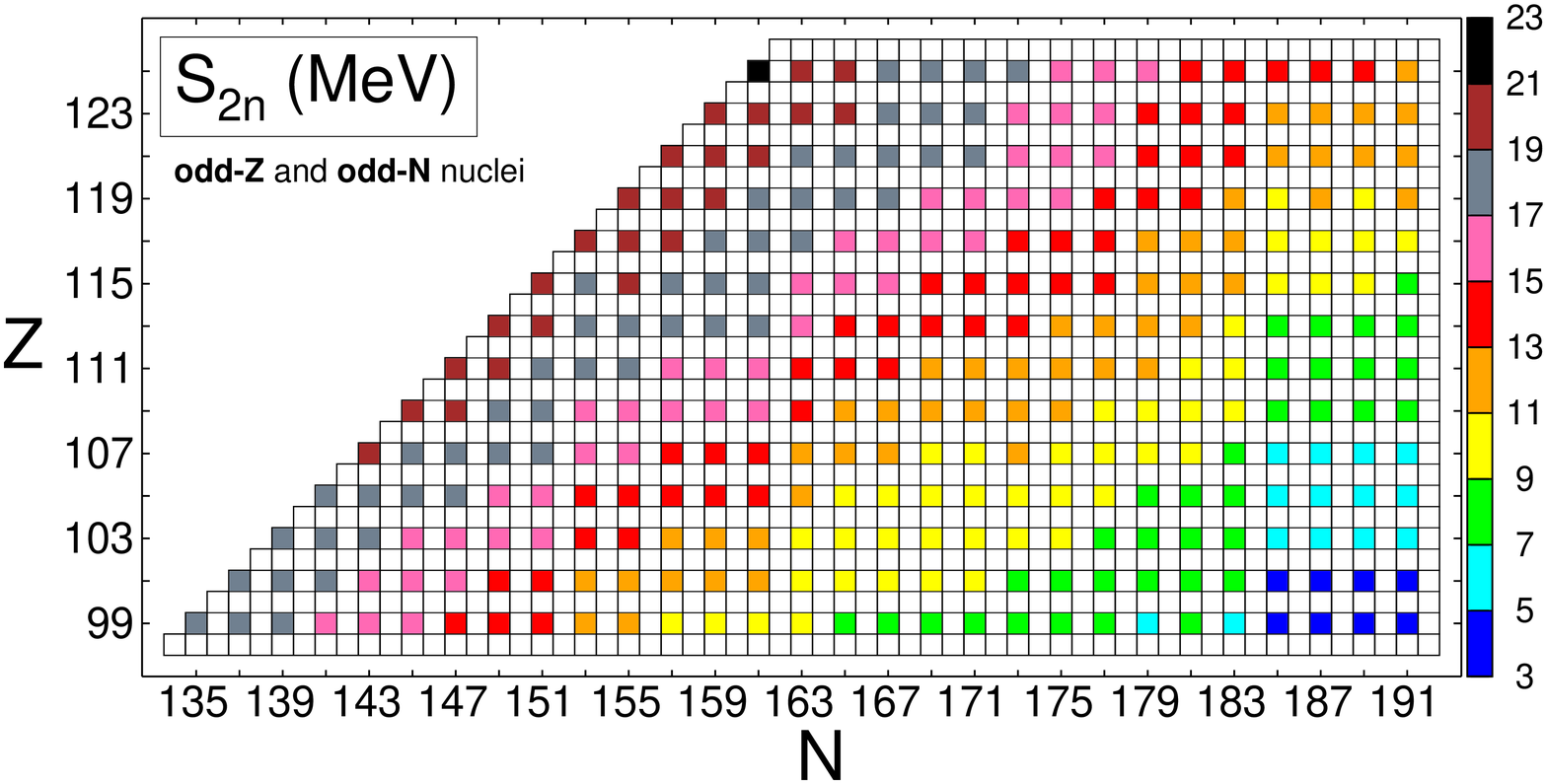}
\caption{Two neutron separation energy $S_{2n}$ (in MeV) in 4 separate groups of nuclei
as a function of proton $Z$ and neutron $N$ numbers.}
\label{S2N}
\end{figure}

Calculated proton, neutron, two-proton and two-neutron separation energies
(in MeV)
are shown in Fig. \ref{SP}, \ref{SN}, \ref{S2P}, \ref{S2N}, respectively.
Their numerical values are given in \mbox{Table 1} in the last four columns.
Separation energies $S_p$ and $S_{2p}$, for a nucleus with $Z$ protons and $N$
 neutrons, were calculated from masses:
\begin{equation}
S_p(Z,N) = -M(Z,N)+M(Z-1,N)+M_p, \qquad {\rm and} \qquad S_{2p}(Z,N) = -M(Z,N)+M(Z-2,N)+2M_p,
\end{equation}
where $M_p$ is the mass of the proton. Similarly, the separation energies of
 one or two neutrons, $S_n$ and $S_{2n}$, were obtained from:
\begin{equation}
S_n(Z,N) = -M(Z,N)+M(Z,N-1)+M_n, \qquad {\rm and} \qquad S_{2n}(Z,N) = -M(Z,N)+M(Z,N-2)+2M_n,
\end{equation}
where $M_n$ is the mass of the neutron.


\subsection{Saddle point properties}

In our saddle-point calculations we used the IWF method on a multidimensional
grid of deformations (nuclear shapes), whereas
the easier, but sometimes misleading, minimization was used only
for the tests. Such (in principle exact) method of the search for saddles
implies that their locations and heights can be more or less different
from those read directly from the maps like in
 \mbox{Fig. \ref{fig5}, \ref{fig6}} below. This is
 mainly due to the energy minimization used in the latter case. Hovewer,
 despite this fact, the potential energy surfaces
 can be very useful for verifying the results obtained from the IWF
 calculations, especially when there are a few saddles of similar height.
 In practice, we have generated and analyzed in detail such maps
 for each nucleus of $1305$ considered.
Therefore, our calculations are extensive and systematic, but rely on
 the detailed and individual analysis.

\begin{figure}[h]
\centering
\includegraphics[scale=0.26]{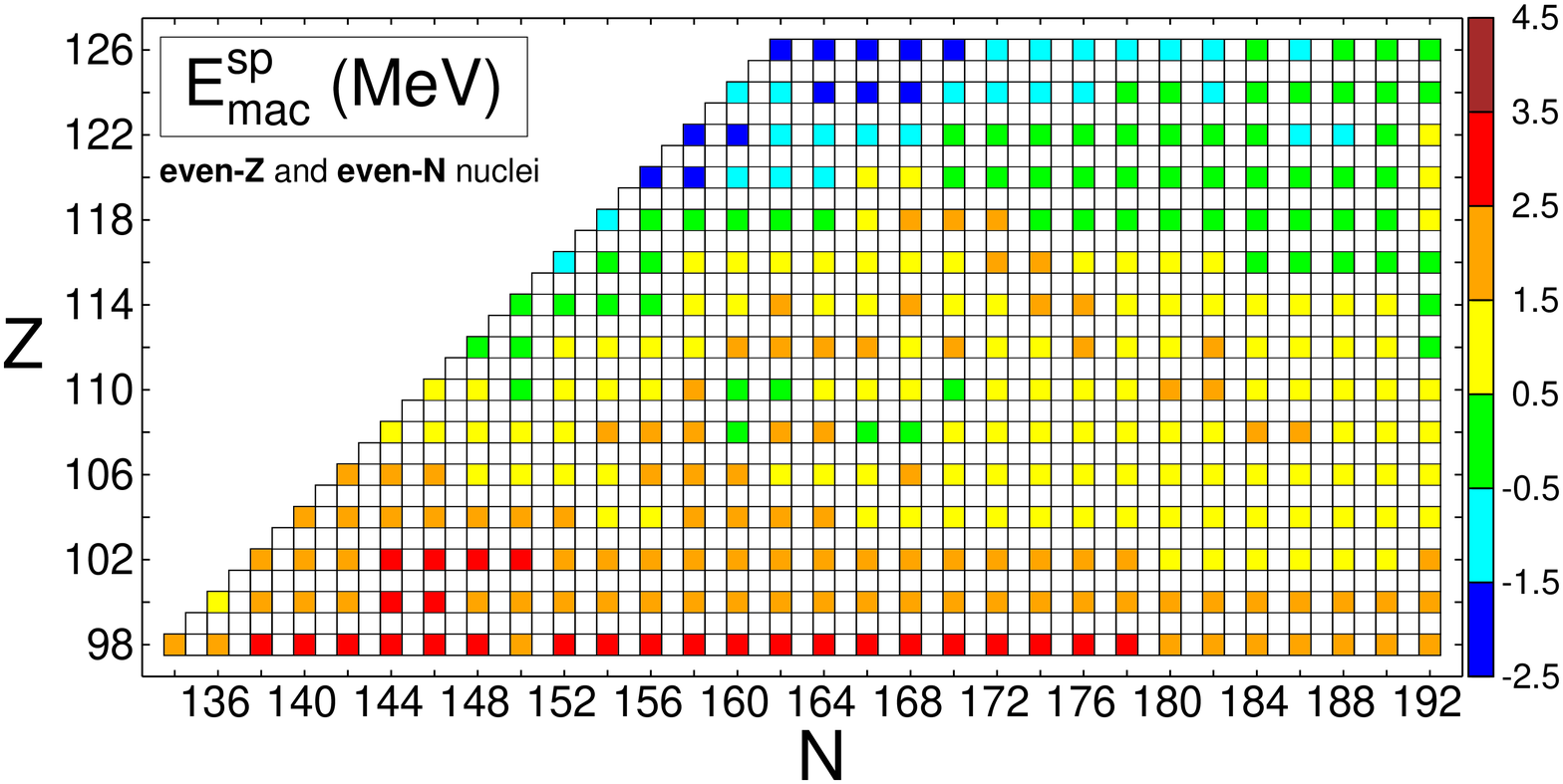}
\includegraphics[scale=0.26]{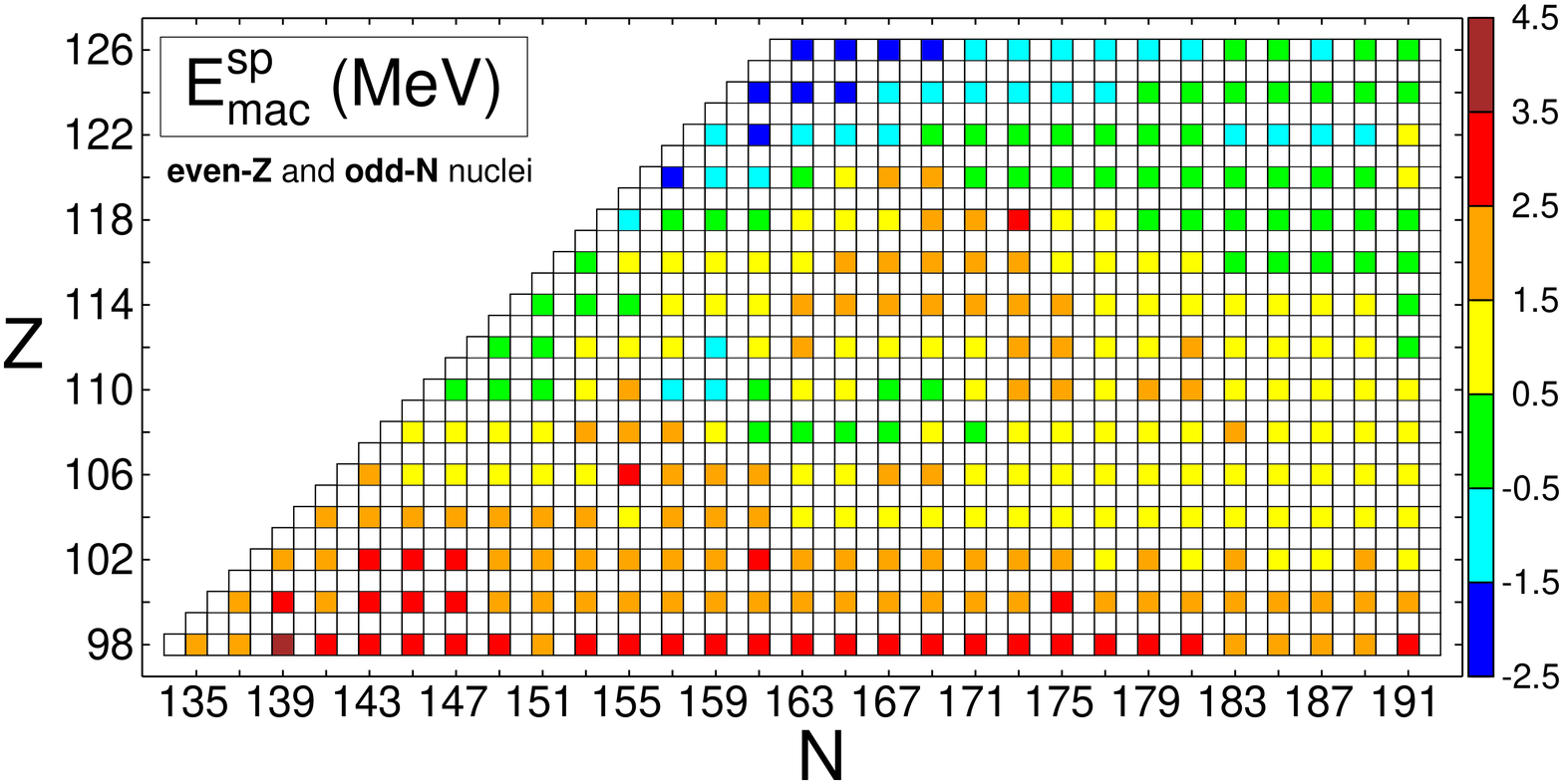}
\includegraphics[scale=0.26]{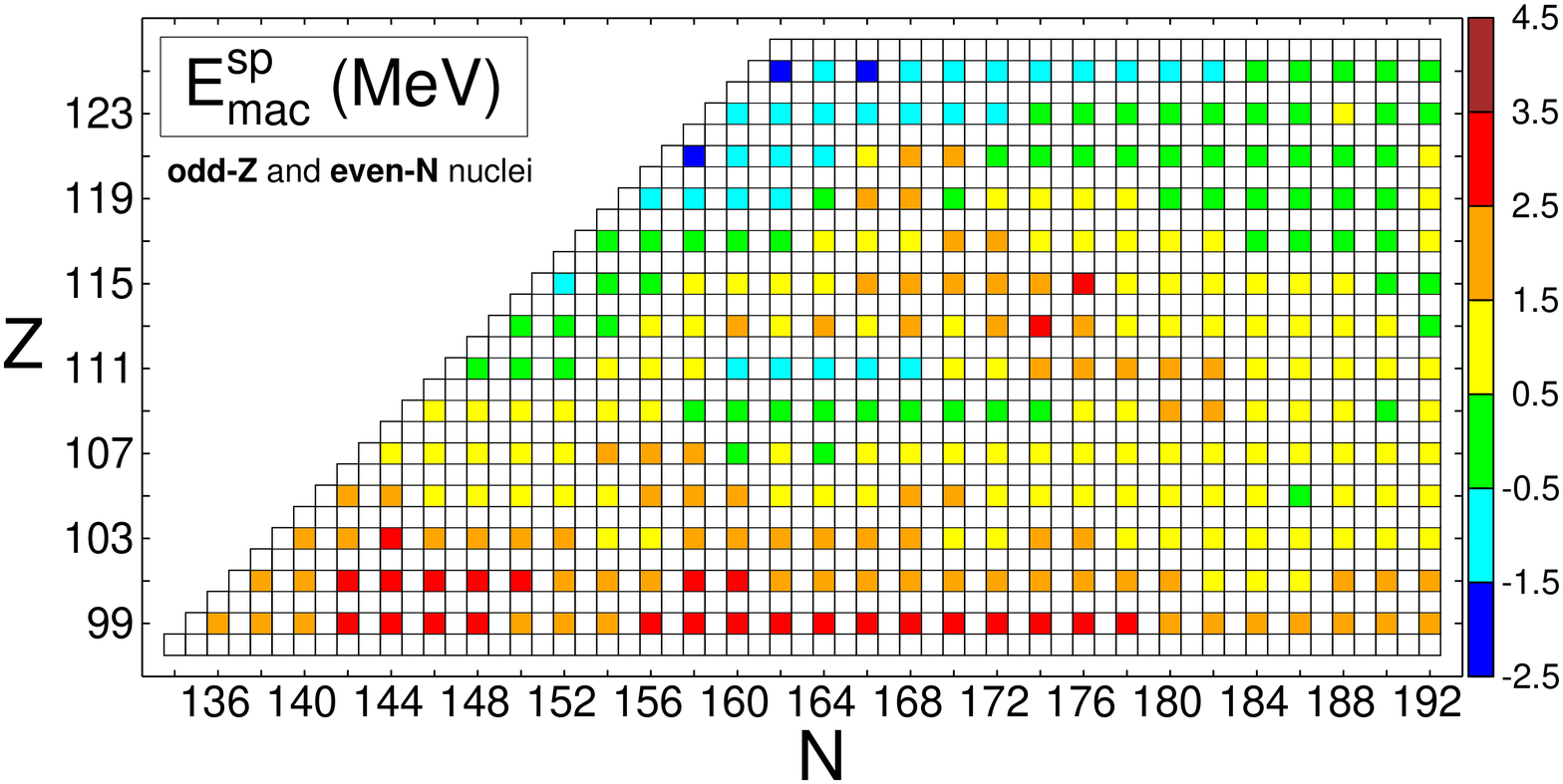}
\includegraphics[scale=0.26]{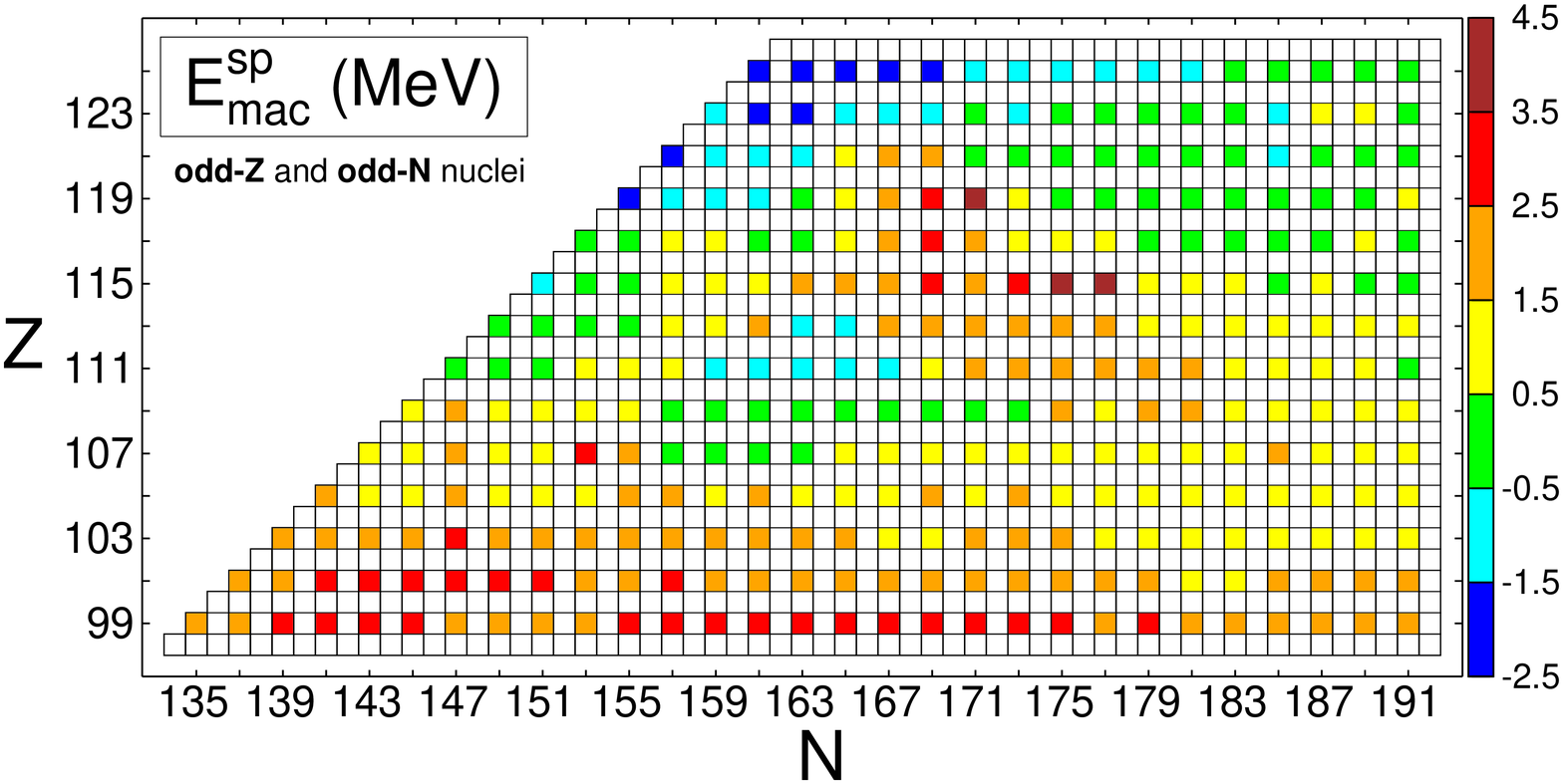}
\caption{As in Fig. \ref{fig1}, but for the calculated saddle points.}
\label{fig12}
\end{figure}

\begin{figure}[h]
\centering
\includegraphics[scale=0.26]{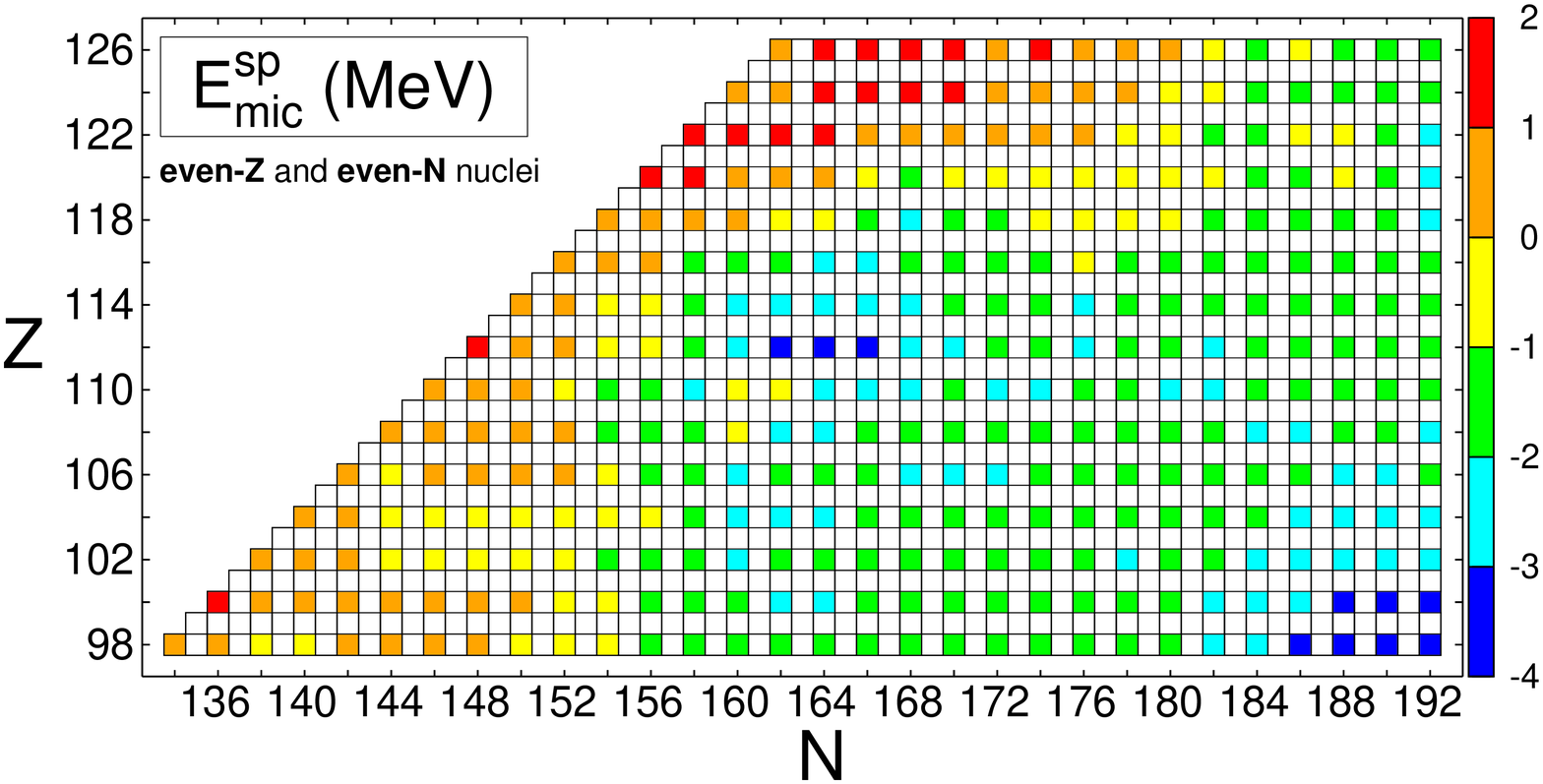}
\includegraphics[scale=0.26]{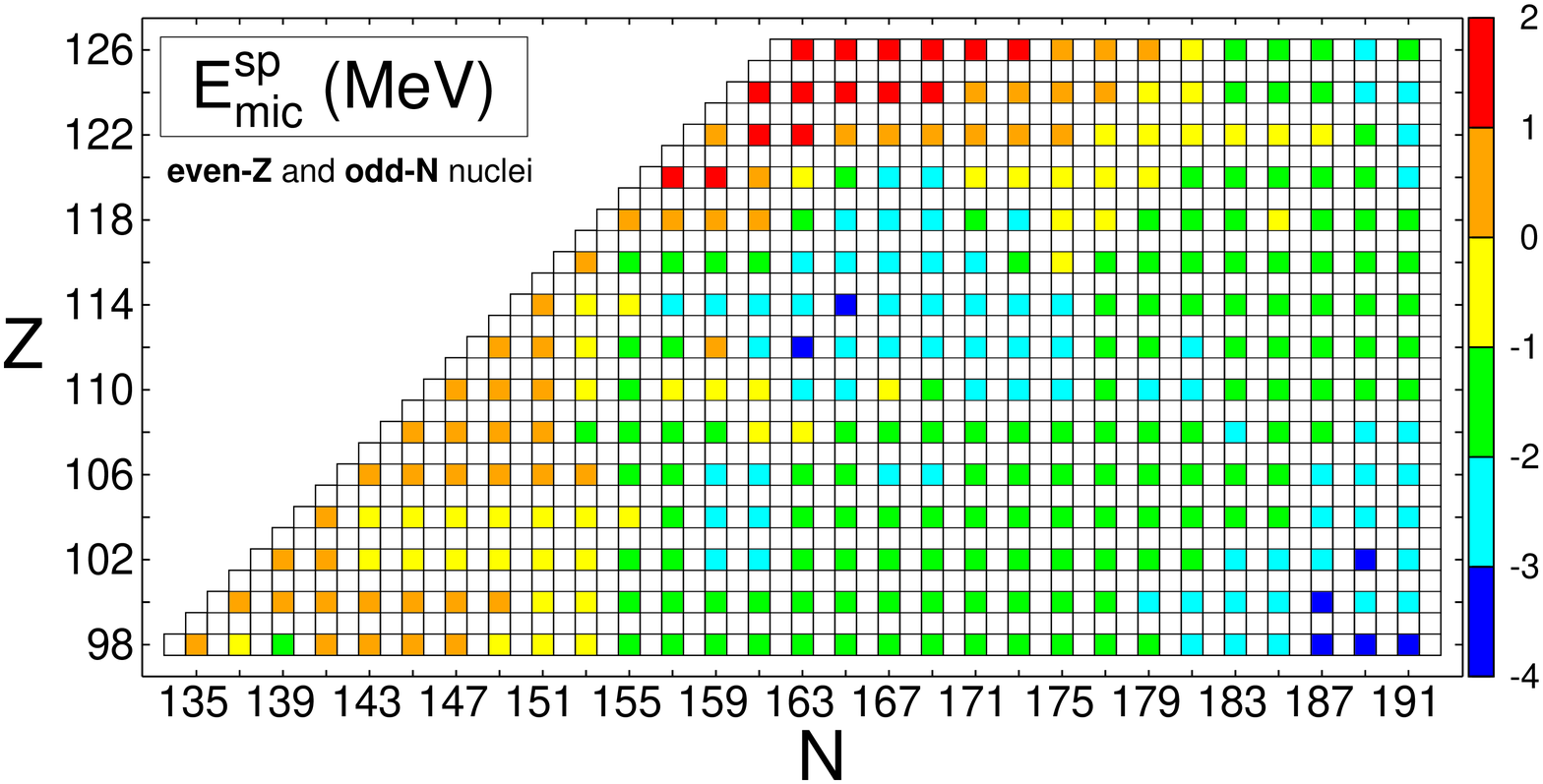}
\includegraphics[scale=0.26]{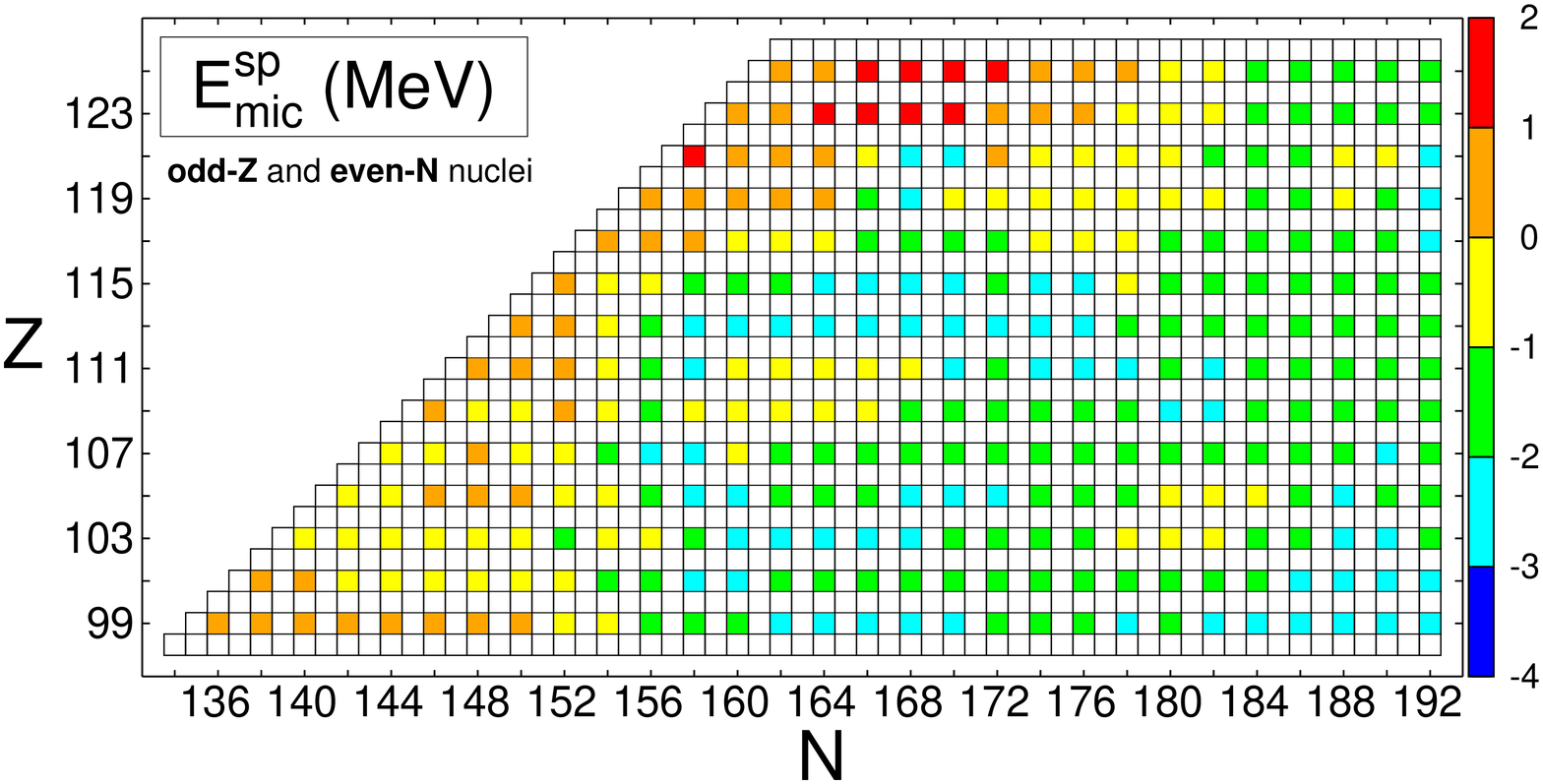}
\includegraphics[scale=0.26]{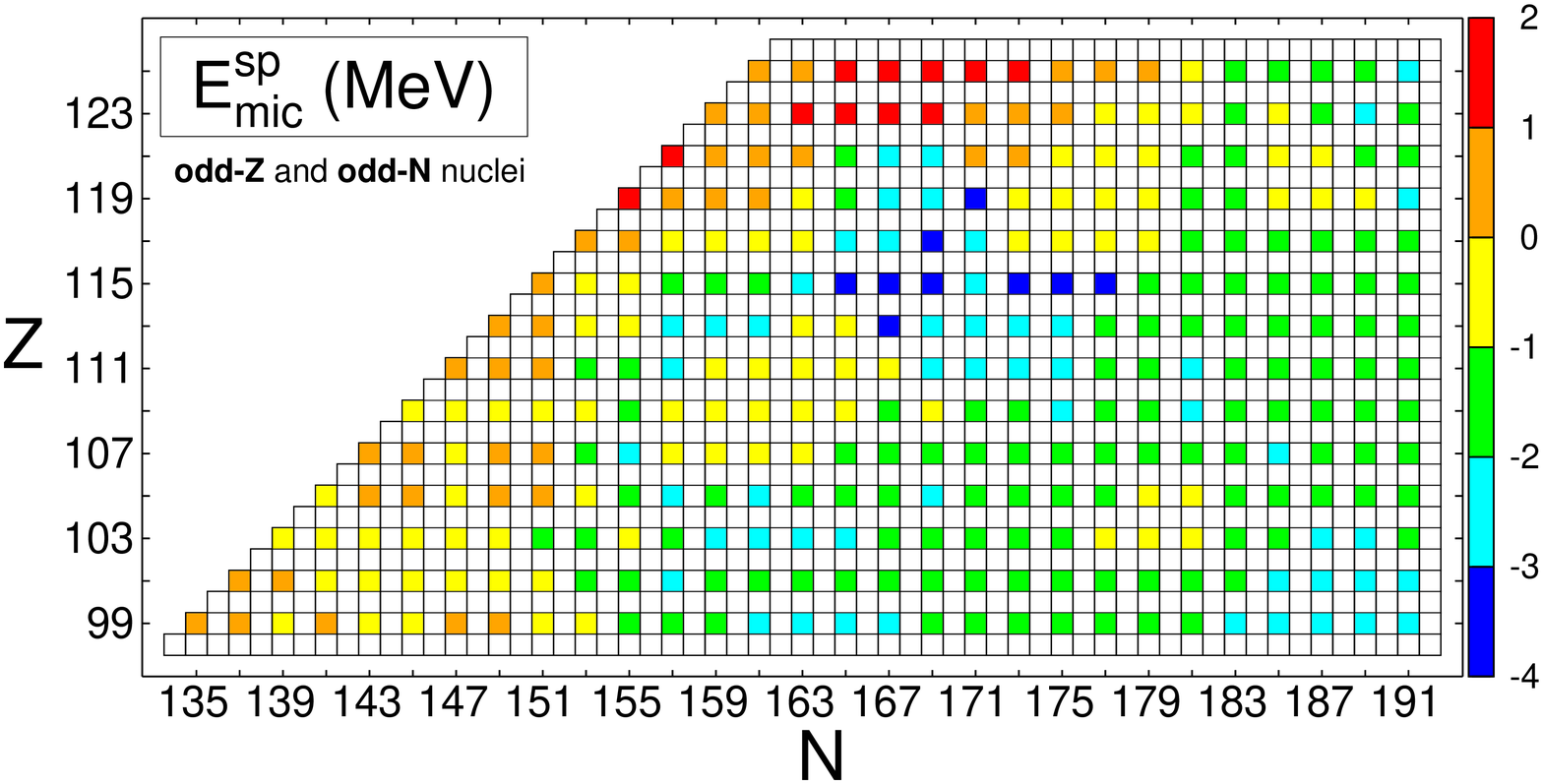}
\caption{As in Fig. \ref{fig2}, but for the calculated saddle points.}
\label{fig13}
\end{figure}

As in the case of the ground-state configuration, the macroscopic energy and
 the microscopic correction at the saddle-point configuration are shown in
 Fig. \ref{fig12} and \ref{fig13}, respectively.
Their numerical values are given in \mbox{Table 2}.
Generally, one can see that the shell effects at the saddle point are much
 weaker than in the ground state, but in many cases not negligible.
The most pronounced minima in $E_{mic}^{sp}$ (the absolute values of $\sim$
 3-4 MeV), were obtained in the vincinity of $Z=115$, $N=170$ and in the very
 exotic region $Z=100$, $N=190$. For the majority of
considered nuclei the microscopic energy at the saddle point is negative
 while the macroscopic one - positive. The main exception from this is the
 region of SDO nuclei.

\subsubsection{Saddle point shapes (deformations)}

\begin{figure}[h]
\centering
\includegraphics[scale=0.26]{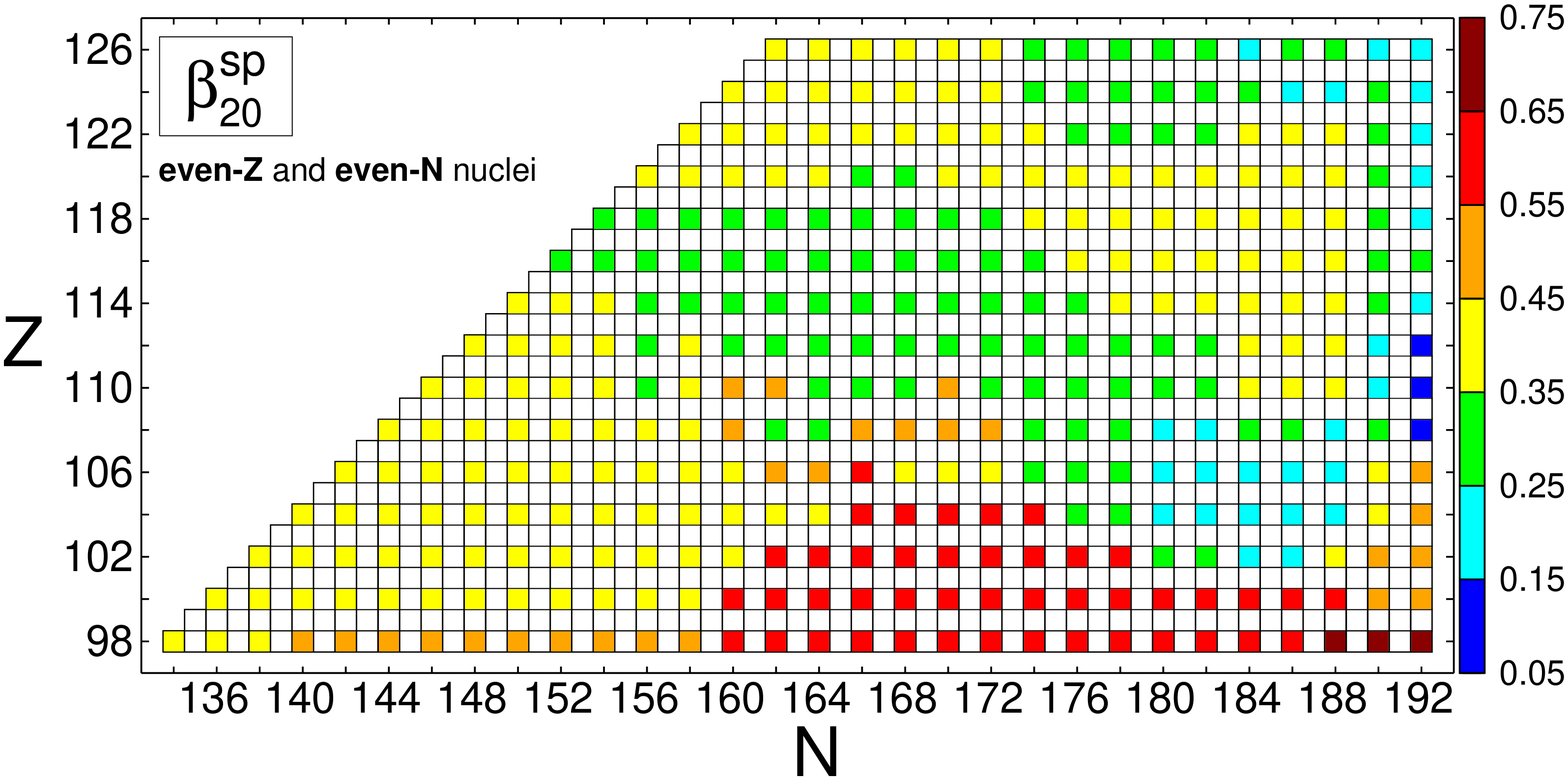}
\includegraphics[scale=0.26]{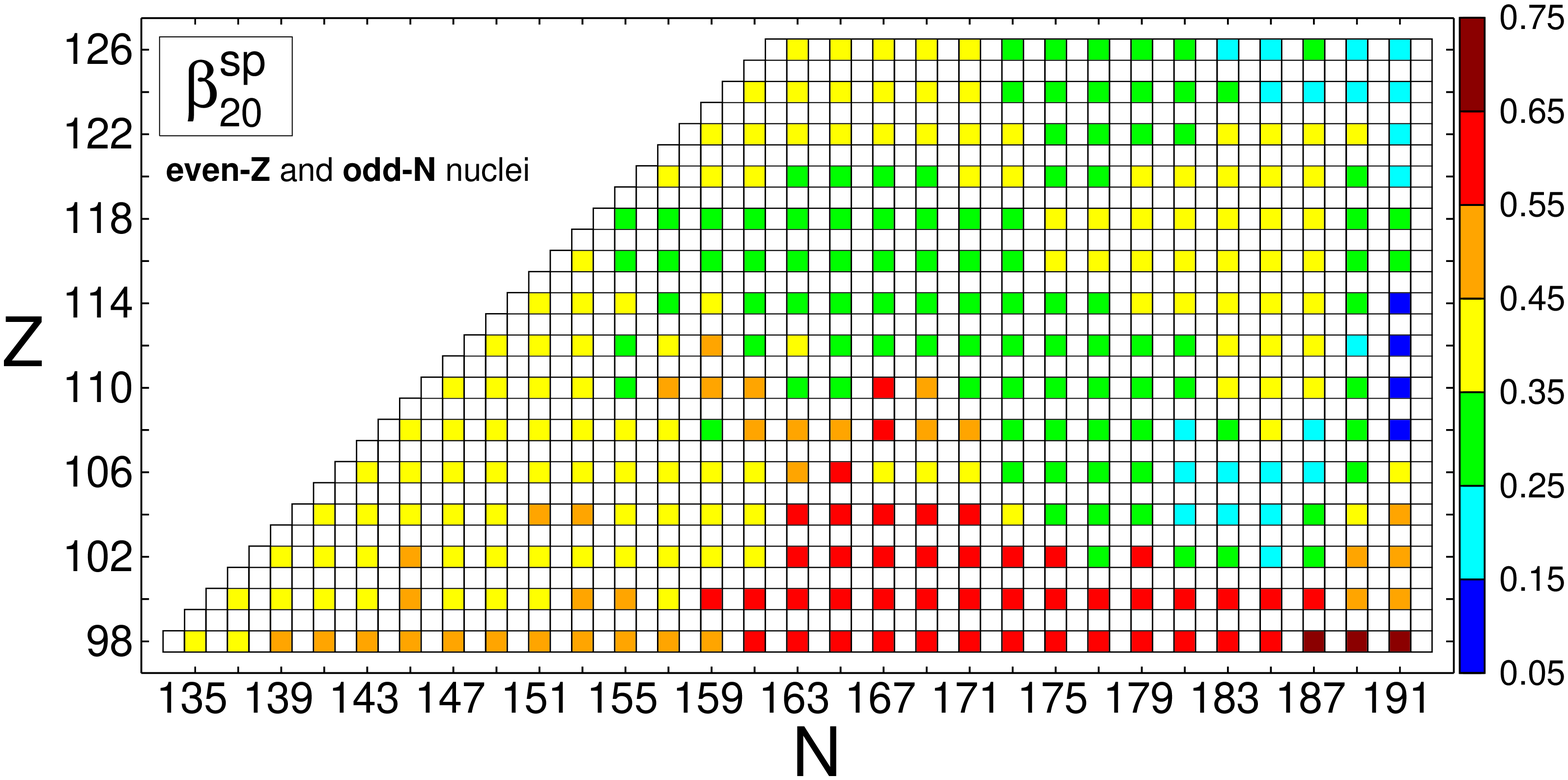}
\includegraphics[scale=0.26]{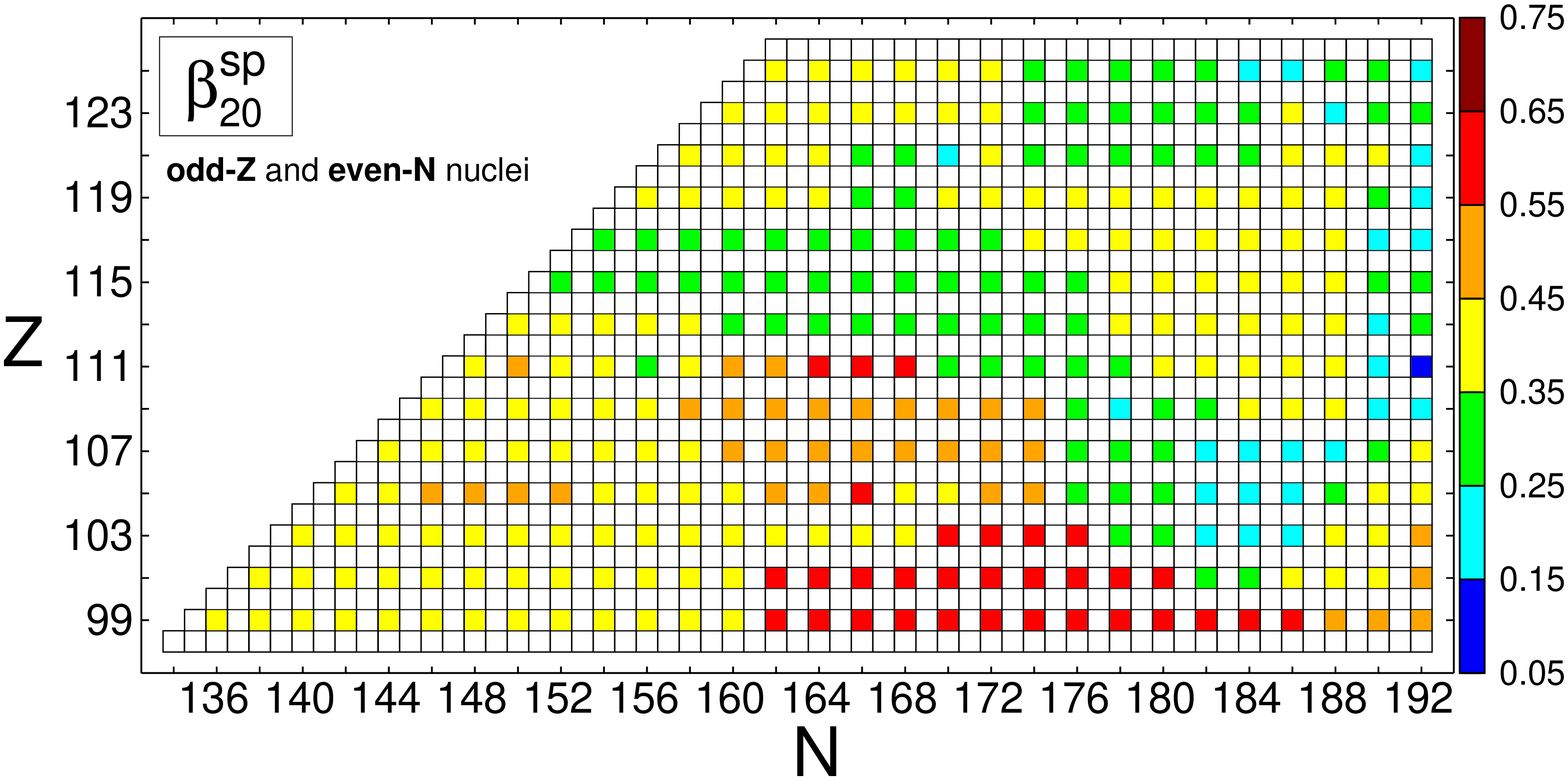}
\includegraphics[scale=0.26]{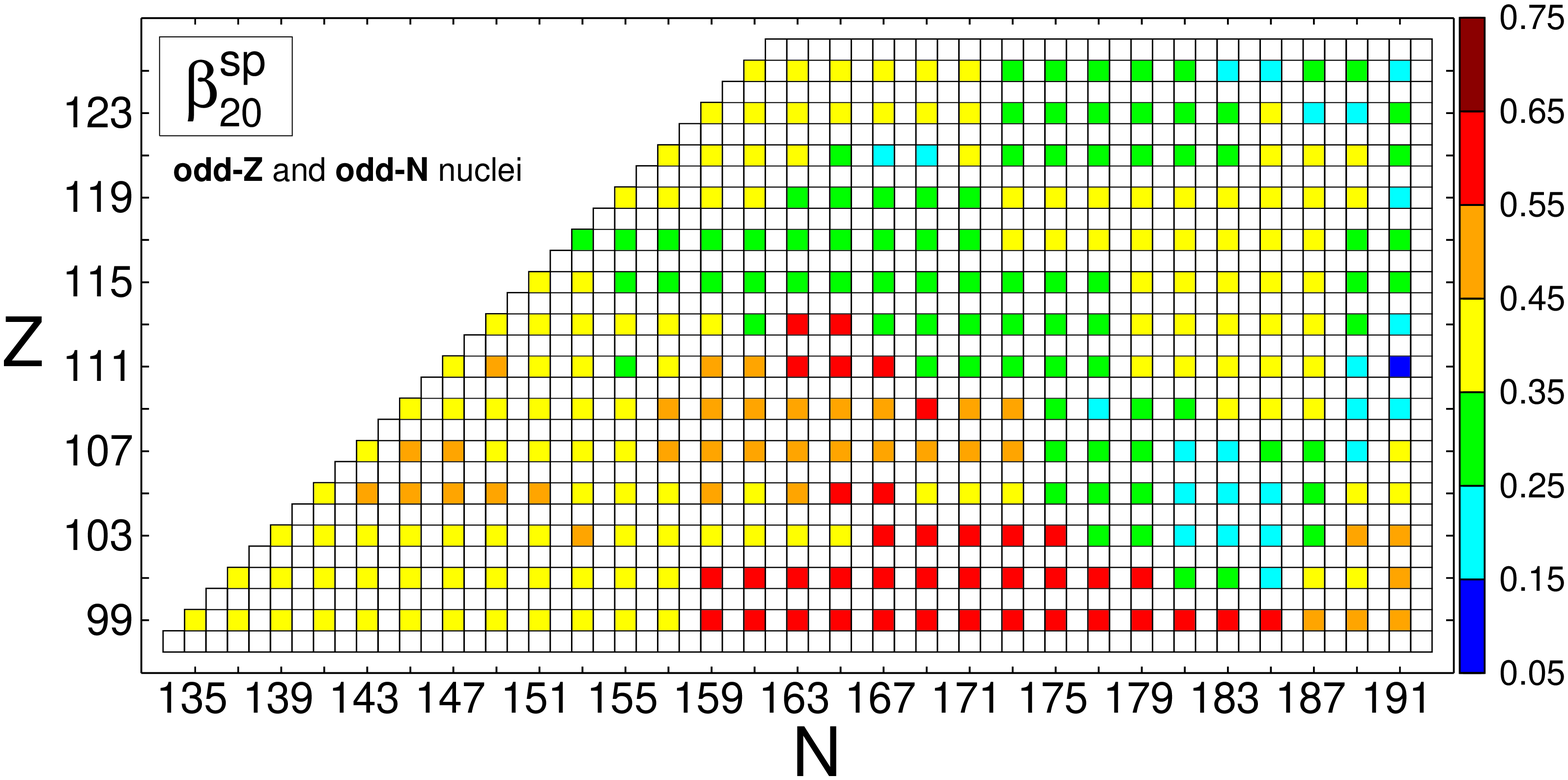}
\caption{As in Fig. \ref{fig4}, but for the calculated saddle points.}
\label{fig14}
\end{figure}

\begin{figure}[h]
\centering
\includegraphics[height=6cm]{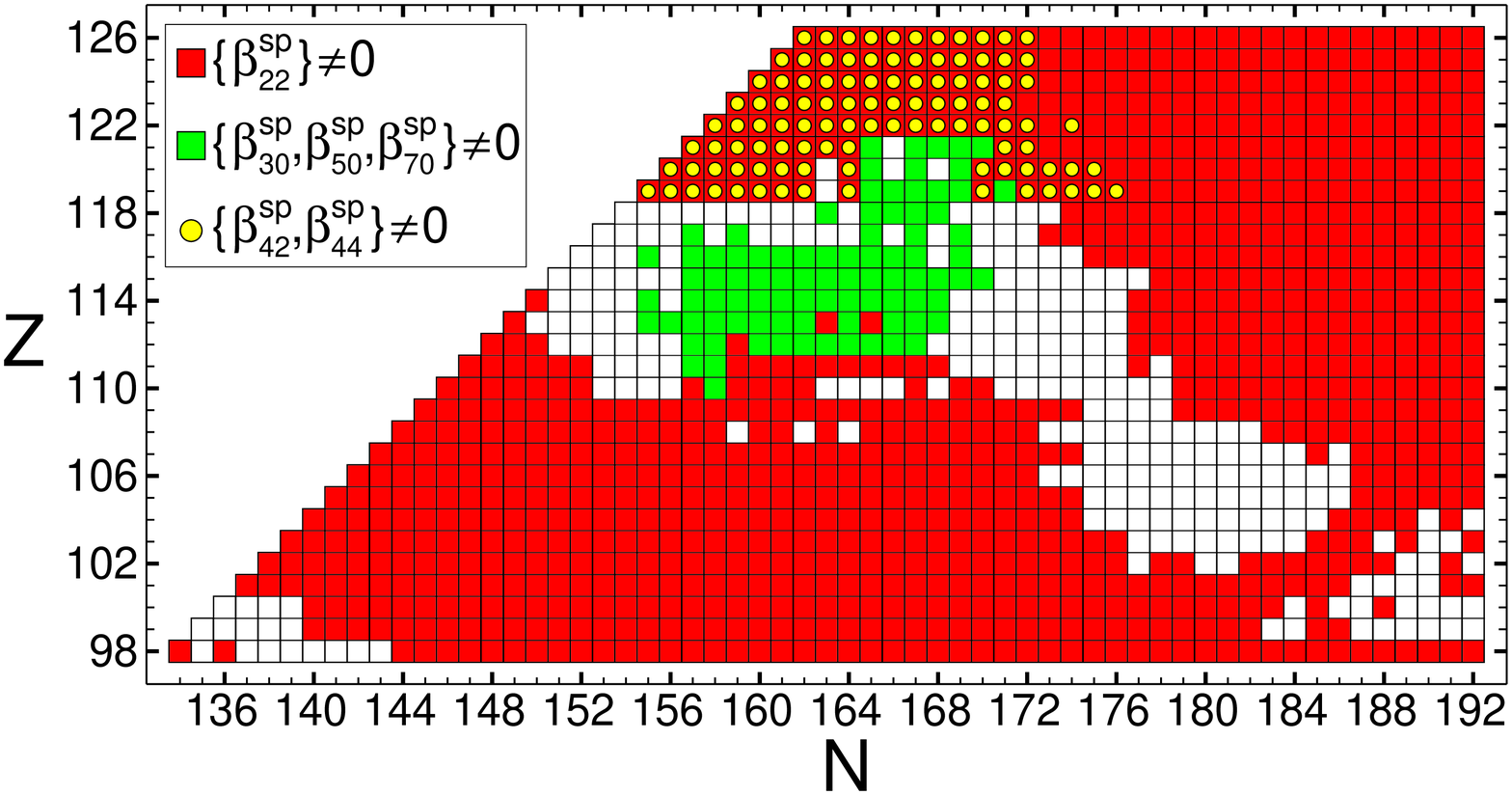}
\caption{Nuclei with $\beta_{22}$ deformation at the saddle point (red squares),
mass-asymmetric fission saddle point (green squares) and
hexadecapole-asymmetric fission saddle point (yellow circles).
White regions in the considered area denotes nuclei with axially and reflection symmetric saddle point.}
\label{fig15}
\end{figure}

The calculated saddle-point quadrupole deformations
$\beta^{sp}_{20}$ are shown in a $Z$ vs. $N$ plot in \mbox{Fig. \ref{fig14}}.
In the vast majority their values are below $0.65$
which corresponds to relatively short fission barriers
and justifies the used shape parametrization (expansion
of the nuclear radius vector onto spherical harmonics).
The calculated saddle-point shapes are mostly triaxial,
 i.e. about $74\%$ of the axially-symmetric saddle points
are noticeably lowered after the inclusion of the non-axial quadrupole
deformation parameter $\beta_{22}$. All nuclei with $Z \geqslant 122$
have exclusively triaxial saddle points.
Moreover, there are many triaxial saddles with rather sizable values of
 deformation $\beta_{22}$, up to $0.34$, see \mbox{Table 2}. This especially
 concerns the region of SDO nuclei where saddle points additionally have
  hexadecapole nonaxiality: $\beta_{42}$ and $\beta_{44}$, see \mbox{Fig. \ref{fig15}}.

Since the saddle-point deformation is a result of the competition
between axial and triaxial saddle-point energies, even their tiny difference may
result in an abrupt change in $\beta^{sp}_{22}$, and also in $\beta^{sp}_{20}$.
 This is indeed reflected in their variation between neighbouring isotopes -
 Table 2.
Stated otherwise: a large $\beta^{sp}_{22}$ does not necessarily correlate with
a large nonaxiality effect on the saddle-point energy (mass).

\subsubsection{Saddle point mass excesses - fission barrier heights}

\begin{figure}[h]
\centering
\includegraphics[scale=0.26]{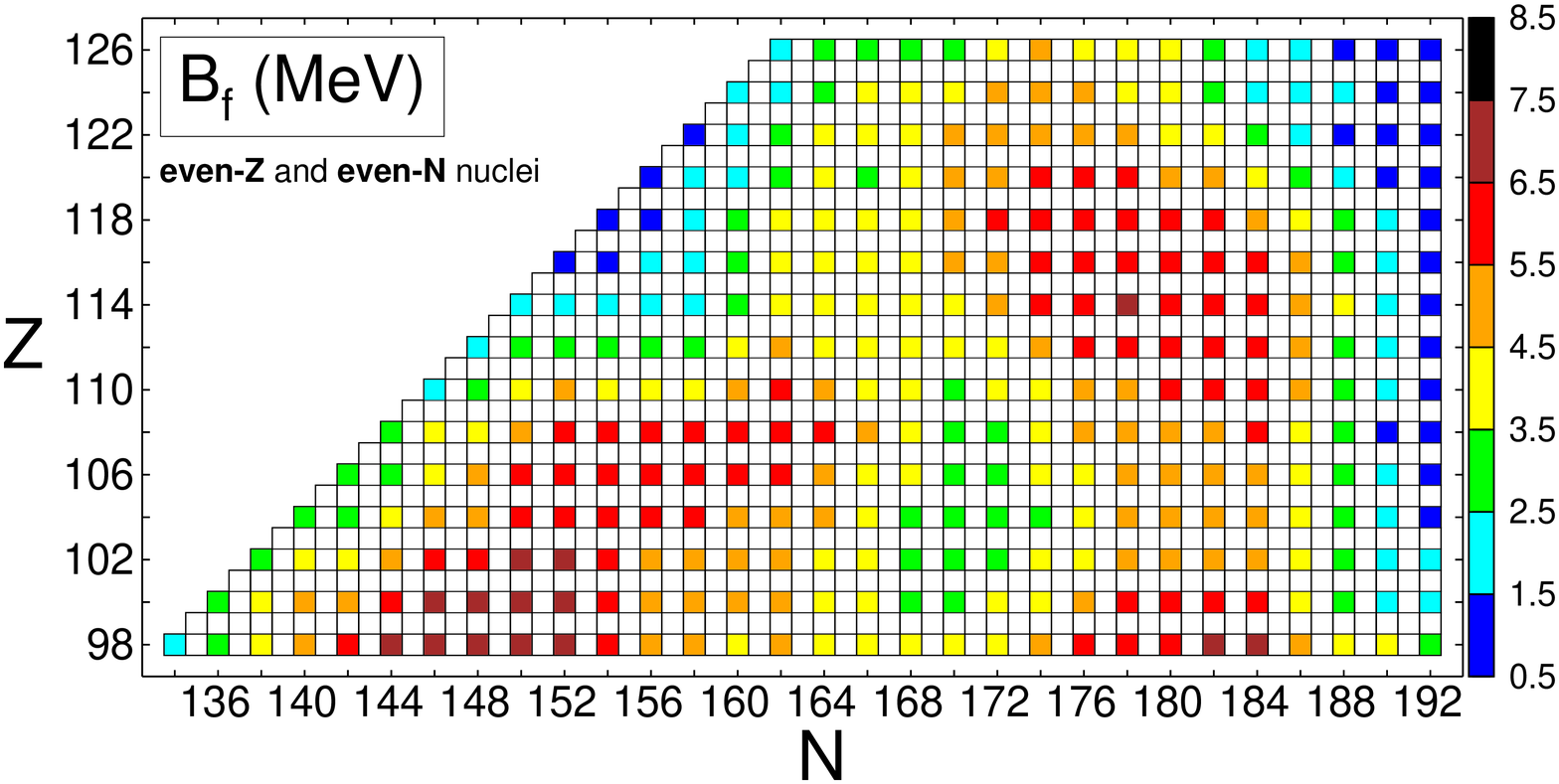}
\includegraphics[scale=0.26]{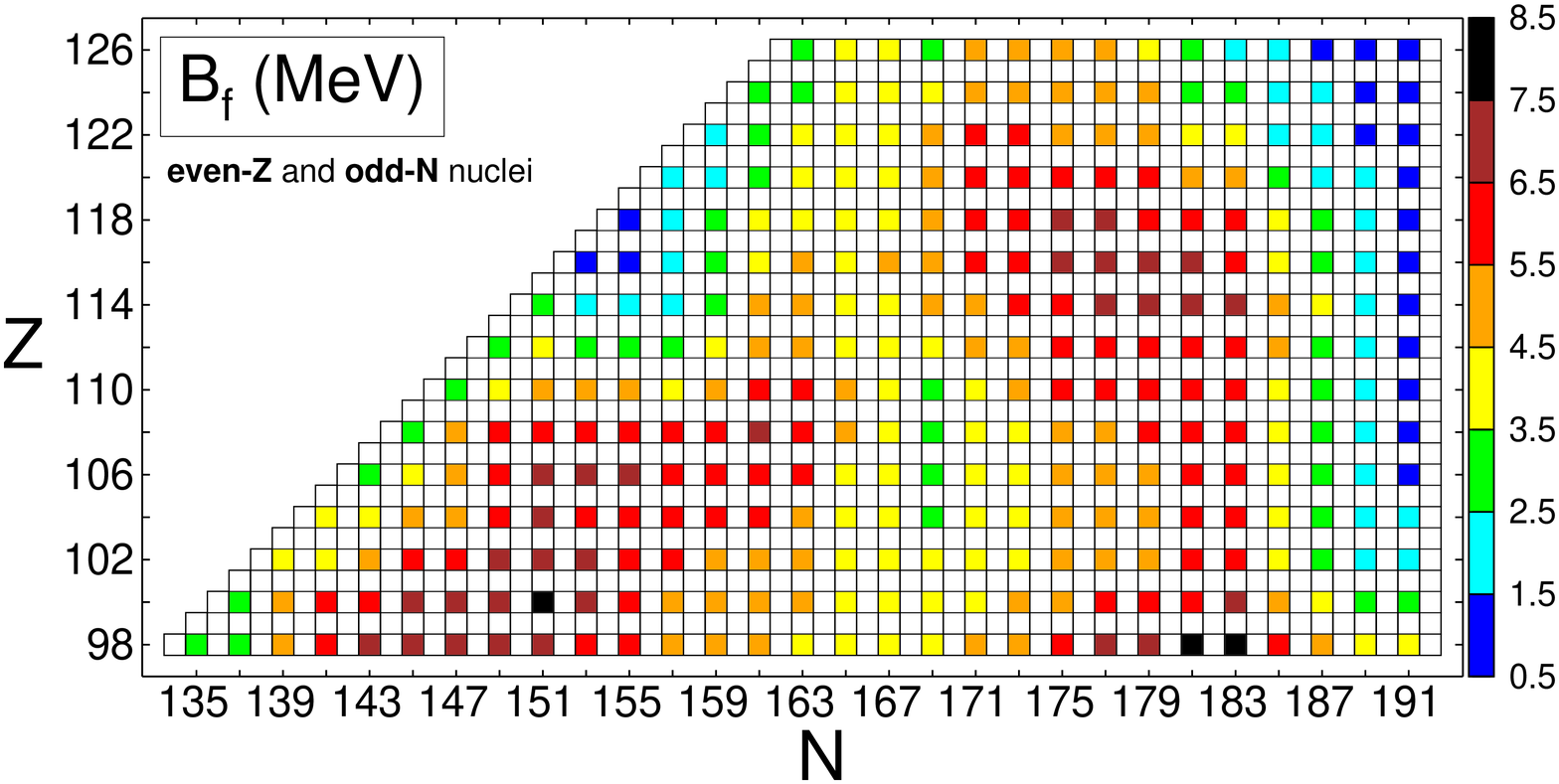}
\includegraphics[scale=0.26]{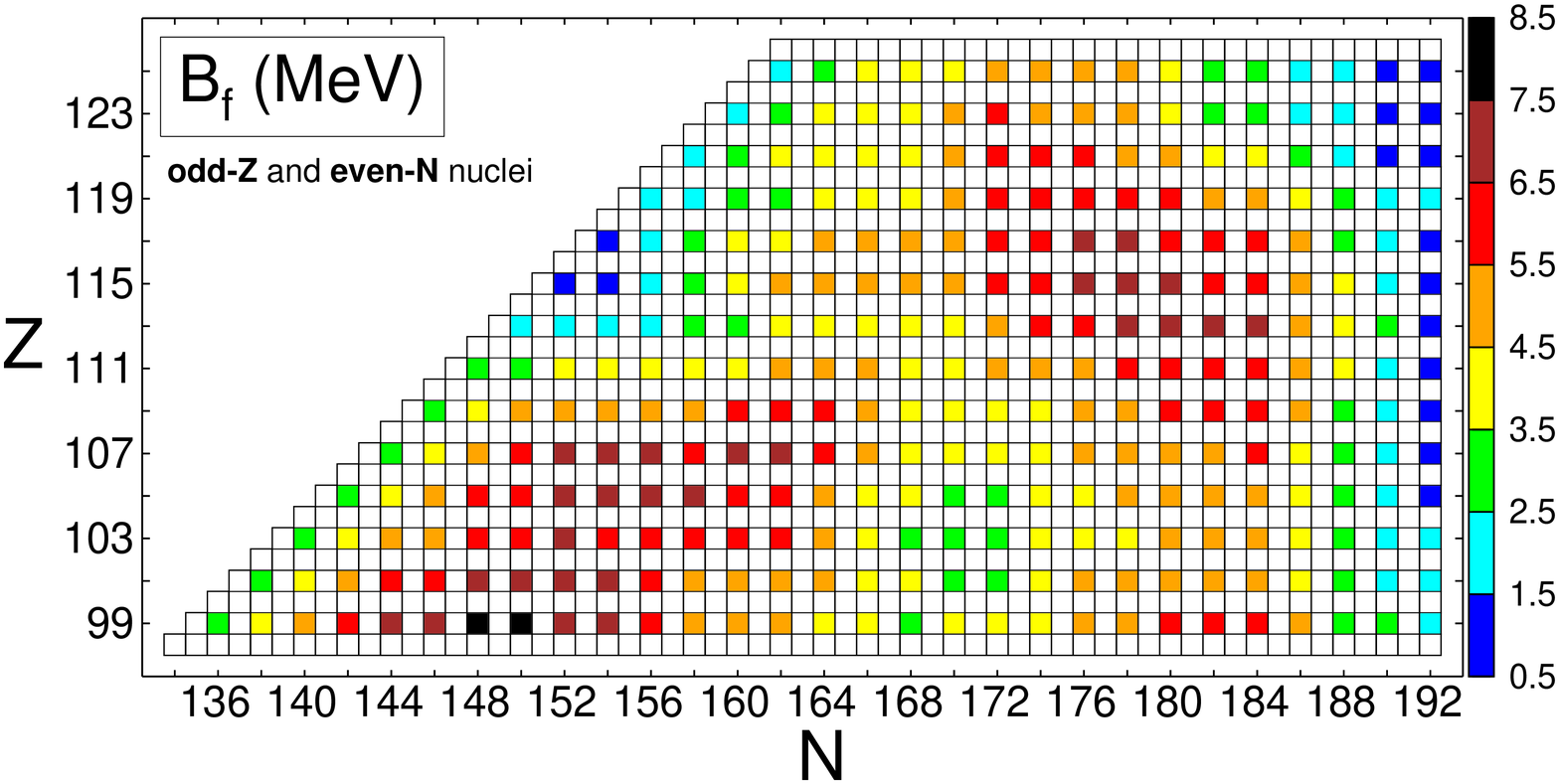}
\includegraphics[scale=0.26]{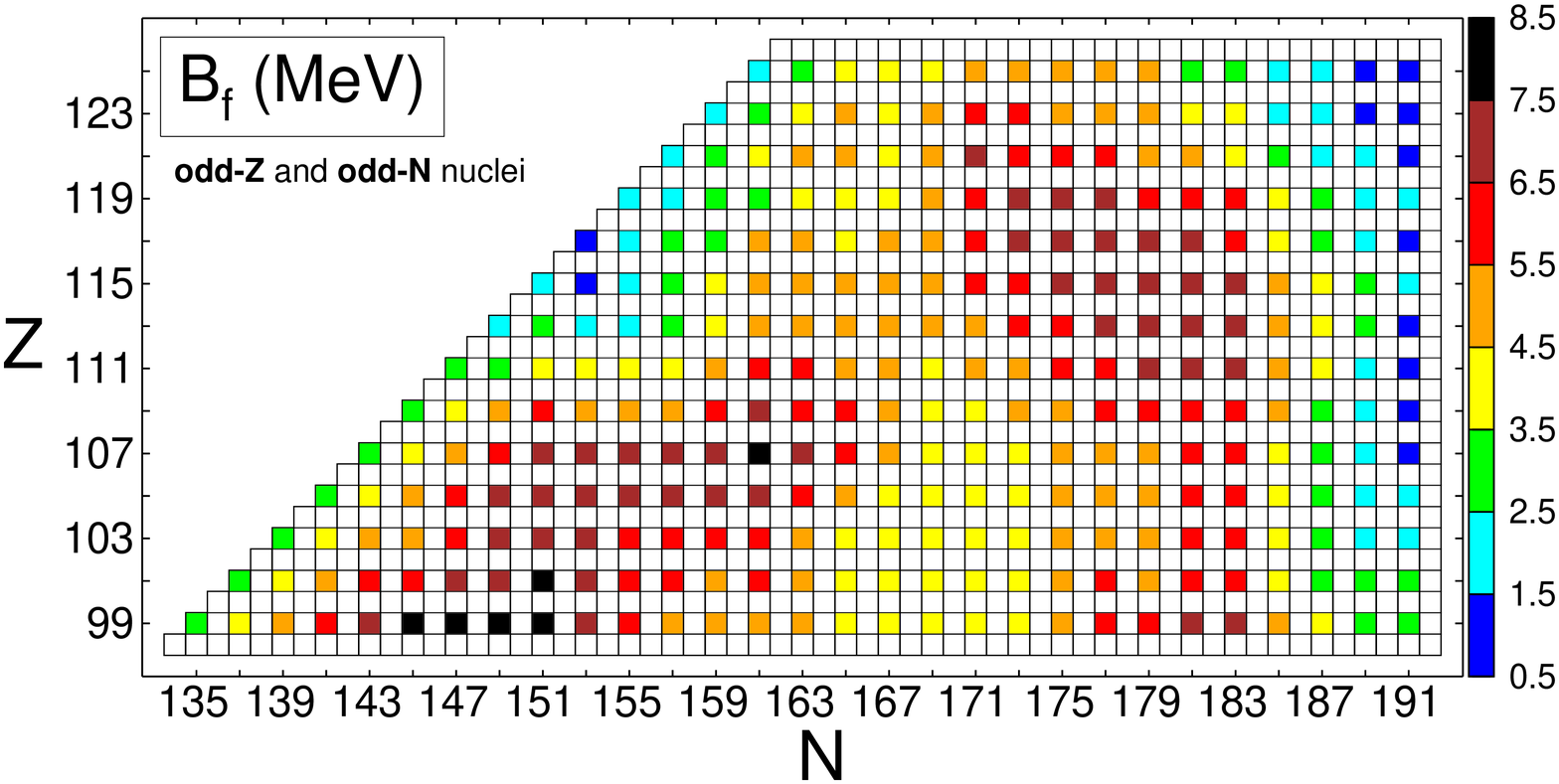}
\caption{Calculated fission-barrier heights $B_f$ (in MeV)
in $4$ separate groups of heavy and superheavy nuclei.}
\label{fig16}
\end{figure}

The fission barrier heights $B_f$, given in the last column of \mbox{Table 2}
 and shown in \mbox{Fig. \ref{fig16}}, are calculated by using the adiabatic
assumption, i.e., they are the smallest possible.
These values may be independently obtained from the \mbox{Tables 1 and 2} as:
\begin{equation}\label{figBf}
B_{f}=  E^{sp}_{tot}-E^{gs}_{tot}=M^{sp}_{th}-M^{gs}_{th}.
\end{equation}
 Such values of $B_f$ are relevant for fission
rates from excited states (thermal rates), with the excitation energy greater
 than the barrier itself. The spontaneous fission rates involve an additional
inertia effect, see e.g. \cite{Smolanczuk1995}.

 The comparisons of calculated $B_f$ values with other recent calculations and
 some empirical estimates are given and discussed in \cite{Kowal2010}, and
 especially in \cite{Brodz2015,Jachimowicz20172}.
 In \cite{Jachimowicz20172} we also discussed the odd-even
 staggering in the calculated fission barriers $B_f$. This effect is quite
 pronounced as can be seen in \mbox{Table 2}. It is related in part to a
 decrease in the BCS pairing gap due to blocking. Our tests of the pairing
 influence on barriers indicate that a possible overestimate of $B_f$ in
 odd-$A$ and odd-odd nuclei, induced by the blocking, should not be larger
 than $0.5$ MeV.

 One can see in \mbox{Fig. \ref{fig16}} that the calculated barrier heights
 $B_f$ in the considered region of nuclei are limited by $8.5$ MeV.
 There are visible maxima in $B_f$ corresponding to:
 ($Z=100$, $N=150$), ($Z=108$, $N=162$) and ($Z=114$, $N=178$), as well as the
 very exotic, beyond the neutron drip-line: ($Z=98$, N=$184$).
 The local minima in $B_f$, prominent for $Z\leq$112, occur near $N=$168 - 170.
  Small fission barriers are obtained at the boundary of the studied region,
 for the smallest and largest values of $N$.
 The first two maxima in $B_f$, related to the corresponding shell effects,
 are reflected in Fig. \ref{fig5} for isotopes of Md, Rf and Mt. Of those,
 fission was observed (experimental half-lives \cite{Hessberger2017} given
 in parentheses) in:
 $^{245}$Md (0.9 ms), $^{259}$Md (1 h 36 m), $^{256}$Rf (6.4 ms),
 $^{265}$Rf (1 m) and $^{277}$Mt (5 ms). Maps for Fl, Og
 and $Z=124$ in Fig. \ref{fig5} (of which $^{286}$Fl fissions with
 $T^{SF}_{1/2}=0.3$ s \cite{Hessberger2017}) indicate a shift of the maximum
 in $B_f(N)$ for larger $Z$: from $N=184$ for $Z\leq 110$ to
  $N\approx 178$ for Fl and Og and $N\approx 172$ for $Z=124$ -
  cf Fig. 9 - 18 in \cite{Jachimowicz20172}.

\section{Results in the actinide region}

 Additional results of our model and methodology are provided in a recent
 work \cite{Jachimowicz2020}
 where fission barrier heights and excitation energies of superdeformed
 isomeric minima in selected actinides were calculated within the above -
 described microscopic-macroscopic method.
 The calculations were made for 75 nuclei, from Ac to Cf,
 for which the experimental estimates of fission barrier heights are known.
 The comparison of our results with these data provides a  test of our model.
 Since most of the actinide nuclei are characterized by a double-humped
 barriers, our calculations were made in this case separately for the first
 and second barrier peaks.
 Determined in this way first and second fission barrier heights,
 as well as excitation energies of superdeformed (SD) minima, were given in
 \cite{Jachimowicz2020}, whereas detailed ground-state, first and second
 saddle, and isomeric minimum properties of those nuclei are collected in
 \mbox{Table \ref{GS_ACTIN} - \ref{SD_ACTIN}}, respectively.

 Properties of the first saddle point, on the basis of which our
 $B_f^{I(th)}$ values in \cite{Jachimowicz2020} were systematically determined,
 are contained in the \mbox{Table \ref{SP1_ACTIN}}.
 Their statistical comparison with available two sets of
 experimental estimates: \cite{Capote2009} and \cite{Smirenkin1993}, gives
 the rms deviation not greater than $0.85$ and \mbox{$0.94$ MeV}, respectively.
 Posisible sources of those discrepanties as well as possible way to their
 reduction were discussed in \cite{Jachimowicz2020}.

  Similar data, but for the second saddle points, are collected in
 \mbox{Table~\ref{SP2_ACTIN}}.
 It is worth mentioning that due to sizable elongations corresponding to the
 location of the second barrier, the range of deformations used in the
 search was significantly larger than specified by Eq. (\ref{7Dgrid}),
 reaching $\beta_{20}=1.5$ - see \cite{Jachimowicz2020} for details.
 Further, it has to be emphasized that in this calculation was introduced
 the additional deformation $\beta_{10}$, corresponding to the spherical
 harmonic $Y_{10}$. For each set of deformations $\beta_{20}$,...,$\beta_{80}$,
  the value of $\beta_{10}$ was fixed by requiring that the center of mass of
  the actual shape would be at zero - thus, in effect, one has
 $\beta_{10}(\beta_{20},...,\beta_{80})$.
  The second saddles in \mbox{Table~\ref{SP2_ACTIN}} and
 \cite{Jachimowicz2020} were found on the original
 grid, {\it without} any additional energy interpolation.
 The comparison between experimental estimates: \cite{Capote2009},
 \cite{Smirenkin1993}, and our second fission barrier heights $B_f^{II(th)}$
 gives the rms deviation: $0.82$ and $0.92$ MeV, respectively.
  The 100-200 keV differences between the present set of the second barriers
 and those for the even-even actinides, which we published previously in
 \cite{Jachimowicz2012}, follow form the present omission of the interpolation;
 the differences between saddle-point deformations here and in
 \cite{Jachimowicz2012} reflect the nonuniqeness of the
  parametrization of the reflection-asymmetric shape, which
 depends on whether $\beta_{10}$ or the translation of the whole shape are
  used to fix the center of mass - see \cite{Kowal2012}.

 The calculated properties of the isomeric (second) super-deformed minima
 in 75 actinide nuclei are given in \mbox{Table \ref{SD_ACTIN}}.
 A comparison of their calculated excitation energies for 28 nuclei for which
 experimental values were taken from \cite{Singh1996} gives the mean deviation
 and the rms error of 0.46 and 0.53 MeV, respectively \cite{Jachimowicz2020}.

\section{Summary}

Using the microscopic-macroscopic Woods-Saxon model we have calculated for the
 nuclei: $Z=98-126$ , $N=134-192$, the
 following ground-state and saddle-point properties:
 masses, macroscopic energies, shell corrections, and deformations;
 from those we derived: the g.s. to g.s. alpha decay energies,
 one- and two-nucleon separation energies and static, adiabatic fission
 barrier heights.


\ack Micha{\l} Kowal was co-financed by the National Science
Centre under Contract No. UMO-2013/08/M/ST2/00257  (LEA COPIGAL)
and by International Atomic Energy Agency (IAEA)
{\it "Recommended Input Parameter Library for Fission Cross Section Calculations (RIPL)"}-(F41033).
\clearpage

\section*{Table 1. Ground state properties}

For the isotopes of the elements $Z=98-126$, tabulates the g.s.
masses, total energies, macroscopic and microscopic
energies, equilibrium deformations, corresponding
$\alpha$-decay g.s. to g.s. $Q$-values
and nucleon separation energies. Note that the equilibrium deformations
  $\beta_{30}$, $\beta_{50}$ and $\beta_{70}$ are non-zero for
   relatively few nuclei.

\begin{center}
\begin{tabular}{lllllll}
$Z$&      The atomic number\\
$N$&      The neutron number\\
$A$&      The mass number\\
$M^{gs}_{th}$& The mass excess in MeV\\
$E^{gs}_{tot}$& The total energy in MeV \\
$E^{gs}_{mac}$& The macroscopic energy in MeV \\
$E^{gs}_{mic}$& The microscopic energy in MeV\\
$\beta^{gs}_{\lambda 0}$ & The g.s. equilibrium deformations\\
$Q^{th}_{\alpha}$& Theoretical $Q$ value of the $\alpha$ decay in MeV \\
$S^{th}_{p}$& Calculated one-proton separation energy in MeV  \\
$S^{th}_{n}$& Calculated one-neutron separation energy in MeV  \\
$S^{th}_{2p}$& Calculated two-proton separation energy in MeV  \\
$S^{th}_{2n}$& Calculated two-neutron separation energy in MeV  \\
\end{tabular}
\end{center}

\datatables
\footnotesize
\LTright=0pt \LTleft=-20pt

\begin{longtable}{|cccccccccccccccccc|}
\caption{Ground state properties of isotopes $Z=98-126$} \label{tab:A}
$N$ & $A$ & $M^{gs}_{th}$ & $E^{gs}_{tot}$ & $E^{gs}_{mac}$ &
$E^{gs}_{mic}$ & $\beta^{ gs}_{20}$ & $\beta^{ gs}_{30}$ &  $\beta^{gs}_{40}$ &
$\beta^{gs}_{50}$  & $\beta^{gs}_{60}$  &  $\beta^{gs}_{70}$  & $\beta^{gs}_{80}$
 &  $Q^{th}_{\alpha}$ & $S^{th}_{p}$ & $S^{th}_{n}$ & $S^{th}_{2p}$ & $S^{th}_{2n}$   \\
 &   & (MeV) & (MeV) & (MeV) & (MeV) &   &   &   &   &  &    &
 & (MeV) & (MeV) & (MeV) & (MeV) & (MeV)   \\ \hline
\multicolumn{18}{|c|}{} \\
\multicolumn{18}{|c|}{Z=98}  \\
 134 & 232 &  59.52 &  0.07  &  1.54  & -1.46 &  0.21 &  0.00 &  0.08 &  0.00 &  0.01 &  0.00 &  0.00 &  9.29 &  1.41 &  9.33 &  2.12 & 17.87 \\
 135 & 233 &  59.29 & -0.68  &  1.89  & -2.56 &  0.20 &  0.00 &  0.09 &  0.00 &  0.01 &  0.00 &  0.00 &  8.98 &  1.51 &  8.30 &  2.42 & 17.63 \\
 136 & 234 &  58.37 & -0.62  &  1.67  & -2.29 &  0.21 &  0.00 &  0.09 &  0.00 &  0.01 &  0.00 &  0.00 &  8.89 &  1.87 &  8.99 &  2.98 & 17.29 \\
 137 & 235 &  58.54 & -1.27  &  1.71  & -2.98 &  0.21 &  0.00 &  0.09 &  0.00 &  0.00 &  0.00 & -0.01 &  8.98 &  1.96 &  7.90 &  3.34 & 16.89 \\
\end{longtable}
\normalsize

\clearpage

\section*{Table 2. Saddle point properties}

For the isotopes of the elements $Z=98-126$, tabulates the saddle
point masses, total energies, macroscopic and microscopic
energies, saddle point deformations as well as static
fission barrier heights. Note that only for relatively few nuclei there
 occur non-zero saddle-point deformations $\beta_{30}$, $\beta_{50}$,
 $\beta_{70}$, or $\beta_{42}$, $\beta_{44}$.

\begin{center}
\begin{tabular}{lllllll}
$Z$& The atomic number\\
$N$& The neutron number\\
$A$& The mass number\\
$M^{sp}_{th}$& The mass excess in MeV\\
$E^{sp}_{th}$& The total energy in MeV \\
$E^{sp}_{mac}$& The macroscopic energy in MeV \\
$E^{sp}_{mic}$& The microscopic energy in MeV\\
$\beta^{sp}_{\lambda \mu}$&  The saddle point deformations\\
$B_f$& The calculated fission-barrier heights in MeV \\
\end{tabular}
\end{center}

\footnotesize

\begin{longtable}{|ccccccccccccccccc|}
\caption{Saddle point properties of isotopes $Z=98-126$}
\label{tab:B}
$N$ & $A$ & $M^{sp}_{th}$ & $E^{sp}_{tot}$ & $E^{sp}_{mac}$ & $E^{sp}_{mic}$ & $\beta^{sp}_{20}$ & $\beta^{sp}_{22}$ &
$\beta^{sp}_{30}$ & $\beta^{sp}_{40}$ & $\beta^{sp}_{42}$ & $\beta^{sp}_{44}$ & $\beta^{sp}_{50}$  &  $\beta^{sp}_{60}$
& $\beta^{sp}_{70}$  &  $\beta^{sp}_{80}$ & $B_{f}$  \\
  &   & (MeV) & (MeV) & (MeV) & (MeV) &   &   &   &   &   &   &   &   &   &   & (MeV)    \\ \hline
\multicolumn{17}{|c|}{} \\
\multicolumn{17}{|c|}{Z=98}  \\
 134 & 232 &  61.80 &  2.35  &  1.68  &  0.67  &  0.39 &  0.04 &  0.00 & -0.02 &  0.00 &  0.00 &  0.00 & -0.01 &  0.00 & -0.02 &  2.28 \\
 135 & 233 &  62.03 &  2.06  &  1.60  &  0.47  &  0.38 &  0.00 &  0.00 & -0.01 &  0.00 &  0.00 &  0.00 & -0.02 &  0.00 & -0.02 &  2.74 \\
 136 & 234 &  61.20 &  2.20  &  1.83  &  0.37  &  0.39 &  0.03 &  0.00 & -0.02 &  0.00 &  0.00 &  0.00 & -0.02 &  0.00 & -0.02 &  2.83 \\
 137 & 235 &  61.99 &  2.18  &  2.30  & -0.12  &  0.43 &  0.00 &  0.00 & -0.04 &  0.00 &  0.00 &  0.00 & -0.02 &  0.00 & -0.01 &  3.45 \\
\hline
\end{longtable}
\normalsize

\clearpage

\section*{Tables 3-6: Static fission properties of selected 75 actinide nuclei with $Z=89-98$}

\noindent
Properties of the ground state minima for nuclei investigated in \cite{Jachimowicz2020}, specified by
$Z$, $N$ (and $A$). $M^{gs}_{th}$ is the calculated mass excess, $E^{ gs}_{tot}$ is the total energy normalized to
the macroscopic energy at the spherical shape. $E^{ gs}_{mac}$ and $E^{gs}_{mic}$
are the macroscopic and microscopic parts of the total energy $E^{gs}_{tot}$, respectively.
$\beta^{gs}_{\lambda 0}$ are values of equilibrium deformations of g.s. minimum.
\begin{longtable}{|cccccccccccccc|}
\caption{Ground state properties of selected actinides $Z=89-98$} \label{GS_ACTIN}
$Z$ & $N$ & $A$ & $M^{gs}_{th}$ & $E^{gs}_{tot}$ & $E^{gs}_{mac}$ &
$E^{gs}_{mic}$ & $\beta^{gs}_{20}$ & $\beta^{gs}_{30}$ &  $\beta^{ gs}_{40}$ &
$\beta^{gs}_{50}$  & $\beta^{gs}_{60}$  &  $\beta^{gs}_{70}$  & $\beta^{gs}_{80}$ \\
  &  &  & (MeV) & (MeV) & (MeV) & (MeV) &   &   &   &   &  &    &  \\ \hline
89 &  137 &  226 & 23.58 & -3.02 & 3.46 & -6.48 &  0.16 &  0.10 &  0.10 &  0.04 &  0.02 &  0.00 & -0.02  \\
89 &  138 &  227 & 25.48 & -2.55 & 3.26 & -5.81 &  0.16 &  0.09 &  0.09 &  0.03 &  0.02 &  0.00 & -0.01  \\
89 &  139 &  228 & 28.25 & -2.72 & 3.13 & -5.86 &  0.18 &  0.02 &  0.11 &  0.00 &  0.03 &  0.00 & -0.02  \\
\hline
90 &  137 &  227 & 25.15 & -2.53 & 3.41 & -5.94 &  0.17 &  0.10 &  0.09 &  0.04 &  0.02 &  0.00 & -0.02  \\
90 &  138 &  228 & 26.62 & -2.13 & 3.13 & -5.26 &  0.18 &  0.00 &  0.11 &  0.00 &  0.03 &  0.00 & -0.01  \\
90 &  139 &  229 & 29.06 & -2.55 & 3.10 & -5.65 &  0.19 & -0.02 &  0.11 & -0.01 &  0.03 &  0.00 & -0.02  \\
90 &  140 &  230 & 30.71 & -2.25 & 2.91 & -5.16 &  0.19 &  0.00 &  0.11 &  0.00 &  0.02 &  0.00 & -0.02  \\
90 &  141 &  231 & 33.33 & -2.76 & 2.81 & -5.57 &  0.20 &  0.00 &  0.10 &  0.00 &  0.01 &  0.00 & -0.02  \\
90 &  142 &  232 & 35.30 & -2.40 & 2.67 & -5.07 &  0.20 &  0.00 &  0.10 &  0.00 &  0.01 &  0.00 & -0.02  \\
90 &  143 &  233 & 38.35 & -2.74 & 2.69 & -5.44 &  0.21 &  0.00 &  0.10 &  0.00 &  0.00 &  0.00 & -0.02  \\
90 &  144 &  234 & 40.47 & -2.51 & 2.50 & -5.01 &  0.21 &  0.00 &  0.09 &  0.00 &  0.00 &  0.00 & -0.02  \\
\hline
91 &  139 &  230 & 31.01 & -3.47 & 3.37 & -6.84 &  0.19 &  0.00 &  0.12 &  0.00 &  0.03 &  0.00 & -0.02  \\
91 &  140 &  231 & 32.60 & -3.14 & 3.21 & -6.34 &  0.20 &  0.00 &  0.11 &  0.00 &  0.02 &  0.00 & -0.02  \\
91 &  141 &  232 & 34.85 & -3.67 & 3.14 & -6.81 &  0.20 &  0.00 &  0.11 &  0.00 &  0.02 &  0.00 & -0.02  \\
91 &  142 &  233 & 36.80 & -3.25 & 3.02 & -6.27 &  0.20 &  0.00 &  0.11 &  0.00 &  0.01 &  0.00 & -0.02  \\
91 &  143 &  234 & 39.55 & -3.55 & 2.97 & -6.53 &  0.21 &  0.00 &  0.11 &  0.00 &  0.01 &  0.00 & -0.02  \\
\hline
92 &  139 &  231 & 33.07 & -2.97 & 3.29 & -6.26 &  0.20 &  0.01 &  0.12 &  0.00 &  0.03 &  0.00 & -0.02  \\
92 &  140 &  232 & 34.30 & -2.65 & 3.10 & -5.75 &  0.20 &  0.00 &  0.11 &  0.00 &  0.02 &  0.00 & -0.02  \\
92 &  141 &  233 & 36.39 & -3.26 & 2.95 & -6.21 &  0.21 &  0.00 &  0.11 &  0.00 &  0.01 &  0.00 & -0.02  \\
92 &  142 &  234 & 37.98 & -2.86 & 2.82 & -5.68 &  0.21 &  0.00 &  0.11 &  0.00 &  0.01 &  0.00 & -0.02  \\
92 &  143 &  235 & 40.57 & -3.23 & 2.84 & -6.06 &  0.21 &  0.00 &  0.11 &  0.00 &  0.00 &  0.00 & -0.02  \\
92 &  144 &  236 & 42.28 & -2.97 & 2.54 & -5.51 &  0.21 &  0.00 &  0.10 &  0.00 &  0.00 &  0.00 & -0.02  \\
92 &  145 &  237 & 45.03 & -3.44 & 2.66 & -6.10 &  0.23 &  0.00 &  0.09 &  0.00 & -0.01 &  0.00 & -0.03  \\
92 &  146 &  238 & 47.15 & -3.04 & 2.47 & -5.50 &  0.23 &  0.00 &  0.09 &  0.00 & -0.01 &  0.00 & -0.03  \\
92 &  147 &  239 & 50.23 & -3.43 & 2.36 & -5.79 &  0.23 &  0.00 &  0.08 &  0.00 & -0.02 &  0.00 & -0.03  \\
92 &  148 &  240 & 52.66 & -2.97 & 2.09 & -5.06 &  0.23 &  0.00 &  0.08 &  0.00 & -0.02 &  0.00 & -0.02  \\
\hline
93 &  140 &  233 & 37.53 & -3.13 & 3.26 & -6.39 &  0.21 &  0.04 &  0.11 &  0.01 &  0.02 & -0.01 & -0.03  \\
93 &  141 &  234 & 39.25 & -3.75 & 3.07 & -6.83 &  0.21 &  0.02 &  0.11 &  0.00 &  0.01 & -0.01 & -0.03  \\
93 &  142 &  235 & 40.70 & -3.40 & 3.01 & -6.41 &  0.21 &  0.03 &  0.11 &  0.00 &  0.01 & -0.01 & -0.03  \\
93 &  143 &  236 & 42.89 & -3.83 & 3.02 & -6.86 &  0.22 &  0.02 &  0.11 &  0.00 &  0.00 & -0.01 & -0.03  \\
93 &  144 &  237 & 44.48 & -3.61 & 3.04 & -6.65 &  0.22 &  0.04 &  0.10 &  0.01 &  0.00 & -0.02 & -0.03  \\
93 &  145 &  238 & 46.70 & -4.28 & 3.32 & -7.60 &  0.23 &  0.06 &  0.10 &  0.01 & -0.01 & -0.02 & -0.04  \\
93 &  146 &  239 & 48.86 & -3.74 & 3.38 & -7.12 &  0.23 &  0.06 &  0.10 &  0.01 & -0.01 & -0.02 & -0.04  \\
\hline
94 &  141 &  235 & 41.78 & -3.28 & 2.64 & -5.91 &  0.21 &  0.00 &  0.10 &  0.00 &  0.01 &  0.00 & -0.02  \\
94 &  142 &  236 & 42.88 & -2.93 & 2.54 & -5.47 &  0.22 &  0.01 &  0.10 &  0.00 &  0.00 &  0.00 & -0.02  \\
94 &  143 &  237 & 44.95 & -3.40 & 2.58 & -5.98 &  0.22 &  0.00 &  0.10 &  0.00 &  0.00 &  0.00 & -0.02  \\
94 &  144 &  238 & 46.16 & -3.21 & 2.41 & -5.62 &  0.22 &  0.00 &  0.09 &  0.00 & -0.01 &  0.00 & -0.02  \\
94 &  145 &  239 & 48.30 & -3.87 & 2.96 & -6.83 &  0.23 &  0.06 &  0.09 &  0.00 & -0.01 & -0.02 & -0.03  \\
94 &  146 &  240 & 50.06 & -3.40 & 2.34 & -5.74 &  0.23 &  0.00 &  0.09 &  0.00 & -0.02 &  0.00 & -0.03  \\
94 &  147 &  241 & 52.66 & -3.87 & 2.21 & -6.08 &  0.24 &  0.00 &  0.08 &  0.00 & -0.02 &  0.00 & -0.02  \\
94 &  148 &  242 & 54.65 & -3.42 & 1.96 & -5.38 &  0.23 &  0.00 &  0.07 &  0.00 & -0.02 &  0.00 & -0.02  \\
94 &  149 &  243 & 57.66 & -3.72 & 1.63 & -5.36 &  0.24 &  0.00 &  0.06 &  0.00 & -0.03 &  0.00 & -0.01  \\
94 &  150 &  244 & 59.80 & -3.38 & 1.60 & -4.98 &  0.24 &  0.00 &  0.05 &  0.00 & -0.03 &  0.00 & -0.01  \\
94 &  151 &  245 & 62.88 & -3.87 & 1.57 & -5.44 &  0.24 &  0.00 &  0.04 &  0.00 & -0.04 &  0.00 &  0.00  \\
94 &  152 &  246 & 65.43 & -3.36 & 1.54 & -4.90 &  0.24 &  0.00 &  0.04 &  0.00 & -0.04 &  0.00 &  0.00  \\
\hline
95 &  144 &  239 & 49.42 & -3.71 & 2.16 & -5.87 &  0.23 &  0.00 &  0.09 &  0.00 & -0.01 &  0.00 & -0.02  \\
95 &  145 &  240 & 51.17 & -4.41 & 2.52 & -6.94 &  0.23 &  0.05 &  0.08 &  0.00 & -0.02 & -0.01 & -0.03  \\
95 &  146 &  241 & 52.81 & -3.98 & 2.16 & -6.14 &  0.24 &  0.00 &  0.08 &  0.00 & -0.02 &  0.00 & -0.02  \\
95 &  147 &  242 & 55.02 & -4.49 & 2.10 & -6.59 &  0.24 &  0.00 &  0.07 &  0.00 & -0.03 &  0.00 & -0.02  \\
95 &  148 &  243 & 56.94 & -4.05 & 1.91 & -5.96 &  0.24 &  0.00 &  0.07 &  0.00 & -0.03 &  0.00 & -0.02  \\
95 &  149 &  244 & 59.59 & -4.37 & 1.64 & -6.01 &  0.24 &  0.00 &  0.06 &  0.00 & -0.03 &  0.00 & -0.01  \\
95 &  150 &  245 & 61.66 & -4.03 & 1.61 & -5.64 &  0.24 &  0.00 &  0.05 &  0.00 & -0.03 &  0.00 & -0.01  \\
95 &  151 &  246 & 64.42 & -4.49 & 1.58 & -6.07 &  0.24 &  0.00 &  0.04 &  0.00 & -0.04 &  0.00 &  0.00  \\
95 &  152 &  247 & 66.92 & -3.96 & 1.55 & -5.51 &  0.24 &  0.00 &  0.04 &  0.00 & -0.04 &  0.00 &  0.00  \\
\hline
96 &  145 &  241 & 53.65 & -4.03 & 2.26 & -6.29 &  0.24 & -0.03 &  0.08 &  0.00 & -0.02 &  0.01 & -0.03  \\
96 &  146 &  242 & 54.88 & -3.67 & 2.02 & -5.69 &  0.24 &  0.00 &  0.08 &  0.00 & -0.02 &  0.00 & -0.02  \\
96 &  147 &  243 & 56.99 & -4.21 & 1.97 & -6.18 &  0.24 &  0.00 &  0.07 &  0.00 & -0.03 &  0.00 & -0.02  \\
96 &  148 &  244 & 58.51 & -3.81 & 1.78 & -5.59 &  0.24 &  0.00 &  0.07 &  0.00 & -0.03 &  0.00 & -0.02  \\
96 &  149 &  245 & 61.01 & -4.22 & 1.57 & -5.78 &  0.24 &  0.00 &  0.05 &  0.00 & -0.03 &  0.00 & -0.01  \\
96 &  150 &  246 & 62.72 & -3.89 & 1.54 & -5.43 &  0.24 &  0.00 &  0.05 &  0.00 & -0.03 &  0.00 & -0.01  \\
96 &  151 &  247 & 65.29 & -4.48 & 1.55 & -6.02 &  0.24 &  0.00 &  0.04 &  0.00 & -0.04 &  0.00 &  0.00  \\
96 &  152 &  248 & 67.44 & -3.96 & 1.53 & -5.49 &  0.24 &  0.00 &  0.03 &  0.00 & -0.04 &  0.00 &  0.00  \\
96 &  153 &  249 & 70.94 & -3.87 & 1.66 & -5.53 &  0.24 &  0.00 &  0.02 &  0.00 & -0.05 &  0.00 &  0.01  \\
96 &  154 &  250 & 73.04 & -3.65 & 1.46 & -5.11 &  0.24 &  0.00 &  0.03 &  0.00 & -0.04 &  0.00 &  0.01  \\
\hline
97 &  147 &  244 & 60.36 & -4.72 & 1.92 & -6.64 &  0.24 &  0.00 &  0.07 &  0.00 & -0.03 &  0.00 & -0.02  \\
97 &  148 &  245 & 61.79 & -4.35 & 1.71 & -6.06 &  0.24 &  0.00 &  0.06 &  0.00 & -0.03 &  0.00 & -0.01  \\
97 &  149 &  246 & 63.88 & -4.82 & 1.57 & -6.40 &  0.24 &  0.00 &  0.05 &  0.00 & -0.04 &  0.00 &  0.00  \\
97 &  150 &  247 & 65.51 & -4.50 & 1.56 & -6.07 &  0.24 &  0.00 &  0.05 &  0.00 & -0.04 &  0.00 &  0.00  \\
97 &  151 &  248 & 67.66 & -5.17 & 1.60 & -6.77 &  0.25 &  0.00 &  0.03 &  0.00 & -0.04 &  0.00 &  0.00  \\
97 &  152 &  249 & 69.77 & -4.63 & 1.60 & -6.23 &  0.25 &  0.00 &  0.03 &  0.00 & -0.04 &  0.00 &  0.00  \\
97 &  153 &  250 & 72.92 & -4.54 & 1.70 & -6.24 &  0.25 &  0.00 &  0.02 &  0.00 & -0.05 &  0.00 &  0.01  \\
\hline
98 &  152 &  250 & 71.28 & -4.53 & 1.58 & -6.11 &  0.25 &  0.00 &  0.03 &  0.00 & -0.05 &  0.00 &  0.00  \\
98 &  153 &  251 & 74.34 & -4.46 & 1.72 & -6.18 &  0.25 &  0.00 &  0.02 &  0.00 & -0.05 &  0.00 &  0.01  \\
98 &  154 &  252 & 76.07 & -4.21 & 1.53 & -5.74 &  0.25 &  0.00 &  0.02 &  0.00 & -0.05 &  0.00 &  0.01  \\
98 &  155 &  253 & 79.41 & -4.12 & 1.74 & -5.86 &  0.25 &  0.00 &  0.01 &  0.00 & -0.05 &  0.00 &  0.02  \\
\hline
\end{longtable}

\clearpage
\noindent
Properties of the first saddle point $sp(I)$ for the selected actinides
 specified by
$Z$, $N$, investigated in \cite{Jachimowicz2020}.
$M^{sp(I)}_{th}$ is the calculated mass excess, $E^{sp(I)}_{tot}$ is the total energy normalized to
the macroscopic energy at the spherical shape. $E^{sp(I)}_{mic}$ and $E^{sp(I)}_{mac}$
are the microscopic and macroscopic parts of the total energy $E^{sp(I)}_{tot}$, respectively.
$\beta^{sp(I)}_{\lambda \mu}$ are first saddle-point deformation parameters.

\begin{longtable}{|ccccccccccc|}
\caption{First saddle point properties of selected actinides $Z=89-98$} \label{SP1_ACTIN}
$Z$ & $N$ & $M^{sp(I)}_{th}$ & $E^{sp(I)}_{tot}$ & $E^{sp(I)}_{mac}$ & $E^{sp(I)}_{mic}$ &
$\beta^{sp(I)}_{20}$ & $\beta^{sp(I)}_{22}$  & $\beta^{sp(I)}_{40}$ & $\beta^{sp(I)}_{60}$ & $\beta^{sp(I)}_{80}$    \\
 &  & (MeV) & (MeV) & (MeV) & (MeV) &
 &  &       &  &    \\
89 & 137 & 27.65 & 1.05 & 3.07 & -2.03 & 0.38 & 0.07 &  0.00 &  0.00 & -0.04 \\
89 & 138 & 29.42 & 1.39 & 3.21 & -1.82 & 0.41 & 0.00 & -0.02 &  0.00 & -0.03 \\
89 & 139 & 32.63 & 1.65 & 3.26 & -1.61 & 0.40 & 0.00 & -0.03 &  0.00 & -0.03 \\
\hline
90 & 137 & 28.89 & 1.21 & 2.87 & -1.66 & 0.37 & 0.07 &  0.00 &  0.00 & -0.04 \\
90 & 138 & 30.19 & 1.44 & 2.85 & -1.41 & 0.39 & 0.05 & -0.01 &  0.00 & -0.03 \\
90 & 139 & 33.23 & 1.62 & 3.28 & -1.66 & 0.40 & 0.06 & -0.03 &  0.00 & -0.03 \\
90 & 140 & 34.69 & 1.73 & 3.58 & -1.85 & 0.42 & 0.00 & -0.04 &  0.00 & -0.03 \\
90 & 141 & 38.11 & 2.02 & 3.76 & -1.74 & 0.43 & 0.04 & -0.05 & -0.01 & -0.01 \\
90 & 142 & 39.85 & 2.15 & 3.75 & -1.60 & 0.43 & 0.03 & -0.05 & -0.01 & -0.01 \\
90 & 143 & 43.56 & 2.47 & 4.05 & -1.59 & 0.45 & 0.05 & -0.05 & -0.01 & -0.01 \\
90 & 144 & 45.50 & 2.52 & 4.42 & -1.90 & 0.46 & 0.04 & -0.06 & -0.01 & -0.01 \\
\hline
91 & 139 & 36.11 & 1.63 & 3.57 & -1.94 & 0.42 & 0.02 & -0.05 & -0.01 & -0.02 \\
91 & 140 & 37.58 & 1.84 & 3.90 & -2.07 & 0.44 & 0.00 & -0.05 & -0.02 & -0.02 \\
91 & 141 & 40.57 & 2.05 & 4.24 & -2.19 & 0.44 & 0.00 & -0.06 & -0.03 & -0.01 \\
91 & 142 & 42.34 & 2.28 & 4.31 & -2.03 & 0.46 & 0.00 & -0.06 & -0.02 & -0.01 \\
91 & 143 & 45.77 & 2.67 & 3.99 & -1.32 & 0.46 & 0.00 & -0.05 & -0.02 &  0.00 \\
\hline
92 & 139 & 37.71 & 1.66 & 3.44 & -1.78 & 0.42 & 0.02 & -0.05 & -0.01 & -0.02 \\
92 & 140 & 38.82 & 1.87 & 3.54 & -1.68 & 0.43 & 0.00 & -0.05 & -0.02 & -0.01 \\
92 & 141 & 41.68 & 2.03 & 3.83 & -1.80 & 0.43 & 0.00 & -0.06 & -0.02 & -0.01 \\
92 & 142 & 43.09 & 2.26 & 3.95 & -1.69 & 0.44 & 0.00 & -0.06 & -0.02 & -0.01 \\
92 & 143 & 46.43 & 2.63 & 4.13 & -1.50 & 0.46 & 0.00 & -0.06 & -0.02 &  0.00 \\
92 & 144 & 47.97 & 2.72 & 4.14 & -1.42 & 0.46 & 0.00 & -0.06 & -0.02 &  0.00 \\
92 & 145 & 51.48 & 3.01 & 4.28 & -1.27 & 0.49 & 0.07 & -0.05 & -0.01 &  0.00 \\
92 & 146 & 53.21 & 3.03 & 4.04 & -1.01 & 0.49 & 0.07 & -0.04 & -0.01 &  0.00 \\
92 & 147 & 56.93 & 3.26 & 3.84 & -0.58 & 0.50 & 0.09 & -0.02 &  0.00 &  0.00 \\
92 & 148 & 58.78 & 3.15 & 4.02 & -0.87 & 0.49 & 0.10 & -0.03 & -0.01 &  0.00 \\
\hline
93 & 140 & 42.68 & 2.02 & 4.55 & -2.53 & 0.49 & 0.00 & -0.07 & -0.02 &  0.00 \\
93 & 141 & 45.35 & 2.35 & 4.72 & -2.38 & 0.49 & 0.00 & -0.07 & -0.03 &  0.01 \\
93 & 142 & 46.60 & 2.49 & 4.74 & -2.26 & 0.49 & 0.00 & -0.07 & -0.03 &  0.01 \\
93 & 143 & 49.68 & 2.96 & 4.03 & -1.08 & 0.49 & 0.06 & -0.05 & -0.01 &  0.00 \\
93 & 144 & 51.02 & 2.92 & 4.68 & -1.75 & 0.48 & 0.00 & -0.07 & -0.03 &  0.01 \\
93 & 145 & 54.11 & 3.14 & 4.01 & -0.87 & 0.50 & 0.08 & -0.04 & -0.01 &  0.00 \\
93 & 146 & 55.84 & 3.23 & 3.84 & -0.61 & 0.50 & 0.09 & -0.03 &  0.00 &  0.00 \\
\hline
94 & 141 & 47.42 & 2.36 & 4.40 & -2.04 & 0.49 & 0.00 & -0.07 & -0.02 &  0.01 \\
94 & 142 & 48.37 & 2.56 & 4.47 & -1.92 & 0.48 & 0.00 & -0.07 & -0.03 &  0.01 \\
94 & 143 & 51.21 & 2.87 & 3.63 & -0.76 & 0.47 & 0.00 & -0.05 & -0.02 &  0.00 \\
94 & 144 & 52.40 & 3.03 & 4.09 & -1.07 & 0.47 & 0.00 & -0.06 & -0.03 &  0.01 \\
94 & 145 & 55.38 & 3.21 & 3.73 & -0.52 & 0.49 & 0.08 & -0.04 & -0.01 &  0.00 \\
94 & 146 & 56.67 & 3.21 & 3.58 & -0.37 & 0.49 & 0.09 & -0.03 & -0.01 &  0.00 \\
94 & 147 & 59.74 & 3.22 & 3.46 & -0.24 & 0.50 & 0.11 & -0.01 &  0.00 &  0.00 \\
94 & 148 & 61.24 & 3.18 & 3.40 & -0.22 & 0.50 & 0.10 & -0.01 &  0.00 &  0.00 \\
94 & 149 & 64.36 & 2.98 & 3.48 & -0.50 & 0.50 & 0.12 &  0.00 &  0.00 &  0.01 \\
94 & 150 & 66.17 & 2.99 & 3.31 & -0.32 & 0.50 & 0.11 &  0.01 &  0.01 &  0.00 \\
94 & 151 & 69.46 & 2.72 & 3.31 & -0.60 & 0.49 & 0.12 &  0.02 &  0.01 &  0.01 \\
94 & 152 & 71.45 & 2.66 & 3.40 & -0.74 & 0.50 & 0.12 &  0.02 &  0.01 &  0.01 \\
\hline
95 & 144 & 56.36 & 3.23 & 3.58 & -0.35 & 0.48 & 0.00 & -0.05 & -0.02 &  0.01 \\
95 & 145 & 58.89 & 3.31 & 3.46 & -0.16 & 0.50 & 0.09 & -0.03 & -0.01 &  0.00 \\
95 & 146 & 60.27 & 3.47 & 3.22 &  0.25 & 0.51 & 0.09 & -0.01 &  0.01 &  0.00 \\
95 & 147 & 62.84 & 3.33 & 3.17 &  0.16 & 0.50 & 0.11 &  0.00 &  0.01 &  0.00 \\
95 & 148 & 64.26 & 3.27 & 3.27 &  0.01 & 0.51 & 0.11 &  0.01 &  0.02 &  0.00 \\
95 & 149 & 67.03 & 3.07 & 3.31 & -0.24 & 0.51 & 0.12 &  0.01 &  0.01 &  0.01 \\
95 & 150 & 68.59 & 2.91 & 3.29 & -0.37 & 0.51 & 0.11 &  0.02 &  0.02 &  0.01 \\
95 & 151 & 71.44 & 2.53 & 3.16 & -0.62 & 0.50 & 0.12 &  0.03 &  0.01 &  0.01 \\
95 & 152 & 73.48 & 2.60 & 3.28 & -0.68 & 0.51 & 0.12 &  0.02 &  0.01 &  0.01 \\
\hline
96 & 145 & 60.98 & 3.30 & 3.18 &  0.12 & 0.49 & 0.09 & -0.03 & -0.01 &  0.00 \\
96 & 146 & 61.84 & 3.29 & 2.90 &  0.40 & 0.50 & 0.09 & -0.01 &  0.00 &  0.00 \\
96 & 147 & 64.33 & 3.14 & 3.00 &  0.14 & 0.50 & 0.12 &  0.00 &  0.00 &  0.00 \\
96 & 148 & 65.42 & 3.10 & 3.11 & -0.01 & 0.50 & 0.13 &  0.00 &  0.00 &  0.00 \\
96 & 149 & 68.11 & 2.88 & 3.02 & -0.14 & 0.51 & 0.12 &  0.02 &  0.01 &  0.01 \\
96 & 150 & 69.40 & 2.79 & 3.06 & -0.27 & 0.50 & 0.13 &  0.02 &  0.01 &  0.01 \\
96 & 151 & 72.27 & 2.51 & 2.86 & -0.36 & 0.48 & 0.13 &  0.03 &  0.00 &  0.01 \\
96 & 152 & 73.82 & 2.42 & 2.99 & -0.57 & 0.49 & 0.13 &  0.03 &  0.01 &  0.01 \\
96 & 153 & 76.96 & 2.15 & 2.94 & -0.79 & 0.48 & 0.13 &  0.04 &  0.01 &  0.01 \\
96 & 154 & 78.76 & 2.08 & 2.97 & -0.90 & 0.50 & 0.12 &  0.04 &  0.01 &  0.01 \\
\hline
97 & 147 & 68.04 & 2.96 & 2.62 &  0.34 & 0.46 & 0.14 &  0.02 & -0.01 &  0.01 \\
97 & 148 & 68.98 & 2.84 & 2.71 &  0.13 & 0.46 & 0.15 &  0.04 &  0.00 &  0.01 \\
97 & 149 & 71.27 & 2.57 & 2.72 & -0.15 & 0.46 & 0.15 &  0.03 &  0.00 &  0.01 \\
97 & 150 & 72.53 & 2.52 & 2.76 & -0.24 & 0.46 & 0.14 &  0.04 &  0.00 &  0.02 \\
97 & 151 & 75.15 & 2.32 & 2.78 & -0.46 & 0.46 & 0.14 &  0.04 &  0.00 &  0.02 \\
97 & 152 & 76.54 & 2.15 & 2.79 & -0.64 & 0.46 & 0.14 &  0.04 &  0.00 &  0.02 \\
97 & 153 & 79.27 & 1.81 & 2.84 & -1.03 & 0.46 & 0.14 &  0.05 &  0.00 &  0.02 \\
\hline
98 & 152 & 77.95 & 2.14 & 2.58 & -0.44 & 0.46 & 0.15 &  0.04 &  0.00 &  0.01 \\
98 & 153 & 80.60 & 1.80 & 2.65 & -0.86 & 0.46 & 0.14 &  0.05 &  0.00 &  0.02 \\
98 & 154 & 82.05 & 1.77 & 2.58 & -0.82 & 0.47 & 0.14 &  0.05 &  0.00 &  0.01 \\
98 & 155 & 85.02 & 1.49 & 2.66 & -1.17 & 0.50 & 0.13 &  0.05 &  0.01 &  0.01 \\
\hline
\end{longtable}

\clearpage
\noindent
Properties of the second saddle point $sp(II)$ for the selected actinides
 specified by $Z$, $N$, investigated in \cite{Jachimowicz2020}.
$M^{sp(II)}_{th}$ is the calculated mass excess, $E^{sp(II)}_{tot}$ is the total energy normalized to
the macroscopic energy at the spherical shape. $E^{sp(II)}_{mic}$ and $E^{sp(II)}_{mac}$
are the microscopic and macroscopic parts of the total energy $E^{sp(II)}_{tot}$, respectively.
$\beta^{sp(II)}_{\lambda 0}$ are second saddle-point deformation parameters
 (with $\beta^{sp(II)}_{10}$ being a function of others - see text).

\begin{longtable}{|cccccccccccccc|}
\caption{Second saddle point properties of selected actinides $Z=89-98$} \label{SP2_ACTIN}
$Z$ & $N$ & $M^{sp(II)}_{th}$ & $E^{sp(II)}_{tot}$ & $E^{sp(II)}_{mac}$ & $E^{sp(II)}_{mic}$ &
$\beta^{sp(II)}_{10}$ & $\beta^{sp(II)}_{20}$  & $\beta^{sp(II)}_{30}$ & $\beta^{sp(II)}_{40}$ &
$\beta^{sp(II)}_{50}$ & $\beta^{sp(II)}_{60}$ & $\beta^{sp(II)}_{70}$ & $\beta^{sp(II)}_{80}$    \\
 &  & (MeV) & (MeV) & (MeV) & (MeV) &
 &  &  &  &
 &  &  &     \\
\hline
89 & 137 & 30.74 &  4.14 &  5.98 & -1.84 & -0.05 & 0.70 & 0.10 &  0.10 &  0.00 & 0.00 &  0.00 &  0.00 \\
89 & 138 & 32.44 &  4.41 &  7.87 & -3.46 & -0.08 & 0.75 & 0.15 &  0.15 &  0.00 & 0.00 &  0.00 & -0.05 \\
89 & 139 & 35.05 &  4.08 &  9.08 & -5.01 & -0.11 & 0.80 & 0.20 &  0.15 &  0.00 & 0.05 &  0.00 &  0.00 \\
\hline
90 & 137 & 31.44 &  3.76 &  5.59 & -1.83 & -0.05 & 0.70 & 0.10 &  0.10 &  0.00 & 0.00 &  0.00 &  0.00 \\
90 & 138 & 32.76 &  4.01 &  7.41 & -3.40 & -0.08 & 0.75 & 0.15 &  0.15 &  0.00 & 0.00 &  0.00 & -0.05 \\
90 & 139 & 35.19 &  3.58 &  8.57 & -4.99 & -0.11 & 0.80 & 0.20 &  0.15 &  0.00 & 0.05 &  0.00 &  0.00 \\
90 & 140 & 36.87 &  3.91 &  7.49 & -3.57 & -0.08 & 0.75 & 0.15 &  0.15 &  0.00 & 0.00 &  0.00 & -0.05 \\
90 & 141 & 39.67 &  3.58 &  7.03 & -3.45 & -0.08 & 0.75 & 0.15 &  0.15 &  0.00 & 0.00 &  0.00 &  0.00 \\
90 & 142 & 41.64 &  3.93 &  7.06 & -3.13 & -0.08 & 0.75 & 0.15 &  0.15 &  0.00 & 0.00 &  0.00 &  0.00 \\
90 & 143 & 44.70 &  3.60 &  7.01 & -3.41 & -0.09 & 0.75 & 0.15 &  0.15 &  0.05 & 0.00 &  0.00 &  0.00 \\
90 & 144 & 46.80 &  3.83 &  7.04 & -3.21 & -0.09 & 0.75 & 0.15 &  0.15 &  0.05 & 0.00 &  0.00 &  0.00 \\
\hline
91 & 139 & 37.81 &  3.33 &  7.65 & -4.32 & -0.11 & 0.80 & 0.20 &  0.10 &  0.05 & 0.00 &  0.00 & -0.05 \\
91 & 140 & 39.51 &  3.77 &  7.13 & -3.36 & -0.08 & 0.80 & 0.15 &  0.15 &  0.00 & 0.05 &  0.00 &  0.00 \\
91 & 141 & 41.90 &  3.38 &  7.38 & -4.00 & -0.08 & 0.75 & 0.15 &  0.20 &  0.00 & 0.00 &  0.00 &  0.00 \\
91 & 142 & 43.75 &  3.70 &  6.52 & -2.82 & -0.09 & 0.75 & 0.15 &  0.15 &  0.05 & 0.00 &  0.00 &  0.00 \\
91 & 143 & 46.41 &  3.31 &  6.55 & -3.24 & -0.09 & 0.75 & 0.15 &  0.15 &  0.05 & 0.00 &  0.00 &  0.00 \\
\hline
92 & 139 & 38.91 &  2.87 &  6.29 & -3.43 & -0.08 & 0.80 & 0.15 &  0.10 &  0.05 & 0.00 &  0.00 & -0.05 \\
92 & 140 & 40.25 &  3.29 &  6.59 & -3.29 & -0.08 & 0.80 & 0.15 &  0.15 &  0.00 & 0.05 &  0.00 &  0.00 \\
92 & 141 & 42.63 &  2.97 &  6.87 & -3.90 & -0.08 & 0.75 & 0.15 &  0.20 &  0.00 & 0.00 &  0.00 &  0.00 \\
92 & 142 & 44.14 &  3.30 &  6.03 & -2.73 & -0.09 & 0.75 & 0.15 &  0.15 &  0.05 & 0.00 &  0.00 &  0.00 \\
92 & 143 & 46.72 &  2.92 &  6.07 & -3.15 & -0.09 & 0.75 & 0.15 &  0.15 &  0.05 & 0.00 &  0.00 &  0.00 \\
92 & 144 & 48.41 &  3.16 &  6.10 & -2.95 & -0.09 & 0.75 & 0.15 &  0.15 &  0.05 & 0.00 &  0.00 &  0.00 \\
92 & 145 & 51.53 &  3.05 &  6.56 & -3.51 & -0.09 & 0.80 & 0.15 &  0.15 &  0.05 & 0.05 &  0.00 &  0.00 \\
92 & 146 & 53.42 &  3.23 &  6.60 & -3.36 & -0.09 & 0.80 & 0.15 &  0.15 &  0.05 & 0.05 &  0.00 &  0.00 \\
92 & 147 & 57.29 &  3.62 &  6.62 & -3.00 & -0.09 & 0.80 & 0.15 &  0.15 &  0.05 & 0.05 &  0.00 &  0.00 \\
92 & 148 & 59.25 &  3.62 &  6.65 & -3.03 & -0.09 & 0.80 & 0.15 &  0.15 &  0.05 & 0.05 &  0.00 &  0.00 \\
\hline
93 & 140 & 43.39 &  2.73 &  5.91 & -3.18 & -0.08 & 0.75 & 0.15 &  0.15 &  0.05 & 0.00 &  0.00 & -0.05 \\
93 & 141 & 45.60 &  2.59 &  5.30 & -2.70 & -0.05 & 0.75 & 0.10 &  0.10 &  0.05 & 0.00 &  0.00 & -0.05 \\
93 & 142 & 46.95 &  2.84 &  6.01 & -3.17 & -0.08 & 0.75 & 0.15 &  0.15 &  0.05 & 0.00 &  0.00 & -0.05 \\
93 & 143 & 49.29 &  2.56 &  6.22 & -3.65 & -0.06 & 0.75 & 0.10 &  0.20 &  0.00 & 0.05 &  0.00 &  0.00 \\
93 & 144 & 50.93 &  2.83 &  5.58 & -2.74 & -0.06 & 0.75 & 0.10 &  0.15 &  0.05 & 0.05 &  0.00 &  0.00 \\
93 & 145 & 53.67 &  2.70 &  6.66 & -3.96 & -0.12 & 0.85 & 0.20 &  0.15 &  0.05 & 0.05 &  0.00 &  0.00 \\
93 & 146 & 55.46 &  2.86 &  6.05 & -3.20 & -0.09 & 0.80 & 0.15 &  0.15 &  0.05 & 0.05 &  0.00 &  0.00 \\
\hline
94 & 141 & 47.15 &  2.09 &  5.80 & -3.71 & -0.08 & 0.75 & 0.15 &  0.20 &  0.00 & 0.00 &  0.00 &  0.00 \\
94 & 142 & 48.20 &  2.39 &  5.50 & -3.10 & -0.08 & 0.75 & 0.15 &  0.15 &  0.05 & 0.00 &  0.00 & -0.05 \\
94 & 143 & 50.43 &  2.08 &  5.06 & -2.97 & -0.09 & 0.75 & 0.15 &  0.15 &  0.05 & 0.00 &  0.00 &  0.00 \\
94 & 144 & 51.70 &  2.34 &  5.10 & -2.77 & -0.09 & 0.75 & 0.15 &  0.15 &  0.05 & 0.00 &  0.00 &  0.00 \\
94 & 145 & 54.31 &  2.14 &  6.03 & -3.89 & -0.12 & 0.85 & 0.20 &  0.15 &  0.05 & 0.05 &  0.00 &  0.00 \\
94 & 146 & 55.76 &  2.31 &  5.64 & -3.33 & -0.12 & 0.85 & 0.20 &  0.15 &  0.05 & 0.00 &  0.00 &  0.00 \\
94 & 147 & 59.19 &  2.66 &  5.53 & -2.87 & -0.09 & 0.80 & 0.15 &  0.15 &  0.05 & 0.05 &  0.00 &  0.00 \\
94 & 148 & 60.74 &  2.67 &  5.73 & -3.05 & -0.12 & 0.85 & 0.20 &  0.15 &  0.05 & 0.00 &  0.00 &  0.00 \\
94 & 149 & 64.46 &  3.08 &  5.14 & -2.06 & -0.09 & 0.80 & 0.15 &  0.15 &  0.05 & 0.00 &  0.00 &  0.00 \\
94 & 150 & 66.15 &  2.97 &  6.38 & -3.41 & -0.12 & 0.85 & 0.20 &  0.20 &  0.05 & 0.00 &  0.00 & -0.05 \\
94 & 151 & 70.02 &  3.27 &  5.19 & -1.92 & -0.09 & 0.80 & 0.15 &  0.15 &  0.05 & 0.00 &  0.00 &  0.00 \\
94 & 152 & 71.93 &  3.14 &  5.88 & -2.74 & -0.10 & 0.80 & 0.15 &  0.15 &  0.10 & 0.00 &  0.00 &  0.00 \\
\hline
95 & 144 & 54.86 &  1.74 &  5.20 & -3.47 & -0.09 & 0.75 & 0.15 &  0.20 &  0.05 & 0.00 &  0.00 &  0.00 \\
95 & 145 & 57.18 &  1.59 &  4.66 & -3.07 & -0.09 & 0.85 & 0.15 &  0.15 &  0.05 & 0.05 &  0.00 &  0.00 \\
95 & 146 & 58.44 &  1.65 &  4.72 & -3.07 & -0.09 & 0.85 & 0.15 &  0.15 &  0.05 & 0.05 &  0.00 &  0.00 \\
95 & 147 & 61.59 &  2.08 &  4.93 & -2.85 & -0.09 & 0.85 & 0.15 &  0.15 &  0.05 & 0.05 &  0.00 & -0.05 \\
95 & 148 & 63.03 &  2.05 &  5.31 & -3.27 & -0.13 & 0.85 & 0.20 &  0.20 &  0.05 & 0.00 &  0.00 &  0.00 \\
95 & 149 & 66.27 &  2.31 &  5.02 & -2.71 & -0.09 & 0.80 & 0.15 &  0.20 &  0.05 & 0.00 &  0.00 &  0.00 \\
95 & 150 & 67.89 &  2.20 &  5.06 & -2.86 & -0.09 & 0.80 & 0.15 &  0.20 &  0.05 & 0.00 &  0.00 &  0.00 \\
95 & 151 & 71.39 &  2.48 &  5.98 & -3.50 & -0.10 & 0.80 & 0.15 &  0.15 &  0.10 & 0.00 & -0.05 & -0.05 \\
95 & 152 & 73.18 &  2.30 &  6.02 & -3.72 & -0.10 & 0.80 & 0.15 &  0.15 &  0.10 & 0.00 & -0.05 & -0.05 \\
\hline
96 & 145 & 58.79 &  1.11 &  4.00 & -2.89 & -0.09 & 0.85 & 0.15 &  0.15 &  0.05 & 0.05 &  0.00 &  0.00 \\
96 & 146 & 59.73 &  1.18 &  4.06 & -2.88 & -0.09 & 0.85 & 0.15 &  0.15 &  0.05 & 0.05 &  0.00 &  0.00 \\
96 & 147 & 62.75 &  1.55 &  4.29 & -2.73 & -0.09 & 0.85 & 0.15 &  0.15 &  0.05 & 0.05 &  0.00 & -0.05 \\
96 & 148 & 63.88 &  1.55 &  4.64 & -3.09 & -0.13 & 0.85 & 0.20 &  0.20 &  0.05 & 0.00 &  0.00 &  0.00 \\
96 & 149 & 67.05 &  1.82 &  4.42 & -2.60 & -0.09 & 0.80 & 0.15 &  0.20 &  0.05 & 0.00 &  0.00 &  0.00 \\
96 & 150 & 68.35 &  1.74 &  5.09 & -3.35 & -0.12 & 0.85 & 0.20 &  0.20 &  0.05 & 0.00 &  0.00 & -0.05 \\
96 & 151 & 71.82 &  2.05 &  4.50 & -2.45 & -0.09 & 0.80 & 0.15 &  0.20 &  0.05 & 0.00 &  0.00 &  0.00 \\
96 & 152 & 73.33 &  1.93 &  4.54 & -2.61 & -0.09 & 0.80 & 0.15 &  0.20 &  0.05 & 0.00 &  0.00 &  0.00 \\
96 & 153 & 76.77 &  1.97 &  5.45 & -3.48 & -0.06 & 0.80 & 0.10 &  0.10 &  0.10 & 0.00 & -0.05 & -0.05 \\
96 & 154 & 78.56 &  1.87 &  5.32 & -3.45 & -0.07 & 0.80 & 0.10 &  0.15 &  0.10 & 0.00 & -0.05 & -0.05 \\
\hline
97 & 147 & 65.78 &  0.70 &  3.62 & -2.92 & -0.09 & 0.85 & 0.15 &  0.15 &  0.05 & 0.05 &  0.00 & -0.05 \\
97 & 148 & 66.86 &  0.72 &  3.69 & -2.96 & -0.10 & 0.85 & 0.15 &  0.20 &  0.05 & 0.05 &  0.00 & -0.05 \\
97 & 149 & 69.66 &  0.96 &  3.80 & -2.84 & -0.09 & 0.80 & 0.15 &  0.20 &  0.05 & 0.00 &  0.00 &  0.00 \\
97 & 150 & 70.89 &  0.88 &  3.85 & -2.97 & -0.09 & 0.80 & 0.15 &  0.20 &  0.05 & 0.00 &  0.00 &  0.00 \\
97 & 151 & 73.88 &  1.05 &  4.65 & -3.60 & -0.07 & 0.80 & 0.10 &  0.15 &  0.10 & 0.00 & -0.05 & -0.05 \\
97 & 152 & 75.30 &  0.90 &  6.36 & -5.46 & -0.02 & 0.95 & 0.05 & -0.05 &  0.00 & 0.00 &  0.00 &  0.00 \\
97 & 153 & 77.96 &  0.50 &  6.05 & -5.55 & -0.06 & 1.00 & 0.10 &  0.00 &  0.05 & 0.00 & -0.05 & -0.05 \\
\hline
98 & 152 & 76.42 &  0.61 &  5.30 & -4.69 & -0.06 & 1.00 & 0.10 &  0.00 &  0.05 & 0.00 & -0.05 & -0.05 \\
98 & 153 & 78.93 &  0.12 &  5.95 & -5.83 & -0.02 & 1.00 & 0.05 & -0.05 &  0.00 & 0.00 &  0.00 &  0.00 \\
98 & 154 & 80.28 &  0.00 &  6.00 & -6.00 & -0.02 & 1.00 & 0.05 & -0.05 &  0.00 & 0.00 &  0.00 &  0.00 \\
98 & 155 & 83.00 & -0.53 &  7.27 & -7.81 & -0.03 & 1.15 & 0.05 & -0.05 &  0.05 & 0.00 & -0.05 & -0.05 \\
\hline
\end{longtable}

\clearpage
\noindent
Properties of the secondary (superdeformed) minima (SD) for the selected
 actinides specified by $Z$, $N$, investigated in \cite{Jachimowicz2020}.
$M^{SD}_{th}$ is the calculated mass excess, $E^{SD}_{tot}$ is the total energy normalized to
the macroscopic energy at the spherical shape. $E^{SD}_{mic}$ and $E^{SD}_{mac}$
are the microscopic and macroscopic parts of the total energy $E^{SD}_{tot}$, respectively.
$E^{*}_{th}$ is obtained by us excitation energy of the SD minimum relative
to the ground state. $\beta^{SD}_{\lambda 0}$ are values of equilibrium
deformation parameters of a SD minimum.

\begin{longtable}{|ccccccccccc|}
\caption{Properties of secondary minima of selected actinides $Z=89-98$}\label{SD_ACTIN}
$Z$ & $N$ & $M^{SD}_{th} $ & $E^{SD}_{tot} $ & $E^{SD}_{mac} $ & $E^{SD}_{mic} $ &
$\beta^{SD}_{20}$  & $\beta^{SD}_{40}$ & $\beta^{SD}_{60}$ & $\beta^{SD}_{80}$ & $E^{*}_{th} $ \\
 & & (MeV) & (MeV) & (MeV) & (MeV) &
 & & & & (MeV) \\
\hline
89 & 137  & 26.63 &  0.03 & 5.90 & -5.87  & 0.51 & -0.08 & -0.01 & -0.02 &  3.05                \\
89 & 138  & 28.26 &  0.24 & 5.29 & -5.05  & 0.55 & -0.04 &  0.02 & -0.03 &  2.78                \\
89 & 139  & 31.26 &  0.29 & 5.84 & -5.55  & 0.60 &  0.05 &  0.07 & -0.01 &  3.01                \\
\hline
90 & 137  & 28.02 &  0.34 & 5.07 & -4.73  & 0.55 & -0.04 &  0.02 & -0.03 &  2.87                \\
90 & 138  & 29.10 &  0.35 & 4.98 & -4.63  & 0.54 & -0.04 &  0.02 & -0.03 &  2.48                \\
90 & 139  & 31.96 &  0.35 & 5.44 & -5.09  & 0.58 &  0.02 &  0.06 & -0.01 &  2.90                \\
90 & 140  & 33.33 &  0.37 & 5.30 & -4.92  & 0.59 &  0.03 &  0.06 & -0.01 &  2.62                \\
90 & 141  & 35.68 & -0.41 & 5.57 & -5.98  & 0.60 &  0.02 &  0.07 & -0.01 &  2.35                \\
90 & 142  & 37.41 & -0.29 & 5.42 & -5.71  & 0.61 &  0.04 &  0.06 & -0.02 &  2.11                \\
90 & 143  & 39.84 & -1.26 & 5.63 & -6.89  & 0.61 &  0.04 &  0.06 & -0.03 &  1.49                \\
90 & 144  & 42.09 & -0.88 & 5.69 & -6.57  & 0.61 &  0.03 &  0.06 & -0.03 &  1.62                \\
\hline
91 & 139  & 34.92 &  0.45 & 5.21 & -4.76  & 0.57 & -0.04 &  0.01 & -0.03 &  3.91                \\
91 & 140  & 36.26 &  0.52 & 4.89 & -4.37  & 0.60 &  0.02 &  0.05 &  0.01 &  3.66                \\
91 & 141  & 38.29 & -0.23 & 5.32 & -5.54  & 0.60 &  0.02 &  0.06 & -0.01 &  3.44                \\
91 & 142  & 39.93 & -0.12 & 5.23 & -5.35  & 0.61 &  0.03 &  0.06 & -0.02 &  3.13                \\
91 & 143  & 41.99 & -1.11 & 5.50 & -6.60  & 0.62 &  0.03 &  0.06 & -0.03 &  2.45                \\
\hline
92 & 139  & 36.48 &  0.43 & 4.82 & -4.39  & 0.60 &  0.01 &  0.05 &  0.00 &  3.41                \\
92 & 140  & 37.40 &  0.44 & 4.58 & -4.14  & 0.61 &  0.03 &  0.05 & -0.01 &  3.10                \\
92 & 141  & 39.25 & -0.41 & 4.67 & -5.07  & 0.62 &  0.04 &  0.05 & -0.02 &  2.86                \\
92 & 142  & 40.54 & -0.30 & 4.74 & -5.04  & 0.62 &  0.03 &  0.05 & -0.02 &  2.57                \\
92 & 143  & 42.51 & -1.29 & 5.07 & -6.36  & 0.62 &  0.03 &  0.06 & -0.03 &  1.94                \\
92 & 144  & 44.33 & -0.92 & 5.10 & -6.02  & 0.62 &  0.03 &  0.06 & -0.03 &  2.05                \\
92 & 145  & 46.95 & -1.52 & 5.38 & -6.90  & 0.64 &  0.02 &  0.06 & -0.03 &  1.92                \\
92 & 146  & 49.09 & -1.10 & 5.27 & -6.37  & 0.63 &  0.02 &  0.06 & -0.02 &  1.94                \\
92 & 147  & 52.25 & -1.41 & 5.60 & -7.01  & 0.65 &  0.00 &  0.05 & -0.02 &  2.02                \\
92 & 148  & 54.69 & -0.94 & 5.38 & -6.32  & 0.65 &  0.01 &  0.05 & -0.02 &  2.04                \\
\hline
93 & 140  & 40.99 &  0.33 & 4.27 & -3.94  & 0.62 &  0.02 &  0.04 &  0.00 &  3.46                \\
93 & 141  & 42.56 & -0.45 & 4.17 & -4.62  & 0.63 &  0.03 &  0.04 & -0.01 &  3.31                \\
93 & 142  & 43.76 & -0.35 & 4.36 & -4.71  & 0.62 &  0.02 &  0.04 & -0.01 &  3.06                \\
93 & 143  & 45.47 & -1.26 & 4.91 & -6.17  & 0.63 &  0.03 &  0.06 & -0.03 &  2.58                \\
93 & 144  & 47.17 & -0.93 & 4.57 & -5.50  & 0.63 &  0.02 &  0.05 & -0.02 &  2.69                \\
93 & 145  & 49.37 & -1.61 & 5.24 & -6.85  & 0.65 &  0.02 &  0.06 & -0.03 &  2.67                \\
93 & 146  & 51.42 & -1.18 & 4.76 & -5.95  & 0.64 &  0.01 &  0.04 & -0.02 &  2.56                \\
\hline
94 & 141  & 44.42 & -0.64 & 3.86 & -4.49  & 0.63 &  0.04 &  0.04 & -0.02 &  2.64                \\
94 & 142  & 45.30 & -0.51 & 4.00 & -4.51  & 0.63 &  0.03 &  0.04 & -0.02 &  2.42                \\
94 & 143  & 46.87 & -1.48 & 4.38 & -5.86  & 0.63 &  0.03 &  0.05 & -0.03 &  1.92                \\
94 & 144  & 48.20 & -1.17 & 4.39 & -5.56  & 0.63 &  0.02 &  0.05 & -0.03 &  2.04                \\
94 & 145  & 50.32 & -1.86 & 4.71 & -6.56  & 0.64 &  0.02 &  0.05 & -0.03 &  2.02                \\
94 & 146  & 52.00 & -1.46 & 4.63 & -6.09  & 0.65 &  0.01 &  0.05 & -0.03 &  1.94                \\
94 & 147  & 54.60 & -1.93 & 4.99 & -6.91  & 0.66 &  0.00 &  0.05 & -0.02 &  1.94                \\
94 & 148  & 56.61 & -1.46 & 4.95 & -6.41  & 0.66 &  0.01 &  0.05 & -0.03 &  1.97                \\
94 & 149  & 59.83 & -1.55 & 5.04 & -6.60  & 0.67 &  0.02 &  0.06 & -0.03 &  2.17                \\
94 & 150  & 61.94 & -1.24 & 4.92 & -6.16  & 0.67 &  0.00 &  0.04 & -0.02 &  2.14                \\
94 & 151  & 65.69 & -1.06 & 4.58 & -5.64  & 0.67 &  0.01 &  0.04 & -0.02 &  2.81                \\
94 & 152  & 67.87 & -0.92 & 4.80 & -5.72  & 0.67 &  0.01 &  0.04 & -0.02 &  2.44                \\
\hline
95 & 144  & 51.61 & -1.51 & 4.00 & -5.51  & 0.64 &  0.02 &  0.04 & -0.02 &  2.19                \\
95 & 145  & 53.36 & -2.22 & 4.31 & -6.53  & 0.65 &  0.01 &  0.05 & -0.03 &  2.19                \\
95 & 146  & 54.91 & -1.89 & 4.29 & -6.18  & 0.65 &  0.01 &  0.04 & -0.02 &  2.10                \\
95 & 147  & 57.04 & -2.47 & 4.65 & -7.12  & 0.67 & -0.01 &  0.04 & -0.02 &  2.02                \\
95 & 148  & 59.01 & -1.97 & 4.56 & -6.53  & 0.67 & -0.01 &  0.04 & -0.02 &  2.07                \\
95 & 149  & 62.00 & -1.96 & 5.09 & -7.05  & 0.69 &  0.01 &  0.06 & -0.04 &  2.41                \\
95 & 150  & 63.89 & -1.79 & 4.39 & -6.18  & 0.68 &  0.00 &  0.03 & -0.02 &  2.23                \\
95 & 151  & 67.28 & -1.64 & 5.01 & -6.64  & 0.69 & -0.02 &  0.02 & -0.02 &  2.86                \\
95 & 152  & 69.35 & -1.54 & 4.44 & -5.98  & 0.68 &  0.00 &  0.03 & -0.02 &  2.43                \\
\hline
96 & 145  & 55.30 & -2.38 & 4.04 & -6.42  & 0.65 &  0.01 &  0.05 & -0.03 &  1.65                \\
96 & 146  & 56.52 & -2.03 & 3.96 & -5.99  & 0.66 &  0.01 &  0.04 & -0.03 &  1.64                \\
96 & 147  & 58.56 & -2.64 & 4.61 & -7.25  & 0.68 & -0.01 &  0.05 & -0.03 &  1.57                \\
96 & 148  & 60.17 & -2.16 & 4.65 & -6.80  & 0.68 & -0.01 &  0.05 & -0.03 &  1.66                \\
96 & 149  & 62.98 & -2.25 & 4.41 & -6.66  & 0.68 &  0.01 &  0.05 & -0.03 &  1.97                \\
96 & 150  & 64.61 & -2.00 & 4.30 & -6.30  & 0.68 & -0.01 &  0.04 & -0.02 &  1.89                \\
96 & 151  & 67.89 & -1.88 & 4.63 & -6.51  & 0.70 & -0.01 &  0.04 & -0.03 &  2.60                \\
96 & 152  & 69.68 & -1.73 & 4.32 & -6.04  & 0.69 &  0.00 &  0.03 & -0.03 &  2.24                \\
96 & 153  & 73.14 & -1.67 & 4.60 & -6.26  & 0.70 & -0.02 &  0.03 & -0.02 &  2.20                \\
96 & 154  & 75.16 & -1.53 & 4.51 & -6.04  & 0.71 & -0.01 &  0.03 & -0.03 &  2.12                \\
\hline
97 & 147  & 61.62 & -3.46 & 4.28 & -7.74  & 0.69 & -0.02 &  0.04 & -0.03 &  1.26                \\
97 & 148  & 63.16 & -2.98 & 4.21 & -7.19  & 0.69 & -0.02 &  0.03 & -0.03 &  1.37                \\
97 & 149  & 65.72 & -2.98 & 4.29 & -7.27  & 0.70 &  0.00 &  0.05 & -0.04 &  1.85                \\
97 & 150  & 67.17 & -2.84 & 4.37 & -7.21  & 0.70 & -0.02 &  0.03 & -0.03 &  1.66                \\
97 & 151  & 69.97 & -2.86 & 5.31 & -8.17  & 0.74 & -0.03 &  0.04 & -0.04 &  2.31                \\
97 & 152  & 71.75 & -2.64 & 5.48 & -8.12  & 0.74 & -0.04 &  0.03 & -0.03 &  1.98                \\
97 & 153  & 74.61 & -2.85 & 6.36 & -9.21  & 0.78 & -0.06 &  0.03 & -0.03 &  1.69                \\
\hline
98 & 152  & 73.11 & -2.70 & 4.01 & -6.71  & 0.71 & -0.02 &  0.03 & -0.03 &  1.83                \\
98 & 153  & 75.98 & -2.82 & 5.74 & -8.56  & 0.78 & -0.06 &  0.03 & -0.03 &  1.63                \\
98 & 154  & 77.66 & -2.63 & 5.16 & -7.79  & 0.76 & -0.04 &  0.03 & -0.03 &  1.58                \\
98 & 155  & 80.47 & -3.06 & 6.78 & -9.84  & 0.83 & -0.08 &  0.03 & -0.03 &  1.06                \\
\hline
\end{longtable}

\clearpage

\noindent
Ground state masses: calculated $M_{gs}^{th}$ and measured $M_{gs}^{exp}$ \cite{Audi2003}.
Calculated first $B_f^{(I)th}$ and second $B_f^{(II)th}$ fission barrier heights compared with two sets of empirical compilations: EXP1 \cite{Smirenkin1993} and EXP2 \cite{Capote2009}, Excitation energy of the SD minimum - $E^{*th}_{II}$ relative to the ground state, experimental values of $E^{*exp}_{II}$ are taken from \cite{Singh1996}.

\begin{table*}[t]
\scriptsize
\caption{Ground state mass excess: calculated $M_{gs}^{th}$ and measured $M_{gs}^{exp}$ \cite{Audi2003}.
Calculated first $B_f^{(I)th}$ and second $B_f^{(II)th}$ fission barrier heights compared with two sets of empirical compilations: EXP1 \cite{Smirenkin1993} and EXP2 \cite{Capote2009}, Excitation energy of the SD minimum - $E^{*th}_{II}$ relative to the ground state, experimental values of $E^{*exp}_{II}$ are taken from \cite{Singh1996}}
 \label{TABLE_TOT}
\centering
\begin{tabular}{|ccc|cc|ccc|cc|ccc|}
\hline
\multicolumn{3}{|c|}  {Nucleus} &
\multicolumn{2}{c|}  {} &
\multicolumn{3}{c|} {} &
\multicolumn{2}{c|} {} &
\multicolumn{3}{c|}  {} \\
\hline
$Z$ & $N$ & $A$ & $M_{gs}^{th}$ & $M_{gs}^{exp}$ & $B_f^{(I)th}$ & $B_f^{(I)EXP1}$ & $B_f^{(I)EXP2}$ & $E^{*th}_{II}$ & $E^{*exp}_{II}$ & $B_f^{(II)th}$ & $B_f^{(II)EXP1}$ & $B_f^{(II)EXP2}$ \\
\hline
89 & 137 & 226 & 23.58 &  24.31 &       4.07 & -   &  -        &  3.05  &  -                   & 7.16 & -   & 7.8  \\
89 & 138 & 227 & 25.48 &  25.85 &       3.94 & -   &  -        &  2.78  &  -                   & 6.96 & -   & 7.4  \\
89 & 139 & 228 & 28.25 &  28.90 &       4.38 & -   &  -        &  3.01  &  -                   & 6.80 & -   & 7.1  \\
\hline
90 & 137 & 227 & 25.15 &  25.81 &       3.74 & 5.9 &  -        &  2.87  &  -                   & 6.30 & 6.6 &  -   \\
90 & 138 & 228 & 26.62 &  26.77 &       3.57 & 6.2 &  -        &  2.48  &  -                   & 6.14 & 6.5 &  -   \\
90 & 139 & 229 & 29.06 &  29.59 &       4.17 & 5.9 &  -        &  2.90  &  -                   & 6.13 & 6.3 &  -   \\
90 & 140 & 230 & 30.71 &  30.86 &       3.98 & 6.1 & 6.1       &  2.62  &  -                   & 6.17 & 6.1 & 6.8  \\
90 & 141 & 231 & 33.33 &  33.82 &       4.78 & 6.0 & 6.0       &  2.35  &  -                   & 6.34 & 6.1 & 6.7  \\
90 & 142 & 232 & 35.31 &  35.45 &       4.55 & 5.8 & 5.8       &  2.11  &  -                   & 6.33 & 6.2 & 6.7  \\
90 & 143 & 233 & 38.36 &  38.73 &       5.21 & 6.1 & 5.1       &  1.49  &  $1.85 (\pm 0.25)$   & 6.35 & 6.3 & 6.65 \\
90 & 144 & 234 & 40.47 &  40.61 &       5.03 & 6.1 &  -        &  1.62  &  -                   & 6.33 & 6.3 &  -   \\
\hline
91 & 139 & 230 & 31.01 &  32.17 &       5.10 & 5.4 & 5.6       &  3.91  &  -                   & 6.81 & 5.4 & 5.8  \\
91 & 140 & 231 & 32.60 &  33.43 &       4.98 & 5.7 & 5.5       &  3.66  &  -                   & 6.91 & 5.7 & 5.5  \\
91 & 141 & 232 & 34.85 &  35.95 &       5.72 & 6.0 & 5.0       &  3.44  &  -                   & 7.05 & 6.1 & 6.4  \\
91 & 142 & 233 & 36.80 &  37.49 &       5.54 & 6.0 & 5.7       &  3.13  &  -                   & 6.95 & 6.0 & 5.8  \\
91 & 143 & 234 & 39.55 &  40.34 &       6.23 &  -  & 6.3       &  2.45  &  -                   & 6.87 & -   & 6.15 \\
\hline
92 & 139 & 231 & 33.07 &  33.81 &       4.64 & 5.2 & 4.4       &  3.41  &  -                   & 5.84 & 5.2 & 5.5  \\
92 & 140 & 232 & 34.30 &  34.61 &       4.52 & 5.4 & 4.9       &  3.10  &  -                   & 5.95 & 5.3 & 5.4  \\
92 & 141 & 233 & 36.40 &  36.92 &       5.29 & 5.7 & 4.35      &  2.86  &  -                   & 6.23 & 5.7 & 5.55 \\
92 & 142 & 234 & 37.98 &  38.15 &       5.12 & 5.9 & 4.8       &  2.57  &  -                   & 6.16 & 5.7 & 5.5  \\
92 & 143 & 235 & 40.57 &  40.92 &       5.86 & 6.0 & 5.25      &  1.94  &  -                   & 6.14 & 5.8 & 6.0  \\
92 & 144 & 236 & 42.28 &  42.45 &       5.69 & 5.6 & 5.0       &  2.05  &  $2.75 (\pm 0.01)$   & 6.13 & 5.6 & 5.67 \\
92 & 145 & 237 & 45.04 &  45.39 &       6.45 & 6.2 & 6.4       &  1.92  &  -                   & 6.49 & 5.9 & 6.15 \\
92 & 146 & 238 & 47.15 &  47.31 &       6.06 & 6.0 & 6.3       &  1.94  &  2.56                & 6.27 & 5.8 & 5.5  \\
92 & 147 & 239 & 50.23 &  50.57 &       6.70 & 6.3 & 6.45      &  2.02  &  -                   & 7.05 & 6.0 & 6.0  \\
92 & 148 & 240 & 52.66 &  52.72 &       6.13 & 6.1 &  -        &  2.04  &  -                   & 6.59 & 5.8 &  -   \\
\hline
93 & 140 & 233 & 37.53 &  37.95 &       5.14 & 5.0 &  -        &  3.46  &  -                   & 5.86 & 5.1 &  -   \\
93 & 141 & 234 & 39.25 &  39.96 &       6.10 & 5.5 &  -        &  3.31  &  -                   & 6.35 & 5.4 &  -   \\
93 & 142 & 235 & 40.71 &  41.04 &       5.89 & 5.5 &  -        &  3.06  &  -                   & 6.24 & 5.5 &  -   \\
93 & 143 & 236 & 42.89 &  43.38 &       6.79 & 5.8 & 5.9       &  2.58  &  -                   & 6.40 & 5.6 & 5.4  \\
93 & 144 & 237 & 44.48 &  44.87 &       6.54 & 5.7 & 6.0       &  2.69  &  $2.80 (\pm 0.40)$   & 6.44 & 5.5 & 5.4  \\
93 & 145 & 238 & 46.70 &  47.46 &       7.41 & 6.0 & 6.5       &  2.67  &  -                   & 6.98 & 5.9 & 5.75 \\
93 & 146 & 239 & 48.87 &  49.31 &       6.98 & 5.8 &  -        &  2.56  &  -                   & 6.60 & 5.4 &  -   \\
\hline
94 & 141 & 235 & 41.78 &  42.18 &       5.64 & 5.7 &  -        &  2.64  &  $3.00 (\pm 0.20)$   & 5.37 & 5.1 &  -   \\
94 & 142 & 236 & 42.88 &  42.90 &       5.49 & 5.7 &  -        &  2.42  & $\sim 3.00$          & 5.32 & 4.5 &  -   \\
94 & 143 & 237 & 44.95 &  45.09 &       6.26 & 5.6 & 5.10      &  1.92  & $ 2.60 (\pm 0.20)$   & 5.48 & 5.4 & 5.15 \\
94 & 144 & 238 & 46.16 &  46.16 &       6.24 & 5.9 & 5.6       &  2.04  & $\sim 2.40$          & 5.55 & 5.2 & 5.1  \\
94 & 145 & 239 & 48.31 &  48.59 &       7.08 & 6.2 & 6.2       &  2.02  & $ 3.10 (\pm 0.20)$   & 6.01 & 5.5 & 5.7  \\
94 & 146 & 240 & 50.06 &  50.13 &       6.61 & 5.8 & 6.05      &  1.94  & $\sim 2.80$          & 5.71 & 5.3 & 5.15 \\
94 & 147 & 241 & 52.66 &  52.96 &       7.08 & 6.2 & 6.15      &  1.94  & $\sim 2.20$          & 6.53 & 5.6 & 5.50 \\
94 & 148 & 242 & 54.65 &  54.72 &       6.60 & 5.7 & 5.85      &  1.97  & $\sim 2.20$          & 6.09 & 5.3 & 5.05 \\
94 & 149 & 243 & 57.66 &  57.76 &       6.70 & 5.9 & 6.05      &  2.17  & $ 1.70 (\pm 0.30)$   & 6.80 & 5.5 & 5.45 \\
94 & 150 & 244 & 59.80 &  59.81 &       6.37 & 5.5 & 5.7       &  2.14  &  -                   & 6.35 & 5.2 & 4.85 \\
94 & 151 & 245 & 62.88 &  63.11 &       6.58 & 5.5 & 5.85      &  2.81  & $ 2.00 (\pm 0.40)$   & 7.13 & 5.4 & 5.25 \\
94 & 152 & 246 & 65.43 &  65.40 &       6.02 & 5.4 &  -        &  2.44  &  -                   & 6.50 & 5.3 &  -   \\
\hline
95 & 144 & 239 & 49.42 &  49.39 &       6.94 & 6.3 & 6.00      &  2.19  & $ 2.50 (\pm 0.20)$   & 5.44 & 4.9 & 5.40 \\
95 & 145 & 240 & 51.17 &  51.51 &       7.72 & 6.4 & 6.10      &  2.19  & $ 3.00 (\pm 0.20)$   & 6.00 & 5.2 & 6.00 \\
95 & 146 & 241 & 52.81 &  52.94 &       7.46 & 6.2 & 6.00      &  2.10  & $\sim 2.20$          & 5.63 & 5.1 & 5.35 \\
95 & 147 & 242 & 55.02 &  55.47 &       7.82 & 6.4 & 6.32      &  2.02  & $ 2.20 (\pm 0.08)$   & 6.57 & 5.4 & 5.78 \\
95 & 148 & 243 & 56.94 &  57.18 &       7.31 & 6.1 & 6.40      &  2.07  & $ 2.30 (\pm 0.20)$   & 6.09 & 5.4 & 5.05 \\
95 & 149 & 244 & 59.59 &  59.88 &       7.44 & 6.2 & 6.25      &  2.41  & $ 2.80 (\pm 0.40)$   & 6.68 & 5.4 & 5.9  \\
95 & 150 & 245 & 61.66 &  61.90 &       6.93 & 6.1 &  -        &  2.23  & $ 2.40 (\pm 0.40)$   & 6.23 & 5.2 &  -   \\
95 & 151 & 246 & 64.42 &  64.99 &       7.02 & 5.8 &  -        &  2.86  & $\sim 2.00$          & 6.98 & 5.0 &  -   \\
95 & 152 & 247 & 66.92 & (67.15)&       6.56 & 5.7 &  -        &  2.43  &  -                   & 6.26 & 4.8 &  -   \\
\hline
96 & 145 & 241 & 53.65 &  53.70 &       7.33 & 6.4 & 7.15      &  1.65  & $\sim 2.30$          & 5.14 & 4.3 & 5.5  \\
96 & 146 & 242 & 54.88 &  54.81 &       6.96 & 6.0 & 6.65      &  1.64  & $ 1.90 (\pm 0.20)$   & 4.85 & 4.0 & 5.0  \\
96 & 147 & 243 & 56.99 &  57.18 &       7.34 & 6.5 & 6.33      &  1.57  & $ 1.90 (\pm 0.30)$   & 5.76 & 4.6 & 5.4  \\
96 & 148 & 244 & 58.51 &  58.45 &       6.91 & 6.1 & 6.18      &  1.66  & $\sim 2.20$          & 5.36 & 4.3 & 5.10 \\
96 & 149 & 245 & 61.01 &  61.00 &       7.10 & 6.3 & 6.35      &  1.97  & $ 2.10 (\pm 0.30)$   & 6.04 & 4.9 & 5.45 \\
96 & 150 & 246 & 62.72 &  62.62 &       6.68 & 6.0 & 6.0       &  1.89  &  -                   & 5.63 & 4.7 & 4.80 \\
96 & 151 & 247 & 65.29 &  65.53 &       6.98 & 6.1 & 6.12      &  2.60  &  -                   & 6.53 & 4.9 & 5.10 \\
96 & 152 & 248 & 67.44 &  67.39 &       6.38 & 5.9 & 5.8       &  2.24  &  -                   & 5.89 & 5.0 & 4.80 \\
96 & 153 & 249 & 70.94 &  70.75 &       6.02 & 5.7 & 5.63      &  2.20  &  -                   & 5.83 & 4.7 & 4.95 \\
96 & 154 & 250 & 73.04 &  72.99 &       5.72 & 5.4 &  -        &  2.12  &  -                   & 5.52 & 4.4 &  -   \\
\hline
97 & 147 & 244 & 60.36 &  60.72 &       7.68 & 6.6 &  -        &  1.26  &  -                   & 5.42 & 4.2 &  -   \\
97 & 148 & 245 & 61.79 &  61.82 &       7.19 & 6.4 &  -        &  1.37  & $\sim 1.56$          & 5.07 & 4.2 &  -   \\
97 & 149 & 246 & 63.88 &  63.97 &       7.40 & 6.5 &  -        &  1.85  &  -                   & 5.78 & 4.7 &  -   \\
97 & 150 & 247 & 65.51 &  65.49 &       7.02 & 6.5 &  -        &  1.66  &  -                   & 5.38 & 4.6 &  -   \\
97 & 151 & 248 & 67.66 & (68.08)&       7.49 & 6.3 &  -        &  2.31  &  -                   & 6.22 & 4.8 &  -   \\
97 & 152 & 249 & 69.77 &  69.85 &       6.77 & 6.1 &  -        &  1.98  &  -                   & 5.53 & 4.5 &  -   \\
97 & 153 & 250 & 72.92 &  72.95 &       6.35 & 6.1 &  -        &  1.69  &  -                   & 5.04 & 4.1 &  -   \\
\hline
98 & 152 & 250 & 71.28 &  71.17 &       6.67 & 5.6 &  -        &  1.83  &  -                   & 5.14 & 3.8 &  -   \\
98 & 153 & 251 & 74.35 &  74.13 &       6.25 & 6.2 &  -        &  1.63  &  -                   & 4.58 & 3.9 &  -   \\
98 & 154 & 252 & 76.08 &  76.03 &       5.97 & 5.3 &  -        &  1.58  &  -                   & 4.21 & 3.5 &  -   \\
98 & 155 & 253 & 79.41 &  79.30 &       5.61 & 5.4 &  -        &  1.06  &  -                   & 3.59 & 3.5 &  -   \\
\hline
\end{tabular}
\end{table*}

\clearpage

\end{document}